\documentclass[11pt]{diss_book}

\usepackage[english]{babel}
\usepackage[dvips]{graphicx}    
\usepackage[dvips]{color}
\usepackage[dvips]{epsfig} 
\usepackage[dvips]{rotating}
\usepackage{latexsym}
\usepackage{amsmath,amssymb}
\usepackage{dcolumn}
\usepackage{fancyhdr}
\usepackage{wrapfig}
\usepackage{floatfig}

\renewcommand{\chaptermark}[1]{
  \markboth{
    \chaptername\ \thechapter. \ #1
    }{}
  }

\newfont{\gros}{cmbx12 at 40pt}

\newcommand{\beq}{\begin{equation}}
\newcommand{\eeq}{\end{equation}}

\newcommand{\beqn}{\begin{equation}}
\newcommand{\beqa}{\begin{eqnarray}}
\newcommand{\eeqa}{\end{eqnarray}}
\newcommand{\bit}{\begin{itemize}}
\newcommand{\eit}{\end{itemize}}

\newcommand{\X}{ {}{\rm x}{} }

\newcommand{\U}  [2]{U_{#1}(#2)}

\newcommand{\bs}{\begin{equation}}
\newcommand{\es}{ \end{equation}}
\newcommand{\bc}{\begin{center}}
\newcommand{\ec}{ \end{center}}
\newcommand{\ba}{\begin{eqnarray}}
\newcommand{\ea}{ \end{eqnarray}}

\def\NP{{Nucl.\ Phys.\ }}

\def\PL{{Phys.\ Lett.\ }}
\def\PR{{Phys.\ Rev.\ }}

\def\PRL{{Phys.\ Rev.\ Lett.\ }}
\def\RMP{{Rev.\ Mod.\ Phys.\ }}

\def\ZP{{Z.\ Phys.\ }}

\def\s#1{\tilde{{\bf #1}}}

\def\MEF{m_{\rm eff}}\def\mef{\ifmmode\MEF\else$\MEF$\fi}

\def\SM{s_{\mu}}\def\xm{\ifmmode\SM\else$\SM$\fi}

\def\lsim{\raise0.3ex\hbox{$<$\kern-0.75em\raise-1.1ex\hbox{$\sim$}}}
\def\gsim{\raise0.3ex\hbox{$>$\kern-0.75em\raise-1.1ex\hbox{$\sim$}}}

\def\PRL{{\em Phys. Rev. Lett.}}

\begin{document}

\thispagestyle{empty}

\title{Percolation and Deconfinement \\
       in SU(2) Gauge Theory\\
       \rule[-1mm]{0mm}{3mm}\\
       \vspace*{-4mm}}
\author{{\bf Santo Fortunato} \\[.2cm]
               \\
         PhD Thesis
                \\[.2cm]
                \\[.2cm]
        Physics Faculty\\[.1cm]
        University of Bielefeld\\
}
\date{~~~~~~~~}
\maketitle

\thispagestyle{empty}
\cleardoublepage

\pagenumbering{Roman}
\setlength{\parskip}{1mm}

\chapter*{Acknowledgements}\label{Acknowledgements}
\addcontentsline{toc}{chapter}{Acknowledgements}
\thispagestyle{empty}
\markboth{\sc\ \ Acknowledgements }{\sc Acknowledgements }

At the end of a long and 
intense experience like a PhD
in physics, there is normally a long 
queue of people to thank.
\vskip0.35cm
My first thanks go to my supervisor, Prof. Dr. Helmut Satz,
for giving me this chance in a delicate moment
of my life in which I had
started to have doubts on myself. 
The challenging tasks he gave me and the 
many discussions we had about the related topics
awaked a passion for physics I thought I had lost.
I am happy to have the possibility to keep working
with him also after the PhD. 
\vskip0.35cm
A key role in my educational process during these three years
has also been played by Prof. Dr. J\"urgen Engels, who introduced
me into the computational aspects of my research work,
providing me most of the tools I needed to get  
numerical results safely and
efficiently.
As far as this is concerned, I would also like to thank 
Dr. Marzia Nardi, Dr. Manfred Oevers and Dr. Piotr Bialas 
for their patience in
assisting me during my early steps in the world of computer
programming.
\vskip0.35cm
I am particularly 
indebted to
Prof. Dr. Dietrich Stauffer, whom I owe most of what I know about
percolation theory and to Prof. Dr. Daniel Gandolfo, who 
let me become acquainted with 
analytical results about percolation which turned out to be 
very useful in my work. 
\vskip0.35cm
I gratefully acknowledge several interesting discussions with Prof.
Dr. Frithjof Karsch, Prof. Dr. Philippe Blanchard, Dr. Sanatan
Digal, Dr. Peter Petreczky, Dr. Tereza Mendes, Dr. Attilio Cucchieri.
The presence of such a high number of experts in 
lattice gauge theory and Monte Carlo simulations has allowed
me to grow very quickly in this field. 
\vskip0.35cm
I would like to express all my gratefulness to the whole 
staff of the Physics Faculty of Bielefeld, for 
the help and support I received in all circumstances.
I thank the secretaries, Gudrun Eickmeyer, Karin Lacey and 
Susi von Reder, for their sympathy and for
facilitating my life especially at the beginning of my stay,
when many things had to be properly arranged.
I thank the younger members of the staff, undergraduate and
PhD students, including some of those who are no longer here, 
for allowing me to get easily integrated in a
reality which is quite different from the Italian one. 
Without them it wouldn't have been 
possible for me to learn quickly a complicated 
language like German, which is of course 
an essential step towards a cultural integration.
I cannot write all names because of 
the limited space, but I would like to 
mention the ones with whom I spent most of my time: Ines Wetzorke,
Daria
Ahrensmeier, Matthias Buse, Peter Schmidt, Olaf 
Kaczmarek, Andreas Peikert, Burkhard Sturm, Christian Legeland,
Manfred Oevers, Sven Stickan, Olaf Leonhard, Markus Dirks.
\vskip1.5cm
Dedico questa tesi alla mia famiglia, che 
ha sempre avuto un ruolo insostituibile nella mia 
vita e nella mia carriera. L'amore e la comprensione dei miei 
cari sono stati essenziali, soprattutto nei momenti difficili
che ho dovuto affrontare. Li ringrazio soprattutto per avermi 
sempre lasciato libero di decidere cosa fare, anche quando 
ci\`o comportava dei sacrifici notevoli per me e per loro, 
come quando ho
deciso di continuare i miei studi all'estero. 
Vorrei chiudere con un ringraziamento speciale per il mio amico e
relatore Prof. Antonio Insolia, per l'amicizia 
e la pazienza che ha dimostrato nel seguire ed 
assecondare le mie decisioni, pensando 
solo a ci\`o che \`e meglio per me e non ad interessi personali. 
\vskip 2cm

$\,\,\,\,\,\,\,\,\,\,\,\,\,\,\,\,\,\,\,\,\,\,\,\,\,\,\,\,\,\,\,\,\,\,\,\,\,\,\,\,\,\,\,\,\,\,\,\,\,\,
\,\,\,\,\,\,\,\,\,\,\,\,\,\,\,\,\,\,\,\,\,\,\,\,\,\,\,\,\,\,\,\,\,\,\,\,\,\,\,\,\,\,\,\,\
\,\,\,\,\,\,\,\,\,\,\,\,\,\,\,\,\,\,\,\,\,\,\,\,\,\,\,\,\,\,\,\,\,\,\,\,\,\,\,\,\,\,\,\,\
\,\,\,\,\,\,\,\,\,\,\,\,\,\,\,\,\,\,\,\,\,\,\,\,\,\,\,\,\,\,\,\,\,\,\,\,\,\,\,\,\,\,\,\,\
\,\,\,\,\,\,\,\,\,\,\,$ 
{\large{\it Santo Fortunato}}

\addcontentsline{toc}{chapter}{Tables of}
\addcontentsline{toc}{section}{Contents}
\tableofcontents


\addcontentsline{toc}{section}{Figures}
\listoffigures

\addcontentsline{toc}{section}{Tables}
\listoftables

\pagestyle{headings}
\pagestyle{fancy}
\parskip3.0ex plus1.2ex minus0.7ex

\setcounter{footnote}{1}
\chapter*{Introduction}
\addcontentsline{toc}{chapter}{Introduction}

\pagenumbering{arabic}
\setcounter{page}{1}

\thispagestyle{empty}
\markboth{\sc\ \ Introduction}{\sc Introduction}


\def\J{$J/\psi$}
\def\j{J/\psi}
\def\X{$\chi$}
\def\x{\chi}
\def\P{$\psi'$}
\def\p{\psi'}
\def\U{$\Upsilon$}
\def\u{\Upsilon}
\def\C{c{\bar c}}
\def\cg{c{\bar c}\!-\!g}
\def\bg{b{\bar b}\!-\!g}
\def\b{b{\bar b}}
\def\q{q{\bar q}}
\def\Q{Q{\bar Q}}
\def\L{\Lambda_{\rm QCD}}
\def\A{$A_{\rm cl}$}
\def\a{A_{\rm cl}}
\def\N{$n_{\rm cl}$}
\def\n{n_{\rm cl}}
\def\S{S_{\rm cl}}
\def\s{s_{\rm cl}}
\def\bb{\bar \beta}

\def\be{\begin{equation}}
\def\ee{\end{equation}}

\def\lsim{\raise0.3ex\hbox{$<$\kern-0.75em\raise-1.1ex\hbox{$\sim$}}}
\def\gsim{\raise0.3ex\hbox{$>$\kern-0.75em\raise-1.1ex\hbox{$\sim$}}}


\def\CMP{{ Comm.\ Math.\ Phys.\ }}
\def\NP{{ Nucl.\ Phys.\ }}
\def\PL{{ Phys.\ Lett.\ }}
\def\PR{{ Phys.\ Rev.\ }}
\def\PRep{{ Phys.\ Rep.\ }}
\def\PRL{{ Phys.\ Rev.\ Lett.\ }}
\def\RMP{{ Rev.\ Mod.\ Phys.\ }}
\def\ZP{{ Z.\ Phys.\ }}

The study of critical phenomena has always been one of the 
most challenging and fascinating topics in physics.
One can give many examples of systems which 
undergo phase transitions, from familiar 
cases like the boiling of
water in a kettle or the 
paramagnetic-ferromagnetic transition of 
iron at the Curie temperature, to 
the more complicated case of 
the transition from hadronic 
matter to quark-gluon plasma which is likely
to be obtained by high-energy heavy-ion experiments 
in the coming years. 
In all cases, one observes big changes 
of some properties of the system caused by
small variations of some parameter (usually the temperature) 
around a particular value of the parameter (critical point).

In spite of the wide
variety of systems in which such phenomena are observed,
one has only two main types of phase transitions: first order 
and continuous (basically second-order) transitions. 
One of the most attractive features is 
the fact that whole {\it classes} of systems, ruled by
dynamics which look very different from each other, 
happen to have the same behaviour at the phase transition.
This is particularly striking for second-order phase transitions, 
as one can define a set of {\it critical indices} 
(exponents, amplitudes' ratios), which rule
the behaviour of the thermodynamic variables near the critical point:
all systems belonging to a class are 
characterized by the same set of critical indices ({\it universality}).
It is not clear which common elements "unify" 
different systems so that they
have the same critical behaviour; however, 
it seems that the number of space dimensions plays an important role.
This connection to {\it geometry} is at the basis of our 
future considerations.

In general, a phase transition corresponds to a change
in the {\it order} of the system. Going 
from a phase to another, the microscopic 
constituent particles of the system "choose" a different way of staying together.
The interesting thing is the fact that the 
order is a {\it macroscopic} feature, while the 
fundamental interactions which are responsible of the 
physics of the system, including the phase change, are
{\it microscopic} interactions between the particles.
How can parts of the system which are far from each other know
about their respective situations, so that they switch all together
to the same state of order?

The usual interpretation of this fact is that
the interplay of the microscopic interactions all
throughout a system at thermal equilibrium gives rise 
to a {\it correlation} between the states of the 
particles. The extent of this correlation
depends on the thermal parameters (i.e. the temperature, eventual external
fields, etc.) and it is 
expressed by the so-called {\it correlation length} $\chi$,
which is the distance over which the fluctuations of 
the microscopic degrees of freedom (position of the atoms, spin orientation, etc.)
are significantly correlated with each other.

The correlation creates thus "ordered" regions 
which drive the behaviour of the whole system.
Because of that, it is natural
to consider such regions as the 
leading characters of the phenomenon 
and describe phase transitions in terms of the 
properties of compound objects. 
The interaction "builds" these objects: 
the phase transition
is related to the {\it geometry} of the resulting structures.

This 
picture has, among other things,
two big advantages. 
First, it would justify 
the connection between 
critical phenomena and  
geometry that we have stressed above.
Second, if the degrees of freedom relevant for the phase change
are the ones of sets of particles, and not of
single particles,
it is likely that
they do not depend on the details of the microscopic interaction, but only
on its gross features (e.g. symmetries): this could explain 
the universality of the critical indices.

On these grounds, it is easy to understand
why several 
attempts have been made to 
find a geometrical description of phase transitions.
The first ideas date back 
to the end of the $40$'s, when
Onsager \cite{onsag} proposed
an interpretation of the $\lambda$-transition in liquid $^{4}He$
based on the behaviour of one-dimensional strings, whose 
size would change dramatically from one phase to the other:
whereas in the superfluid phase only finite strings are present,
at the critical point infinite strings appear.

This kind of picture 
is analogous to the well known phenomenon 
of {\it percolation} \cite{stauff, grimm}, which takes place
when geometrical {\it clusters}, formed by 
elementary objects of some system, stick to each other giving rise to
an infinite network, that spans the whole system.
Here, criticality is reached when the {\it density} 
of the elementary objects is sufficiently high. 
The onset of 
percolation marks a distinction between two different phases
of the system, characterized by the presence or the absence 
of an infinite cluster.
The percolation phenomenon
turns out to have astonishing analogies with
ordinary second order thermal phase transitions.
In particular, the behaviour of the
percolation variables at criticality is also
described by simple power laws, with relative exponents; 
the values of the exponents, related to each other by simple scaling relations,
are fixed only by the number of space dimensions of the system
at study, regardless of its structure and of the special type
of percolation process one considers.
  
For these reasons, percolation seems to be an ideal framework
for the geometrical description of phase transitions
we are looking for. One could try to map the thermal
transition into a geometrical percolation transition.
In order to do that, one must require that 
the two critical thresholds coincide, and that 
the thermal variables can be expressed in terms 
of corresponding percolation quantities.

The first studies in this direction started at the
beginning of the $70$'s, and concentrated on the Ising model.
The main problem was to 
look for a suitable cluster definition.
The first structures which were investigated 
were the ordinary magnetic domains, i.e. 
clusters formed by nearest-neighbouring 
spins of the same sign. In two dimensions
such clusters happen indeed to percolate at the thermal
critical temperature $T_c$ \cite{connap}. Nevertheless, the values
of the critical exponents differ from the 
corresponding Ising values \cite{sykes}. Besides,
in three dimensions, 
the magnetic domains of 
the spins oriented like the magnetization percolate at any temperature;
the domains formed by the spins opposite to the magnetization
percolate for $T\,{\geq}\,T_p$, 
with $T_p\,{\neq}\,T_c$ \cite{krumb}.

The problem was solved when 
it became clear that, to define the `physical' 
islands of a thermal system, one
must take into account correctly the degree of correlation
between the spins. The size 
of the ordinary magnetic domains, in fact,
happens to be too large because
of purely geometrical effects,
which operate independently of
the spins' correlation. In order to get rid
of these disturbing effects, Coniglio and Klein 
introduced a {\it bond probability} $p=1-\exp(-2J/kT)$ ($J$
is the Ising coupling, $T$ the temperature).
The new islands are {\it site-bond} clusters, i.e.
clusters formed by nearest neighbouring like-signed
spins, which are connected with a probability $p$, and not
always like in the previous definition ($p=1$).
These clusters had actually been introduced some years before
by Fortuin and Kasteleyn. They had shown
that, by means of such objects, one can 
reformulate the Ising model as a geometrical model \cite{fort}. 
This result indicates that these apparently
artificial structures are strictly related to
the Ising dynamics.
Coniglio and Klein showed that the new clusters
percolate at the thermal threshold 
and that the percolation exponents coincide with the
Ising exponents \cite{conkl}.
 
So, it is possible to describe the 
paramagnetic-ferromagnetic transition
of the Ising model as a percolation transition
of suitably defined clusters. The paramagnetic-ferromagnetic transition
is due to the {\it spontaneous breaking}
of the $Z(2)$ symmetry of the Ising Hamiltonian, i.e., the symmetry
under inversion of the spins. 
The spontaneous breaking of the $Z(2)$ symmetry is also
responsible of the 
confinement-deconfinement transition
in $SU(2)$ pure gauge theory. 
Because of that, it was conjectured that 
$SU(2)$ has the same critical behaviour
of the Ising model \cite{svetitsky}, i.e., 
it undergoes a second order phase transition with Ising exponents,
as it was successively confirmed by 
lattice simulations \cite{engles}.

It is then natural to see whether the $SU(2)$ confinement-deconfinement
phase transition can be described as a percolation transition 
like for the Ising model: this is the aim of this work.

The analogue of the spin variable in $SU(2)$ pure gauge theory
is the {\it Polyakov loop} $L$, 
a real number which is a well
defined function of the gauge fields. The deconfined 
region is the ordered phase of the system, characterized
by a non-vanishing lattice average of the 
Polyakov loop. In this way, regions 
of the space where the Polyakov loop has the same sign
can be viewed as local "bubbles" of deconfinement.
In each of these regions, in fact, the average 
of the Polyakov loop is necessarily non-zero.
If we put a test colour charge into a bubble,
it will be free to move within the portion of space
occupied by the bubble. But to have a real 
deconfined phase, the test charge must be able to move freely
all throughout the system, so that there must
be bubbles whose size is of the same order of the 
volume of the system.
A working percolation picture would support
the proposal of
such a mechanism for the deconfinement transition.

The question is, again, what clusters 
to choose. From what we have said, it is simple 
to deduce that
the clusters must be formed 
by sites at which 
the Polyakov loops have the same sign. But it is not 
clear if and how we can extract
the other necessary ingredient 
for the cluster building, namely the 
bond probability.

The search of the right bond probability 
is affected by two problems:
\begin{itemize}
\item{The Polyakov loop is not a two-valued 
variable like the spin in the Ising model
but a continuous one;
its values vary in a range that, with the normalization
convention we use, is $[-1,1]$.}
\item{The $SU(2)$ Lagrangian is a function of the gauge fields
which cannot in general  
be expressed only in terms
of the Polyakov loop $L$.}

\end{itemize}

The first point 
led us to investigate
{\it continuous spin models}, i.e. 
models where the spin is a continuous variable,
to check whether the Coniglio-Klein
result can be extended to such more general cases.
We began by analyzing the continuous spin Ising model,
which is an Ising model with continuous spins.
We will see that, in this case,
an equivalent percolation picture can be obtained
by introducing a bond weight which is similar to the Coniglio-Klein one,
with the difference that it contains an explicit dependence
on the spins connected by the bond. This {\it local} bond
probability solves the first of the two 
afore-mentioned problems.
Besides, the result can be further extended to 
models with several spin-spin interactions,
if ferromagnetic. We will also show that
eventual spin
distribution functions and self-interaction terms
do not influence the percolation picture. Finally,
we will analyze $O(n)$ spin models and find again that
their critical behaviour can be
easily described by means of cluster percolation.

The second difficulty is hard to overcome. 
In fact, it seems clear that 
the percolation picture of a model
is strictly related to the {\it interactions} 
of the model. In particular, a bond
is associated to each spin-spin interaction, with a probability 
which depends on the value of the coupling strength of the
interaction. 
But, if the $SU(2)$ Lagrangian is not simply a function
of $L$, we cannot
exactly specify how the "gauge spins", i. e., the Polyakov loops,
interact with each other. It seems then
impossible to derive rigorously the corresponding 
percolation scheme.

However, we can try to solve the problem
by using suitable approximations. 
The best thing to do is to 
try to
approximate $SU(2)$ pure gauge theory
by means of an {\it effective theory},
hoping that 
the effective model admits a percolation picture.

We shall first exploit a
strong coupling expansion 
derived by Green and Karsch \cite{Ka}, 
which shows that the partition function of $SU(2)$
can be reduced to the partition function
of one of the continuous spin models we have analyzed.
This approximation is valid
only in the strong coupling limit,
more precisely in the cases $N_{\tau}=1,2$
($N_{\tau}$=number of lattice spacings in the temperature
direction).
We will analyze the case $N_{\tau}=2$, both
in two and in three space dimensions, and 
show that the percolation picture derived by the effective theory
describes well the thermal transition of $SU(2)$.

Next, we will try to find a procedure which
can be also applied to the more
interesting weak coupling case.
This time we shall construct the effective theory
starting not from the $SU(2)$ Lagrangian, but from
the Polyakov loop configurations.
Actually we shall consider the Ising-projected configurations,
i.e. the distributions of the {\it signs} of the Polyakov loops.
This is done assuming that the $Z(2)$ symmetry is the 
only relevant feature at the basis of the critical behaviour.

We will essentially look for a model which can reproduce
the Ising-projected Polyakov loop configurations. 
The effective model must be necessarily chosen inside the
group of spin models for which a working percolation
picture exists.
Our ansatz will be an Ising-like model 
with just ferromagnetic spin-spin interactions, 
to which the Coniglio-Klein result can be trivially 
extended by associating a bond to each coupling.
The couplings of the effective theory are calculated 
following a method used in Monte Carlo renormalization group studies
of field theories \cite{okawa, oka2}.
We will examine $SU(2)$ in $3+1$ dimensions, 
for $N_{\tau}=2$ and $N_{\tau}=4$.
The results will be shown to be
satisfactory in both cases.

Our results are entirely obtained by means
of lattice Monte Carlo simulations  
of the various models we have studied. 
We have always used workstations
except for 
some lenghty $SU(2)$ simulations which were 
performed on a Cray $T3E$ ($ZAM$, J\"ulich). 

This work is
structured as follows.
Chapter 1 is devoted to a presentation of the
main concepts of percolation theory with a special
attention to numerical techniques. 
In Chapter 2 we focus on the analogies
between percolation and thermal phase transitions,
which lead to the percolation formulation of 
the Ising model of Coniglio and Klein.
Chapter 3 collects all percolation studies 
on continuous spin models that we have mentioned above.
In Chapter 4 we show the results for
$SU(2)$ pure gauge theory.
Finally, the conclusions
of our investigation are drawn.
In Appendix A we present 
the procedure we have adopted to 
perform the so-called {\it cluster labeling}, 
i.e.
the identification 
of the cluster configurations.

\chapter{Introduction to Percolation Theory}\label{Introduction to Percolation Theory}

\setcounter{footnote}{1}
\thispagestyle{empty}


\def\J{$J/\psi$}
\def\j{J/\psi}
\def\X{$\chi$}
\def\x{\chi}
\def\P{$\psi'$}
\def\p{\psi'}
\def\U{$\Upsilon$}
\def\u{\Upsilon}
\def\C{c{\bar c}}
\def\cg{c{\bar c}\!-\!g}
\def\bg{b{\bar b}\!-\!g}
\def\b{b{\bar b}}
\def\q{q{\bar q}}
\def\Q{Q{\bar Q}}
\def\L{\Lambda_{\rm QCD}}
\def\A{$A_{\rm cl}$}
\def\a{A_{\rm cl}}
\def\N{$n_{\rm cl}$}
\def\n{n_{\rm cl}}
\def\S{S_{\rm cl}}
\def\s{s_{\rm cl}}
\def\bb{\bar \beta}

\def\be{\begin{equation}}
\def\ee{\end{equation}}

\def\lsim{\raise0.3ex\hbox{$<$\kern-0.75em\raise-1.1ex\hbox{$\sim$}}}
\def\gsim{\raise0.3ex\hbox{$>$\kern-0.75em\raise-1.1ex\hbox{$\sim$}}}


\def\CMP{{ Comm.\ Math.\ Phys.\ }}
\def\NP{{ Nucl.\ Phys.\ }}
\def\PL{{ Phys.\ Lett.\ }}
\def\PR{{ Phys.\ Rev.\ }}
\def\PRep{{ Phys.\ Rep.\ }}
\def\PRL{{ Phys.\ Rev.\ Lett.\ }}
\def\RMP{{ Rev.\ Mod.\ Phys.\ }}
\def\ZP{{ Z.\ Phys.\ }}

\section{Definition of the problem}\label{paolo}

\vspace*{-1.5ex}
\begin{wrapfigure}{o}{8.8cm}
\begin{center}
\vspace*{-0.5cm}
\hspace*{-0.2cm}\epsfig{file=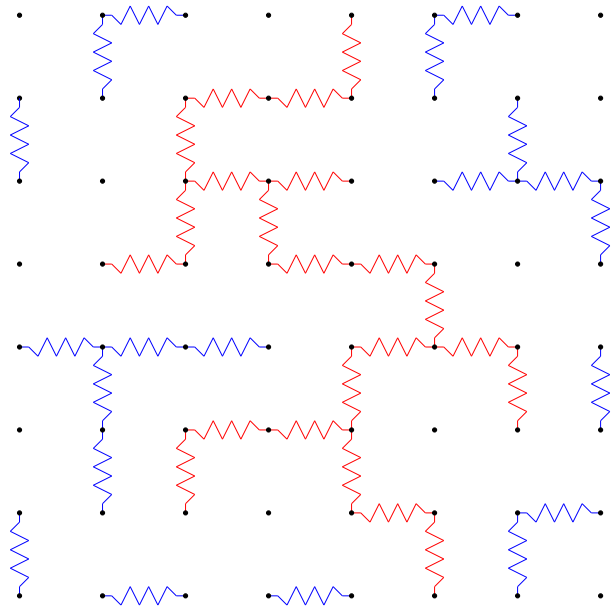,  width=8cm}
\caption[Scheme of a random resistor network]
{\label{randnet}{Scheme of a two-dimensional random resistor network. The
spanning structure formed by the resistors (marked in red) allows
electric current to flow all through the material.}}
\end{center}
\end{wrapfigure}
\vspace*{-0.5ex}
Let us suppose to have a piece of some material $X$
which is given by the mixture of two different substances $A$ and $B$.
Substance $A$ is a metal, substance $B$ an insulator. 
One could ask oneself  
whether the material $X$ is insulating or conducting. 
Fig. \ref{randnet} schematizes the situation, assuming for simplicity
our system to be two-dimensional. The geometry 
of the sample $X$ is the one of a regular square lattice,
represented by the black points. 
If we assume that the mixing process is disordered,
we can visualize the presence of 
the metal by distributing randomly resistors between 
pairs of nearest neighbouring sites. If we set a 
voltage between the upper and the lower edge of our sample, electric current
will flow through the substance if the resistors form 
a connected structure from top to bottom (red path in the figure).
Let $p$ be the concentration 
of the metal in the substance. Our problem can be
reformulated in the following way: what is the minimum value of 
$p$ which is necessary to have 
a connected bridge of resistors all through the lattice?

The system we have presented here is a    
{\it random resistor network}, and represents 
only one of the
many applications of
{\it percolation theory} \cite{stauff,grimm}. The original problem which gave rise to
this theory was studied by Flory and Stockmayer \cite{flory} during
the Second World War. They  
had a set of
small branching molecules and increased the number of
chemical bonds between them. In this way 
larger and larger macromolecules are formed. 
At some stage it may happen
that the chemical bonds form a structure which spans the whole system
({\it gelation}). 

Nowadays the set of problems which can be modelled by using percolation 
theory is big and various: diffusion in
disordered media \cite{disord}, critical behaviour  
of systems undergoing second order phase transitions (the topic of this work),
fractality \cite{mandelvi}, spread of
epidemics or fires in large orchards \cite{frischcox}, 
stock market fluctuations \cite{stock}.
In this chapter we want to introduce the percolation 
problem and illustrate its main features. 

Suppose to have some infinite periodic lattice\footnote{
We remark that the percolation phenomenon does not 
require a lattice structure, but it can be 
also studied on continuous manifolds. 
However, since our work is centered on lattice systems, 
we will disregard continuum percolation. The interested reader 
is invited to look at \cite{contperc}.}    
in $d$
dimensions. For simplicity, we consider here a two-dimensional 
square lattice. 
We start by distributing randomly objects on the lattice,
something like placing pawns on a chessboard.
At this stage we have two possibilities: we can place our pawns
on the edges of the lattice, or on its vertices.
If we work on the edges we have the so-called {\it bond percolation}: 
our random resistor network is an example of it. 
If we instead place our pawns on the sites
we are in the {\it site percolation} case.
Other choices are allowed, but they
are given by combinations of
site and bond percolation (for example one can
use edges and sites together). 
Every bond model
may be reformulated as a site model on 
a different lattice \cite{bondsite}, but the converse is false. Therefore
site models are more general than bond models and in what follows we will
deal essentially with the former ones. 
We assume that 
an edge (site) is occupied with
some probability $p$ ($0\,{\leq}\,p\,{\leq}\,1$), independently of
the other edges (sites). 
To complete the picture we only need  
to establish a rule to form compound structures ({\it clusters})
out of our pawns. Percolation theory
deals with the properties of the clusters thus formed. 

If we increase the probability $p$, the clusters 
at the beginning will increase in number and size. 
Successively most of them will stick to each other
to form bigger and bigger clusters until, 
for some value $p_c$ of the
occupation probability, an infinite spanning structure is formed
({\it percolating cluster}). 
Further increases of $p$ lead to an increase of the size
of the percolating cluster which slowly embodies the remaining
ones until, for $p=1$, it 
invades every edge (site) of the lattice.

\begin{figure}[h]
\begin{center}
\epsfig{file=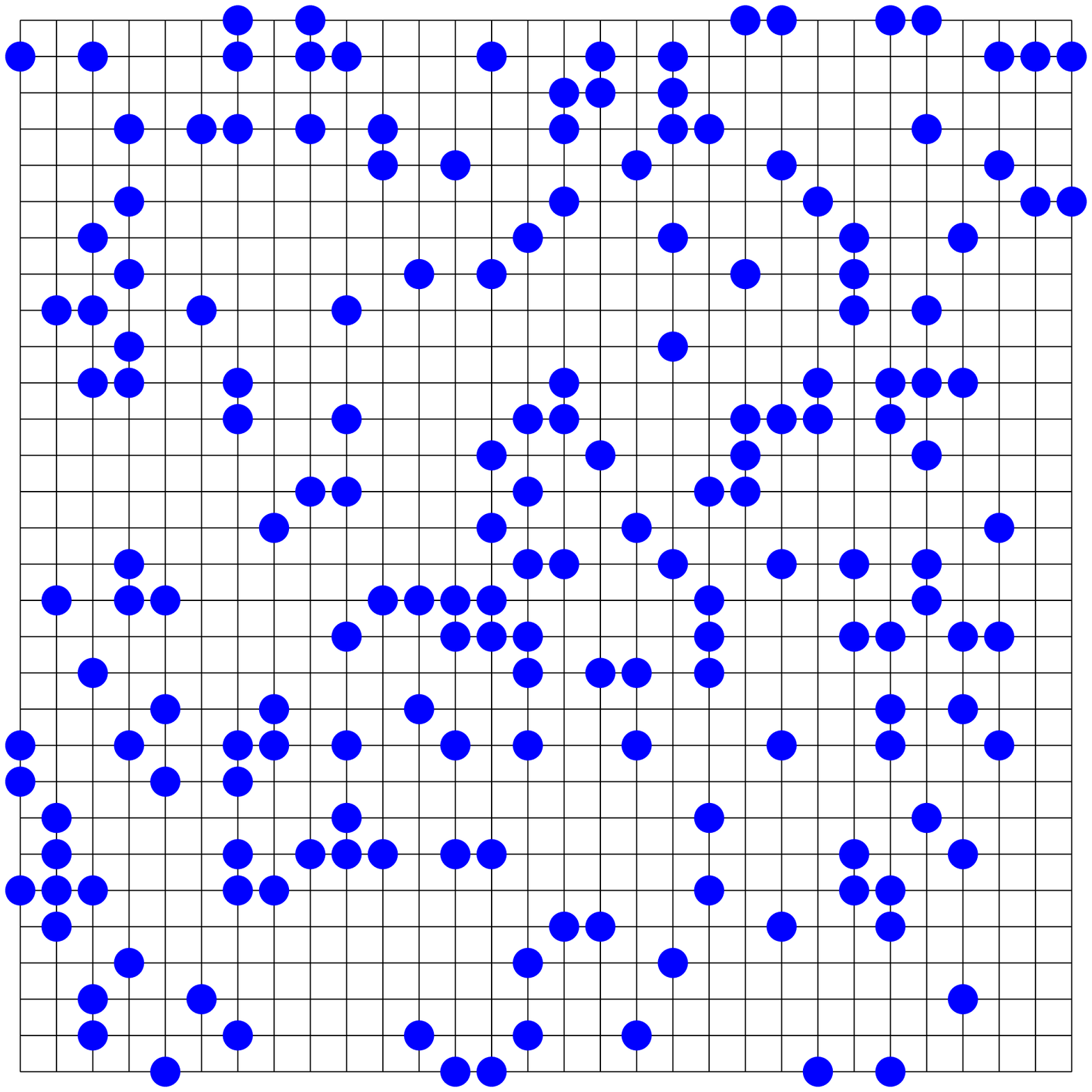,width=4.8cm}
\epsfig{file=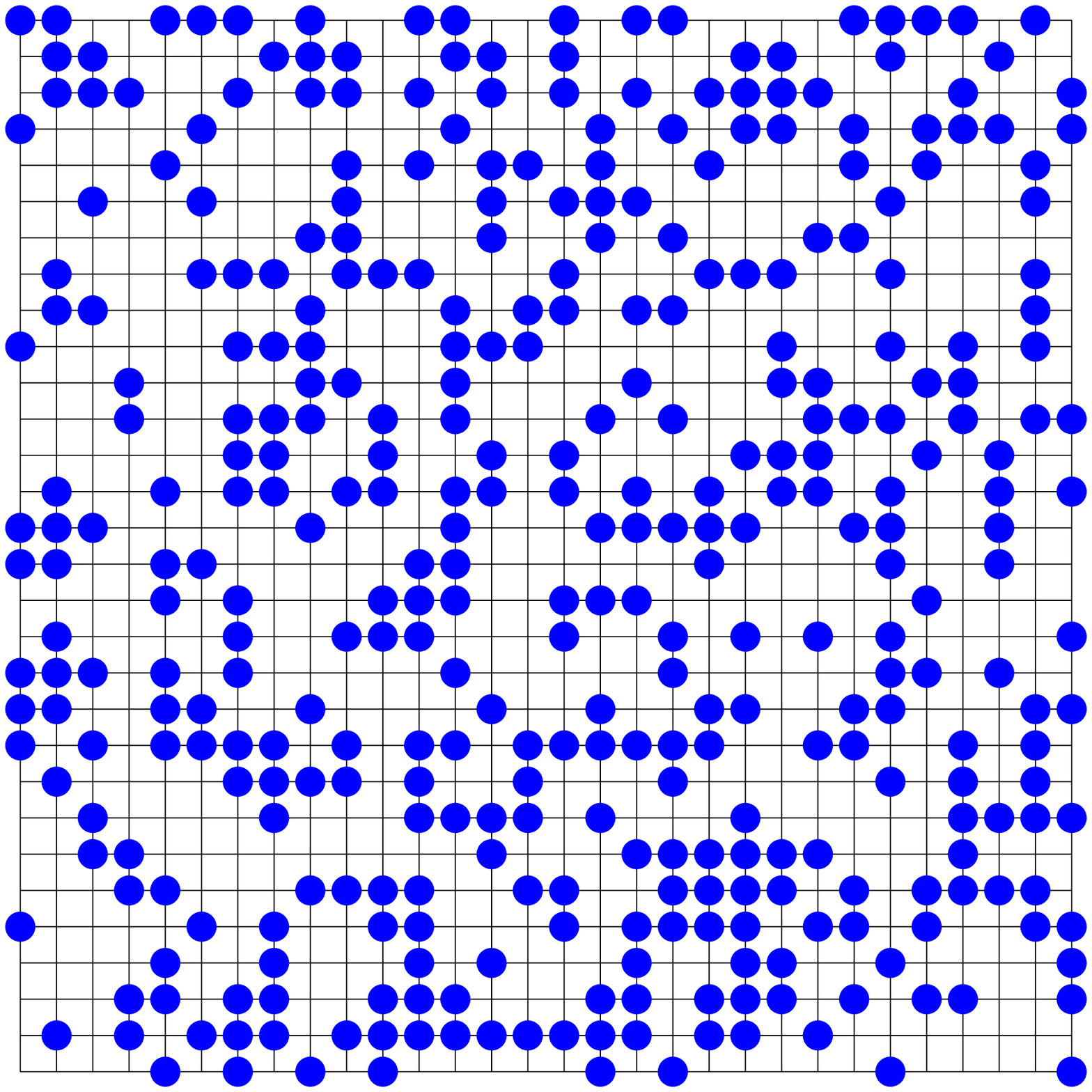,width=4.8cm}
\epsfig{file=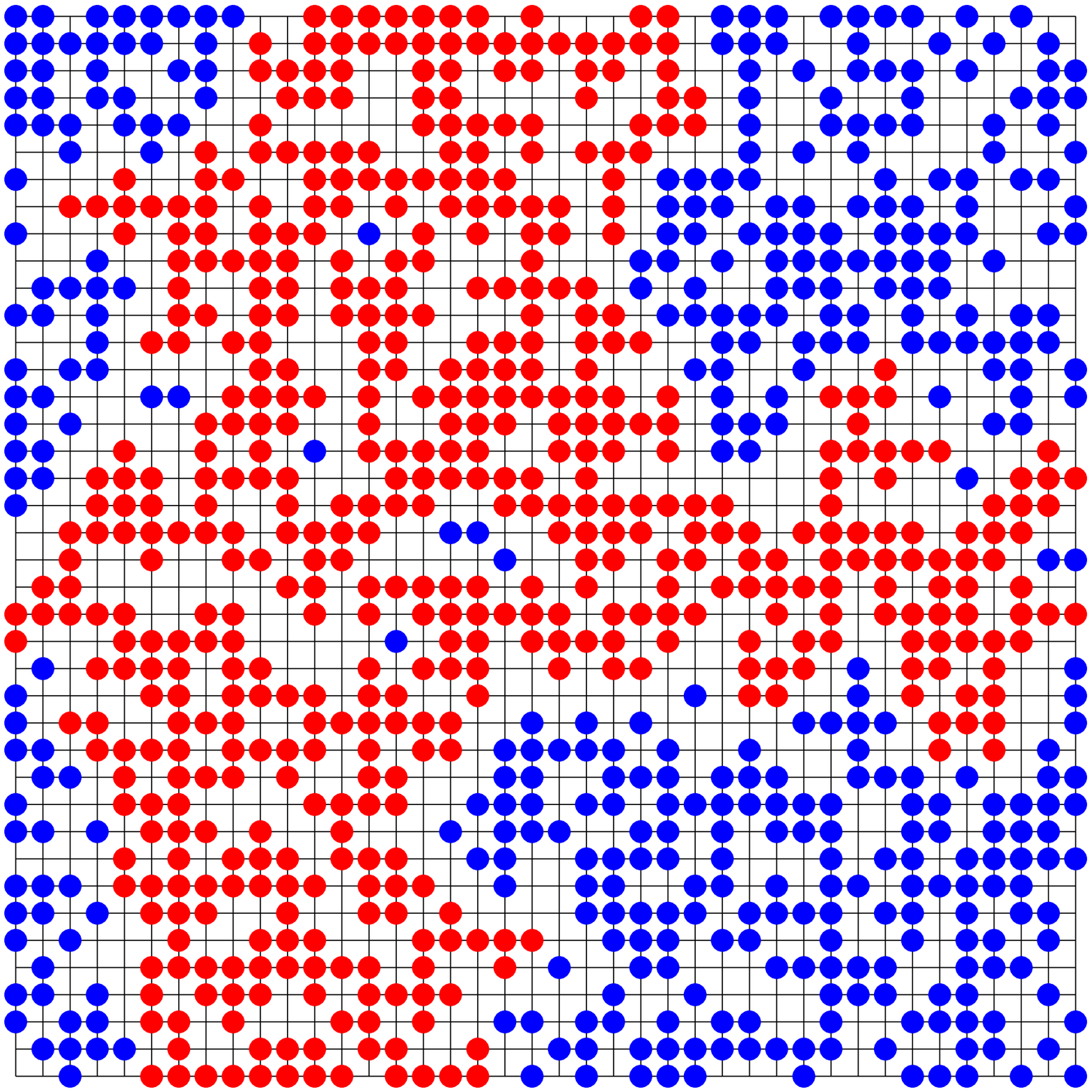,width=4.8cm}
\vskip0.2cm
\begin{picture}(0,-30)
    \put(-145.2,-3){\begin{minipage}[t]{7.4cm}{(a)}
    \end{minipage}}
    \put(-5,-3){\begin{minipage}[t]{7.4cm}{(b)}
    \end{minipage}}
    \put(135,-3){\begin{minipage}[t]{7.4cm}{(c)}
    \end{minipage}}
\end{picture}
\vskip0.3cm
\caption[Pure site percolation on a square lattice]
{\label{RP_2D}{Pure site percolation on a 2-dimensional square lattice.
          In (a) the density of occupied sites is low and the
          clusters small. In (b) the density is increased and the
          corresponding clusters are larger. For a still higher
          density many clusters stick
          together to form a spanning structure (red cluster in (c)).}}
\end{center}
\end{figure}

Fig. \ref{RP_2D} shows three ``pictures'' of this phenomenon 
for the so-called pure site percolation case, for which 
two nearest neighbouring sites always belong to the same cluster.
Fig. \ref{RP_2D}a shows a lattice configuration corresponding to a small value 
of $p$, in Fig. \ref{RP_2D}b $p$ is higher but below
$p_c$ and in Fig. \ref{RP_2D}c $p$ is slightly above $p_c$.

Particularly interesting is what happens for values
of $p$ near $p_c$. The aspects related to that are called
{\it critical phenomena} and we will focus mainly on that.
Indeed, at the percolation threshold $p_c$ 
a sort of {\it phase transition} takes place, because
our system changes dramatically its behaviour at 
one particular value of a continuously varying parameter.
For an occupation probability ${p_c}-\epsilon$ ($\epsilon$ is
an arbitrarily small numer) there is no percolating cluster, for
${p_c}+\epsilon$ there is (at least) one. 

\begin{figure}[htb]
\begin{center}
\epsfig{file=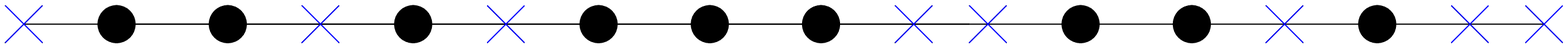,width=10cm}
\caption[Site percolation on a 1-dimensional linear chain]
{\label{1dimperc}{Site percolation on a 1-dimensional linear chain. 
Nearest-neighbouring black circles form the clusters. The 
crosses indicate vacancies, which separate 
the clusters from each other. Percolation 
can take place only if all sites are occupied ($p=1$).}}
\end{center}
\end{figure}

\begin{figure}[htb]
\begin{center}
\epsfig{file=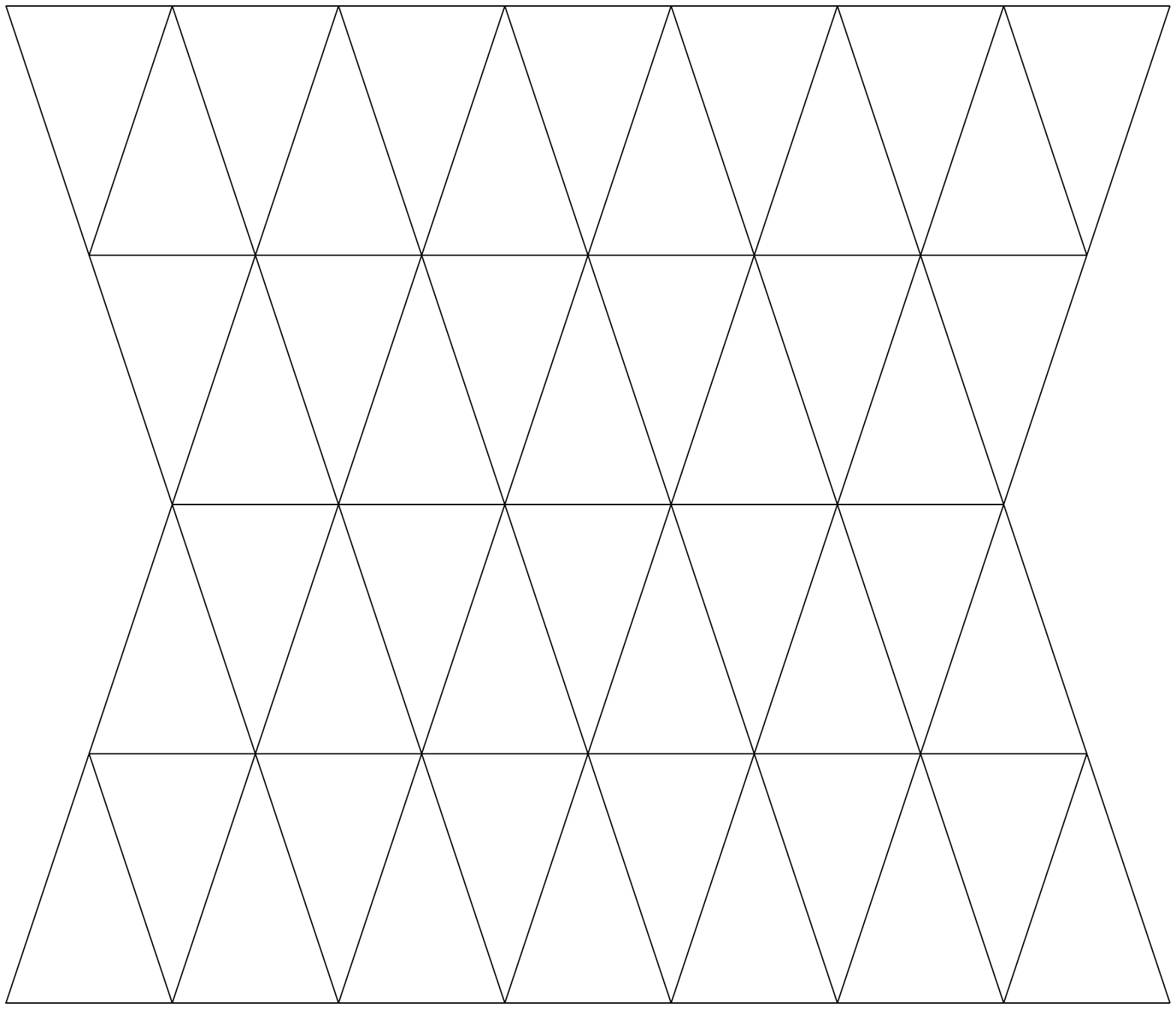,width=5cm}
\caption[Scheme of a triangular lattice]{Scheme of a triangular lattice. If we 
consider as sites the centers of the triangles we get
the so-called honeycomb lattice.}
\label{triang}
\end{center}
\end{figure}

We have defined the percolation process on
a regular lattice in $d$ dimensions. It is easy to see
that $d$ must be at least 2 in order to 
have a critical phenomenon. Let us suppose that $d=1$.
Our system can be represented by an infinitely long
linear chain, as shown in Fig. \ref{1dimperc}.
\vskip0.5cm
The black circles in the figure represent the occupied 
sites. If the occupation probability $p$ is smaller than 1, 
there will be holes along the chain. 
But a spanning cluster
in this special case must include all sites, therefore there can be 
percolation only for $p=1$.
There is no separation in two phases, and that 
makes the one-dimensional case not as interesting as 
the higher-dimensional
ones. We shall thus always assume that $d{\geq}2$.
The lattice structures on which we can play our percolation
game are not restricted to the
simple square (cubic) ones: we can use as well triangular,
honeycomb lattices (Fig. \ref{triang}). 
Besides, we can use the
same structure in different ways, like in the case of the
simple 3-dimensional cubic lattice, from which 
we can get three lattices: we can consider as sites
just the vertices of the cubic cells, 
the vertices plus the centers of the cubes ({\it
body centered cubic} or {\it bcc} lattice), 
or the vertices plus the centers of 
the six faces of each cube ({\it face centered cubic} or {\it fcc} lattice).

Because of the different lattice structures, the 
critical values of the occupation probability $p_c$ will be
in general different in each case. In Table \ref{critthr}  
we have listed the values corresponding to
the most studied systems. We notice that, for a fixed lattice structure,
$p_c$ gets smaller
the higher the dimension $d$ of the lattice.
\vskip1cm
\begin{table}[h]
\begin{center}
\begin{tabular}{|c|c|c|}
\hline$\vphantom{\displaystyle\frac{1}{1}}$
Lattice & Site & Bond\\
\hline
\hline$\vphantom{\displaystyle\frac{1}{1}}$
$d=2$ honeycomb & 0.6962 & 1-2$sin(\pi/18)$\\
\hline$\vphantom{\displaystyle\frac{1}{1}}$
$d=2$ square & 0.592746 & 1/2\\
\hline$\vphantom{\displaystyle\frac{1}{1}}$
$d=2$ triangular & 1/2&2$sin(\pi/18)$\\
\hline$\vphantom{\displaystyle\frac{1}{1}}$
$d=3$ simple cubic & 0.31160 & 0.2488\\
\hline$\vphantom{\displaystyle\frac{1}{1}}$
$d=3$ bcc & 0.246 & 0.1803\\
\hline$\vphantom{\displaystyle\frac{1}{1}}$
$d=3$ fcc & 0.198 & 0.119\\
\hline$\vphantom{\displaystyle\frac{1}{1}}$
$d=4$ hypercubic & 0.197 & 0.1601\\
\hline$\vphantom{\displaystyle\frac{1}{1}}$
$d=5$ hypercubic & 0.141 & 0.1182\\
\hline$\vphantom{\displaystyle\frac{1}{1}}$
$d=6$ hypercubic & 0.107 & 0.0942\\
\hline$\vphantom{\displaystyle\frac{1}{1}}$
$d=7$ hypercubic & 0.089 & 0.0787\\
\hline
\end{tabular}
\caption[Percolation thresholds for various lattices]{\label{critthr} Selected percolation thresholds for various lattices.}
\end{center}
\end{table}
\clearpage

\section{Cluster Size}\label{filippo}

\subsection{Cluster Distribution}\label{andrea}

Once we have defined the problem, we have to see how it is possible to
study the percolation phenomenon quantitatively. 
Percolation is a {\it random}
process, because random is the way in which we occupy the sites (bonds) of the 
lattice. If we repeat the procedure over and over we will 
have clusters of different sizes and shapes
and therefore it makes sense to study the {\it averages } of 
quantities related to the clusters.
In order to do that, we must study the {\it statistics }
of these clusters. 

In general we define as size $s$ of a cluster the number 
of sites (bonds) belonging to it. It is interesting 
to see how the clusters are distributed according to their size.
This information is expressed by a function $n_s$, which depends both
on $s$ and on the density $p$. We define
$n_s$ as the {\it number
of clusters of size s per lattice site}, according
to the following formula
\begin{equation}
\label{nspec}
n_s\,=\,\lim_{V{\rightarrow}\infty}\frac{{N_V}(s)}{V},
\end{equation}
where $V$ is the volume (number of sites) of a finite lattice and
${N_V}(s)$ the number of clusters of size $s$ on that lattice.
\begin{figure}[h]
\begin{center}
\epsfig{file=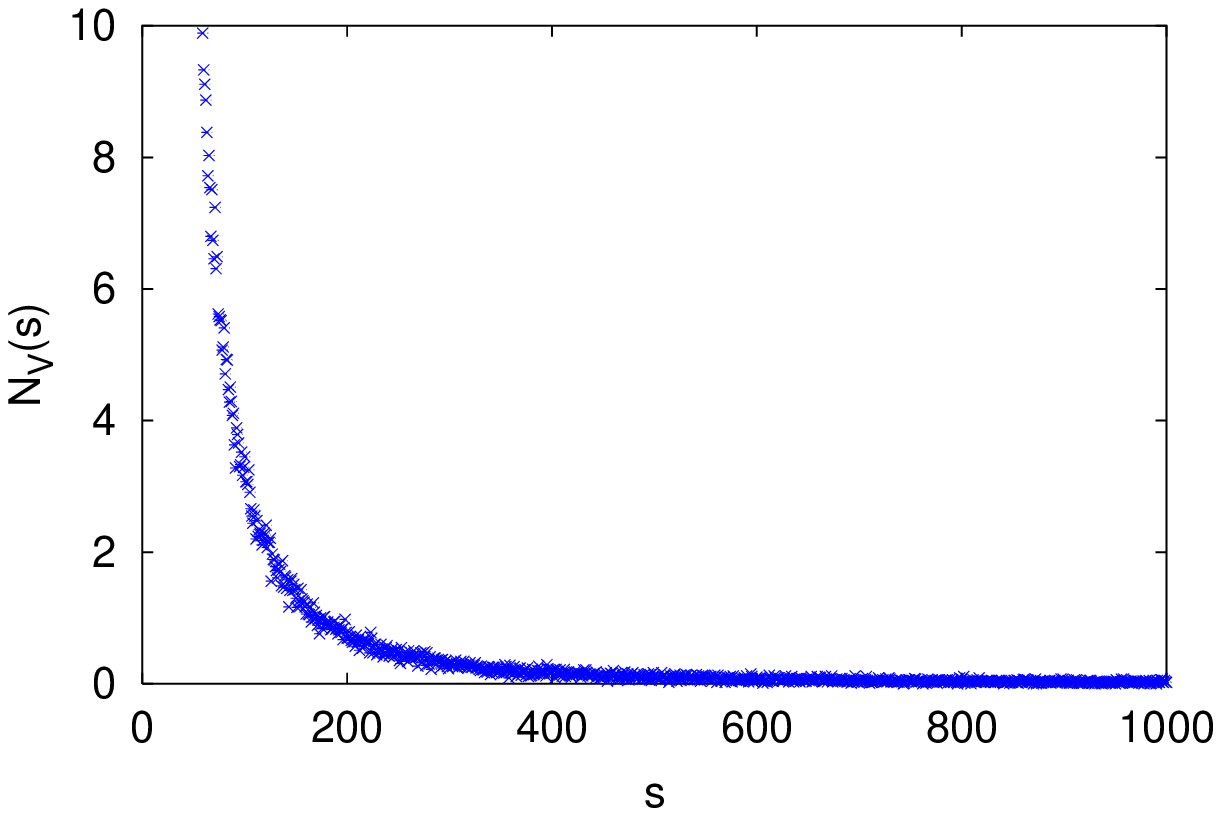,  width=7.9cm}
\epsfig{file=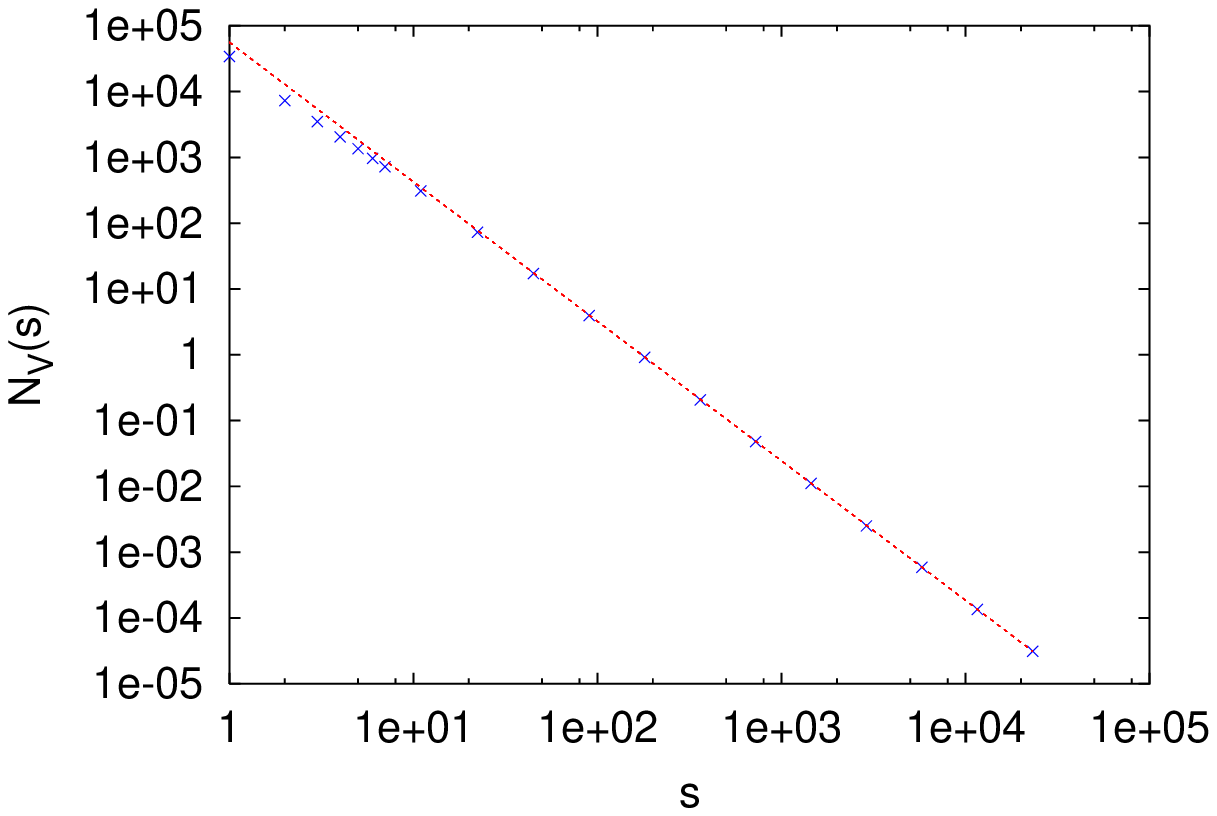,  width=7.9cm}
\vskip0.2cm
\begin{picture}(0,-30)
    \put(-109.9,2){\begin{minipage}[t]{8.4cm}{(a)}
    \end{minipage}}
    \put(123.7,2){\begin{minipage}[t]{8.4cm}{(b)}
    \end{minipage}}
\end{picture}
\caption[Cluster distribution for site percolation on a simple cubic lattice]
{(a) Cluster distribution 
for site percolation on a $100^3$ simple cubic lattice in correspondence of the
critical density $p_c=0.3116$. (b) Log-log plot of the cluster distribution  
shown in (a). 
The data are stored in bins to reduce 
the fluctuations. The slope of the straight line 
gives an approximated estimate of the critical exponent $\tau$.}
\label{clusdistr}
\end{center}
\end{figure}

It is generally found that,
near the critical density $p_c$ and for sufficiently big values 
of the size $s$, the distribution 
$n_s$ has the scaling form:
\begin{equation}
\label{ns}
n_s\,{\propto}\,\,s^{-\tau}f[(p-p_c)\,s^{\sigma}],
\end{equation}
where $f$ is a function to be 
determined in each specific case and $\tau$, $\sigma$ are
{\it critical exponents}. 
The function $f(z)$, however, 
has some general features: it is basically 
constant for $|z|\,{\ll}\,1$ and it decays  
rapidly for $|z|\,{\gg}\,1$. That means that, for a fixed value
of the density $p$,
$n_s$ will be appreciably different from zero for those values
of the size $s$ for which
\begin{equation}
\label{cros}
s\,<\,|p-p_c|^{-1/{\sigma}}.
\end{equation}
For $p\,=\,p_c$ the 
distribution is a simple power law:
\begin{equation}
\label{ns2}
n_s\,{\propto}\,\,s^{-\tau}\,.
\end{equation}
Fig. \ref{clusdistr}a shows the
cluster number distribution for pure site percolation
on a cubic lattice at the 
critical threshold $p_c\,=\,0.3116$. 
The lattice size is $100^3$ and 
we have analyzed 100 samples in order to get a 
satisfactory statistics. 
The values on the $y$ axis are the 
unrenormalized cluster numbers $N_V(s)$. 
We can see the 
main features of the cluster distribution, in particular
the rapid decrease with the size $s$. 
To check whether $n_s$ has really the power law behaviour 
of Eq. (\ref{ns2}), we have plotted our distribution in log-log scale.
To obtain a good quality of the plot we have tried 
to reduce the fluctuations which are visible
in Fig. \ref{clusdistr}a. An efficient method 
to do that consists in  
dividing the $s$ axis in bins and calculating the
average of $n_sV$ in each bin. 
The result can be seen in Fig. \ref{clusdistr}b, 
where all our data are represented by few 
points: they look rather stable. 
Eq. (\ref{ns2}) is valid only for
big values of $s$, therefore we have excluded the 
points corresponding to low values of $s$ ($s\,\leq\,20$)
and performed
a linear best fit on the remaining ones.
The straight line we have drawn 
is in good agreement with the data points,
which confirms the correctedness
of Eq. (\ref{ns2}). 
The slope of the straight line is 2.13, which 
is a fair approximation of 
the exponent $\tau$ for this 
system ($\tau=2.18$).

\subsection{Average Cluster Size}\label{giovanni}

If we know the cluster distribution function $n_s$, we 
may ask ourselves how big on average a cluster is. 
We must be careful in specifying
what we exactly mean by "average" in this case.
Let us suppose that we point randomly 
to a lattice site and want to know how big
the cluster to which that site belongs is.
If the size of the cluster is $s$, the number of
clusters of that size (per site) is $n_s$. 
Therefore, the quantity ${n_s}s$ is just the probability
of picking up a site belonging to one of those clusters.
On the other hand the probability 
that a site of the lattice taken at random belongs to
any finite cluster is given by
\begin{equation}\label{probns}
\sum_{s}\,{n_{s}s}
\end{equation}
(the sum excludes the eventual percolating cluster).
So, if we hit some occupied site
of the lattice, the probability $w_s$ that it belongs 
to a cluster of size $s$ is
given by
\begin{equation}\label{probws}
w_s\,=\,\frac{n_{s}s}{\sum_{\bar{s}}{n_{\bar{s}}\bar{s}}}
\end{equation}
Our procedure will thus lead us to the 
following definition of 
{\it average cluster size }$S$:
\begin{equation}\label{defS}
  S\,=\,\sum\limits_{s}\,w_s\,s\,=\,\frac{\sum_{s} {{n_{s}s^2}}}{\sum_{\bar{s}}{n_{\bar{s}}\bar{s}}}~.
\end{equation}
\vskip0.5cm
\begin{figure}[h]
\begin{center}
\epsfig{file=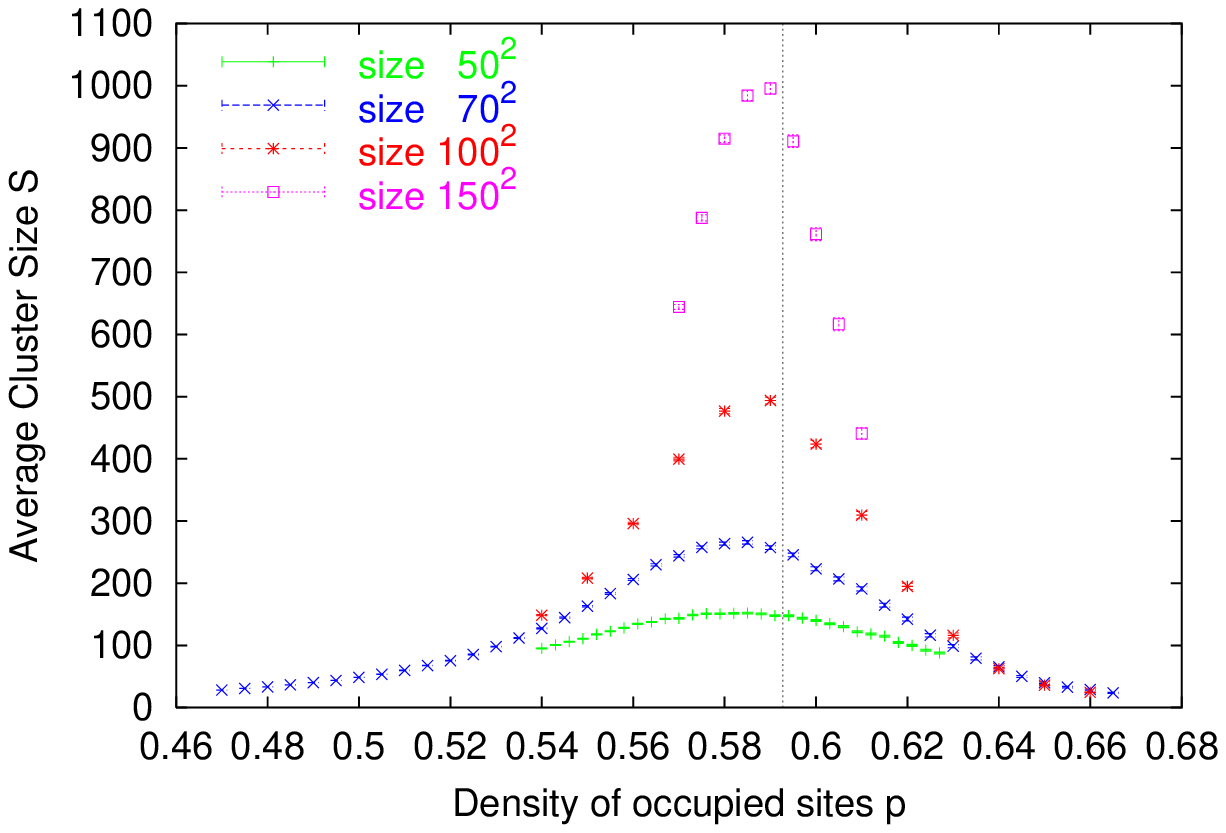,  width=7.8cm}
\epsfig{file=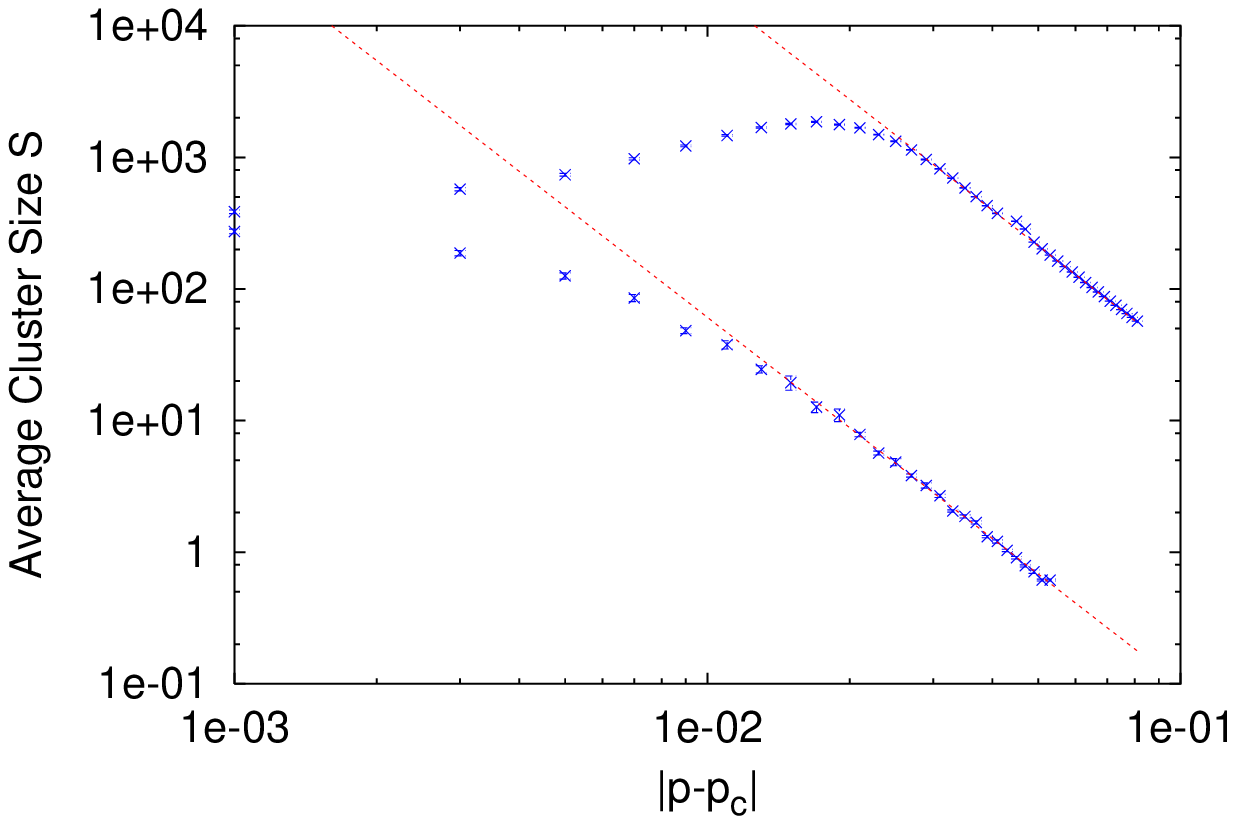,  width=8.0cm}
\vskip0.2cm
\begin{picture}(0,-30)
    \put(-109.9,2){\begin{minipage}[t]{7.4cm}{(a)}
    \end{minipage}}
    \put(120,2){\begin{minipage}[t]{7.4cm}{(b)}
    \end{minipage}}
\end{picture}
\caption[Average cluster size $S$ for pure site percolation on a square lattice]
{(a) Average cluster size $S$ as a function of 
the density $p$ for pure site percolation on a 2-dimensional square lattice.
The curves correspond to four different lattice sizes and  
peak near the infinite volume threshold $p_c=0.592746$, which is
represented by the dotted line. (b) Average cluster size $S$ as a function of 
 $|p-p_c|$ for pure site percolation on a $300^2$ square lattice.
In the logarithmic scale of the plot the scaling behaviour
of $S$ is clearly indicated by the two 
straight lines, which correspond to the different branches of
the curve around the peak.}
\label{avcl2dRP}
\end{center}
\end{figure}
If the sums included the eventual percolating cluster,  
$S$ would become infinite above the critical threshold.
In this way instead the average cluster size is divergent only
at the critical density $p_c$. Besides, its behaviour near 
$p_c$ is again expressed by a power law:
\begin{equation}\label{Satpc}
S\,{\propto}\,\,|p-p_c|^{-\gamma}
\end{equation}
where $\gamma$ is another critical exponent. The behaviour of $S$ as a function
of $p$ is illustrated in Fig. \ref{avcl2dRP}a, where we present the results
of simulations for pure site percolation on a square lattice in correspondence
of different lattice sizes. The divergence of $S$ can be seen 
through the peaks of the curves, which become higher and narrower
the larger is the size of the lattice. Besides,
increasing the lattice volume, the position of the peaks approaches 
the critical point of the geometrical transition (dotted line). 
To check the scaling behaviour of $S$ expressed by Eq. (\ref{Satpc})
we use other data relative to  
pure site percolation on a square lattice. In general, 
scaling relations are clearer for big volumes because 
the effects due to the finite size of the 
lattice are small (see Section \ref{markus}). In Fig. \ref{avcl2dRP}b we have plotted 
$S$ as a function of $|p-p_c|$ for a $300^2$ lattice.
The branches of the curve to the right and to the left 
of the peak are represented by the two straight lines in the figure.
They are approximately parallel, which confirms the fact that both
branches have a power law behaviour with the same exponent 
$\gamma$ as in (\ref{Satpc}). 
Actually the condition of 
best parallelism of the two lines is in general obtained for
a value of $p_c$ which is slightly different from the 
infinite volume one also for relatively large lattices:
that shows that the infinite volume limit
is a condition that is hard to simulate even using
modern supercomputers. 

\subsection{Percolation Strength}\label{marco}

In introducing the average cluster size $S$ we stressed the fact
that to evaluate this variable we don't need any information
about the eventual percolating cluster. But such information is
of course very important for a thorough understanding of the
percolation phenomenon. We thus introduce 
another variable, the {\it percolation strength } $P$, defined as
the {\it probability that an arbitrarily chosen site of the lattice
belongs to the spanning cluster}. $P$ is then basically the fraction
of the lattice volume which is occupied by the percolating cluster.
On an infinite lattice $P$ 
is zero for any density $p$
below the critical value $p_c$ (no
percolation), and a number between zero and one
above $p_c$. $P$ is the {\it order parameter }
of the percolation transition, as its value allows us to distinguish
the two phases of the system.
\vskip0.5cm
\begin{figure}[h]
\begin{center}
\epsfig{file=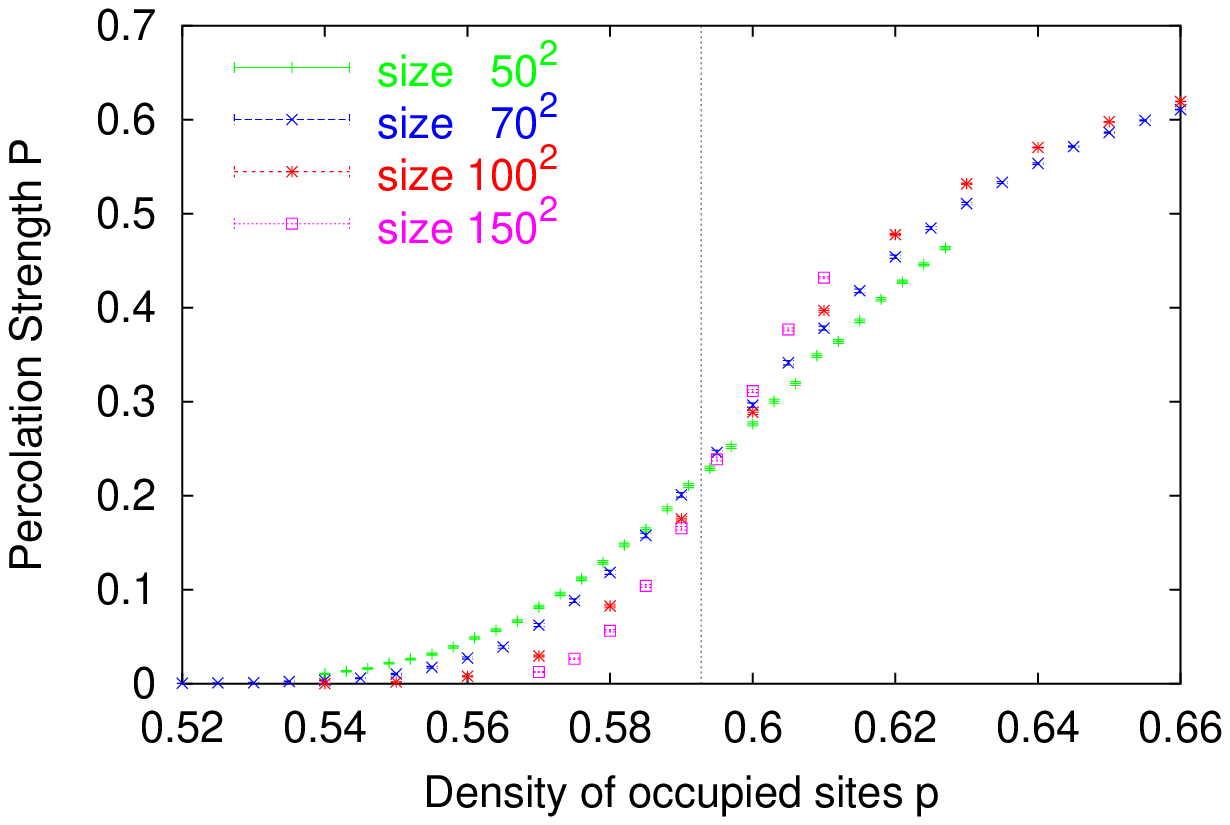,  width=7.8cm}
\epsfig{file=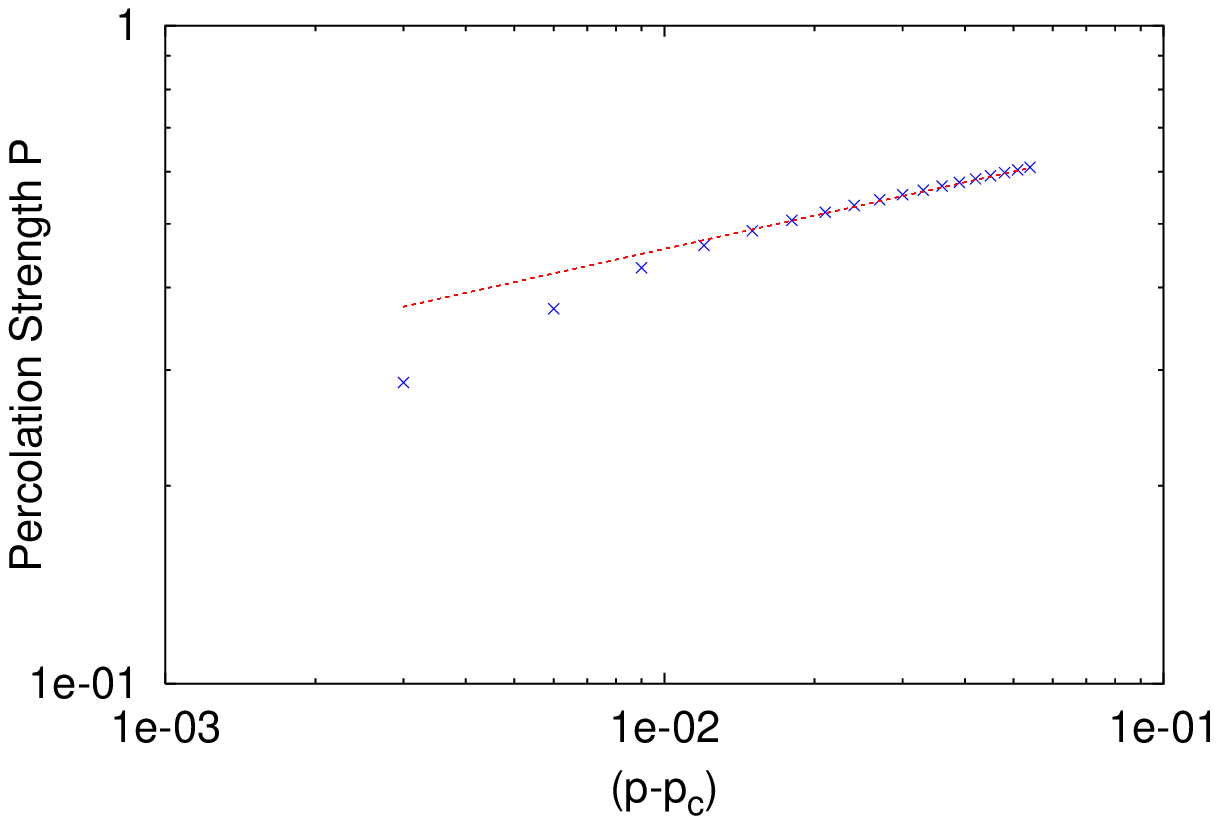,  width=8.0cm}
\vskip0.2cm
\begin{picture}(0,-30)
    \put(-109.9,2){\begin{minipage}[t]{7.4cm}{(a)}
    \end{minipage}}
    \put(117,2){\begin{minipage}[t]{7.4cm}{(b)}
    \end{minipage}}
\end{picture}
\caption[Percolation strength $P$ for pure site percolation on a square lattice]
{(a) Percolation strength $P$ as a function of 
the density $p$ for pure
site percolation on a square lattice.
The lattice sizes are the same as in 
Fig. \ref{avcl2dRP}a.
The tail of the curves to the left of the
critical threshold (dotted line) 
is smaller the 
greater the lattice size. (b) Percolation strength $P$ 
as a function of 
$(p-p_c)$ for pure site percolation on a $600^2$ square lattice.
Excluding the closest values of $p$ to $p_c$, for which
the results 
are strongly affected by the finite size of the lattice,
our data points follow approximately  
a straight line, which 
confirms the scaling behaviour of Eq. (\ref{Ppower}).}
\label{Pstreng}
\end{center}
\end{figure}
Near the critical density $p_c$ the behaviour of the 
percolation strength as a function of the density $p$ is 
again expressed by a power law:
\begin{equation}
\label{Ppower}
P\,\propto\,(p-p_c)^{\beta},
\end{equation}
relation which is obviously valid for $p>p_c$. 
Fig. \ref{Pstreng}a shows the $P$ curves 
corresponding to the $S$ curves 
of Fig. \ref{avcl2dRP}a. The finite size of our lattices
allows percolation to occur also at values 
of $p$ which are smaller than $p_c$, but 
the tails of the $P$ curves
to the left of $p_c$ get smaller the bigger the lattice size is.
In Fig. \ref{Pstreng}b we show a plot in logarithmic scale
of the percolation strength as a function of $p$
for $600^2$ lattice.
Disregarding the closest points to the threshold,
which feel strongly the effects of the finite size of the system 
(see Section \ref{markus}),
the scaling behaviour of Eq. (\ref{Ppower})
is clearly represented by the straight line to the right of the figure.
\vskip1cm

\section{Cluster Structure}\label{giuseppe}

\subsection{Perimeter of a Cluster}\label{irene}

\begin{figure}[h]
\begin{center}
\epsfig{file=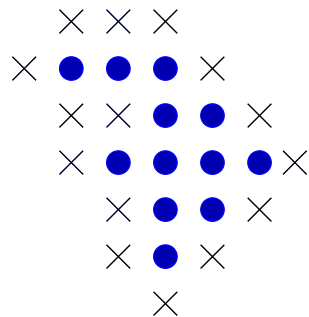,  width=4cm}
\caption[Perimeter of a small cluster]{Perimeter of a small cluster. We see that  
the number of sites of the perimeter
is of the same order as the size of the cluster. This fact is 
also valid for large clusters.}
\label{perim}
\end{center}
\end{figure}

Most of what we have discussed so far has to do with the {\it size }
of the clusters. But there are also
other aspects that can be studied. In particular, we can 
examine the {\it cluster structure}, which can 
let us know the {\it geometrical properties}
of our objects. For example, how can we define the {\it perimeter}
of a cluster? The easiest thing to think of is the number of
empty sites neighbouring a cluster. In Fig. \ref{perim} the crosses
around the cluster mark its perimeter 
according to this definition. 
If we count the sites of the perimeter of Fig. \ref{perim} we find that
they are approximately as many as the sites of the 
cluster (15 vs 12).
\vspace*{.2ex}
However, from  
geometry we know that, in a $d$-dimensional space, 
the perimeter of an object of
linear dimension $L$ is proportional to $L^{d-1}$, 
while its volume is proportional to $L^{d}$: the ratio 
perimeter/volume goes then like $L^{-1}$.
We might object that this fact is due to the 
small size of the cluster we have taken in
our example, and that going to larger 
structures we would recover the right behaviour.
As strange as it may seem, this objection is not correct.
We should not forget that our clusters are 
{\it random} structures; because of that, large clusters have
in general {\it holes} in their body (like the holes in 
a Swiss cheese). The empty sites of these holes contribute to 
the perimeter as well. We can take as example the 
big spanning cluster of Fig. \ref{RP_2D}c. 
There are more than forty holes in it, some
of which are so big that other clusters are contained in them.
On these grounds it isn't surprising that 
even the perimeter of large clusters is proportional to their
size. One could still say that the real perimeter is only the 
external one, i.e., it is given by the empty sites  
surrounding the cluster, excluding the contribution of eventual 
inner holes. But even in this case, the result remains valid.
We can easily convince ourselves in the case of site percolation 
on a simple cubic lattice. If we take a density $p$ between 
$0.4$ and $0.6$, we have percolation
both for the occupied and for the empty sites of the cube. In fact,
both the density of occupied sites $p$ and 
the one of empty sites $1-p$ are above the critical threshold ($p_c=0.3116$).
Nearly every occupied (empty) site belongs to 
the infinite network of occupied (empty) sites. 
Thus everywhere in the lattice, each occupied site has 
with high probability at least one neighbour belonging to the infinite
cluster of empty sites. 
Such empty site 
contributes to the external perimeter,
since inner holes are, of course, disconnected from 
the infinite network. This simple example
shows clearly that
the perimeter of a cluster is proportional to its size $s$ and not 
to $s^{(d-1)/d}$.

\subsection{Cluster Radius and Fractal Dimension}\label{patty}

To examine the cluster structure 
it is also important to define the linear dimension of the 
cluster, i.e., its {\it radius}. To define the radius of 
such complicated objects may not be that easy. The need to focus
on some features of the cluster geometry instead of others may
lead to different definitions. We will define 
the radius $R_s$ of a cluster of size $s$ through 
\begin{equation}
\label{radius}
{R_s}^2=\sum_{\bf i=1}^{s}\frac{|{\bf r_i}-{\bf r_0}|^2}{s},
\end{equation}
where
\begin{equation}
\label{cenmass}
{\bf r_0}=\sum_{\bf i=1}^{s}\frac{{\bf r_i}}{s},
\end{equation}
is the position of the {\it center of mass} of the cluster
and ${\bf r_i}$ 
the coordinates of the site {\bf i}. 
If we relate $R_s$ to the average distance between two cluster sites 
we get the formula:
\begin{equation}
\label{radiustwo}
{R_s}^2=\sum_{\bf i,j}\frac{|{\bf r_i}-{\bf r_j}|^2}{2s^2}.
\end{equation}
(We put the origin of the coordinates at the cluster centre-of-mass.)
It is interesting to check whether the radius $R_s$ of a cluster
is related in some
simple way to the cluster size $s$. One finds that for 
large values of $s$ the following simple power law  
is valid
\begin{equation}
\label{fractal}
R_s\,\propto\,s^{1/D}.
\end{equation}
The number $D$ is called
{\it fractal dimension}. An interesting feature of Eq. (\ref{fractal})
is the fact that $D$ varies with the density $p$. In particular, it may 
take non integer values. 
To evaluate the fractal dimension $D$ in correspondence
of some density $p$ we just need to test the scaling relation
(\ref{fractal}). However, there is a special 
case in which $D$ is relatively easy to determine. 

\begin{figure}[h]
\begin{center}
\epsfig{file=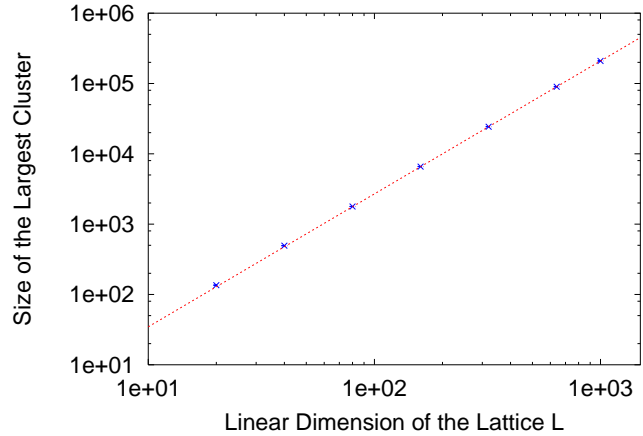,  width=10cm}
\caption[Determination of the fractal dimension]
{Test of the scaling relation (\ref{fractalpc}) for 2-dimensional
site percolation on a square lattice. Except the 
points corresponding to the smaller lattice sizes, 
our data points fall remarkably well on a straight line. The slope 
is the fractal dimension $D$ at $p_c$.}
\label{scalD}
\end{center}
\end{figure}

In fact, 
at the critical density $p_c$, the radius of the largest clusters 
on a lattice of linear dimension $L$ is 
with good approximation just $L$. On big lattices
one can thus write
\begin{equation}
\label{fractalpc}
s\,\propto\,L^{D},
\end{equation}
being $s$ the size of the largest cluster.
Fig. \ref{scalD} illustrates a numerical test of Eq. (\ref{fractalpc}).
The clusters are again the ones of pure site percolation
on a square lattice.
We have drawn our data points on a log-log plot, and it is 
clear that, apart from little deviations for the smaller lattices, the behaviour 
expressed by Eq. (\ref{fractalpc}) is correct. The slope of the 
straight line is $1.89(1)$, in good agreement with the 
exact value $91/48=1.896$. 
Since $D$ is smaller than the space dimension $d$ of the system,
we say that large clusters at criticality are 
{\it fractal objects}. This is not true at higher densities.
One can easily argue that, for $p\,{\approx}\,1$, large clusters
do not present holes in their body and therefore 
they are `real' objects, i.e., $s\,{\propto}\,{R_s}^d$. One finds that
this result is more general, namely
\begin{equation}
\label{Dabove}
D(p>p_c)\,=\,d 
\end{equation}
So, there is a jump in the value of the fractal dimension
when one goes from $p_c$ to $p\,>\,p_c$. 
Large clusters have again the 
same fractal dimension at any $p\,<\,p_c$.
In general
\begin{equation}
\label{Deq}
D(p<p_c)\,<\,D(p=p_c)\,<\,D(p>p_c)\,=\,d. 
\end{equation}
What we have just said about the fractal dimension allows us
to illustrate an important point that we have 
on purpose neglected till now.
We have so far spoken of "percolating cluster",
assuming that, at $p\,\geq\,p_c$, only 
one spanning network can be formed.
This fact is not at all trivial, and it has been
a source of hot debates inside the percolation community.
Nowadays the situation
seems to be clear and we 
present it here, without going into the details. 
One has to distinguish
two cases:
\begin{itemize}
\item{$p\,>\,p_c$. The fractal dimension
of a percolating cluster is, from Eq. (\ref{Dabove}),
equal to the number of space dimensions of the lattice. 
That means that
their density inside the lattice is {\it finite}, no matter how small, i.e.
the clusters cover a finite fraction of the 
whole lattice. Starting from this, it was proved rigorously that
there can be {\it only one} percolating cluster \cite{aizen}.}
\item{$p\,=\,p_c$. In this case, as we have seen,
the fractal dimension 
of a percolating cluster is smaller than $d$. The relative density of such
a cluster inside the lattice is {\it zero}, like the density of a straight
line on a plane. This would allow, in principle, the existence
of several percolating clusters at $p_c$. 
Aizenman proved 
that there is indeed a small
but finite probability to have more that one spanning cluster, 
even in two and three space dimensions \cite{aiz}.}
\end{itemize}
On the grounds of these results,
we shall keep assuming that there is
a single percolating cluster,
meaning a spanning cluster with a finite density.

\subsection{Correlation Length}\label{nancy}
 
If we take a site of a cluster, the probability that 
an occupied site put at some distance $r$ from the first one
belongs to the same cluster is 
non-negligeable as long as $r$ is of the same order of the
cluster radius $R_s$. The average value of this probability
is the {\it correlation function} $g(r)$. If we sum 
$g(r)$ over all distances
$r$, we get the average number of sites connected to some occupied
site of the lattice. The equivalence of $\sum_{r}g(r)$ and the 
average cluster size $S$ is clear. So, in general:
\begin{equation}
\label{eqSgr}
pS\,=\,\sum_{s}n_ss^2\,=\,p\sum_{r}g(r),
\end{equation}
relation that is valid for $p\,<\,p_c$ because, 
above $p_c$, $g(r)$ would take into account the spanning cluster as well,
whereas $S$ excludes it. 
Eq. (\ref{eqSgr}) can, however, be extended also
to the region $p\,>\,p_c$. For that it is enough to
subtract the contribution of the 
spanning cluster from the definition of the
correlation function $g(r)$.
The probability $p_{inf}$ that an occupied site $s_0$ taken at random belongs to
the infinite cluster is
given by $P/p$, where $P$ is the percolation
strength $P$. In fact, let $S_p$ be the size of the 
infinite cluster, $N$ the number of occupied sites and 
$V$ the lattice volume. 
The probability $p_{inf}$ is given by
\begin{equation}
\label{pinf}
p_{inf}\,=\,\frac{S_p}{N}\,=\,\frac{S_p}{V}\,\frac{V}{N}\,\equiv\,P\,\frac{1}{p}.
\end{equation}
Now,
the probability that another randomly selected site
$s_r$ (occupied or not), distant $r$ from $s_0$, belongs as well to the 
infinite cluster is simply given by $p_{inf}\,P\,=\,P^2/p$. 
The contribution of the spanning cluster to the correlation
function is thus $P^2/p$.
In this way, if
we replace $g(r)$ by $g(r)-P^2/p$, 
we get
\begin{equation}
\label{eqfinal}
p\sum_{r}[g(r)-P^2/p]\,=\,p\sum_{r}g(r)-p\sum_{r}P^2/p\,=\,\sum_{s}n_ss^2-
P^2V\,=\,\sum_{s'}n_{s'}{s'}^2\,=\,pS
\end{equation}
(the sum over $s'$ runs over
non-percolating clusters),
which is the generalization of Eq. (\ref{eqSgr}) for any value
of the occupation probability $p$.

We define the {\it correlation} or {\it connectivity length} $\xi$ 
as some average distance of two sites belonging to the same cluster:
\begin{equation}
\label{Corlen}
{\xi}^2=\frac{\sum_{r}r^2g(r)}{\sum_{r}g(r)},
\end{equation}
The sum over $r$ can be written as a sum over the cluster size $s$
following this reasoning. If we point to an occupied site of the lattice,
the probability $g(r)$ will be zero for all sites which do not 
belong to the same cluster. So, we have basically to perform
a sum only within each cluster and average over all clusters
of the lattice. Now we have 
to express Eq. (\ref{Corlen}) in terms of
$s$-quantities. Let us take at random a site $i$ 
of the lattice. Supposing it belongs to a cluster of
size $s$, we have
\begin{equation}
\label{umsch}
\sum_{r}g(r)=p\sum_{s}{\frac{1}{s}}\sum_{i}\sum_{j}|{\bf r_i}-{\bf r_j}|^2{n_s}s
\end{equation}
where the indices $i$ and $j$ run over all sites of the cluster. 
The probability that any site belongs to a cluster of size $s$ is $n_ss$,
and that weighs the distance $|{\bf r_i}-{\bf r_j}|^2$ in our 
equation. The second sum (divided by $s$) corresponds to averaging over the 
site $i$ picked up at the beginning. 
From Eq. (\ref{radiustwo}) we get
\begin{equation}
\label{substit}
\sum_{\bf i,j}{|{\bf r_i}-{\bf r_j}|^2}=2{R_s}^2{s^2}.
\end{equation}
by which we can write
\begin{equation}
\label{final}
\sum_{r}g(r)=p\sum_{s}2{R_s}^2{n_s}s^2~.
\end{equation}
The denominator of Eq. (\ref{Corlen}) can be easily
rewritten using Eq. (\ref{eqSgr}), so that we finally obtain
\begin{equation}
\label{Corlenrad}
{\xi}^2=\frac{\sum_{s}2{R_s}^2{n_s}s^2}{\sum_{s}n_ss^2}.
\end{equation}
Eq. (\ref{Corlenrad}) shows that
the correlation length is basically determined 
by those clusters which give the main contribution to the
average size $S$: $\xi$ is essentially the average radius
of those clusters. 
Approaching the critical density, 
the correlation length as well
as $S$ are thus divergent
at $p_c$. From what we have said
it is not surprising that, for $p\,{\approx}\,p_c$, 
also $\xi$ has a power law behaviour,
\begin{equation}
\label{Corlennu}
{\xi}\propto|p-p_c|^{-\nu}
\end{equation}
with $\nu$ as critical exponent.
There is, however, much more than that. It is rather easy to argue that
all divergencies we have encountered so far 
are also due to the clusters which are responsible for the divergencies
of the average size $S$ and the correlation length $\xi$.
For all variables, indeed, a key role is played
by the cluster number distribution $n_s$, which is explicitly
or implicitly present in all our definitions. 
We have actually seen above (Eq. (\ref{cros}))
that there is a sort of {\it cutoff} for the size of
the clusters for which $n_s$
is non negligible: 
the properties of these clusters {\it determine the critical behaviour 
of the percolation phenomenon}. In particular, the divergence of the
correlation length is at the basis of the scaling laws we have met 
up to now, as we will explain more in detail in the next section. 
If the behaviour of all variables we have introduced
is determined by the few properties of some special clusters, 
it is easy to deduce that
the corresponding critical exponents, 
which fix the functional dependence on $p$
of the variables at criticality, are somehow related to each other.
The distribution 
$n_s$ at criticality is ruled by the two 
critical exponents $\tau$ and $\sigma$ (see Eq. (\ref{ns})), 
so that we expect that all other exponents are simple combinations
of $\tau$ and $\sigma$. That turns out to be true: below we show how
one can calculate all exponents starting from the two fundamental ones
\begin{equation}
\label{scalrel}
\alpha=2-\frac{\tau-1}{\sigma},\,\,\,\,\,\,\,\,\,\beta=\frac{\tau-2}{\sigma},\,\,\,\,\,\,\,\,\,
\gamma=\frac{3-\tau}{\sigma},\,\,\,\,\,\,\,\,\, \nu=\frac{\tau-1}{{\sigma}d},
\,\,\,\,\,\,\,\,\,D(p=p_c)=\frac{1}{\sigma\nu}=\frac{d}{\tau-1}.
\end{equation}
(We indicate with $d$ the space dimension of the lattice.)
If we play a bit with Eqs. (\ref{scalrel})  
we can derive other useful relations: particularly important is 
\begin{equation}
\label{hyper}
2\,\frac{\beta}{\nu}\,+\,\frac{\gamma}{\nu}\,=\,d.
\end{equation}
The relations containing the dimension $d$ are called
{\it hyperscaling relations}. It is believed that 
the hyperscaling relations are valid only for values 
of $d$ satisfying $d\,\leq\,d_u$, for some
$d_u$ called the {\it upper critical dimension}.
When $d\,\geq\,d_u$, one finds that 
the percolation process behaves roughly in the same manner
as percolation on an infinite regular tree, like
the {\it Bethe lattice}. 
The values of the critical exponents
for this problem are analytically 
known: $\tau=5/2$, $\sigma=1/2$, $\nu=1/2$.
We can ask ourselves 
for which value of $d$ the 
hyperscaling relation
\begin{equation}
\label{upper}
\nu\,=\,\frac{\tau-1}{{\sigma}d}
\end{equation}
is satisfied by such exponents.
According to what we have said, 
the solution is just the upper critical dimension $d_u$.
From Eq. (\ref{upper}) one obtains $d_u\,=\,6$.

The fact that the space dimension $d$ of the lattice is 
present in Eqs. (\ref{scalrel}) means that 
the scaling relations are well-defined once we fix the value of 
$d$, independently of the type of percolation 
(site, bond) and (or) the lattice structure we are studying. 
It is actually remarkable that the dimension $d$ seems to fix not only
the scaling relations (\ref{scalrel}) but even the values
of the single exponents. This property is called
{\it universality} and so far all tests which 
have been performed, both analytically and numerically, haven't
found exceptions to it.
In Table \ref{percritexp} we have reported 
the values of the critical exponents for several values of 
the space dimension $d$. 
\newpage
\begin{table}
\begin{center}
\begin{tabular}{|c|c|c|c|c|}
\hline$\vphantom{\displaystyle\frac{1}{1}}$
Exponent & d=2 & d=3\footnotemark[3] & d=4 & d=5 \\
\hline
\hline$\vphantom{\displaystyle\frac{1}{1}}$
$\alpha$ & -2/3 & -0.6295(53) &-0.72 & -0.86\\
\hline$\vphantom{\displaystyle\frac{1}{1}}$
$\beta$ & 5/36 &0.4181(8) & 0.64 & 0.84\\
\hline$\vphantom{\displaystyle\frac{1}{1}}$
$\gamma$ & 43/18 & 1.793(4) & 1.44 & 1.18 \\
\hline$\vphantom{\displaystyle\frac{1}{1}}$
$\nu$ & 4/3& 0.8765(17) & 0.68 & 0.57\\
\hline$\vphantom{\displaystyle\frac{1}{1}}$
$\sigma$ & 36/91 &0.4522(9) & 0.48 & 0.49\\
\hline$\vphantom{\displaystyle\frac{1}{1}}$
$\tau$ & 187/91 & 2.18906(8) & 2.31 & 2.41\\
\hline$\vphantom{\displaystyle\frac{1}{1}}$
$D(p=p_c)$ & 91/48 &2.5230(2) & 3.06 & 3.54\\
\hline$\vphantom{\displaystyle\frac{1}{1}}$
$D(p<p_c)$& 1.56\footnotemark[4]&2 & 12/5 & 2.8\\
\hline$\vphantom{\displaystyle\frac{1}{1}}$
$D(p>p_c)$& 2 &3 & 4 & 5\\
\hline
\end{tabular}
\caption[Percolation critical exponents in $d$ dimensions]
{Percolation critical exponents in $d$ dimensions.}
\label{percritexp}
\end{center}
\end{table}
\footnotetext[3]{The values of the critical indices 
for $d\,=\,3$ are taken from a recent study of random percolation
on a simple cubic lattice \cite{parisi}.}
\footnotetext[4]{One could
wonder why we have given a numerical estimate of $D(p<p_c)$ 
in two dimensions, whereas for $d=3, 4$
analytical results are known. 
The percolation clusters below $p_c$ belong to the universality class of
{\it lattice animals}. In
1980 Parisi and Sourlas showed that the d-dimensional lattice animal
problem corresponds to a $(d-2)$-dimensional different problem, solvable 
in one and two dimensions: that is why exact results are known for 
$d=3$ and $d=4$.}

\section{Real Space Renormalization}\label{paul}

As we have seen up to now, the behaviour of 
all percolation variables at criticality 
is described by simple power laws. 
Apart from the simplicity
of their form, power laws have a remarkable property:
they are {\it scale free}.
To understand this feature, 
we take the simple function $f(x)=x^{1/2}$, and focus
on two intervals of the $x$-axis, namely $[1,2]$ and $[10,20]$. 
The ratio of the extremes of the intervals is the same ($2:1=20:10=2$)
in both cases: the corresponding ratios of the values of the 
function is also the same (${2^{1/2}}:1={20^{1/2}}:{10^{1/2}}={2^{1/2}}$). 
That means that if we perform a change of scale,
from $x$ to $x'=ax$, the $y$-axis will be
correspondingly rescaled, and the curve will look identical
after the transformation. That does not happen if we use, for example,
an exponential function. In fact, taking $g(x)=e^x$ and the same intervals
of our example, we would find two different ratios
for the values of the 
function at the extremes
of the ranges ($e^2:e^1=e\,{\neq}\,e^{20}:e^{10}=e^{10}$): if we go from a range 
to another through a scale change, the function will
look different after the transformation. 
In this sense we say that there is no characteristic 
length for a phenomenon described by a power law: it will look
identical in each scale. 

We have stressed in the previous section that 
the correlation length $\xi$ is the characteristic length
of the percolation phenomenon, expressing the average 
radius of those clusters which give the greatest contribution
to the percolation variables. So, at some density $p$, the value
of $\xi$ fixes the scale   
of the phenomenon: the (large) clusters of radius $R_s$ smaller than $\xi$
determine the percolation variables. The correlation
length thus divides all clusters in two distinct categories. At
the critical density $p_c$, $\xi$ becomes infinite. Therefore, in
a sense, there are no longer fundamental distinctions
between two large clusters $A$ 
and $B$ at criticality, even if 
$A$ is much bigger/smaller in size than $B$. 
If we take out a medium size piece of a big lattice, 
the linear dimensions of the 
lattice and of the piece 
are both much smaller than $\xi$ at $p_c$. The 
original lattice and its part will then be similar as far as 
their average properties are concerned. 
A nice example of this is represented by 
Fig. \ref{scalD}: the average size 
of the largest cluster for all lattice 
sizes above $100^2$ scales clearly with the linear dimension
$L$, which means 
that all lattices are basically equivalent to each other.
In this respect, the lattice $100^2$ 
contains all the information that can be extracted by
$1000^2$, $10000^2$, etc. Going from 
a lattice size $A$ to $B$ we just need 
to rescale properly the values of the 
variables in $A$ to obtain the values we would
measure in $B$. 
This feature is called {\it self-similarity} 
at the critical point and, 
according 
to what we have said at the beginning of this section,
it naturally leads to the power law behaviour of the 
percolation variables. 

\begin{figure}[h]
\begin{center}
\epsfig{file=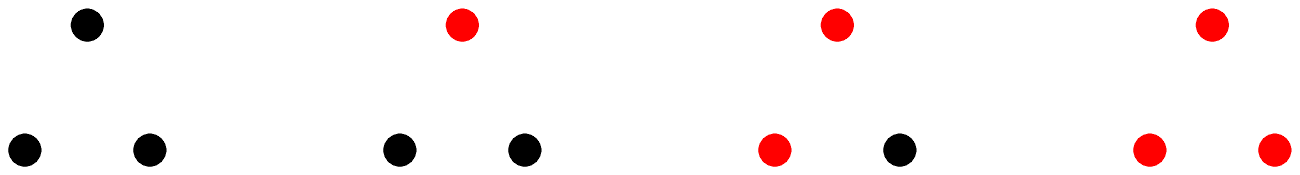,width=11cm}
\vskip0.4cm
\begin{picture}(0,-30)
    \put(-153.4,-3){\begin{minipage}[t]{6cm}{$(1-p)^3$}
    \end{minipage}}
    \put(-67,-3){\begin{minipage}[t]{6cm}{$3p\,(1-p)^2$}
    \end{minipage}}
    \put(23.5,-3){\begin{minipage}[t]{6cm}{$3p^2\,(1-p)$}
    \end{minipage}}
    \put(134.5,-3){\begin{minipage}[t]{6cm}{$p^3$}
    \end{minipage}}
\end{picture}
\vskip0.6cm
\caption[Possible "percolation states" of a triangular cell]
{Possible states of a group of three sites. In black we 
mark the free sites, in red the occupied ones. Apart
from irrelevant permutations, there are
only four different combinations. 
Under each scheme
we have written the corresponding probability. }
\label{unblock}
\end{center}
\end{figure}

Self-similarity is the basis of 
the {\it renormalization group} treatment of percolation. 
This approach was historically first applied to
thermal phase transitions by K. G. Wilson \cite{wilson}
to justify the scaling assumptions and to calculate 
the critical exponents. We will briefly present 
the extension to percolation, introduced 
by Reynolds et al. \cite{reynolds,rey2}. 
It is based on the so-called 
{\it real space renormalization}, by which one performs
transformations on the position coordinates
in ordinary space. 
The first step consists in {\it blocking} the lattice,
i.e. dividing the sites of the lattice into groups or
{\it blocks}, and then replacing each block by just one single site.
\begin{wrapfigure}{o}{8.5cm}
\begin{center}
\vspace*{-0.5cm}
\hspace*{0.2cm}\epsfig{file=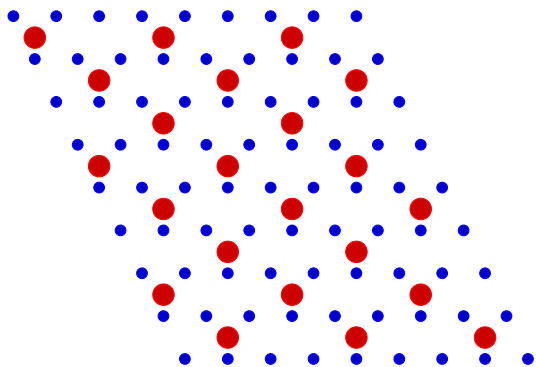,  width=7cm}
\caption[Real space renormalization on a triangular lattice]
{Real space renormalization on a triangular
lattice (blue structure). The new sites, marked in red,
replace the triangles which surround them. The new lattice,
which is still triangular,
has therefore one third of the sites of the original one.}
\label{block}
\end{center}
\end{wrapfigure}
Fig. \ref{block} shows an example of 
this operation on the 2-dimensional triangular lattice.
We block the sites in triangles and replace them by 
the red sites put in the center of each triangle.
One of the requirements of the blocking procedure is 
that one must get the same lattice structure after any transformation.
In our case
we clearly see that the new structure we have formed 
is again a triangular lattice, and it contains 
one third of the sites there were at the beginning.
In order to complete the transformation, we must 
decide which of the new sites are occupied and which
are not. We need that the renormalized lattice keeps
some essential
features of the old one, because the latter is the system we 
want to analyze. That means that the status of
each new site (occupied, free) must be related to the 
status of the three sites it 
replaces.
There is
no unique way of doing that. If we take a group of three sites,
we can get four possible configurations, since we may have 
zero, one, two or three occupied sites
 (Fig. \ref{unblock}).
What we want to keep
is the essential physics of percolation of the 
initial configuration. Since percolation
involves the formation of an infinite connected network, by which one gets
across the whole lattice, a sensible choice could be to
define a cell as occupied if and only if it 
contains a set of sites such that the cell `percolates'. 
\vskip0.2cm
As we can see in Fig. \ref{unblock}, 
the first and the second schemes are not percolating,
and the relative super-site will be set free,
the last two are percolating and the relative
super-site will be occupied. 
Since the 
occurrence of the four triangular schemes of Fig. \ref{unblock}
is a different function of $p$ in each case, the density 
of the blocked configuration will be in general 
some $p'\,{\neq}\,p$. 
In our example it is easy to calculate $p'$: it is just the probability
for a triangular block of the unblocked lattice to be either the third
or the fourth triangle of Fig. \ref{unblock}.
The probability for a triangle to have two occupied sites is 
$3p^2\,(1-p)$, to have three $p^3$. Then
\begin{equation}
\label{pblock}
p'\,=\,3p^2\,(1-p)+p^3
\end{equation}
At $p\,=\,p_c$
we expect our operation
to be basically equivalent to a rescaling of the structures
of the original lattice and, because of 
self-similarity, 
\begin{equation}
\label{pcrit}
p'\,=\,3p^2\,(1-p)+p^3\,=\,p\,=\,p_c
\end{equation}
The equation $p'=p$ has three solutions: 0, 1/2, 1.
Discarding the trivial 0 and 1, we find $p_c=1/2$, which is
indeed the exact value of the percolation threshold
on a two dimensional triangular lattice (see Table \ref{critthr}). 

Moreover, 
by means of the renormalization group approach,
we can evaluate the critical exponents.
If we start from a density
$p$ close to $p_c$, the correlation length $\xi$ 
of the initial
configuration is much bigger than the linear dimension $b$
of the blocks (in our case $b=\sqrt{3}$). That means that 
the blocking introduces 
changes only at a scale which is 
by far smaller than $\xi$. The correlation length of the 
renormalized configuration $\xi'$ has thus
the same functional dependence of $\xi$, i.e.
\begin{equation}
\label{xipr}
\xi'=\,c\,\,|p'-p_c|^{-\nu}
\end{equation}
with the same constant $c$ and exponent $\nu$ of $\xi$. Moreover,
since all the lengths of the initial system are 
rescaled   
by a factor $b$, we have $\xi'=\xi/b$, which establishes
the following relation between $p$, $p'$ and $\nu$
\begin{equation}
\label{nuscale}
b\,|p'-p_c|^{-\nu}\,=\,|p-p_c|^{-\nu},
\end{equation}
from which we derive
\begin{equation}
\label{nulog}
\frac{1}{\nu}=\frac{\log[(p'-p_c)/(p-p_c)]}{\log\,b}=\frac{\log{\left(\frac{dp'}{dp}\right)_{p_c}}}{\log\,b},
\end{equation}
where the last step is justified since we assume that both $p$
and $p'$ are very close to $p_c$.
In our case, knowing the function $p'$ from Eq. (\ref{pcrit}) and
$p_c=1/2$, we get finally
\begin{equation}
\label{nures}
\nu=\frac{\log(\sqrt{3})}{\log(3/2)}=1.355,
\end{equation}
which is a good approximation of the exact value $4/3$.

We have then shown the power of the renormalization group
approach. We must admit, however, that 
we have chosen a particularly suitable example, and that 
the agreement between the values derived in this way 
and the exact values is seldom as good as in our case.
As we have said, in fact, there is some 
freedom in the procedure that leads to the 
renormalized configurations: we may choose several
ways of blocking the lattice, and 
the rule to establish which of the sites of the 
renormalized lattice are occupied and which are free is not fixed either. 
In general, each of the possible ways we may adopt to 
renormalize the initial lattice leads to different results, which
could also be rather far from the exact ones.
The original assumption that, around the critical point, we can 
`rescale' the lattice structures by 
simply replacing groups of sites by single super-sites is indeed
quite strong and not
completely legitimate. It is easy to convince oneself that, for
instance, 
a cluster of the initial lattice could be broken into pieces 
in the renormalized lattice or, vice versa, separate clusters 
can be fused together after the blocking transformation. 
Since the crucial feature is the fact that
the {\it average} properties of the initial configuration are 
not changed, sometimes we 
can be lucky enough to choose
a procedure that induces a sort of compensation
of these two effects: our case of the triangular
lattice is an example of that. However, generally speaking,
renormalizing a configuration involves 
correlations between sites at a block distance $b$ from each other.
After the transformations, in fact, the relative super-sites
can become neighbours and form structures. 
But, if we want to preserve the initial cluster
distribution after any transformation, 
we must forbid that new structures are formed or that some of the
old ones disappear. Reynolds et al. showed that using large cells
one can reduce very much this drawback and get quite precise results for
several systems \cite{rey2}.

From what we have said it emerges that 
blocking the lattice does not only imply   
a new occupation density $p'$ for the 
sites of the renormalized configuration, but 
also some {\it probability} $x$ that neighbouring sites are connected
to each other. This probability is introduced
to eliminate correlations among sites which are not neighbour 
in the initial configuration.
If we start from a pure site percolation problem,
we will thus end up with a {\it site-bond} one. 
Repeating the
transformation over and over, longer range correlations will
be introduced, and, in order to cancel them, the number of parameters
which characterize the percolation system after any transformation 
will increase. But, around criticality,
as long as the range of the 
correlation between sites can be, it will be always negligible 
compared to the (basically)
infinite correlation length and, following the
same reasoning of our example, we deduce that the exponent 
$\nu$ is the same for all the percolation systems mapped
onto each other by renormalization transformations. 
Analogously, if
we consider other percolation variables, 
instead of the correlation length $\xi$,
it is easy to show that each critical exponent is not changed by 
blocking transformations.
At some stage,
one finds that the set of parameters does not vary after performing 
blocking transformations. In an ideal 
parameter space where all percolation systems
can be represented by points according to the 
values of $p$, $x$, etc., the final set of parameters 
represents the 
so-called {\it fixed point } of the
renormalization transformation. 
If one starts from 
some percolation system at criticality in $d$
space dimensions, successive
blockings will lead to the {\it 
same} fixed point. From what we have said,
a consequence of that is the fact that
the critical exponents are the same
for all percolation systems in $d$ dimensions,
which explains
the {\it universality} for the percolation phenomenon.
In so far, we can say that the local differences between 
the various
percolation systems can be smoothed out by 
means of renormalization transformations without changing 
the main features (e.g. the exponents) of the phenomenon.
These features remain unchanged all the way up to the 
fixed point and are, because of that, equal for all possible
systems. 

\section{Finite Size Scaling}\label{markus}

As we said at the beginning, the percolation
problem is relatively old. The simplicity of its
formulation and the 
useful symmetries of several lattices
have allowed to derive a number of results
by means of rigorous analytic proofs: the 
demonstration that the critical density 
for bond percolation 
on a square lattice is $1/2$ is probably the most
spectacular achievement \cite{kesten}.
Indeed, percolation as a critical phenomenon makes sense only
on an infinite lattice, and such an ideal system can be
properly handled by probability theory, which is 
at the basis of the proofs we mentioned. 
Many features of 
apparently simple systems are, however, still out of reach. For instance,
nobody could so far find an analytical expression 
for the value
of the critical density ($p_c=0.592746$)
for the site percolation problem 
on a square lattice. 

The study of percolation systems received
new impulse since fast computers became available. Monte Carlo simulations
are, in fact, a powerful tool to analyze 
complex systems. The data we have shown in our plots so far 
have been derived by means of this numerical approach. 
Computer simulations are {\it experiments}: 
they reproduce the system  
one wants to study by creating a big number of
possible copies of it, and obtain the results 
by averaging the values corresponding to the 
different configurations. In this way, one can 
investigate any system with a degree of accuracy 
which depends mainly on the computer time one invests in
the project. The results can then be made, in principle,  
arbitrarily precise. 

By simulating a system, however, we are forced
to use {\it finite lattices}. The infinite lattice,
which is the ideal system we would like to investigate,
remains an unreachable limit for a computer,
as big as it can be. The rapid evolution of 
fast machines in the last years has allowed
to push the size of the lattices 
that can be realistically studied up to 
values which were unthinkable only ten years before.
Any progress in this direction is always welcome, but the 
fundamental problem of computer simulations is always the same:
how can we extrapolate the infinite volume results out of 
values relative to finite lattices?
One could assume that a huge lattice is already
`infinite' in the sense that the difference between
the critical indices (threshold, exponents) 
that one can derive from it 
and the exact ones is smaller than the degree of accuracy
we want to reach. For some random percolation systems 
this turns out to be a good assumption. 
But in most cases, especially
when one wishes to perform percolation studies on interacting
systems, it does not work. Because of the dynamics, in fact,
the simulations are by far more time consuming than
for ordinary random percolation, and the largest lattices
one can use for the latter 
are out of reach. As a matter of fact, there is 
a way to extract the required infinite volume 
information out of 
values calculated on finite lattices: instead of using 
a single lattice size, one has to take several ones, and exploit 
the {\it scaling} behaviour of the percolation systems.
This procedure is called {\it finite size scaling} and 
is usually applied to all systems which
undergo second order phase transitions.
In this section we shall describe finite size scaling, 
focusing in particular on the 
techniques we adopted to extract the final results 
for the systems we have investigated all through this work.

If we take a look at the plots we have presented in this chapter,
we can already see a number of characteristic
{\it finite size effects},
i.e. features due to the finite size of the system. In Fig. 
\ref{avcl2dRP}a, for example the divergence 
of the average cluster size $S$ becomes a finite peak,
which gets sharper and higher for bigger systems; 
the corresponding percolation strength curves (Fig. \ref{Pstreng}a)
show a little tail 
to the left of the critical point, whereas on an infinite 
lattice $P=0$ for $p<p_c$. The main reason of these 
perturbations 
is obviously the finite number of
sites of the lattice, which
introduces a cut-off for 
the upper size of the clusters. 
Another problem 
is the fact that
the configuration looks different at the boundaries 
of the system than far from them. We can see it in
Fig. \ref{RP_2D}c: the edges of the lattice
cut the clusters close to them. This  
factor can be considerably reduced by using {\it periodic boundary
conditions}, i.e., by connecting opposite sides
(surfaces) of the lattice to each other 
in some way, so that each site is always surrounded 
by other sites. Such a trick is regularly adopted in
simulating systems on the lattice but in all
our cluster analyses we will dispense with it ({\it free boundaries}).

We have seen that the scaling laws are 
effective already for rather small lattices. 
In the previous section we pointed out   
that self similarity at the critical point is 
responsible of that. From
renormalization
group theory it is possible
to find out what the scaling laws look like
on finite lattices. In general, if a variable $\cal O$ is 
supposed to scale as $|p-p_c|^{-\rho}$, on a finite lattice 
of linear  
dimension $L$ at 
a density $p$ close to $p_c$, one observes the following behaviour:
\begin{equation}
{\cal O}(p-p_c,L) \;=\; L^{\rho/\nu}\,Q_{\cal O}\Big[\Big(\frac{p-p_c}{p_c}
\Big)\,L^{1/\nu}, g_iL^{y_i}\Big]\,,
\label{scalas}
\end{equation}
where $\nu$ is the critical exponent we have already met and  
$\,Q_{\cal O}\,$ is a function related 
to the variable $\cal O$ whose form does not depend on the dimension $L$ of the
lattice. 
Besides, one could 
have an eventual dependence 
on other parameters, which we indicate
by $g_i$: $y_i$ are the exponents correspondent to these other parameters.
The further dependence of $\cal O$ on $g_i$ is the 
main source of the so-called
{\it corrections to scaling}, since it modifies
the otherwise simple scaling assumption expressed by Eq. (\ref{scalas}).
Such perturbations are sometimes relevant and one
should take them into account. However, for
all the systems 
we have investigated in this work, 
we will disregard them \footnote{This point will 
be discussed more extensively in the summary.}.
We shall thus always make use of the simple
scaling assumption
\begin{equation}
{\cal O}(p-p_c,L) \;=\; L^{\rho/\nu}\,Q_{\cal O}\Big[\Big(\frac{p-p_c}{p_c}\Big)\,L^{1/\nu}\Big]\,.
\label{simpscal}
\end{equation}
Eq. (\ref{simpscal}) shows that the 
infinite volume information we look for ($p_c$ and 
the exponent $\rho$) is 
`hidden' in the  
finite size results: we have only to extract it 
in some clever way. 
At the critical density $p_c$, Eq. (\ref{simpscal})
becomes
\begin{equation}
{\cal O}(L)_{p_c} \;=\; L^{\rho/\nu}\,Q_{\cal O}[0]\,.
\label{ScalO}
\end{equation}
We notice 
that there is no $L$-dependence
in the values of the function
$Q_{\cal O}$. By plotting ${\cal O}$ as a 
function of $L$ at $p_c$,
we can then obtain the exponents' ratio
$\rho/\nu$ directly from the slope of the 
data points in a log-log plot.
If we have an idea of where 
the critical point could be, e.g. 
from the positions of the peaks 
of the average cluster size curves, we can
evaluate $\cal O$ at different values 
of $p$ for several lattices and 
check for which value of the density we get 
the best $\chi^2$ for the simple linear fit in 
the log-log plot. In this way we would be able
to evaluate $p_c$ as well. The errors on $p_c$ and on the
exponents are calculated by determining
the $p$-range containing $p_c$ such that for each value of $p$
one still gets a good $\chi^2$ for the scaling fit\footnote{A common
criterion is that the value of the $\chi^2$ must be within the $95\,\%$
confidence level.}. 

\begin{figure}[h]
\begin{center}
\epsfig{file=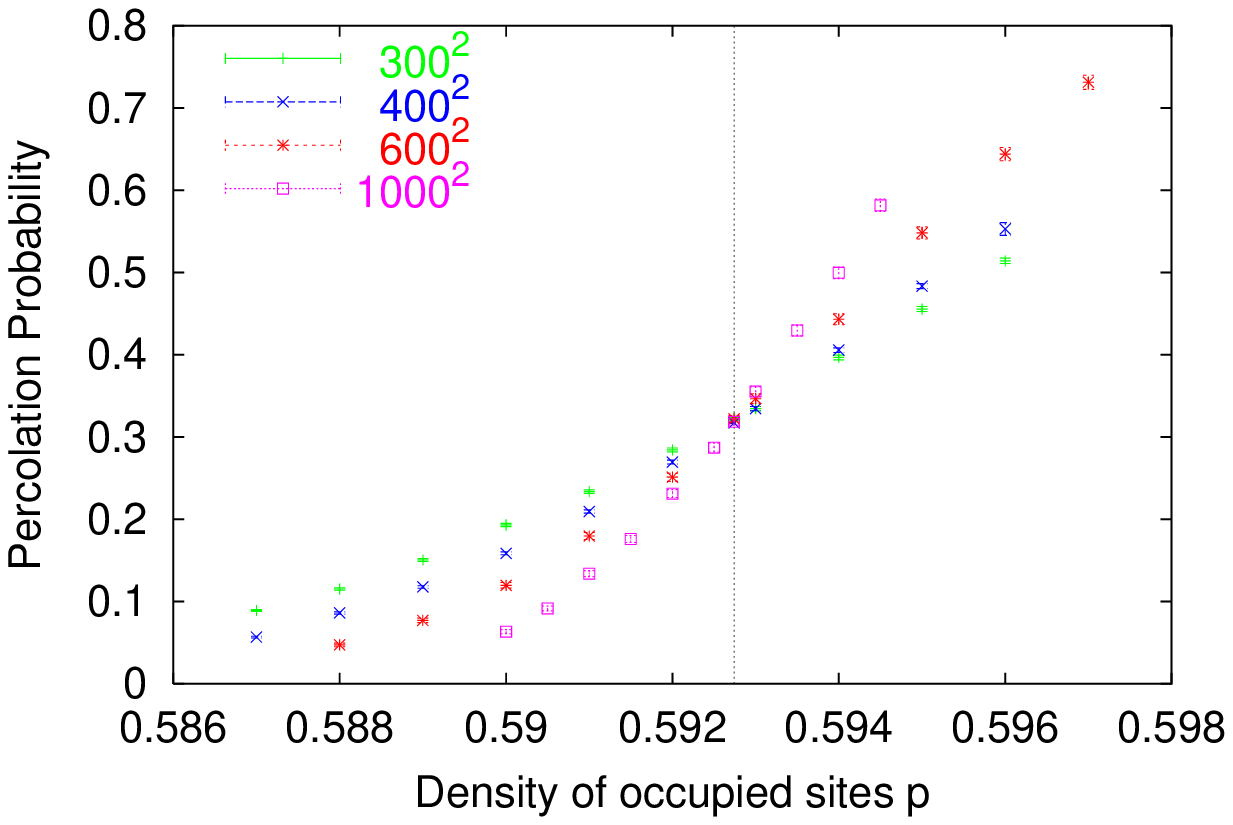,  width=7.9cm}
\epsfig{file=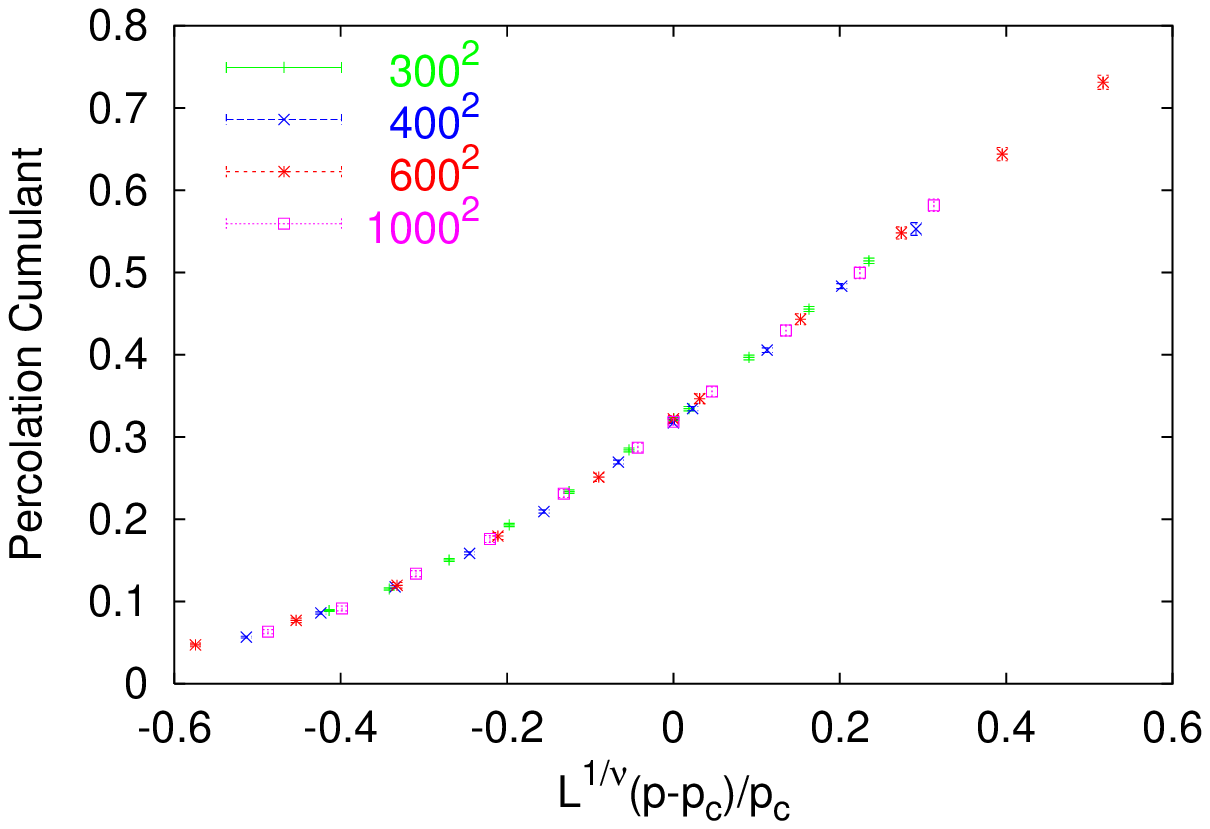,  width=7.9cm}
\vskip0.2cm
\begin{picture}(0,-30)
    \put(-109.6,2){\begin{minipage}[t]{7.4cm}{(a)}
    \end{minipage}}
    \put(121.6,2){\begin{minipage}[t]{7.4cm}{(b)}
    \end{minipage}}
\end{picture}
\caption[Percolation cumulant for pure
site percolation on a square lattice]
{(a) Percolation cumulant as a function of 
the density $p$ for pure
site percolation on a square lattice.
The curves cross remarkably well at the same point, 
in excellent agreement with the infinite volume 
threshold, whose value is marked by the dotted line.
(b) The data points in (a) are plotted  
as a function of 
$(\frac{p-p_c}{p_c})L^{1/{\nu}}$. Both 
$p_c$ and $\nu$ for this problem
are well known 
from the literature: 
$p_c=0.592746$  
and $\nu=4/3$. 
All data points fall on the same curve, which is the scaling 
curve $\Phi(X)$ of equation (\ref{Pprob}) for our system. 
}
\label{Pcros}
\end{center}
\end{figure}

As a matter of fact, there is 
also another method 
to determine quite precisely the critical point 
of the percolation transition.
Because of the finite size of the lattices 
we may find spanning clusters
at any value of the density $p$ of occupied sites, in
particular also for $p<p_c$. For the same reason 
there may be lattice configurations at densities above
the critical threshold for which percolation does not occur.
The probability of finding a spanning 
cluster on a finite lattice of linear dimension $L$ at a density $p$
is a well defined function $\Pi$, which we call {\it  
percolation cumulant} \cite{bind}: for $p\,{\approx}\,p_c$ and 
big values of $L$, it has the following behaviour  
\begin{equation}
\label{Pprob}
\Pi\,=\,\Phi\Big[\Big(\frac{p-p_c}{p_c}\Big)\,L^{1/{\nu}}\Big].
\end{equation}
We recognize the typical functional dependence
of an observable $\cal O$ given by Eq. (\ref{simpscal})
with $\rho=0$. 
The function $\Pi$ is not a real variable for percolation 
because it has a non-trivial meaning only
on finite lattices. On an infinite lattice it reduces itself to a
step function: it is zero for $p<p_c$
and one $p\,>\,p_c$. 
Nevertheless the special features of 
$\Pi$ make it a powerful tool to extract information about
critical properties of the percolation phenomenon. 
In particular, for $p=p_c$,
$\Pi=\Phi(0)$ for any value of $L$. That means that if we calculate 
the percolation cumulant as a function of $p$ for different lattice sizes, 
all curves will cross in correspondence of the critical density $p_c$
(Fig. \ref{Pcros}a).
Besides, if we replot the different curves as a function of 
$X=(\frac{p-p_c}{p_c})L^{1/{\nu}}$, the result must be just the function
$\Phi(X)$ for each lattice size and all curves will fall
on top of each other (Fig. \ref{Pcros}b).
This represents a good technique to determine
the critical point and we will adopt  
this method all through our calculations.
Once we have determined the position of the critical threshold $p_c$ 
with its error $\sigma$, we examine the range 
$[p_c-\sigma, p_c+\sigma]$: by exploiting the scaling
of the percolation variables one can 
see how the exponents's ratios $\rho/\nu$ of Eq. (\ref{ScalO})
vary with $p$ within the range. In this way one determines
the errors on $\rho/\nu$.
  
To get the scaling function $\Phi(X)$,
one needs also to know the value of the exponent $\nu$: 
if we don't know this value we can evaluate it by
making some guesses until we find the best scaling for 
the percolation cumulant curves. This method can indeed help
to restrict the range of possible values of $\nu$; unfortunately
$\Phi$ varies quite slowly with $\nu$ and 
the errors on its value can be rather big (${\approx}\,5\%$ in 
some of our investigations).

An alternative way of evaluating the exponent
$\nu$ consists in determining, for a given 
lattice of dimension $L$, the so-called
{\it pseudocritical point}. Looking at Fig. \ref{avcl2dRP}a 
we notice that the peaks are not centered
at the same value of $p$. In fact, because of the 
finite size, each lattice `feels' itself at criticality
when the correlation length $\xi$ reaches the dimension
of the lattice. Since, around the critical point,
$\xi$ varies with $p$ according to 
Eq. (\ref{Corlennu}), the condition $\xi\,{\approx}\,L$ is 
reached for the density $\bar{p}\,(L)$ for which
\begin{equation}
\label{nuleng}
|\bar{p}(L)-p_c|\,\,{\propto}\,\,L^{-1/\nu}
\end{equation}
The value of $\bar p$ is called pseudocritical point: 
we stress that it 
depends
on the linear dimension $L$ of the lattice.
Plotting in logarithmic scale the values
of the `distances' from the critical point $|\bar{p}(L)-p_c|$
as a function of $L$, we should then get a straight line,
whose slope gives $1/\nu$.
If we have a precise value for the critical density $p_c$, 
Eq. (\ref{nuleng}) allows us then to derive the 
exponent $\nu$. 
Anyway, if our determination of the critical
point is not accurate enough but we have 
data in correspondence of several lattice sizes, we could obtain
$p_c$ by considering it a parameter of 
the fit like $1/\nu$ and the proportionality constant
of the power law of Eq. (\ref{nuleng}). 
This method leads to more precise estimates of $\nu$
than the ones got by means of the scaling of the percolation cumulant;
because of that, in our studies we shall 
determine the percolation exponent $\nu$ from the scaling
of the pseudocritical points.

\chapter{Percolation and Critical Behaviour in the Ising
Model}\label{Percolation and Critical 
Behaviour in the Ising Model}


\setcounter{footnote}{1}
\thispagestyle{empty}


\def\J{$J/\psi$}
\def\j{J/\psi}
\def\X{$\chi$}
\def\x{\chi}
\def\P{$\psi'$}
\def\p{\psi'}
\def\U{$\Upsilon$}
\def\u{\Upsilon}
\def\C{c{\bar c}}
\def\cg{c{\bar c}\!-\!g}
\def\bg{b{\bar b}\!-\!g}
\def\b{b{\bar b}}
\def\q{q{\bar q}}
\def\Q{Q{\bar Q}}
\def\L{\Lambda_{\rm QCD}}
\def\A{$A_{\rm cl}$}
\def\a{A_{\rm cl}}
\def\N{$n_{\rm cl}$}
\def\n{n_{\rm cl}}
\def\S{S_{\rm cl}}
\def\s{s_{\rm cl}}
\def\bb{\bar \beta}

\def\be{\begin{equation}}
\def\ee{\end{equation}}

\def\lsim{\raise0.3ex\hbox{$<$\kern-0.75em\raise-1.1ex\hbox{$\sim$}}}
\def\gsim{\raise0.3ex\hbox{$>$\kern-0.75em\raise-1.1ex\hbox{$\sim$}}}


\def\CMP{{ Comm.\ Math.\ Phys.\ }}
\def\NP{{ Nucl.\ Phys.\ }}
\def\PL{{ Phys.\ Lett.\ }}
\def\PR{{ Phys.\ Rev.\ }}
\def\PRep{{ Phys.\ Rep.\ }}
\def\PRL{{ Phys.\ Rev.\ Lett.\ }}
\def\RMP{{ Rev.\ Mod.\ Phys.\ }}
\def\ZP{{ Z.\ Phys.\ }}

\section{Critical Behaviour}\label{marzia}

In this section we shall introduce 
the formal definition of phase transition and 
point out the main aspects
related to it.

In general, if we have
a system at a temperature $T$,
from its Hamiltonian $\cal H$ one defines
the {\it partition function} ${\cal Z}(T)$ as follows
\begin{equation}
{\cal Z}(T)=\sum_{\{n\}}{e^{{-\beta}{\cal H}}},\,\,\, \,\, \beta=\frac{1}{kT},
\label{partf}
\end{equation}
where $\sum_{\{n\}}$ runs over all
possible states of the system. From ${\cal Z}(T)$ one 
can derive all the {\it thermodynamic potentials}
of the system, which give us the whole thermal information.
In particular, one defines the {\it free energy} ${\cal F}(T)$
\begin{equation}
{\cal F}(T)=-{\frac{1}{\beta}}\log\,{\cal Z}(T).
\label{freee}
\end{equation}
By means of the free energy 
one usually classifies phase transitions in two
main categories:
\begin{itemize}
\item {\it first order} phase transitions, if the first derivative of the free energy 
 ${\cal F}$ as a function of $T$ is discontinuous;

\item
{\it continuous} phase transitions, if the $n^{th}$ derivative of the free energy 
${\cal F}$ as a function of $T$ is discontinuous, all the previous n-1 derivatives 
are continuous
($n=2,3, etc.$). 
\end{itemize}

For first order phase transitions, the discontinuity 
of the first derivative of the free energy implies the discontinuity of the 
{\it energy density} $\epsilon$,
\begin{equation}
\epsilon=\frac{E}{V}=\frac{{\cal F}+T\,\frac{\partial {\cal F}}{\partial T}}{V}
\label{latheat}
\end{equation}
($E$ and $V$ are the
energy and the volume of the system,
respectively).
Because of that, once we reach the critical temperature $T_c$ by heating
or cooling our system, we need to extract (or add) some energy ({\it latent heat})
for the system to pass to the other phase, and during this process
the temperature does not vary. This is exactly what happens during
the water-ice transition: the latent heat $L\,{\simeq}\,334Jg^{-1}$
is the energy released when $H_2O$ molecules neatly
pack themselves into a face-centered cubic lattice, rather
than wandering around.

The most famous example of a continuous 
phase transition is the conversion of iron from paramagnetic to ferromagnetic form at the
Curie temperature ${T_c}=1043^0 K$. At $T>T_c$ iron is paramagnetic, i.e.,
it is not magnetized in absence of an external magnetic field; for $T<T_c$
the material acquires a {\it spontaneous magnetization} ${\bf m}$. The
magnetization as a function of $T$ is continuous and 
the energy density changes as well smoothly. 
The phase change is thus continuous. 

In this work we will deal with
second order phase transitions, therefore we 
shall briefly introduce here the main features of these special
phenomena. 

A common feature of phase transitions is 
the existence of a variable $\Phi$, 
called {\it order parameter}:
$\Phi$ is defined in each point $\bf {\vec r}$
of the volume occupied by the system and its
average value allows to identify the phase of the system. 
For the 
paramagnetic-ferromagnetic transition we have mentioned,
$\Phi$ is just the magnetization ${\bf m}$, which is zero
in the paramagnetic phase and non-zero in the ferromagnetic one.
To express the relationship between two points of the
system at various distances, one defines the 
{\it two-point correlation function}:
\begin{equation}
\label{corfun}
G^{(2)}(r)\equiv \langle \Phi({\bf 0})\cdot\Phi({\bf {\vec r}})\rangle. 
\end{equation}
The brackets indicate thermal averaging, i.e., over many
configurations at some temperature $T$.
In the ordered phase, $G^{(2)}(r)$ 
includes the contribution of the 
non-vanishing average of the 
order parameter $|\langle\Phi\rangle|^2$.
In order to determine
the {\it fluctuations} of $\Phi$ with respect to
its average value, one needs to subtract
that contribution. Therefore one
often uses the {\it connected two-point correlation function}
\begin{equation}
\label{concorfun}
{G_c}^{(2)}(r)\equiv \langle \Phi({\bf 0})\cdot\Phi({\bf {\vec r}})\rangle-
|\langle\Phi\rangle|^2. 
\end{equation}
Experimental evidence leads to the following form of 
the function ${G_c}^{(2)}(r)$ for $T$
close to the 
critical temperature $T_c$:
\begin{equation}
\label{corclos}
{G_c}^{(2)}(r)\sim e^{-r/{\xi}}.
\end{equation}
The length $\xi$ is the {\it correlation length} of the system. 
It expresses the distance within which 
the fluctuations of the order parameter 
are important.
For
second order phase transitions at
$T\,{\approx}\,T_c$,
\begin{equation}
\label{corlng}
\xi\,\sim\,|T-T_c|^{-\nu}
\end{equation}
where $\nu$ is a critical exponent. The correlation length is thus
divergent at $T_c$, which
means that at the critical point
large scale fluctuations of the order parameter occur.
Because of that big dynamical structures 
are generated, though the interactions 
within the system are short-ranged. 

The divergence of $\xi$ at criticality 
is described by
a simple power law.  Actually it turns out that the behaviour of all
variables around the critical point is described by 
simple power laws and corresponding
critical exponents (Table \ref{ta1}).
\vskip0.5cm
\begin{table}[h]
\begin{center}
{\large
\begin{tabular}[t]{|c|cll|}\hline
${\color{red}\alpha}$ & ${\color{blue}c_H \propto \alpha^{-1}\left
( \left(|T-T_c|/T_c\right)^{-\alpha}-1\right)}$~, & $T \rightarrow T_c,$&$H=0$ \\
${\color{red}\beta}$ & ${\color{blue}m \propto \left(T_c-T\right)^\beta}$~,& $T \rightarrow T_c^{-},$&$H=0$ \\
${\color{red}\gamma}$ & ${\color{blue}{\chi} \propto |T-T_c|^{-\gamma}}$~,& $T \rightarrow T_c,$& $H=0$ \\
${\color{red}\delta}$ & ${\color{blue}m \propto H^{1/{\delta}}}$~,& $ T=T_c,$&
$H \rightarrow 0$ \\
${\color{red}\eta}$ & ${\color{blue}G_{c}^{(2)}(r) \propto r^{2-d-\eta}}$~,& $ T=T_c,$& $H=0$ \\
${\color{red}\nu}$ & ${\color{blue}{\xi} \propto |T-T_c|^{-\nu}}$~,& $T \rightarrow T_c,$& $H=0$ \\
\hline
\end{tabular}}
\vskip0.5cm
\caption[Behaviour of thermal variables at criticality]
{\label{ta1} Behaviour at criticality of the main variables 
that characterize a system which undergoes a second order phase transition. 
We indicate by $c_H$ the specific heat, by $m$ the order parameter,
by $\chi$ the susceptibility. 
The presence of another degree of freedom 
besides the temperature $T$, like 
a (small) external field (labeled by $H$), leads to 
other interesting power laws when $H\,{\rightarrow}\,0$. In the
first column we have listed the relative critical exponents. The $d$
present in the expression of $G_{c}^{(2)}(r)$ is the space dimension of 
the system. 
}  
\end{center}
\end{table}

Some laws are valid both to the right and to the 
left of the critical point;  
the values of the relative proportionality constants, or
{\it amplitudes}, are in general different for the 
two branches of the function, whereas the exponent
is the same. 
From Table \ref{ta1} we see that there are altogether six exponents.
Nevertheless they are not independent of each other, but
related by some simple scaling laws
\begin{equation}
\label{scalrelat}
\alpha+2\beta+\gamma=2,\,\,\,\,\,\,\,\,\,\alpha+\beta(\delta+1)=2,\,\,\,\,\,\,\,\,\,
(2-\eta)\nu=\gamma,\,\,\,\,\,\,\,\,\, \nu\,d=2-\alpha,
\end{equation}
so that there are only two independent exponents.
One of the most interesting aspects of second order 
phase transitions is the so-called {\it universality}, i.e.,
the fact that systems which can be very different from each other 
share
the same set of critical indices (exponents and some amplitudes' ratios). 
One can thus subdivide all systems into {\it classes},
each of them being identified by a set of critical indices. 

The divergence of the correlation length
at $T_c$ implies self-similarity at the critical point
and opens the way to analogous arguments as those we have
presented in Section \ref{paul}. 
Real space renormalization \cite{Wils}
gives an account of the scaling behaviour 
and the universality of the critical indices;
moreover, starting
from a general ansatz for the 
free energy, 
it allows to derive 
the scaling relations (\ref{scalrelat})
and the values of the exponents. 

\section{Percolation vs Second Order Thermal Phase Transitions}\label{norita}

The onset of percolation marks a 
borderline between two different {\it geometrical phases}
of the system: on one side we never have a spanning cluster,
on the other we always have one. 
When we introduced the percolation problem, we
stressed the fact that each site of the lattice is 
occupied with a probability $p$ {\it independently 
of the other sites}. There is 
{\it no communication} between different
sites, which is in contrast to 
what one has in real systems, whose
costituents normally interact with each other. 

Nevertheless, the geometrical transition 
of a percolation system has many
features in common with the thermal phase transitions
we have dealt with in the previous section.
We summarize here the most important ones:
{\color{red}
\begin{itemize}

\item{\color{black} In the neighbourhood of the critical point, both the 
      percolation and
      the thermal variables vary according to 
          power laws:

      \begin{center}
        \begin{tabular}{rclcrcl}
          $P$ & $ {\propto}$ & $(p-p_c)^{\beta}$  & 
          \makebox[5mm][]{} &
          $m$ & $ {\propto}$ & $(T_c-T)^{\beta}$  \\ 
          & &  {\makebox[30mm][r]{{\it for} $p>p_c$}} & &
          & &  {\makebox[30mm][r]{{\it for} $T<T_c$}}  \\ \\
          $S$ & $ {\propto}$ & $|p-p_c|^{-\gamma}$  &  &
          ${\chi}$ & $ {\propto}$&$ |T_c-T|^{-\gamma}$    \\
        \end{tabular}
      \end{center}}
      
\item{\color{black}Simple scaling relations are valid, some of which, like the 
      {\it hyperscaling} relation
      \begin{center}
        \begin{equation}
         \label{Hyper}
          d~=~{{\gamma}\over{\nu}}+2~{{\beta}\over{\nu}}\,\,\,\,\,\,\,\,\,\,\,\,\,\,\,\,\,\,
        \end{equation}
        \end{center}
      ( $d$= number of space dimensions), are identical for both kinds 
    of systems;}

\item{\color{black}Universality of the critical indices.}

\end{itemize}
}

Indeed, for several `physical' systems, it is possible 
to single out some cluster-like structures: the magnetic
domains of a piece of iron are a clear example. 
The interplay of such structures in correspondence
of different states of the system 
can be quite 
interesting. In the 
case of the magnetic domains 
of iron, for instance, one 
observes that they
grow by lowering the 
temperature of the sample until they fuse into 
macroscopic structures below the Curie point.
One could ask oneself whether 
the critical behaviour of a system 
could be described in terms of the 
properties of some `physical clusters'. 
The growth of correlations approaching the critical point
would be represented by the growth of the size of the
clusters.
Moreover, the spontaneous 
order appearing below the critical temperature 
$T_c$ could be related to the 
formation of an infinite cluster, which would
map the thermal transition into a geometrical percolation 
transition.
Cluster-like pictures of phase transitions have been discussed
since about 1940. Their
first applications concerned the liquid-gas transition:
the droplet model proposed by Fisher was already 
able to make quantitative predictions and 
the size distribution of its `droplets' 
is very close to the cluster size distribution 
in percolation theory \cite{fisher}.

In general, one establishes the following
correspondence between the 
thermal properties of the model and 
the geometrical features of the `physical' droplets:
\begin{itemize}
\item{they diverge at the thermal critical point;}
\item{the connectedness length diverges as the 
thermal correlation length (same exponent);}
\item{the percolation strength $P$ near the threshold
varies like the order parameter $m$
of the model (same exponent);}
\item{the average cluster size $S$ diverges as the 
physical susceptibility $\chi$ (same exponent).}
\end{itemize}

By turning a thermal system into
a percolation one, we introduce
a new feature with respect to 
the simple geometrical problem we introduced
in the previous chapter: {\it the sites of the 
lattice are no longer independent of each other},
because of the interaction.
This may lead to 
different cluster distributions compared to
the ones of random percolation\footnote{In this case one usually
speaks of {\it correlated percolation}.}. 
The distribution of the `physical' clusters depends on the 
temperature of the system
and on its dynamics. 
If the percolation transition 
takes place at a temperature $T_p\,{\neq}\,T_c$,
the thermal correlation length of the
system, $\xi_{th}$, is finite at $T_p$. That means that
two sites of the lattice separated by a distance $r\,>\,\xi_{th}(T_p)$
will be uncorrelated; the large clusters
which are responsible of the singularities of the 
percolation variables are then 
basically formed by randomly distributed (occupied) sites,
and for this reason they will carry the exponents of 
random percolation. We can easily convince ourselves
by simple space renormalization arguments. At $T_p$,
 $\xi_{th}$ is finite but the percolation correlation length,
$\xi_{p}$, is infinite. By applying successive blocking
transformations, at some stage we will
have reduced $\xi_{th}$ to very small values, i.e., 
the sites of the renormalized configuration 
will be all uncorrelated. But $\xi_{p}$ remains infinite
and, for percolation purposes, the renormalized 
system is characterized by the same exponents of the original one,
as we have explained in Section \ref{paul}. Since the sites 
of the final configuration are uncorrelated, the 
percolation exponents must be the random percolation ones.
Numerical analyses 
performed on several systems have confirmed 
that without exception. 

On the other hand, if
$T_p\,=\,T_c$, each site has a non vanishing  
correlation on any other, and the properties of all clusters
of the system, including the largest ones, will be influenced
by this correlation. 
Therefore, the
exponents describing the  
percolation variables 
need not be the ones 
of random percolation, and may be related to the 
exponents describing the singularities of thermal variables
at criticality: in particular, their
values could coincide.
 
The early attempts to explore quantitatively 
this possibility date back to the 70's,
and the first system to be investigated was the Ising model.
 
\section{The Ising Model}\label{claudio} 

The Ising model is by far the simplest of all 
spin systems. Suppose we 
have a regular lattice 
in $d$ space dimensions and 
place two-valued {\it spins} at 
each lattice site. 
The Ising model is characterized by the following 
Hamiltonian:
\begin{equation}
\label{haIs}
{\cal H}=-J\sum_{ij}s_is_j-H\sum_{i}s_i,
\end{equation}
where $J (>0)$ is the coupling 
of the interaction between  
nearest neighbouring spins $s_i$ and $s_j$
and $H$ an external field. The values of the 
spins are conventionally 
taken to be +1 (up) and -1 (down).
\begin{wrapfigure}{r}{8cm}
\begin{center}
\vspace*{-0.5cm}
\hspace*{0.2cm}\epsfig{file=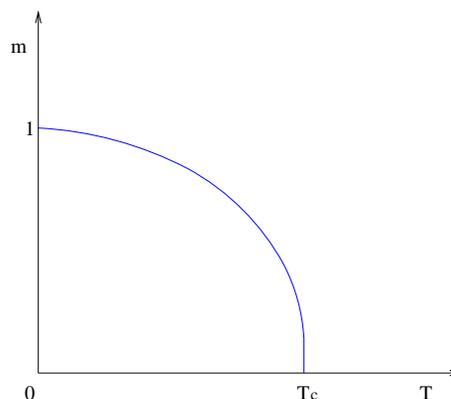,  width=6cm}
\caption[Behaviour of the specific magnetization of 
the Ising model as a function of the 
temperature $T$]
{Behaviour of the specific magnetization of 
the Ising model as a function of the 
temperature $T$. Below the critical temperature $T_c$,
$m$ is different from zero, that is the system 
`chooses' one of the 
two equivalent directions for the spins.}
\label{magnIs}
\end{center}
\end{wrapfigure}
For $H=0$ and space dimension $d\,{\geq}\,2$, 
a macroscopic system ruled by 
(\ref{haIs}) undergoes a second order 
phase transition, going from a high temperature phase without 
spin alignment to a low temperature 
phase with spin alignment. In particular, at $T=0$, all spins
point to the same direction, either up or down. 
In this way the state of the 
system at low temperatures breaks the global symmetry 
under spin inversion enjoyed by the 
Hamiltonian ({\it spontaneous symmetry breaking}).
The {\it order parameter} of the Ising model is 
the lattice average of the spin variable 
$s$, or {\it specific magnetization} $m$:
\begin{equation}
\label{magn}
m\,=\,\frac{1}{V}\sum_{i}s_i,
\end{equation}
being $V$ the lattice volume (number of sites of the lattice).
Fig. \ref{magnIs} shows the behaviour of 
$m$ as a function of the temperature $T$. 

The Ising model without external field was 
exactly solved  
in two dimensions by Onsager \cite{onsager}. 
The behaviour of the thermodynamic potentials
close to the critical temperature is thus known analytically and 
the values of the critical exponents are exactly determined. 
In three dimensions no rigorous solution has been found so far,
and all of what is known about it comes from numerical 
analyses, like high- and low-temperature expansions 
and Monte Carlo simulations. However, the simplicity of the 
system is such that most aspects can be investigated 
with remarkable precision. In Table \ref{ta2} we put the values 
of the critical exponents of the Ising model in two and three
space dimensions, because we will often refer to 
them for comparisons all along this work.
\vskip1cm
\begin{table}[h]
\begin{center}
{\large
\begin{tabular}[t]{|c|c|c|c|c|c|c|}\hline
 &$\alpha$ & $\beta$&$\gamma$&$\delta$&$\eta$&$\nu$\\
\hline
2D & 0&1/8&7/4&15&1/4&1\\
\hline
3D &0.1118(30)&0.3265(4)&1.2353(25)&4.783(16)&0.0374(12)&0.6294(10)\\
\hline
\end{tabular}}
\vskip0.5cm
\caption[Critical exponents of the Ising model in two and three dimensions]
{\label{ta2} Critical exponents of the Ising model in two and three dimensions. For the latter ones
we report the recent numerical evaluations given in \cite{parisi}.} 
\end{center}
\end{table}

\begin{wrapfigure}{l}{9cm}
\begin{center}
\vspace*{-0.5cm}
\hspace*{0.2cm}\epsfig{file=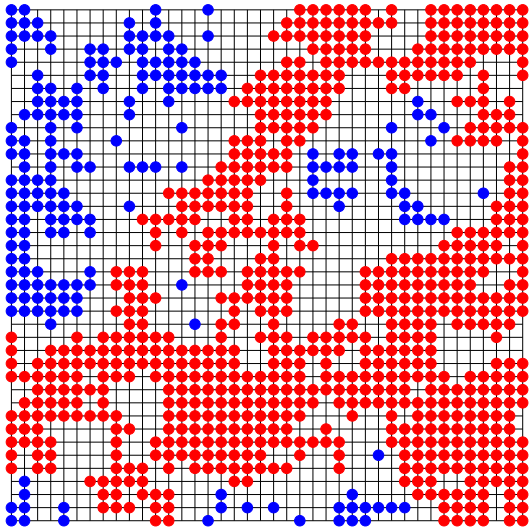,  width=7cm}
\caption[Configuration of the 2D Ising model near the critical temperature $T_c$]
{Ising model with no external field in two dimensions.
The figure shows a typical configuration near the critical temperature
$T_c$. We have marked all up spins with 
balls. It is visible the tendency of the 
spins to clusterize, because of the interaction. If we 
treat the spins up like pawns in a percolation game 
and we form clusters according to the pure site percolation 
scheme, there is a spanning network (red cluster in the figure).}
\label{Isconf}
\end{center}
\end{wrapfigure}

If we take a configuration of the Ising model around $T_c$, it will look like 
in Fig. \ref{Isconf}.
\vspace*{0.5ex}
Because of the spin alignment at low temperatures, the system 
will be finally dominated by spins of one type (up or down). 
If we think of these spins as occupied sites 
in a percolation picture, the clusters formed by 
the aligned spins will increase their size the lower the temperature is,
and at a certain value $T_p$, there will be an infinite network.
The easiest thing to think of is 
to consider as clusters all structures formed by 
nearest neighbouring spins of the same sign, which
is the pure site percolation scheme we have 
very often discussed in the previous chapter.
If we adopt this scheme, the configuration of
Fig. \ref{Isconf} presents a spanning cluster, represented
by the red structure. The first
percolation studies on the Ising model
indeed focused on these clusters. 

In the two-dimensional Ising model, topological considerations 
imply that the percolation transition of the pure site 
clusters and the thermal critical point must coincide \cite{connap}.
It thus became interesting to study the 
behaviour of the percolation variables around criticality to
determine the critical exponents. As we have seen in Chapter 1,
the percolation variables near the critical point vary
as power laws of the `reduced' density of occupied sites $p-p_c$. 
That is valid for pure random percolation. 
If we analyze the clusters formed by interacting 
systems, their features do not depend on the density alone,
and the dynamics plays a major role. For example, if we take the Ising model
at temperatures above $T_c$, 
there will be as many spins up as spins down. So, the density of 
occupied sites (considering either the up or the down spins) remains 
constant above $T_c$. But the features of 
the clusters change if we go from $T=T_c$, where the correlations
between spins are long-ranged, to $T\rightarrow\infty$, where 
spins are uncorrelated.
It turns out that, for `dynamical' clusters, 
the percolation variables vary as simple power laws
of the reduced temperature $T-T_c$, like the thermal ones. 
In the case of the two dimensional Ising model, the average cluster size $S$ 
of the pure site clusters was shown to diverge as
\begin{equation}
\label{syk}
S\,\propto\,|T-T_c|^{-\gamma_p},
\end{equation}
where $\gamma_p=1.91$ \cite{sykes}.
The result, derived by means
of series expansions, is not 
in agreement with the thermal value 
for the susceptibility exponent found by Onsager ($\gamma=7/4$).
Besides, in three dimensions, the spins which
are favoured by the onset of magnetization 
form an infinite network at any temperature, whereas 
the unfavoured spins happen to 
percolate at temperatures higher than 
$T_p$, with $T_p\,{\approx}\,0.96\,T_c$ \cite{krumb}.

So, if it is at all possible to describe the thermal 
phase transition of the Ising model as a percolation 
transition, one must look for a different cluster definition 
than the pure site one.

\section{The Random Cluster Model}\label{mancio} 

At the beginning of the 70's,
contemporary to the research activities mentioned in the previous section,
Fortuin and Kasteleyn \cite{fort} introduced a correlated 
bond-percolation model (the {\it Random Cluster Model}) indexed by
a parameter $q$, and proved identities relating the partition function
and connectedness probabilities in this model to the partition 
function and correlation functions of the 
$q$-state Potts model ($q=2,3,...$). The $q$-state Potts model
has the following Hamiltonian:
\begin{equation}
\label{potts}
{\cal H}\,=\,J_P\sum_{ij}(1-\delta_{\sigma_i\sigma_j}),
\end{equation}
where $J_P(>0)$ is the coupling and the $\sigma$'s represent
the spin variable of the model, which can take $q$ different values.
For $q=2$ it is easy to see that the Hamiltonian (\ref{potts})
is equivalent to the one of an Ising model, whose coupling $J_I={J_P}/2$.
In this section, however, we will keep the Potts notation
because it 
simplifies the mathematical expressions.

To give a feeling of the work of 
Fortuin and Kasteleyn, we will 
show that
their Random Cluster Model
and the $q$-state Potts model 
are equivalent to each other.
In particular, we will see that the partition function of the 
$q$-state Potts model can be rewritten in purely geometrical 
terms, as sum over cluster configurations. 
The clusters are built in the following way: taking 
two nearest neighbouring spins $\sigma_i$ and $\sigma_j$, if 
$\sigma_i\neq\sigma_j$
they are {\it always}
disjoint;
if $\sigma_i=\sigma_j$, they are joined together with
a temperature dependent probability $p_{ij}=1-\exp(-J_P/kT)$.
So, the Fortuin-Kasteleyn clusters are {\it site-bond} clusters:
once we have a spin configuration, we need to
distribute bonds with
the probability $p_{ij}$ among nearest neighbouring spins of the same value
to build the clusters. A cluster configuration will be therefore
completely determined by a spin configuration $\{\sigma\}$ and 
a bond configuration $\{n\}$ superimposed to the former.
For the bond variables $n_{ij}$, we assign $n_{ij}=0$ (open bond) and 
$n_{ij}=1$ (closed bond).

We won't follow the original 
Fortuin-Kasteleyn derivation, because it is too 
technical, but a simplified version
proposed by Sokal and Edwards \cite{sok}.
Given a lattice with Potts spins $\sigma_i=1,..., q$ on the sites
and bond variables $n_{ij}$ on the edges (links), we define 
the joint probability of a certain cluster configuration (spins + bonds) as
\begin{equation}
\label{p1}
P(\sigma,n)={\cal Z}^{-1}\prod_{<ij>}[(1-p_{ij})\delta_{n_{ij},0}+
p_{ij}\delta_{\sigma_{i}\sigma_{j}}\delta_{n_{ij},1}],
\end{equation}
with
\begin{equation}
\label{p2}
{\cal Z}=\sum_{\sigma}\sum_{n}\prod_{<ij>}[(1-p_{ij})\delta_{n_{ij},0}+
p_{ij}\delta_{\sigma_{i}\sigma_{j}}\delta_{n_{ij},1}].
\end{equation}
This is the so-called FKSW
model (Fortuin-Kasteleyn-Swendsen-Wang), 
which 
has, a priori, nothing to do with the dynamics of a spin model.
If we sum over all bond configurations we get
\begin{eqnarray}
\label{p3}
P(\sigma)&=&\sum_{n}P(\sigma,n)\nonumber\\
&=&{\cal Z}^{-1}\prod_{<i,j>}\,\sum_{n_{ij}=0}^{1}[(1-p_{ij})\delta_{n_{ij},0}+
p_{ij}\delta_{\sigma_{i}\sigma_{j}}\delta_{n_{ij},1}]\nonumber\\
&=&{\cal Z}^{-1}\prod_{<i,j>}[(1-p_{ij})+p_{ij}\delta_{\sigma_{i}\sigma_{j}}]\nonumber\\
&=&{\cal Z}^{-1}\exp\Big[-\frac{J_P}{kT}\sum_{<ij>}(1-\delta_{\sigma_i\sigma_j})\Big]\nonumber\\
&=&{\cal Z}^{-1}\exp\Big[-\frac{\cal H(\sigma)}{kT}\Big],
\end{eqnarray}
where ${\cal H(\sigma)}$ is the Hamiltonian (\ref{potts}).
Now we have got rid of the bonds. $P(\sigma)$ is 
the probability associated to the spin configuration $\{\sigma\}$ 
in the FKSW model. 
If we sum over all spin configurations, we obviously obtain
\begin{equation}
\label{p4bis}
\sum\limits_{\sigma}\,P(\sigma)\,=\,1.
\end{equation}
From Eqs. (\ref{p3}) and (\ref{p4bis}) one gets
\begin{equation}
\label{p4}
{\cal Z}\,=\,\sum_{\sigma}\exp\Big[-\frac{\cal H(\sigma)}{kT}\Big].
\end{equation}
We have then found that the partition function of the FKSW model
coincides with the one of the Potts model (see Eq. (\ref{potts})).
The expression of the probability $P(\sigma)$ we have derived 
in Eq. (\ref{p3}) is just the Boltzmann probability 
to have the spin configuration $\{\sigma\}$ 
in a system ruled by the Potts dynamics.
We conclude that, after integrating out the bond configurations,
the FKSW model
is equivalent to the Potts model.
 
Next, we want 
to see what 
happens if we reduce the FKSW model
to a bond model, 
by eliminating the spin degrees of freedom.
For that, we start again from 
Eq. (\ref{p1}) and sum over all spin configurations. We obtain
\begin{eqnarray}
\label{p5}
P(n)&=&\sum_{\sigma}P(\sigma,n)\nonumber\\
&=&{\cal Z}^{-1}\sum_{\sigma}\Big[\prod_{<ij>,n_{ij}=1}p_{ij}\delta_{\sigma_i\sigma_j}\prod_{<ij>,n_{ij}=0}
(1-p_{ij})\Big].
\end{eqnarray}
In the last expression all the terms in the sum 
with a closed bond between two spins in distinct states will vanish
(they are not allowed by definition), so if we denote by $\sigma^{n}$ a spin
configuration compatible with the restriction for two spins to be parallel if 
connected by a closed bond, we get
\begin{equation}
\label{p6}
P(n)={\cal Z}^{-1}\sum_{\sigma^{n}}\Big[\prod_{<ij>,n_{ij}=1}p_{ij}\prod_{<ij>,n_{ij}=0}(1-p_{ij})\Big].
\end{equation}
The terms in the sum are now independent of the spin configuration. Given
the bond configuration, the sum just counts the number of compatible
spin configurations. Defining as a cluster each set of bond-connected spins, we get
\begin{equation}
\label{p7}
P(n)={\cal Z}^{-1}\prod_{<ij>,n_{ij}=1}p_{ij}\prod_{<ij>,n_{ij}=0}(1-p_{ij})q^{c(n)},
\end{equation}
where $c(n)$ is the number of clusters of the given 
bond configuration $n$. Again, we have the normalization
\begin{equation}
\label{p7bis}
\sum\limits_{n}\,P(n)\,=\,1,
\end{equation}
so that
\begin{equation}
\label{p8}
{\cal Z}=\sum_{n}\Big[\prod_{<ij>,n_{ij}=1}p_{ij}\prod_{<ij>,n_{ij}=0}(1-p_{ij})q^{c(n)}\Big].
\end{equation}
The (\ref{p8}) is just the partition function of the Random Cluster 
model introduced by Fortuin and Kasteleyn, which is then equivalent to
the FKSW model when the spins are integrated out.

Summarizing the results we have derived 
so far, we can say that the 
Potts and the Fortuin-Kasteleyn 
models are nothing but the FKSW model when one
eliminates the bonds or the spins, respectively.
Consequently, the Potts model is equivalent to the one
of Fortuin and Kasteleyn.

The site-bond clusters we have used look like 
artificial structures, because 
the bond probability breaks existing geometrical
connections between the spins. Nevertheless, on the grounds of the result
we have just presented, it seems that such "artificial structures"
have a close relationship to the dynamics of the Potts model.
A confirmation of such relationship is 
represented by the fact that the Fortuin-Kasteleyn clusters
can be used 
to implement a non-local Monte Carlo update 
of the Potts model.
This algorithm was proposed by
Swendsen and Wang \cite{swend} 
and it reduces considerably
the problem of {\it critical slowing down},
which makes the simulations around the critical
point very lengthy with traditional local methods.
We conclude the section describing this algorithm.

As we have already said, in order to identify the clusters,
we have a superposition of a spin configuration $\{\sigma\}$ 
and a 
bond configuration $\{n\}$. But if we take a spin configuration
$\{\sigma\}$, it will not be compatible with each 
bond configuration $\{n\}$. So, the probability
to have $\{\sigma\}$ {\it and} $\{n\}$
is not simply given 
by the product of 
the probability of having $\{\sigma\}$
by the probability of having $\{n\}$ independently, but it is 
a more involved expression which requires
the introduction of the concept of 
{\it joint probability}.
If we have two events $A$ and $B$,
one defines joint probability $P(A|B)$ as 
the probability of the event $A$ {\it given}
the event $B$.
According to this definition one gets, trivially
\begin{equation}
\label{p9}
P(\sigma, n) = P(\sigma | n)\,P(n) = P(n | \sigma)\,P(\sigma) 
\end{equation}
Now we can calculate the conditional probabilities 
to get from a bond configuration to a spin configuration
and vice versa:
\begin{eqnarray}
\label{p10}
P(n | \sigma)&=&\frac{P(\sigma, n)}{P(\sigma)}\nonumber\\
&=&\frac{\prod\limits_{<i,j>}[(1-p_{ij})\delta_{n_{ij},0}+
p_{ij}\delta_{\sigma_{i}\sigma_{j}}\delta_{n_{ij},1}]}
{\exp[-H(\sigma)/kT]}\nonumber\\
&=&\frac{\prod\limits_{<i,j>,\sigma_i=\sigma_j}[(1-p_{ij})\delta_{n_{ij},0}+
p_{ij}\delta_{n_{ij},1}]\prod\limits_{<i,j>,\sigma_i\neq\sigma_j}[(1-p_{ij})\delta_{n_{ij},0}]}
{\exp[-\sum_{ij}\frac{J_P}{kT}(1-\delta_{{\sigma_i}{\sigma_j}})]}\nonumber\\
&=&\frac{\prod\limits_{<i,j>,\sigma_i=\sigma_j}[(1-p_{ij})\delta_{n_{ij},0}+
p_{ij}\delta_{n_{ij},1}]\prod\limits_{<i,j>,\sigma_i\neq\sigma_j}\delta_{n_{ij},0}
\prod\limits_{<i,j>,\sigma_i\neq\sigma_j}\exp[-\frac
{J_P}{kT}]}{\prod\limits_{<i,j>,\sigma_i\neq\sigma_j}\exp[-\frac
{J_P}{kT}]}
\nonumber\\
&=&\prod_{<i,j>,\sigma_i=\sigma_j}[(1-p_{ij})\delta_{n_{ij},0}+
p_{ij}\delta_{n_{ij},1}]\prod_{<i,j>,\sigma_i\neq\sigma_j}\delta_{n_{ij},0}
\end{eqnarray}
is the probability to obtain the bond configuration 
$\{n\}$ given the spin configuration $\{\sigma\}$. In the case
$\sigma_i\neq\sigma_j$ only open bonds are allowed; in the case
$\sigma_i=\sigma_j$ a closed bond is put with a probability $p_{ij}$
and an open bond with probability $1-p_{ij}$.
\begin{eqnarray}
\label{p11}
P(\sigma | n)&=&\frac{P(\sigma, n)}{P(n)}\nonumber\\
&=&\frac{\prod\limits_{<i,j>}[(1-p_{ij})\delta_{n_{ij},0}+
p_{ij}\delta_{\sigma_{i}\sigma_{j}}\delta_{n_{ij},1}]}
{\prod\limits_{<ij>,n_{ij}=1}p_{ij}\prod\limits_{<ij>,n_{ij}=0}(1-p_{ij})q^{c(n)}}\nonumber\\
&=&q^{-c(n)}\frac{\prod\limits_{<i,j>,n_{ij}=1}p_{ij}\delta_{\sigma_{i}\sigma_{j}}
\prod\limits_{<i,j>,n_{ij}=0}(1-p_{ij})}
{\prod\limits_{<ij>,n_{ij}=1}p_{ij}\prod\limits_{<ij>,n_{ij}=0}(1-p_{ij})}\nonumber\\
&=&q^{-c(n)}\prod_{<ij>,n_{ij}=1}\delta_{\sigma_{i}\sigma_{j}}
\end{eqnarray}
is the probability to obtain the spin configuration $\{\sigma\}$
given the bond configuration $\{n\}$. In order to have compatibility,
the spin configuration $\{\sigma\}$ must be one of the configurations
which can be obtained by flipping the spins of 
the $c(n)$ clusters formed by the bond $\{n\}$, under the condition
that the flipped spins 
within a 
cluster
take the same value $q$. 
\begin{figure}[h]
\begin{center}
\epsfig{file=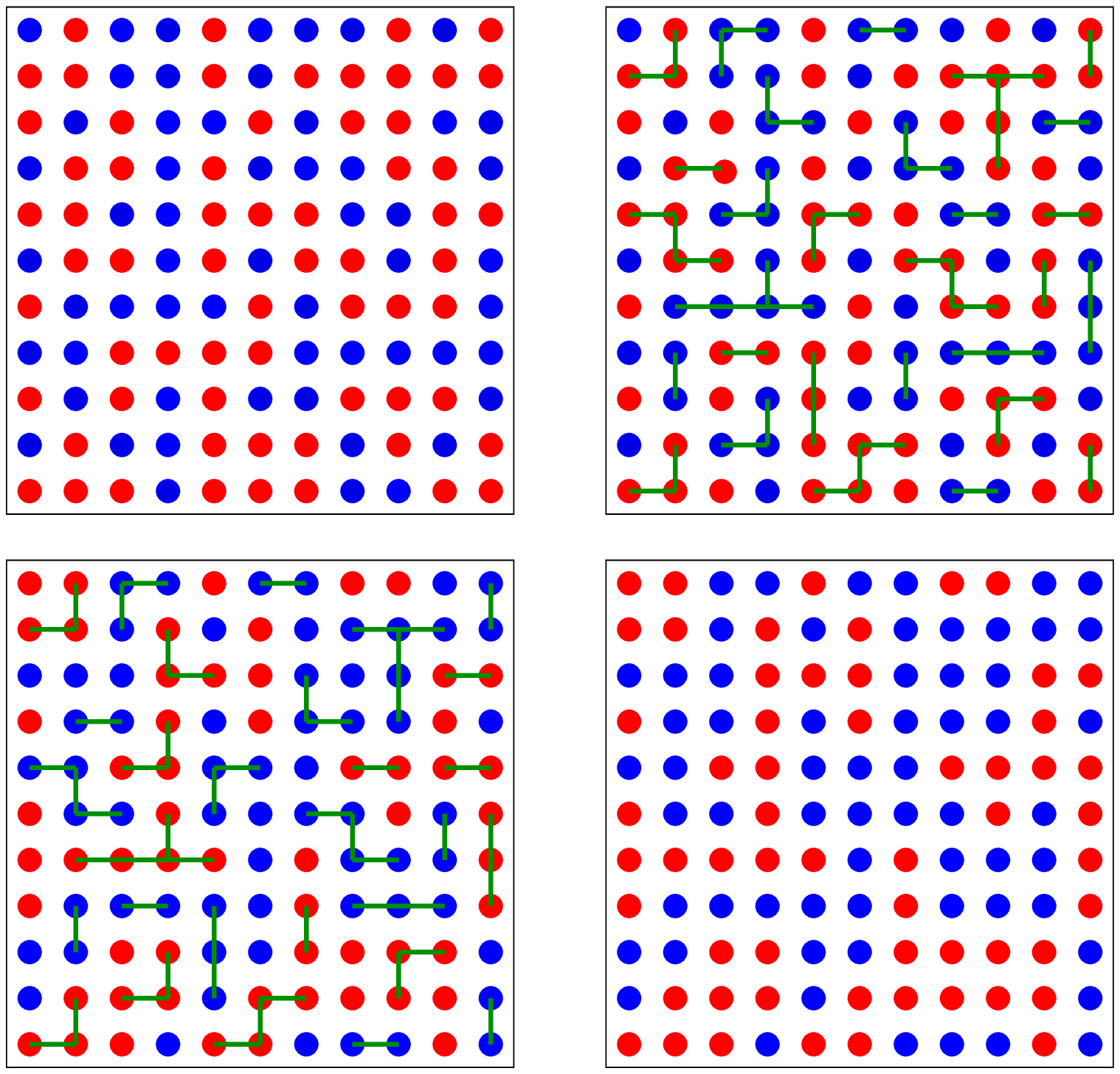,width=10cm}
\caption[Scheme of the Swendsen-Wang cluster update for the 2D Ising model]
{Scheme of the Swendsen-Wang cluster update for the 2D Ising model. 
The two possible spin values are 
labeled by the two colors, blue and red,
of the sites. One starts
from some spin configuration (top left); bonds 
between nearest neighbouring spins of the same color are 
distributed with probability $p_{ij}$ (green links
in the top right diagram); the color of all sites 
which are bond-connected to each other 
(including isolated sites) is set 
to blue or red with equal probability, provided
the color remains uniform within each cluster (lower left); 
taking the bonds away one obtains a new spin configuration (lower right).}
\label{swenwa}
\end{center}
\end{figure}

The Swendsen-Wang cluster update is based exactly on this 
procedure (see Fig. \ref{swenwa}). We can divide it in two steps:
\begin{itemize}
\item{Take a spin configuration and distribute bonds 
between nearest neighbouring spins of the same value
with the probability $p_{ij}=1-\exp(-J_P/kT)$;} 
\item{Set the values of all spins belonging to each
cluster of bond-connected sites 
to one of the possible $q$ values with equal probability.} 
\end{itemize}
It is clear that the algorithm respects the accessibility 
criterium, i.e. any 
spin configuration can be produced provided 
we update the system a sufficiently
large number of times. The probability 
$P_{\sigma,\sigma^{\prime}}$ of getting from the spin configuration
$\{\sigma\}$ to $\{\sigma^{\prime}\}$ is given by:
\begin{equation}
\label{p12}
P_{\sigma,\sigma^{\prime}} = \sum_{n}P(\sigma | n)\, P(n | \sigma^{\prime})
\end{equation}
It is easy to show that 
 $P_{\sigma,\sigma^{\prime}}$ satisfies 
the detailed balance condition, so that 
the algorithm indeed produces a Markov chain which
leads the system to the canonical equilibrium
distribution of the Potts model.

\section{Percolation of Fortuin-Kasteleyn clusters}\label{mancia}

As we have seen, the pure site-clusters
of the Ising model
do not allow to map
the thermal transition 
into a geometrical percolation transition.
The known properties
of the Ising site-clusters suggest
that they
are too big to describe 
the critical behaviour of the Ising model.
The reason is that there are two
contributions
to the Ising clusters: one is due 
to the correlations, and the other is due to purely geometric 
effects. The latter becomes evident 
in the limit of infinite temperature. In this case there are no correlations
but the cluster size is different from zero. In fact, since the density
of occupied sites is $1/2$, they tend to form clusters
just because they happen to be close to each other; in the 3-dimensional
Ising model there is even a spanning pure-site network 
at $T\rightarrow\infty$, because
the critical density of 3D random percolation
is $0.3116$, well below $1/2$. 

It is thus necessary to reduce the size of the 
clusters in some way. We notice that 
the Fortuin-Kasteleyn clusters are indeed smaller
than the pure-site ones. In particular,
the bond probability $p_{ij}=1-\exp(-J_P/kT)$
varies strongly with the temperature $T$,
going from the value $1$ at $T=0$ to 
the value $0$ at $T\rightarrow\infty$, which expresses
the absence of correlation between the sites
that are, therefore, all disjoint. 
Moreover, from the previous section,
it turns out that these clusters have a close relationship with
the dynamics of the $q$-state Potts model (Ising model for $q=2$).
For all that they might be good candidates 
for the droplets we are looking for.

\begin{figure}[h]
\begin{center}
\epsfig{file=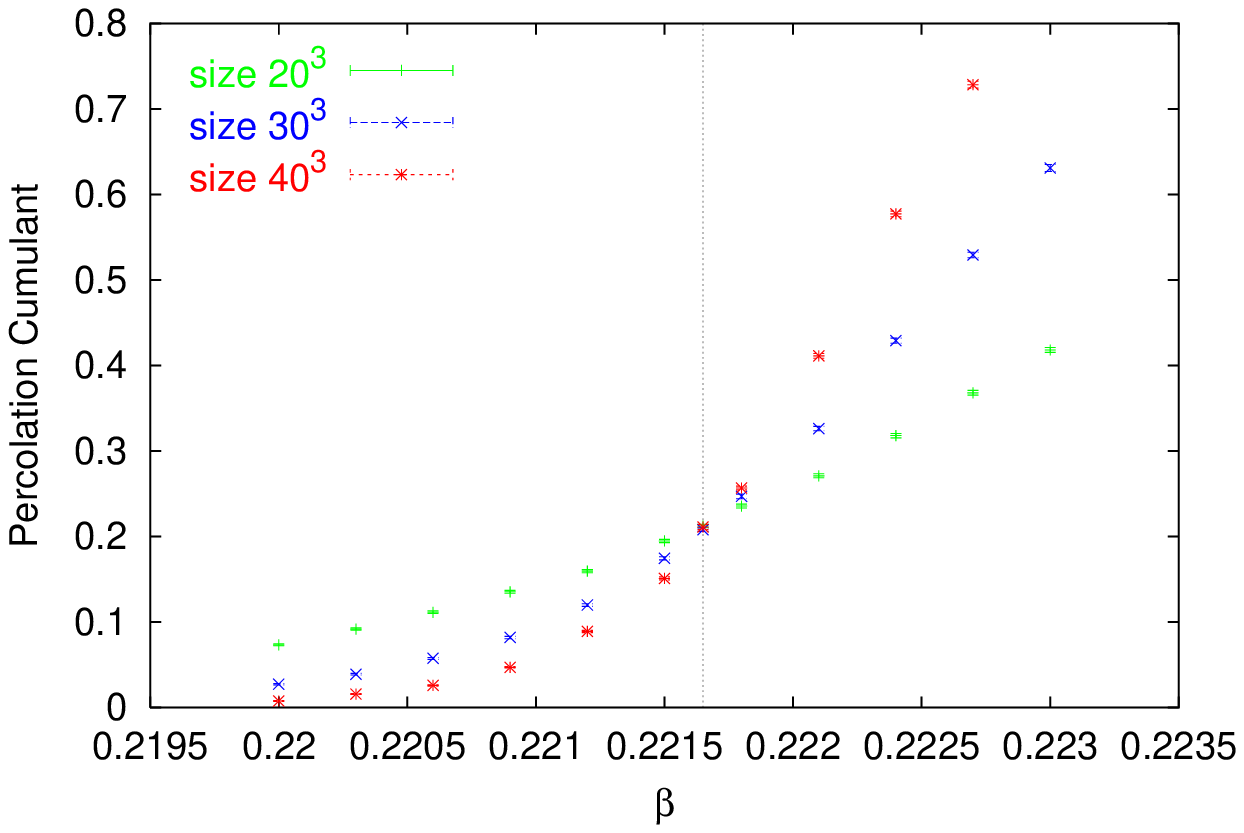,  width=7.9cm}
\epsfig{file=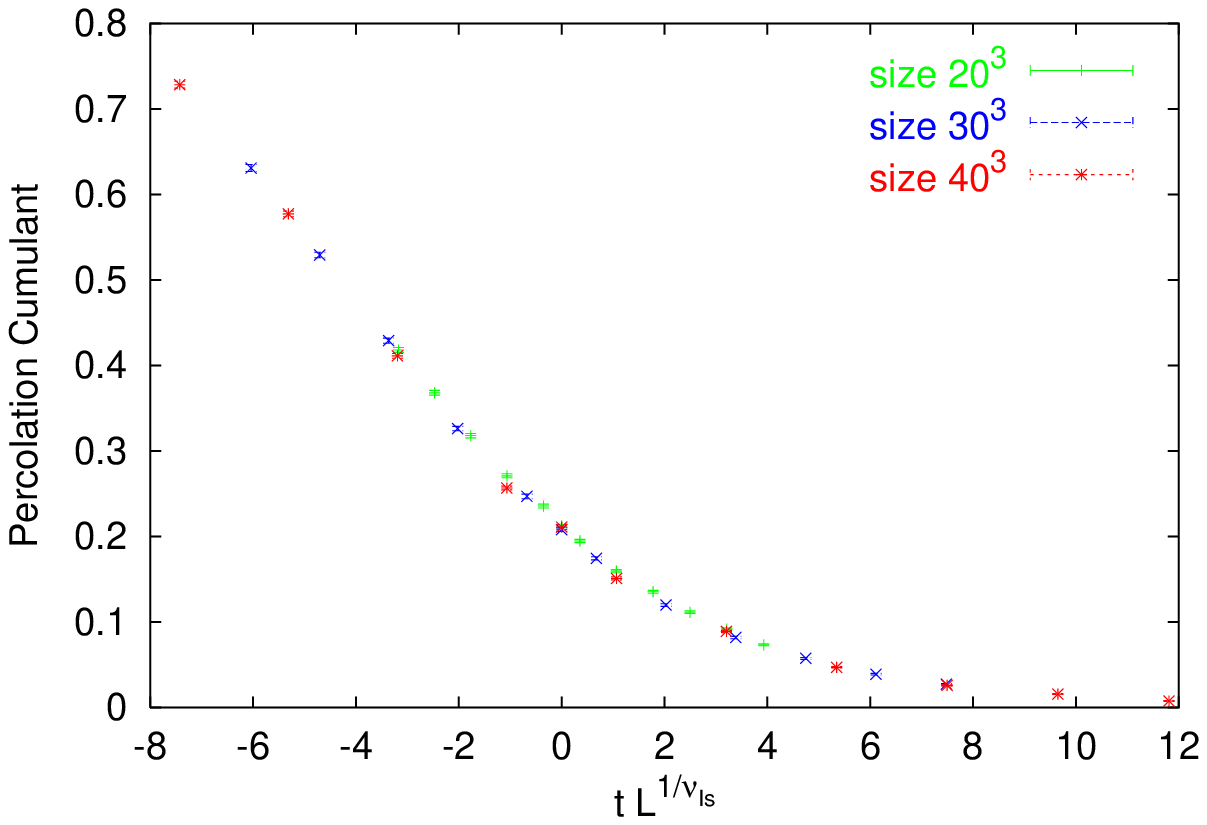,  width=7.9cm}
\vskip0.2cm
\begin{picture}(0,-30)
    \put(-112,2){\begin{minipage}[t]{7.4cm}{(a)}
    \end{minipage}}
    \put(117.7,2){\begin{minipage}[t]{7.4cm}{(b)}
    \end{minipage}}
\end{picture}
\caption[Percolation cumulant as a function of 
$\beta=J/kT$ for Fortuin-Kasteleyn clusters 
of the 3D Ising model]
{(a) Percolation cumulant as a function of 
$\beta=J/kT$ for Fortuin-Kasteleyn clusters 
of the 3-dimensional Ising model. The curves, corresponding
to three different lattice sizes, cross remarkably well
at the thermal critical point, represented by the dashed
line in the plot. (b) Rescaled percolation cumulants
taking as a variable on the X-axis the expression
$tL^{1/{\nu}}$ ($t=(T-T_c)/T_c$, $L$ is the lattice side), where $\nu$ is set to
the 3D Ising model value $\nu_{Is}=0.6294$ (see Table \ref{ta2}). The curves fall on top
of each other, so that $\nu_{perc}=\nu_{Is}$.}
\label{conifig}
\end{center}
\end{figure}

A. Coniglio and W. Klein \cite{conkl}
showed that the Fortuin-Kasteleyn clusters
really have the required properties
of the physical droplets, i. e. they 
percolate at the thermal critical point 
and the geometrical critical exponents  
coincide with the thermal ones. This result,
obtained independently of the Fortuin-Kasteleyn work,
is analytical and is valid for 
any space dimension $d\geq{2}$ and 
any lattice geometry, as long as it is 
homogeneous (Fig. \ref{conifig}).
In the Ising notation, the bond weight of 
Fortuin and Kasteleyn is 
\begin{equation}
\label{bon2k}
p_{ij}=1-\exp(-\frac{2J}{kT})
\end{equation}
(see Section \ref{mancio}).
\vskip0.5cm
For the 2-dimensional Ising model, the 
result of Coniglio and Klein leads to an apparent
paradox. In fact we have seen that 
the pure site clusters
percolate at the thermal threshold.
On the other hand, the Fortuin-Kasteleyn clusters,
which are smaller than the 
pure site ones, form as well an infinite network
at the Ising critical point. From this 
fact, which is indeed unexpected but legitimate,
it follows that site-bond clusters built using 
a bond probability $p$ such that 
$p_{ij}<p<1$ will also give rise to a spanning
cluster at the thermal critical point, as their
size is intermediate between the size of 
the pure site clusters and the one of the Fortuin-Kasteleyn 
clusters.
However, the geometrical critical exponents 
relative to the
percolation transition of these intermediate clusters
are different from the thermal ones, with which 
they coincide {\it only} if $p=p_{ij}$.
This fact shows the key role played by the bond weight
$p_{ij}$.

\section{The Kert\'esz Line}\label{manu}

So far we have been dealing with 
the Ising model in absence of an external
magnetic field. The reason of that is clear: the Ising
model shows critical behaviour
in the usual sense only if 
$H=0$. That means that introducing 
an external field $H$, none of the thermodynamic 
potentials will exhibit discontinuities 
of any kind, because the partition function
is analytical. This result, already
proved by 
Yang and Lee \cite{yalee}, inserts itself
in a quite old debate concerning 
phase transitions. It has been known for a long time
that phases separated by a line of
first order phase transitions can be connected
without thermodynamic singularities 
when using paths around the critical endpoint.
This was discovered experimentally by Andrews (1869)
and explained by the Van der Waals theory of 
liquid and gaseous states (1873). 
Because of that, it was suggested
that, along the `continuous' paths, something
interesting may happen, in spite of the 
absence of standard thermodynamic singularities.
This is indeed true, and is strictly related 
to the droplet description of phase transitions that
we have discussed in this chapter.

\begin{figure}[h]
\begin{center}
\epsfig{file=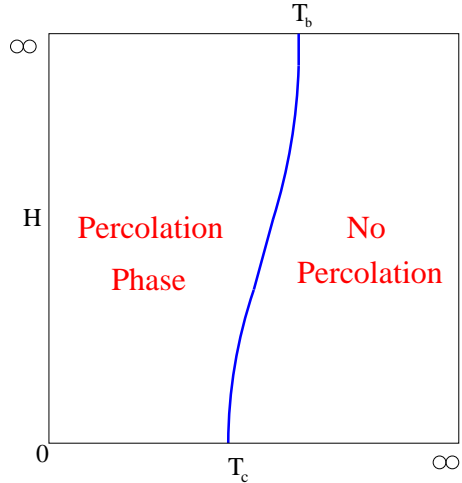,  width=6cm}
\caption[Scheme of the Kert\'esz line]
{Kert\'esz line. For $H=0$,
$T_p$ is equal to the Ising critical point;
for $H\rightarrow\infty$, $T_p$ tends
to the endpoint $T_b$, solution
of Eq. (\ref{limva}).}
\label{kert}
\end{center}
\end{figure}

The Fortuin-Kasteleyn clusters 
are perfectly defined also in the presence of a magnetic
field. 
Because of the field, the system has
a non vanishing magnetization $m$ parallel to the direction of the field
for any
value of the temperature $T$.
However, for $T\rightarrow\infty$, $m{\rightarrow}\,0$. 
For $T=0$, $m=1$ again. 
This suggests that also in this case,
for a fixed value of the field $H$, the clusters
will form an infinite network at some temperature 
$T_p(H)$. Varying
the intensity of the field
one gets a curve $T_p(H)$, which is called {\it Kert\'esz line} \cite{kertesz}.
We have plotted it schematically in Fig. \ref{kert}.
\vskip0.1cm
When the field $H=0$, we obviously get 
the thermal threshold of the Ising model.
When $H\rightarrow\infty$, at any temperature 
$T$ the lattice spins  
will be all aligned with the field. 
The bonds will be then distributed among all
pairs of nearest neighbouring spins, and 
the site-bond problem turns in a pure 
bond percolation problem. The geometric transition will then take place 
for that value of the temperature $T_b$ for which
the probability $p_{ij}$ equals the critical 
density $p_B(d)$ of random bond percolation
in $d$ dimensions:
\begin{eqnarray}
\label{limva}
p_{ij}&=&p_B(d)\nonumber\\
1-\exp(-\frac{2J}{kT_b})&=&p_B(d)\nonumber\\
\log[1-p_B(d)]&=&-\frac{2J}{kT_b}\nonumber\\
T_b&=&-\frac{2J}{k\,\log[1-p_B(d)]}
\end{eqnarray}
So, we have a whole curve whose 
points are percolation
points, with the 
usual singular behaviour of 
cluster-related quantities, though 
the corresponding thermal variables are continuous.

One can ask oneself how `physical' the  
Kert\'esz line is. We have already seen
that some definitions of 
clusters may lead to behaviours which have nothing
to do with the critical behaviour of the system: one example 
is represented by the pure site-clusters of the Ising model.
In the same way, we could 
conclude that the Fortuin-Kasteleyn clusters are
not the `physical droplets' of the system
if we switch on a magnetic field, and that we have
to look for an appropriate definition.
Swendsen and Wang \cite{swendwa} proposed to introduce
a {\it ghost spin} oriented parallel 
to the magnetic field. This ghost spin is 
connected to each spin (oriented like the field)
with a probability $p_H=1-\exp(-2H/kT)$,
formally similar to the Fortuin-Kasteleyn bond weight.
Since, for $H\,{\neq}\,0$, such probability is 
non-zero, no matter how small, spins arbitrarily far from
each other will be connected together through
the ghost spin, giving rise 
to a loose infinite network. Therefore, as long as 
$H\,{\neq}\,0$, at any temperature there will be percolation
in this general sense, with the sites 
being not directly but indirectly connected.
That seems to provide the desired mapping 
to the thermal counterpart, 
in which there is always a non-zero magnetization
and no divergences. An indirect confirmation of that
is given by the fact that, 
by means of this general definition of clusters,
it is possible to implement a cluster update 
which leads to the canonical equilibrium distribution
of the Ising (Potts) model with an external field. 

The success of the Swendsen-Wang definition of clusters
does not imply that we can simply forget the Kert\'esz line
or treat it like an artificial construction.
In fact, it turns out to have some remarkable properties.
\begin{figure}[htb]
\begin{center}
\epsfig{file=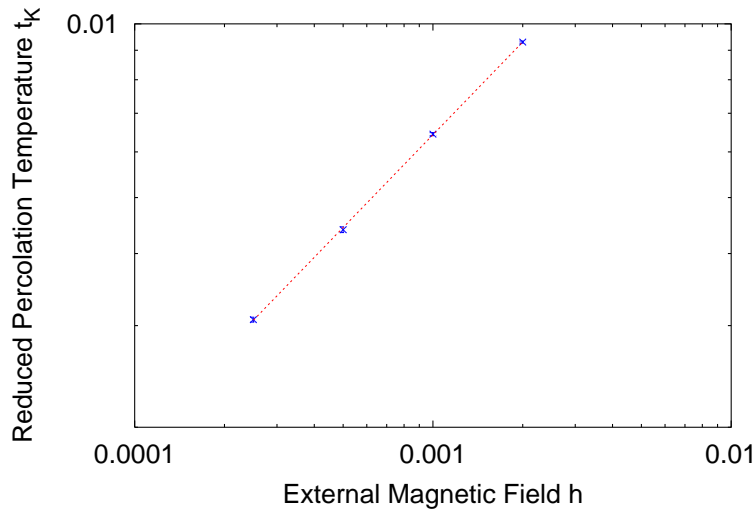,width=10cm}
\caption[Kert\'esz line of the 2D Ising model for small values of the 
external field $h$]
{\label{miaker}{Kert\'esz line for the 2D Ising model for small values of the 
external field $h$.}}
\end{center}
\end{figure}
Fig. \ref{miaker} shows some preliminary results
of an investigation we are carrying on. The system is 
the 2D Ising model, and we calculated few points of the Kert\'esz line
for very small values of the external field $h=H/J$ (see Eq. (\ref{haIs})).
The variable on the $y$-axis is the reduced percolation temperature
$t_K\,=\,(T_K-T_c)/T_c$, 
where $T_c$ is as usual the critical temperature
of the Ising model without field.
In the logarithmic scale of the plot the data points fall remarkably well 
on a straight line. We thus conclude that, for the Kert\'esz line,
\begin{equation}
\label{miak}
t_K\,\propto\,h^{\kappa},\,\,\,\,\,\,\,\,for\,\,h\,{\rightarrow}\,0.
\end{equation}
From the slope of the straight line in the plot we obtain
$\kappa\,=\,0.534(3)$. This result might have an interesting connection
with some thermal properties of the Ising model with external field. 
In fact, even if the susceptibility $\chi$ is not divergent
at any $T$ when $H\,\neq\,0$, it has anyhow a well defined peak.
From the renormalization group ansatz for the free energy of thermal systems
undergoing second order phase transitions,  
it is possible
to determine in general how the position of the 
susceptibility peak $t_{\chi}$ is related to the 
magnetic field $h$, when $h\,{\rightarrow}\,0$. It turns out that
\begin{equation}
\label{betadelta}
t_{\chi}\,\propto\,h^{1/({\beta\delta})},\,\,\,\,\,\,\,\,for
\,\,h\,{\rightarrow}\,0,
\end{equation}
where $\beta$ and $\delta$ are the 
critical exponents we have introduced in Section \ref{marzia}.
For the 2D Ising model $1/({\beta\delta})\,=\,0.533$, which coincides
with our estimate of the Kert\'esz exponent $\kappa$.
That could mean that there is a relationship 
between the two curves. In particular,
it would be interesting to check whether they overlap, at least
for small values of $h$\footnote{For $h\rightarrow\infty$, the position of the 
susceptibility peak 
$t_{\chi}\rightarrow\infty$, whereas we have seen that 
the Kert\'esz line has an endpoint, given by Eq. (\ref{limva}).
So, the two curves will certainly differ for sufficiently 
high values of $h$. Nevertheless, one could introduce a 
dependence on the field $h$ into the Coniglio-Klein factor. 
Simple expressions like $1-\exp[-2\beta(1+h)]$, for example,
would still lead to the same power law behaviour of Eq. (\ref{miak}) with
the same exponent $\kappa$ we have found, because for 
the small $h$-values of the points we have 
considered, the difference 
from the Coniglio-Klein factor is negligible.  
On the other hand, the Kert\'esz line obtained by using 
the new factor tends to infinity for $h\,\rightarrow\infty$, which might
allow a global comparison with the thermal curve of the 
susceptibility peaks.}.
Work in this direction is still in progress.
 
If we take the cluster number distribution $n_s$
of the general Swendsen-Wang droplets, it behaves 
differently on the two 
sides of the Kert\'esz line \cite{kertesz}.
On the low-temperature side
\begin{equation}
\label{drop}
\log\,n_s\,{\propto}\,-\frac{2H}{kT}s-{\Gamma}s^{2/3},
\end{equation}
where $\Gamma$ is a surface tension term; instead, on the 
high-temperature side, there is no surface tension and one has
\begin{equation}
\label{drop2}
\log\,n_s\,{\propto}\,-\frac{2H}{kT}s-const{\cdot}s,
\end{equation}
Similar percolation-type singularities 
appear when one studies
the behaviour of the Fortuin-Kasteleyn droplets
around the Kert\'esz line.
According 
to some numerical investigations \cite{adler},
there seems to be evidence that Taylor
expansions of the free energy 
as a function of $H$ or $T$ have a different 
convergence behaviour (i.e. radius) on the two sides
of this line. That might
be related to the geometrical singularities 
we have just mentioned and could 
represent an argument for a 
generalization of the definition of 
phase change, not exclusively based 
on standard singularities of the 
thermodynamic potentials.

\chapter{Percolation and Magnetization in Continuous Spin Models}\label{Percolation and Magnetization
in Continuous Spin Models}
\setcounter{footnote}{1}
\thispagestyle{empty}


\def\J{$J/\psi$}
\def\j{J/\psi}
\def\X{$\chi$}
\def\x{\chi}
\def\P{$\psi'$}
\def\p{\psi'}
\def\U{$\Upsilon$}
\def\u{\Upsilon}
\def\C{c{\bar c}}
\def\cg{c{\bar c}\!-\!g}
\def\bg{b{\bar b}\!-\!g}
\def\b{b{\bar b}}
\def\q{q{\bar q}}
\def\Q{Q{\bar Q}}
\def\L{\Lambda_{\rm QCD}}
\def\A{$A_{\rm cl}$}
\def\a{A_{\rm cl}}
\def\N{$n_{\rm cl}$}
\def\n{n_{\rm cl}}
\def\S{S_{\rm cl}}
\def\s{s_{\rm cl}}
\def\bb{\bar \beta}

\def\be{\begin{equation}}
\def\ee{\end{equation}}

\def\lsim{\raise0.3ex\hbox{$<$\kern-0.75em\raise-1.1ex\hbox{$\sim$}}}
\def\gsim{\raise0.3ex\hbox{$>$\kern-0.75em\raise-1.1ex\hbox{$\sim$}}}


\def\CMP{{ Comm.\ Math.\ Phys.\ }}
\def\NP{{ Nucl.\ Phys.\ }}
\def\PL{{ Phys.\ Lett.\ }}
\def\PR{{ Phys.\ Rev.\ }}
\def\PRep{{ Phys.\ Rep.\ }}
\def\PRL{{ Phys.\ Rev.\ Lett.\ }}
\def\RMP{{ Rev.\ Mod.\ Phys.\ }}
\def\ZP{{ Z.\ Phys.\ }}

Starting from this chapter we shall  
present the results of our investigations. 
We will initially
try to extend the Coniglio-Klein result to  
models characterized by continuous 
spin variables, which represents a first step towards
the definition of a percolation picture 
for lattice field theory.

\section{The Continuous Spin Ising Model}\label{manuele}

The easiest thing to start with is just 
to take the Ising model without external field, and replace its 
two-valued spins by continuous variables. 
The Hamiltonian is again given by
\begin{equation}
\label{haCIs}
{\cal H}\,=\,-J\,\sum_{ij}S_iS_j,
\end{equation}
with the sum over nearest neighbours, but the spin $S_{i}$ can now take
all values within some range, which 
we assume to be $[-1,+1]$.
This model is the classical continuous spin Ising model introduced 
by Griffiths \cite{Gr}, who studied its 
behaviour in two space dimensions. 
In \cite{Gr}, Griffiths deduced that this model
has the same critical behaviour of the 
Ising model, i. e. it undergoes a second order
phase transition with the
magnetization as order parameter, and its exponents are in the Ising
universality class.
For practical reasons, it is convenient to rewrite 
the spin variable $S$ in the following way
\begin{equation}
S\,=\,sign(S)\,\sigma,
\end{equation}
separating the sign from the amplitude $\sigma$ (e.g. the absolute value) of the spin.
The Hamiltonian (\ref{haCIs}) satisfies a $Z(2)$ global symmetry, i. e. 
it remains invariant after a simultaneous sign change of all spins 
of the system. This symmetry 
will play an important role all through our studies and it implies
that the signs of the spins are equally distributed in the canonical
ensemble of the system. 
In contrast, the amplitudes can in general be weighted
in different ways by choosing a distribution function $f(\sigma)$.
Therefore, the partition function of 
the continuous spin Ising model has the following general form:
\begin{equation}
Z(T)\, =\,  \prod_i \int_{0}^{1}
 d\sigma_i f(\sigma_i)~ \exp\{ \kappa
\sum_{\langle i,j \rangle} S_iS_j\},
\label{partCIs}
\end{equation}
where $\kappa\equiv J/kT$. In the model studied by Griffiths,
$f(\sigma)=1\,\, \forall\,\sigma$. 
We will begin by studying this special case, 
but we will see that our result is valid
also for the more general expression (\ref{partCIs}).

We have carried on a detailed numerical study of 
the model on a simple square lattice. The Monte Carlo update
method we have used is a version of the Wolff algorithm \cite{wolff}
adapted to our system. We briefly describe this algorithm,
that we will often use, in the case of the Ising model. 

The Wolff algorithm is a cluster update which improves
the Swendsen-Wang procedure we have illustrated in Section 
\ref{mancio}.
Starting from a randomly chosen spin $S_0$, one 
visits all nearest neighbours of the same sign
as $S_0$ and 
connects them to it with probability $p=1-\exp(-2J/kT)$. 
Repeating
iteratively this
procedure with newly added spins in the cluster,
at some stage 
no more neighbours will fulfill the above compatibility condition.
Flipping all spins of the cluster one gets a new spin configuration.
It turns out that this
dynamics verifies the detailed balance condition, i.e. it samples the Gibbs
distribution of the Ising model (see \cite{wolff}).
The analogies with the 
Swendsen-Wang method are clear. The Wolff cluster
is constructed in the same way as the 
Fortuin-Kasteleyn-Swendsen-Wang clusters, being the bond probability
the same in both cases. But with the Wolff 
method one flips a single cluster at a time,
a feature that succeeds in eliminating the old
problem of critical slowing down of Monte Carlo simulations.

Because of its effectiveness, we tried to 
implement a Wolff-like cluster update 
for our system, exploiting its analogies 
with the Ising model. We basically repeat the Wolff procedure,
but adopting for the bond probability the expression below 
\begin{equation}
p(i,j)=1-\exp(-\frac{2J}{kT}\sigma_i\sigma_j),
\label{genCK}
\end{equation}
which explicitly depends on the spin amplitudes.
If we simply flip the spins, the dynamics is no longer
ergodic, as the spin amplitudes would remain unchanged.
So, the cluster flipping must be supplemented by a local update method
(like Metropolis or heat bath), in order to respect the 
accessibility criterium. We chose to alternate 
heat bath and Wolff steps. The proof that the resulting
update fulfills both ergodicity criteria and the
detailed balance condition will be omitted here
since it follows closely the derivations that can be found in
\cite{Wo2,Ha,Cha3,Cha4}.

Our version of the Wolff algorithm for 
the continuous Ising model suggests that 
the Fortuin-Kasteleyn clusters in this case
should probably be built as usual,  
the only difference being represented by the local
bond probability (\ref{genCK}). 

To check whether these clusters are indeed 
the physical droplets we are looking for, 
we have performed extensive simulations of our model, 
choosing six different
lattice sizes, namely $64^2$, $96^2$, $128^2$,
$160^2$, $200^2$ and $300^2$.
Our update step consisted of one heat bath sweep for the 
spin amplitudes and three 
Wolff flippings for the signs, which turned out 
to be a good compromise to reduce sensibly the correlation
of the data without making the move be too much time-consuming.
The thermal quantities are the energy density 
\begin{equation}
\label{densen}
\epsilon\,=\,\frac{\sum_{ij}S_iS_j}{V}
\end{equation}
($V$ is the lattice volume),
and the magnetization  
\begin{equation}
\label{densmagn}
m\,=\,\frac{|\sum_{i}S_i|}{V},
\end{equation}
where the absolute value is necessary to take into account
the two equivalent directions of the spins.

As far as the percolation
variables are concerned, after grouping 
all spins into clusters by means of the 
Hoshen and Kopelman labeling (see Appendix A), we 
measure the percolation strength $P$ and the average 
cluster size $S$, as defined in Sections \ref{giovanni} and \ref{marco}.
For the cluster labeling we have used free boundary conditions.
We say that a cluster percolates if it spans the lattice in both 
directions, that is if it touches all four sides of the lattice.
This choice was made
to avoid the possibility that, due to the finite lattice size,
one could find more than one percolating cluster, making 
ambiguous the evaluation of our variables\footnote{In three dimensions 
        even this definition of spanning cluster
        does not exclude the possibility of having more than one
        of such clusters for the same configuration. 
        Nevertheless the occurrence of such cases 
        is so rare that we can safely ignore them.}.   
The three fundamental features  
we have just 
mentioned, i. e. the  
Hoshen-Kopelman algorithm, the use of free boundary 
conditions and the definition of percolating 
cluster in all directions, will 
be always present in 
our percolation investigations, unless stated otherwise.

The statistical errors of all variables were determined 
by using the Jackknife method \cite{jack} with ten bins
of data: such method will be  
applied in all our studies.
The quantities of interest were measured every five updates
for any temperature and lattice size. That makes both
the percolation and the thermal variables basically uncorrelated.
 
After some preliminary scans of our program for several 
values of the temperature $\kappa$ ($\kappa=J/kT$), we focused on
the $\kappa$-range between $1.07$ and $1.11$, where 
the transition seems to take place. 
The number of iterations for each run goes from 
20000 (for $\kappa$ values close to the extremes of the range) to 50000  
(around the center of the range).
The thermal results have been interpolated by means
of the density of state method ($DSM$) \cite{DSM}, which contributes
to reduce the errors relative to the data points. 
We shall regularly apply this method to study 
thermal phase transitions. Unfortunately
the $DSM$ fails if one tries to interpolate 
the percolation data, because the 
probability of having a given cluster configuration
must take into account not only the 
distribution of the spins, which
is weighted by the Hamiltonian of the model, 
but also the distribution of the bonds.
Besides,
for the percolation quantities, 
standard interpolation methods (like cubic spline)
do not help to improve the situation
because of the fluctuations of the data at criticality.
Therefore we used directly the data points to extract the
critical indices. 

\begin{figure}[h]
\begin{center}
\epsfig{file=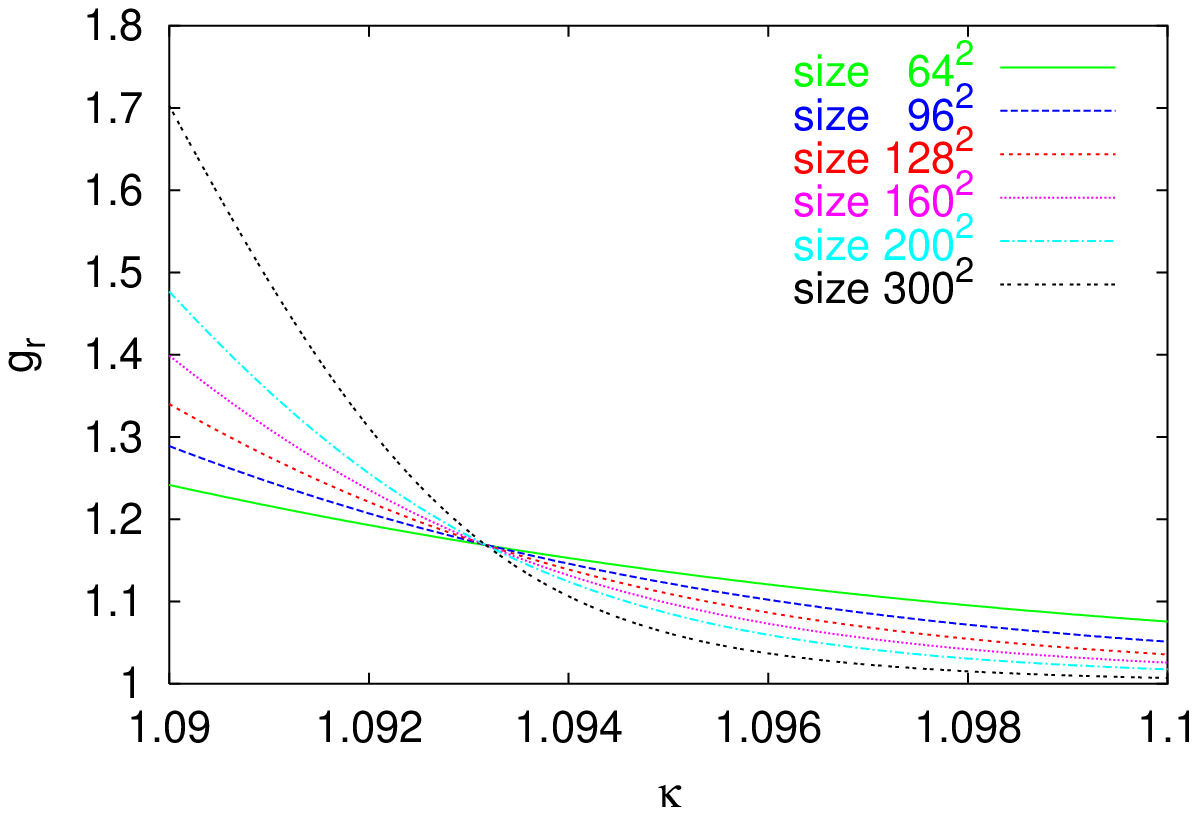,  width=8cm}
\caption[Binder cumulant as a function of $\kappa=J/kT$ for the
classical continuous Ising model of Griffiths]
{Classical continuous Ising model of Griffiths.
Binder cumulant as a function of $\kappa$ for six lattice sizes.
}
\label{grCIs}
\vskip0.1cm
\epsfig{file=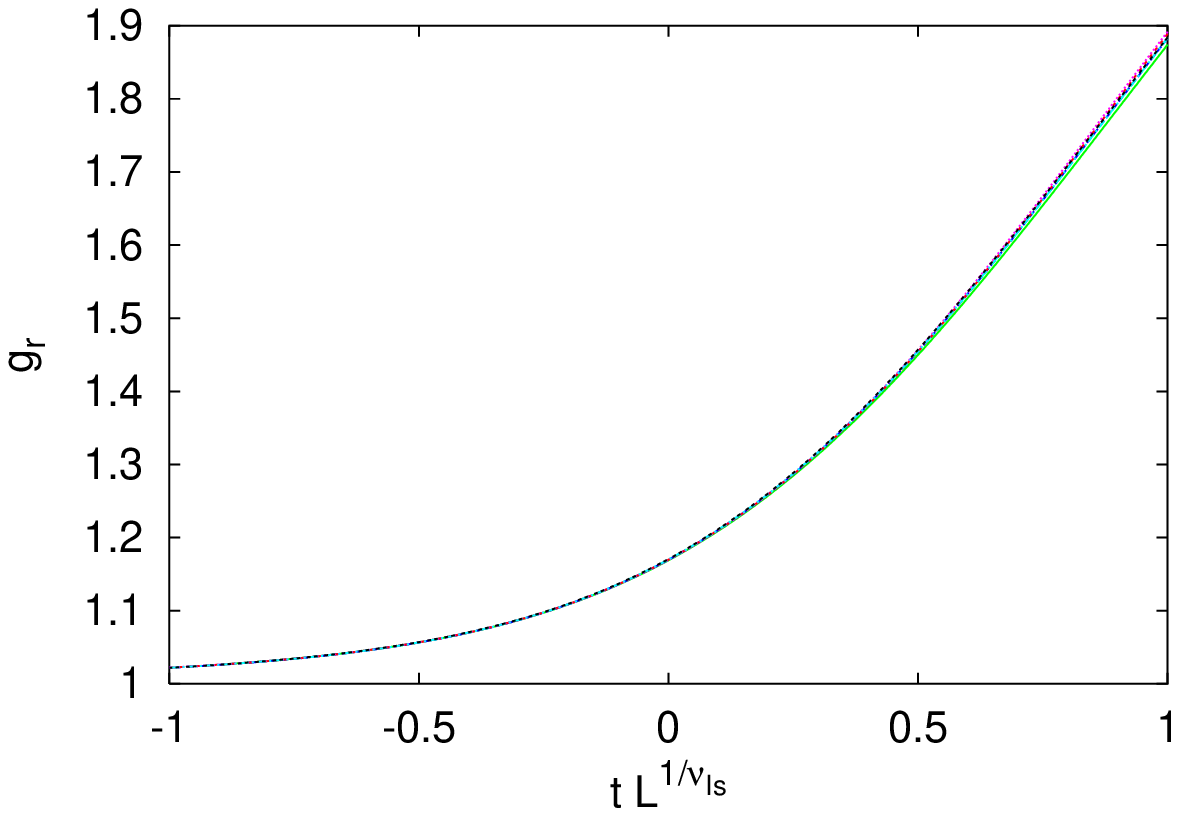,  width=8cm}
\caption[Rescaling of the Binder cumulant curves of Fig. \ref{grCIs}]
{
Rescaling of the Binder cumulant curves
shown in Fig. \ref{grCIs}. We took ${\kappa_{crit}}=1.0932$
and for the exponent $\nu$ the 2D Ising value $\nu_{Is}=1$. }
\label{ScgrCIs}
\end{center}
\end{figure}

To locate the critical point of the thermal transition 
we used the Binder cumulant\footnote{In all figures of this
work showing the Binder cumulant
we will just plot the ratio ${\langle{m^4}\rangle}/{{\langle{m^2}\rangle}^2}$; 
that allows to separate neatly  
the Binder and the percolation cumulant in the same figure,
which provides a visual comparison of the critical thresholds.} 
\begin{equation}
\label{cummag}
  g_r=3-\frac{\langle{m^4}\rangle}{{\langle{m^2}\rangle}^2}~.
\end{equation}
Fig. \ref{grCIs} 
shows $g_r$ as a function of $\kappa$ for the different
lattice sizes we used. The lines cross remarkably well 
at the same point, which suggests that also in our case $g_r$ 
is a scaling function. As a numerical proof we replot the lines
as a function of $tL^{1/\nu}$ ($t=(T-T_c)/T_c$, $L$ is the lattice side), choosing
for the exponent $\nu$ the 2D Ising value $1$. The plot (Fig. \ref{ScgrCIs})
shows that indeed $g_r$ is a scaling function 
with the critical exponent 
$\nu$ equal to the 2D Ising one.

\begin{figure}[t]
\begin{center}
\epsfig{file=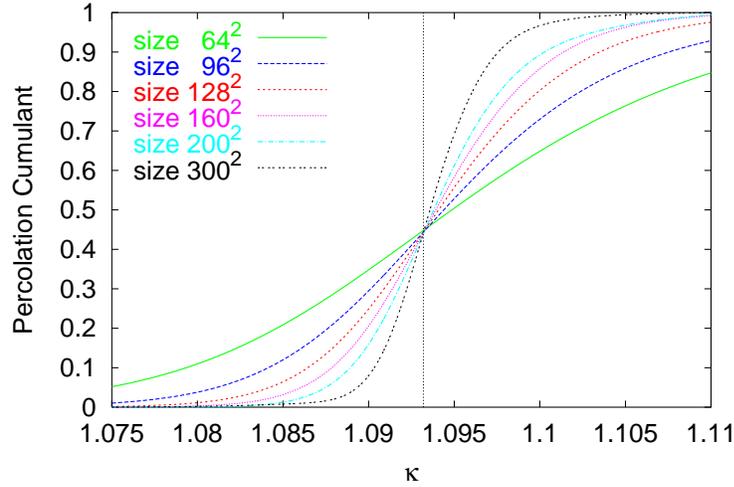,  width=11cm}
\caption[Percolation cumulant as a function of $\kappa=J/kT$
for the classical continuous Ising model of Griffiths]
{Classical continuous Ising model of Griffiths.
Percolation cumulant as a function of $\kappa$ for six
lattice sizes. The dashed line indicates the 
thermal critical threshold.}
\label{CISpr}
\end{center}
\end{figure}

To find the critical point of the percolation transition
we use the percolation cumulant introduced in Section \ref{markus}.
The results can be seen in Fig. \ref{CISpr}.
The agreement between the thermal threshold
and the geometrical one is excellent.

For the evaluation of the exponents we used 
standard finite size scaling techniques (see Section \ref{markus}).
To obtain the thermal exponents we adopted the
$\chi^2$ method \cite{engles}, a procedure that we will
apply to most of the models we are interested in. 
The results we got are reported in Table
 \ref{ISK}, from which 
it is clear that the critical exponents 
of the two transitions agree with each other
and with the 2D Ising model values.
\vskip0.2cm
\begin{table}[h]
  \begin{center}{
      \begin{tabular}{|c||c|c|c|c|}
\hline
  &      Critical point  & $\beta/{\nu}$ & ${\gamma}/{\nu}$ & $\nu$\\ \hline\hline
$\vphantom{\displaystyle\frac{1}{1}}$  Thermal results &$1.09312^{+0.00012}_{-0.00008}$&$
  0.128^{+0.005}_{-0.006}$&$1.745^{+0.007}_{-0.007}$&$1.01^{+0.01}_{-0.02}$\\  
\hline $\vphantom{\displaystyle\frac{1}{1}}$ 
  Percolation results &$1.09320^{+0.00008}_{-0.00008}$& $ 0.130^{+0.008}_{-0.010}$
  &$ 1.753^{+0.006}_{-0.006}$
&$0.98^{+0.03}_{-0.02}$ \\ \hline
$\vphantom{\displaystyle\frac{1}{1}}$2D Ising values& & $1/8\,=\,0.125$&$7/4\,=\,1.75$&1\\ \hline
      \end{tabular}
      }
\caption[Thermal and percolation critical indices for the 
classical continuous Ising model of Griffiths]
{\label{ISK} Thermal and percolation critical indices for the 
classical continuous Ising model of Griffiths.} 
  \end{center}

\end{table}
So far we have investigated
a relatively simple case, namely a model with the 
uniform amplitudes distribution $f(\sigma)=1$. 
One can ask whether the result remains valid for the 
general ansatz (\ref{partCIs}).
As a matter of fact, the distribution $f(\sigma)$ plays an important
role as far as the thermal properties 
of the system are concerned; in particular, it may influence
the order of the phase transition.
For this reason, since we want to study models with continuous
transitions, the choice
of the function $f(\sigma)$ is not arbitrary. It can be proved that it
must obey certain regularity conditions, which are not very restrictive,
however \cite{ellis}. Here we consider
the following form for $f(\sigma)$:
\begin{equation}
f(\sigma)=\sqrt{1-{\sigma}^2}
\label{3eq}
\end{equation}
which is the Haar measure of the $SU(2)$ group. 
We have made this choice
because our final target is to define a percolation
picture for $SU(2)$ gauge theory, and 
the function (\ref{3eq})  
appears quite often in formal 
expressions of this theory, like series expansions. 

It is reasonable to 
presume that the bond weight we need 
to define 
our clusters is determined 
by the Hamiltonian of the system, and not by eventual
distribution functions. That is why we tried to test 
the same cluster definition we adopted 
in the previous case. So, our droplets will be again
clusters of like-signed nearest neighbouring spins 
bound to each other with the probability (\ref{genCK}).
 
We have carried on a complete numerical investigation of the model,
performing simulations on 
four lattice sizes, $64^2$, $128^2$, $160^2$ and $240^2$.
Our algorithm consists in heat bath steps for the update of the
spin amplitudes followed by Wolff-like cluster updates for the flipping
of the signs. That is basically the same method as used before, 
although the heat bath procedure is slightly modified to
take into account the presence of the distribution function
$f(\sigma)$: the procedure
is analogous
as the heat bath algorithm of Creutz for 
$SU(2)$ gauge theory \cite{creu}. Also in this case, the proof 
of the detailed balance condition 
is simply obtained from the results in
\cite{Wo2} - \cite{Cha4}.
Again, we alternated one heat bath sweep
and three Wolff flippings and took the measurements every five updates: 
that
makes negligeable the correlation of all quantities. 

\begin{figure}[h]
\begin{center}
\epsfig{file=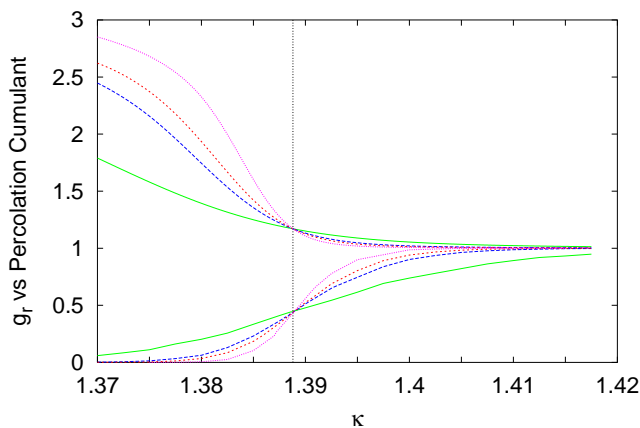,  width=9.5cm}
\caption[Comparison of the thermal and the geometrical
critical point for the 
continuous Ising model with spin amplitudes distribution $f(\sigma)=\sqrt{1-{\sigma}^2}$]
{Comparison of the thermal and the geometrical
critical point for the continuous Ising model
with the distribution (\ref{3eq}).}
\label{CoCuCIs}
\end{center}
\end{figure}

Fig. \ref{CoCuCIs} shows a comparison between the Binder cumulant
$g_r(\kappa)$ and the percolation cumulant, both
as functions of the temperature variable $\kappa$, for different lattice
sizes. The agreement between the two thresholds is excellent.

\begin{figure}[h]
\begin{center}
\epsfig{file=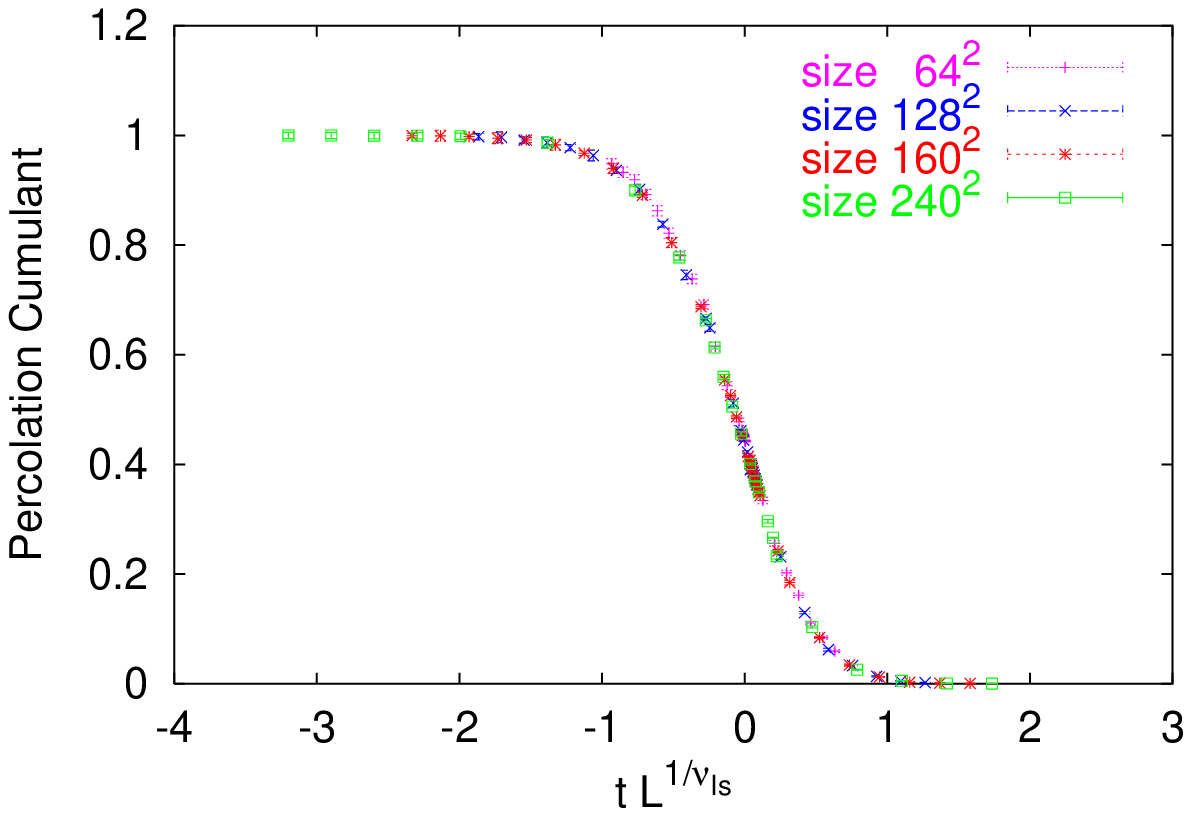,  width=13cm}
\caption[Rescaled percolation cumulant 
using the 2D Ising exponent $\nu_{Is}=1$]
{Continuous Ising model with the distribution (\ref{3eq}).
Rescaled percolation cumulant for four lattice sizes,
using the 2D Ising exponent $\nu_{Is}=1$.
The errors on the data points are smaller than the size of the symbols in the plot.
}
\label{ScprobIsCIS}
\vskip1cm
\epsfig{file=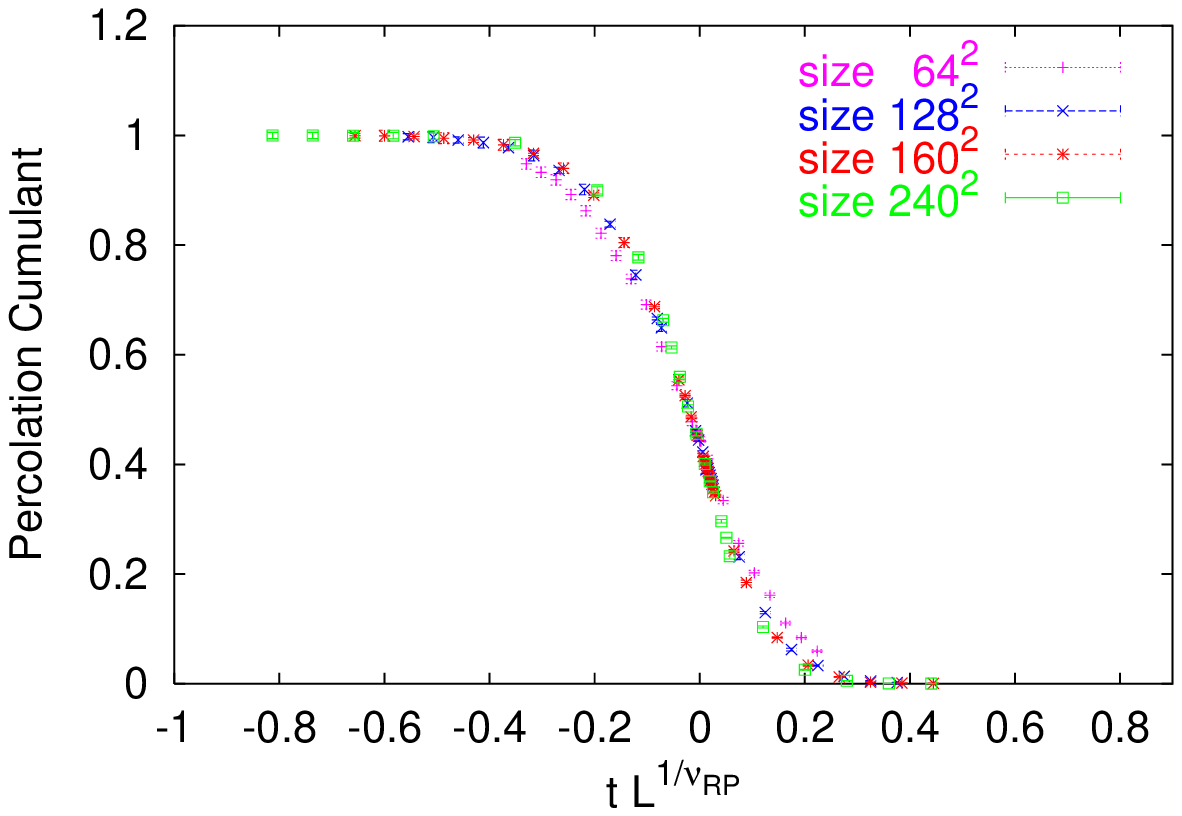,  width=13cm}
\caption[Rescaled percolation cumulant 
using the 2D random percolation exponent $\nu_{RP}=4/3$]
{Continuous Ising model with the distribution (\ref{3eq}).
Rescaled percolation cumulant for four lattice sizes,
using the 2D random percolation exponent $\nu_{RP}=4/3$.
The errors on the data points are smaller than the size of the symbols in the plot.}
\label{ScprobRPCIS}
\end{center}
\end{figure}
\clearpage

We can also get an estimate for the 
percolation critical exponent $\nu$, by rescaling the 
percolation cumulant curves as a function of 
$tL^{1/\nu}$. Figs. \ref{ScprobIsCIS}
and \ref{ScprobRPCIS} show the rescaled curves:
$\kappa_{crit}=1.3888$ and for $\nu$
we have taken the random percolation
value $\nu_{RP}=4/3$ and Ising one $\nu_{Is}=1$, respectively.
It is clear that 
the curves scale 
for $\nu=\nu_{Is}$ and do not
for $\nu=\nu_{RP}$.
To determine
the critical
exponents' ratios
${\beta}/{\nu}$ and ${\gamma}/{\nu}$, 
we have performed high-statistics simulations around the critical
point, with the number of measurements for each value of the coupling
varying from 50000 to 100000.
We have listed the results in Table \ref{IsCNTdis}. 
It is evident that the percolation behaviour
coincides fully with the thermal critical behaviour. This conclusion is
likely to hold in general for the admissable spin distribution
functions.
\vskip0.5cm
\begin{table}[h]
  \begin{center}{
      \begin{tabular}{|c||c|c|c|c|}
\hline
  &      Critical point  & $\beta/{\nu}$ & ${\gamma}/{\nu}$ & $\nu$\\ \hline\hline
$\vphantom{\displaystyle\frac{1}{1}}$  Thermal results &$1.3887^{+0.0002}_{-0.0001}$&$
  0.128^{+0.007}_{-0.010}$&$1.754^{+0.007}_{-0.008}$&$0.99^{+0.03}_{-0.02}$\\  
\hline $\vphantom{\displaystyle\frac{1}{1}}$ 
  Percolation results &$1.3888^{+0.0002}_{-0.0003}$& $ 0.121^{+0.008}_{-0.006}$
  &$ 1.745^{+0.011}_{-0.007}$
&$1.01^{+0.02}_{-0.03}$ \\ \hline
$\vphantom{\displaystyle\frac{1}{1}}$2D Ising values& & $1/8\,=\,0.125$&$7/4\,=\,1.75$&1\\ \hline
      \end{tabular}
      }
\caption[Thermal and percolation critical indices for the 
continuous Ising model corresponding 
to the amplitudes distribution $f(\sigma)=\sqrt{1-{\sigma}^2}$]
{\label{IsCNTdis} Thermal and percolation critical indices for the 
continuous Ising model with 
the amplitude distribution (\ref{3eq}).} 
  \end{center}

\end{table}

\section{Extension to Generalized Continuous Ising-like Models}\label{pippo} 

We shall now address the question whether the introduction of additional 
longer range spin-spin interactions still allows a description of the thermal transition
in terms of percolation. This will turn out to be very useful
in our attempt to
define suitable clusters in $SU(2)$ gauge theory.
Besides, we will examine the effects of eventual 
self-interaction terms, and show that 
they don't play any role in the cluster definition.

\begin{figure}[h]
\begin{center}
\epsfig{file=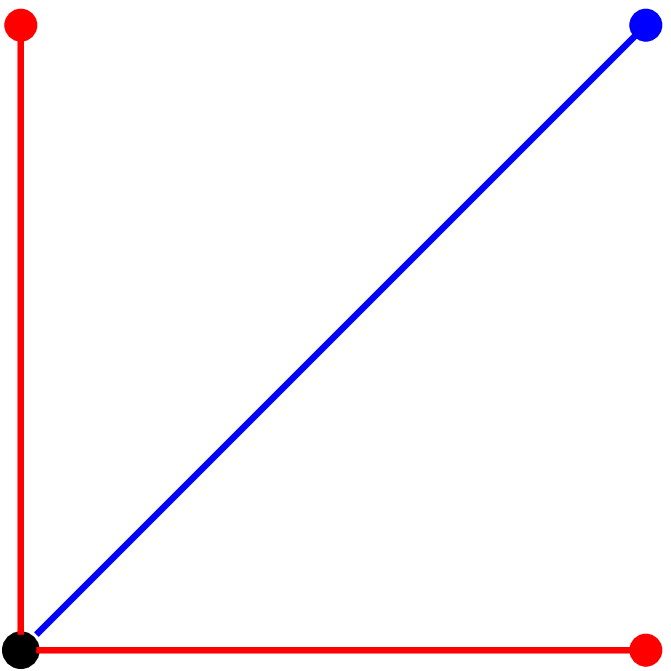,  width=5cm}
\epsfig{file=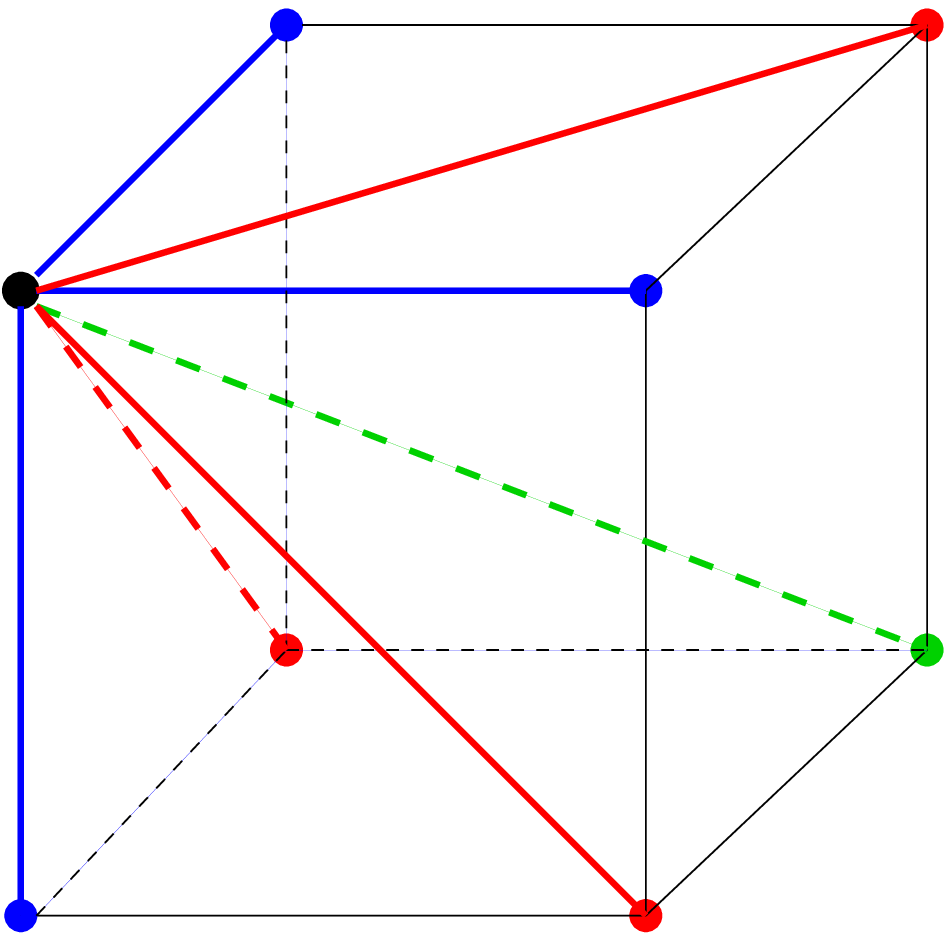,  width=7cm}
\vskip0.5cm
\begin{picture}(0,-50)
    \put(-107,2){\begin{minipage}[t]{8.4cm}{(a)}
    \end{minipage}}
    \put(89,2){\begin{minipage}[t]{8.4cm}{(b)}
    \end{minipage}}
\end{picture}

\caption[Scheme of the spin-spin interactions in the 
continuous spin models $A$, $B$ and $C$]
{Scheme of the spin-spin interactions in the
models we have studied. The figures 
indicate the interactions of the spin
represented by the black circle with its neighbours. 
Lines of the same color are associated to the
same coupling.  a) Model A.
b) Models B and C.}
\label{cubi}
\end{center}
\end{figure}

Our study  
is still based on continuous spin Ising models, in which the
individual spins $s_i$ at each lattice site can take on all values in the
finite range $[-1,1]$. Since these models are more general
than the ones characterized by 
discrete valued spins, the results 
can be then trivially extended to the latter ones.
Here we will consider three more general models of 
this type; $d$ denotes the space dimension:

\medskip

\noindent
A) $d=2$, nearest-neighbour (NN) and diagonal
   next-nearest-neighbour (NTN) interaction (Fig. \ref{cubi}a);

\medskip

\noindent
B) $d=3$, NN and two types of NTN interactions (see
   Fig. \ref{cubi}b);

\medskip

\noindent
C) as case B), but including an additional self-interaction term
   proportional to $S_i^2 ~ \forall~i$.
\vskip0.1cm

The couplings $J_i$ relative to the 
spin-spin interactions are all positive
(ferromagnetic).  
In each case, we will assume a uniform 
distribution for the spin amplitudes. This has only 
practical reasons, since it 
simplifies the numerical analysis, but, according to the
results of the previous section, it does not 
affect the generality
of our conclusions.

\subsection{Model A: Next-to-Nearest Neighbour Interactions}\label{maria}

We have now two terms, with a Hamiltonian of the form
\begin{equation}
{\cal H} = -J_{NN} \sum_{\langle i,j \rangle}^{NN} S_iS_j -
J_{NTN} \sum_{\langle i,j \rangle}^{NTN} S_iS_j
\label{fa}
\end{equation}
where the first sum describes nearest-neighbour and the second
diagonal next-to-nearest neighbour interactions (Fig. \ref{cubi}a). Since longer
range interactions are generally weaker, we have fixed the ratio
between the two couplings at $J_{NN}/J_{NTN}=10$; however, we do not
believe that our results depend on the choice of the couplings, as long
as both are ferromagnetic.

To define clusters, we now extend the
Coniglio-Klein method and define for each two spins $i,j$ of the same
sign, for NN as well as NTN, a bond probability
\begin{equation}
p_B^x(i,j) = 1 - \exp ( - 2 \kappa_x \sigma_i \sigma_j),
\label{ffeq}
\end{equation}
where $x$ specifies $\kappa_{NN}=J_{NN}/kT$ and $\kappa_{NTN}=J_{NTN}/kT$,
respectively. This hypothesis seems to us the most natural, and we will
test it in the following B and C models.

We have studied model A using two different Monte Carlo algorithms, in
order to test if a Wolff-type algorithm can also be applied in the 
presence of NTN interactions. The first is the standard Metropolis 
update, while the second alternates
heat bath steps and a generalized Wolff flipping, for which the
clusters are formed taking into account both interactions. The
generalization of the cluster update is trivial. After several runs,
some with high statistics, we found excellent agreement with the
Metropolis results in all cases. So, the mixed algorithm 
with heat bath and Wolff flippings appears to remain viable also
in the presence of more than the standard NN interaction. Subsequently
we have therefore used this mixed algorithm. The update
alternates like before one heat bath sweep and three Wolff
flippings. The lattice sizes 
ranged from $100^2$ to $400^2$. 
We measured our variables every $5$ updates for 
the smaller lattice sizes and every $10$ for the 
larger ones, keeping these 
numbers fixed at any temperature.
All variables of interest turn out to be basically uncorrelated. 
We accumulated
up to 50000 measures for temperatures close to the critical point.

\begin{figure}[h]
\begin{center}
\epsfig{file=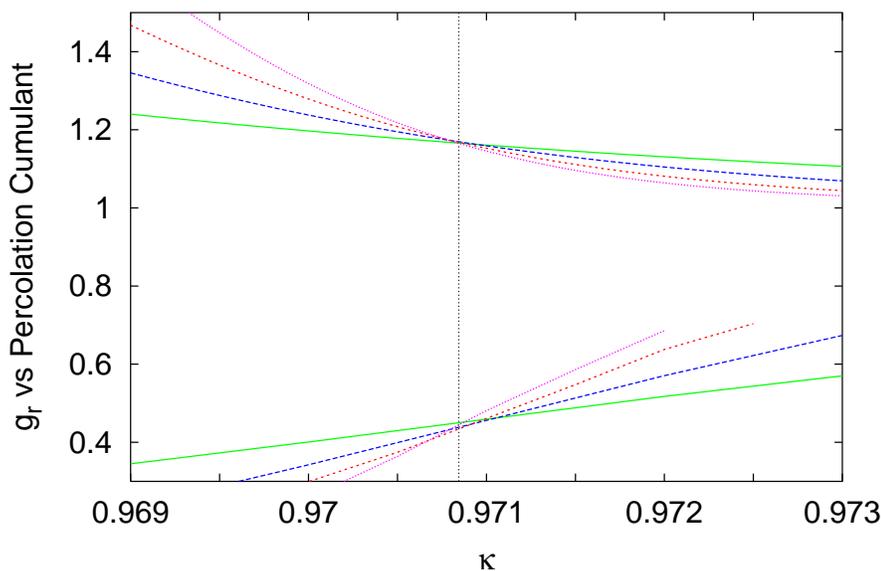,  width=13cm}
\caption[Comparison of the thermal and the geometrical
critical point for Model A]
{Comparison of the thermal and the geometrical
critical point for Model A obtained respectively from the Binder cumulant
$g_r$ and the percolation cumulant.}
\label{Comp_kff}
\end{center}
\end{figure}

\begin{figure}[h]
\begin{center}
\epsfig{file=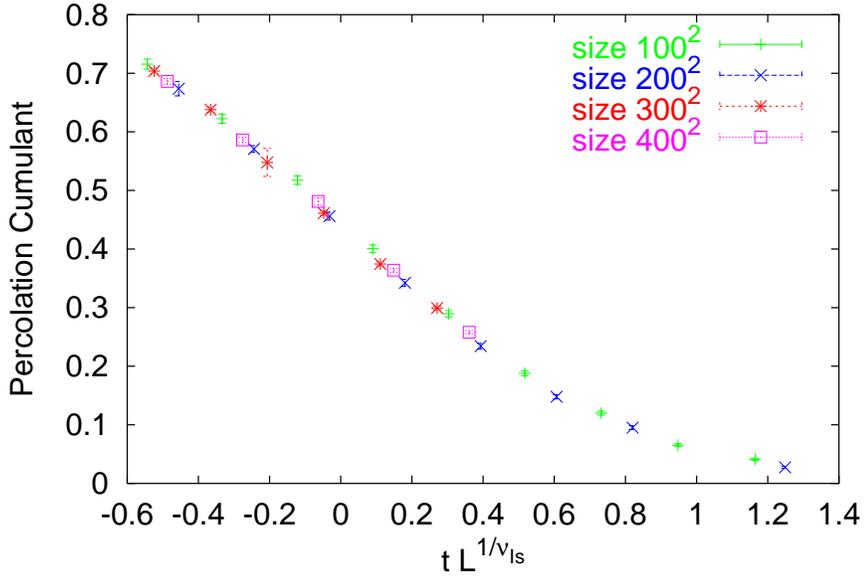,  width=13cm}
\caption[Rescaled percolation cumulant curves 
for model A,
using the 2D Ising exponent $\nu_{Is}=1$]
{Rescaled percolation cumulant curves 
for model A,
using the 2D Ising exponent $\nu_{Is}=1$.
The errors on the data points are smaller than the size of the symbols in the plot.}
\label{Scff}
\end{center}
\end{figure}

We present again the comparison between
percolation and Binder cumulants, in order to test that the
critical points coincide (Fig. \ref{Comp_kff}).
The crossing point
of the percolation curves looks less defined than
the thermal one because we used a simple linear interpolation of the data.
Anyhow, simulations of the model at the 
thermal threshold lead to 
values of the percolation cumulant that, within errors, are the same
for all lattice sizes. 
We then rescale the
percolation cumulant, using the critical $\kappa$ determined in Fig. \ref{Comp_kff}
together with the 
two main options
for the exponent $\nu$, that is the value of 2D random percolation
and the one of the 2D Ising model.
In Fig. \ref{Scff} we show the rescaling done using
$\nu_{Is}$: the curves fall clearly on top of each other.

The determination of the two exponents ratios $\beta/\nu$ and $\gamma/\nu$
confirms that indeed the exponents of
our geometrical islands belong to the 2D Ising universality class
(Table \ref{fftab}).
\vskip0.5cm
\begin{table}[h]

  \begin{center}
\begin{tabular}{|c||c|c|c|c|}
\hline
& Critical point & $\beta/\nu$ & $\gamma/\nu$ & $\nu$\\\hline\hline
$\vphantom{\displaystyle\frac{1}{1}}$ 
Thermal results & 0.9707$^{+0.0003}_{-0.0002}$ & 0.124$^{+0.007}_{-0.005}$
& 1.747$^{+0.009}_{-0.007}$
 & 0.993$^{+0.014}_{-0.010}$ \\
\hline$\vphantom{\displaystyle\frac{1}{1}}$
Percolation results& 0.9708$^{+0.0002}_{-0.0002}$ & 0.129$^{+0.008}_{-0.009}$
& 1.752$^{+0.009}_{-0.011}$ & 1.005$^{+0.012}_{-0.020}$ \\
\hline$\vphantom{\displaystyle\frac{1}{1}}$
2D Ising Model &                   & $1/8\,=\,0.125$&$7/4\,=\,1.75$
 & 1 \\
\hline
\end{tabular}
\caption[Thermal and percolation critical indices for model A]
{\label{fftab} Thermal and percolation critical indices for model A, 
compared to those of
the 2D Ising model.}
\end{center}
\end{table}

\subsection{Model B: Extension to Three Dimensions}\label{giulia}

We now go one step further and repeat the study 
for a $d=3$ model with three different interactions
(Fig. \ref{cubi}b).

To fix the model, we have to specify the ratios of the
nearest-neighbour coupling $J_{NN}$ to $J_{NTN}$ and
$J_{diag}$. We chose them to be $10:2$ and $10:1$, respectively.
Our calculations are performed on lattices ranging from 
$12^3$ to $40^3$.

\begin{figure}[h]
\begin{center}
\epsfig{file=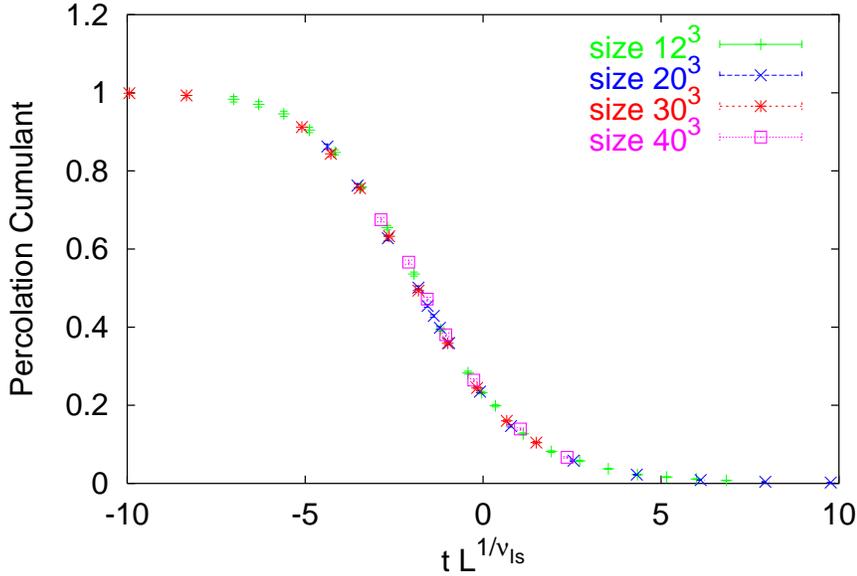,  width=13cm}
\caption[Rescaled percolation cumulant curves 
for model B,
using the 3D Ising exponent $\nu_{Is}=0.6294$.]
{Rescaled percolation cumulant curves 
for model B,
using the 3D Ising exponent $\nu_{Is}=0.6294$.
The errors on the data points are smaller than the size of the symbols in the plot.}
\label{Scalfff}
\end{center}
\end{figure}

Also here we have first compared the results from a mixed algorithm
of the same kind as for the previous case to those from a standard
Metropolis algorithm; again, the agreement turns 
out to be very good. The heat bath sweeps and the Wolff flippings are 
in the ratio $1:3$. We measured our variables every $5$ updates
for any temperature and lattice size. The 
percolation variables are not correlated, whereas
the thermal ones show a correlation which 
is, however, rather small (the autocorrelation time $\tau$
is of about $2-3$ for the magnetization on the 
$40^3$ lattice near criticality). The number of measurements we took
varies from $20000$ to $40000$.
\vskip-0.4cm
\begin{figure}[h]
\begin{center}
\epsfig{file=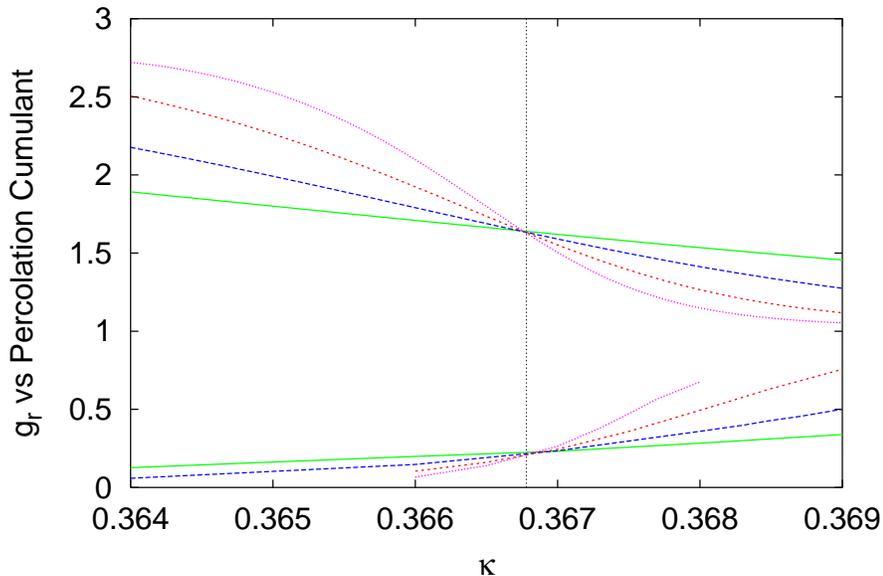,  width=13cm}
\caption[Comparison of the thermal and the geometrical
critical point for Model B]
{Comparison of the thermal and the geometrical
critical point for Model B obtained respectively from the Binder cumulant
$g_r$ and the percolation cumulant.}
\label{Compfff}
\end{center}
\end{figure}
\vskip-0.4cm
Figs. \ref{Compfff} and \ref{Scalfff} then show the comparison of the thresholds
and the scaling of the percolation probability. As before, the
correspondence between percolation and thermal variables is evident
(Table \ref{tafff}).
\begin{table}[h]
  \begin{center}
\begin{tabular}{|c||c|c|c|c|}
\hline
& Critical point & $\beta/\nu$ & $\gamma/\nu$ & $\nu$\\\hline\hline
$\vphantom{\displaystyle\frac{1}{1}}$ 
Thermal results & 0.36677$^{+0.00010}_{-0.00008}$ & 0.530$^{+0.012}_{-0.018}$
& 1.943$^{+0.019}_{-0.008}$
 & 0.640$^{+0.012}_{-0.018}$ \\
\hline$\vphantom{\displaystyle\frac{1}{1}}$
Percolation results& 0.36673$^{+0.00012}_{-0.00010}$ & 0.528$^{+0.012}_{-0.015}$
& 1.975$^{+0.010}_{-0.015}$ & 0.632$^{+0.010}_{-0.015}$ \\
\hline$\vphantom{\displaystyle\frac{1}{1}}$
3D Ising Model &          & 0.5187(14)   
& 1.963(7)
 & 0.6294(10) \\
\hline
\end{tabular}
\caption[Thermal and percolation critical indices for model B]
{\label{tafff} Thermal and percolation critical indices for model B, 
compared to those of
the 3D Ising model.}
\end{center}
\end{table}

\subsection{Model C: Adding Self-Interactions}\label{marina}

From what we have seen up to now, it seems to be clear that the correct
cluster definition can readily be extended to models including several
(ferromagnetic) spin-spin interactions. However, such terms are not the
only possible interactions in a model with Z(2) symmetry and a
continuous transition. There could be anti-ferromagnetic spin-spin
couplings as well as multispin terms, coupling an even number of
spins greater than two (four, six, etc.). Moreover, since the spins
are continuous, the presence of self-interaction terms is possible,
determined by $S^2$, $S^4$, etc. The treatment for antiferromagnetic
and multispin couplings so far remains an open question. In contrast,
self-interactions are not expected to play a role in the cluster
building, since such terms do not relate different spins. Therefore, we
test a cluster definition ignoring any self-interaction term.

We thus consider in Model C the same interactions as in Model B, plus a
term proportional to $J_0 \sum_i S_i^2$. We chose a negative value for
the self-interaction coupling $J_0$; this is the more interesting
case since the corresponding interaction tries to resist the approach
of the system to the ground state at low temperatures ($\sigma=1$
everywhere). The ratios of the NN coupling to the others were
chosen as $J_{NN}:J_{NTN}:J_{diag}:|J_0|= 6:2:1:2$.

We first verify the viability of the mixed algorithm.
The check was successful so that we 
could apply the algorithm for our purposes.
The update consists again in one heat bath sweep and 
three Wolff flippings. In order to 
eliminate the correlation of the data we measured our quantities every 
$40$ updates. We collected up to 70000 measurements 
for temperatures close to criticality.

\begin{figure}[h]
\begin{center}
\epsfig{file=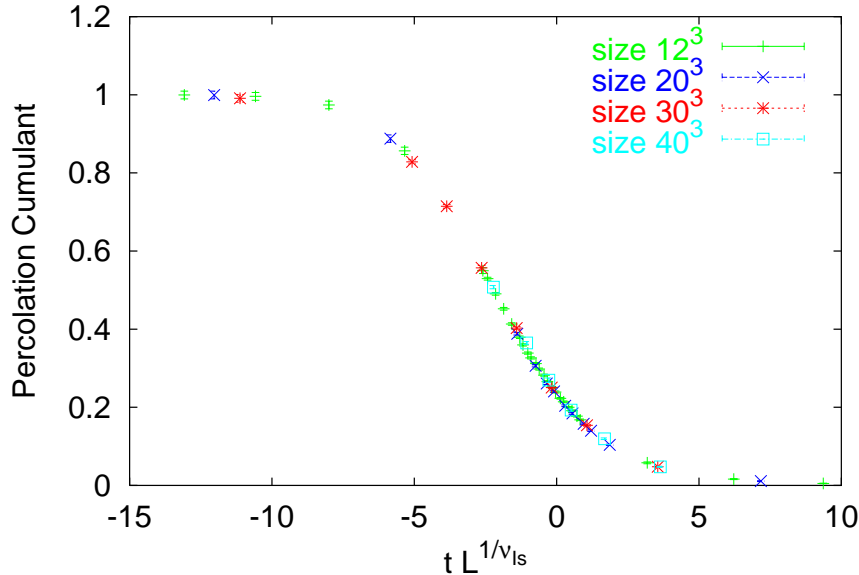,  width=13cm}
\caption[Rescaled percolation cumulant curves 
for model C,
using the 3D Ising exponent $\nu_{Is}=0.6294$]
{Rescaled percolation cumulant curves 
for model C,
using the 3D Ising exponent $\nu_{Is}=0.6294$.
The errors on the data points are smaller than the size of the symbols in the plot.}
\label{ScaSI}
\end{center}
\end{figure}

The critical points were determined by
means of the cumulants. In Fig. \ref{ScaSI} 
we present the rescaling of the percolation
cumulant curves in correspondence of the 3D
Ising exponent $\nu_{Is}=0.6294$. 
The scaling function can be clearly seen. Successively we
have determined the 
critical exponents (Table \ref{taSI}). It is evident that percolation and the
thermal transition again fall into the same universality class.
\vskip0.5cm
\begin{table}[h]
  \begin{center}
\begin{tabular}{|c||c|c|c|c|}
\hline
& Critical point & $\beta/\nu$ & $\gamma/\nu$ & $\nu$\\\hline\hline
$\vphantom{\displaystyle\frac{1}{1}}$ 
Thermal results & 0.3004$^{+0.0002}_{-0.0001}$ & 0.513$^{+0.012}_{-0.010}$
& 1.963$^{+0.014}_{-0.009}$
 & 0.626$^{+0.011}_{-0.010}$ \\
\hline$\vphantom{\displaystyle\frac{1}{1}}$
Percolation results&0.3005$^{+0.0001}_{-0.0001}$  & 0.524$^{+0.010}_{-0.011}$
&1.975$^{+0.008}_{-0.009}$  & 0.636$^{+0.011}_{-0.017}$ \\
\hline$\vphantom{\displaystyle\frac{1}{1}}$
3D Ising Model &       & 0.5187(14)
& 1.963(7)
 & 0.6294(10) \\
\hline
\end{tabular}
\caption[Thermal and percolation critical indices for model C]
{\label{taSI} Thermal and percolation critical indices for model C,
compared to those of
the 3D Ising model.}
\end{center}
\end{table}

We have shown that the equivalence of cluster percolation and spin 
ordering in the description of critical behaviour in the continuous
spin Ising model can be extended to a rather wide class of theories.
In particular, it remains valid also in the presence of more than 
just nearest neighbour interactions, if ferromagnetic, 
and of spin distribution functions.
Moreover, the introduction of self-energy contributions does not 
affect the equivalence.

\section{Cluster Percolation in O(n) Spin Models}\label{marica}

An interesting extension of the 
Coniglio-Klein result concerns 
the $O(n)$ spin models. 

The $O(n)$ spin models with no 
external magnetic field have the following Hamiltonian:
\begin{equation}
\label{onham}
H=-J\sum_{\langle{{\bf i},{\bf j}}\rangle}{\bf s_i s_j},
\end{equation}
where $i$ and $j$ 
are nearest-neighbour sites on a $d$-dimensional 
hypercubic lattice, and 
${\bf s_i}$ is an $n$-component unit vector at site $i$ ($J\,>\,0$ is the 
coupling). The partition function
of these models at the temperature $T$ is 
\begin{equation}
Z(T) = \int{\cal D}[{\bf s}] \exp\{ \beta \sum_{\langle {{\bf i},{\bf j}} \rangle}{\bf s_i s_j} \} \label{1}
\end{equation}
where $\beta=J/kT$ and the integral is extended over all spin configurations
$\{{\bf s}\}$ of the system.

In three space dimensions, 
such models undergo a second order phase transition
due to the spontaneous breaking of the continuous rotational
symmetry of their Hamiltonian.
The $O(n)$ models are very interesting: some physical systems
in condensed matter physics are directly
associated to them. 
The three-dimensional $O(3)$ model is the low-temperature 
effective model for a bidimensional quantum antiferromagnet
\cite{mano}. The $O(2)$ model in three dimensions is known to be
in the same universality class as superfluid ${^4}He$.
$O(n)$ models are also very useful to 
study relativistic field theories.
The $O(4)$ model in three dimensions has
been conjectured to be in the same universality class 
as the finite-temperature chiral phase transition of $QCD$ with
two flavours massless quarks \cite{wilczeck}.

Numerical simulations of $O(n)$ models became much 
quicker and more 
effective after U. Wolff introduced his Monte Carlo
cluster update \cite{wolff}. 
We have already described it 
in the particular case of the Ising model (see Section \ref{manuele}).
As a matter of fact, the Wolff update
was devised for $O(n)$ spin models, of which
the Ising model is a special case (for $n=1$).

The procedure, as we have said, consists in flipping
all spins of a cluster which is built in some way.
For details of the flipping procedure, see \cite{wolff}.
Here we are interested in the way to build up the clusters.
We can split this procedure in two steps:

a) choose a random n-component unit vector {\bf r};

b) bind
together pairs of nearest-neighbouring sites $i$, $j$ with
the probability 
\begin{equation}
\label{bonWo}
p(i,j)=1-\exp\{min[0,-2{\beta}({\bf s_i}{\cdot}{\bf r})
({\bf s_j}{\cdot}{\bf r})]\}.
\end{equation}
From this prescription it follows that if 
the two spins at two nearest-neighbouring 
sites $i$ and $j$ are such that their projections onto the random vector 
{\bf r} are of opposite signs, they will never belong
to the same cluster ($p(i,j)=0$). The random vector {\bf r}, therefore, divides
the spin space in two hemispheres, separating the spins
which have a positive projection onto it from the ones which have a negative
projection. The Wolff clusters are made out of spins which all lie
either in the one or in the other hemisphere.
In this respect, we can again speak of 'up' and 'down' spins, like
for the Ising model. In addition to that, the bond probability is
local, since it depends explicitly on the spin vectors 
${\bf s_i}$ and ${\bf s_j}$, and not only on the temperature
like the Fortuin-Kasteleyn factor.

The analogies with the Ising
model are however clear,
motivating the attempt to study the percolation properties of these
clusters.

Indeed, for $O(2)$ and $O(3)$, it was analytically proven that 
the Wolff 
clusters percolate at the thermal critical point \cite{cha01,cha02}.
Nevertheless, in \cite{cha01,cha02} nothing about the 
relationship between the critical exponents
was said.
 
We have investigated numerically the 3-dimensional
$O(2)$ and $O(4)$ models performing computer simulations for
several lattice sizes. 
The Monte Carlo update was performed by the Wolff algorithm. 
At the end of an iteration, 
the percolation strength $P$ and the average cluster size $S$ were
measured. This has been done for each of the models using
two different approaches.

The {\bf first approach} is the traditional one, based on 
a complete analysis of the lattice configuration.
Once we have the configuration we want to analyze, we 
build Wolff clusters until all spins are set into clusters.
We assign to $P$ the value zero if there is no percolating cluster,
the ratio between the size of the percolating cluster
and the lattice volume otherwise.
We calculate $S$ using the standard formula (\ref{defS}).
We say that a cluster percolates if it spans
the lattice from a face to the opposite one in each of
the three directions $x$, $y$, $z$.
In this approach we have used as usual free boundary conditions.

      The {\bf second approach} is based on a single-cluster
      analysis. Basically one studies the percolation
      properties of the cluster built during the update
      procedure. For the cluster building we 
      have considered periodic boundary conditions.
      Suppose that $s_c$ is the size of the cluster we built. 
      If it percolates,
      we assign value one to the strength $P$
      and zero to the size $S$; otherwise, we write
      zero for $P$ and $s_c$ for $S$.
      These definitions of $P$ and $S$ look different from 
      the standard definitions we have introduced above, but it is easy to see
      that they are instead equivalent to them.
      
      In fact, we build the cluster starting from a lattice site taken at
      random. In this way, the probability that 
      the cluster percolates (expressed by the new $P$)
      coincides with the probability that a site taken at random belongs
      to the percolating cluster (standard definition of $P$).
      As far as the average cluster size is concerned, we can repeat the same
      reasoning: the probability that the cluster we built is
      a non-percolating cluster of size $s_c$ is just 
      the probability $w_{s_c}$ that a randomly taken lattice site
      belongs to a non-percolating cluster of size $s_c$; $w_{s_c}$ is given by 
      \begin{equation}
        w_{s_c}\,=\,n_{s_c}\,{s_c}\,\,.
        \label{probw}
      \end{equation}
      Because of that, whenever we get a non-vanishing size 
      $s_c$, such value will be weighted by the probability 
      $w_{s_c}$ in the final average $S$, which is then given by the following
      formula:
      \begin{equation}
        S\,=\,\sum_{s_c}w_{s_c}s_c\,=\,\sum_{s_c}n_{s_c}\,{s_c}^2,
        \label{tereS}
      \end{equation}
      where the sum runs over the non-percolating clusters.
      We notice that Eq. (\ref{tereS}) coincides with 
      Eq. (\ref{defS}), apart from the denominator $\sum_{s}n_{s}\,s$,
      which is just the density of the sites belonging to finite clusters.
      Since this term does not contribute to the divergence
      of the average cluster size, the power law behaviour
      of the two $S$'s at criticality is identical, 
      so that the critical exponent $\gamma$ is the same in both cases.
      
      As we have said, in the second approach we 
      select a single cluster at a time from the whole configuration. Because of that
      we have now some freedom of choosing the 
      definition of percolating cluster, as we do not risk, like in the first
      approach, to find more spanning structures.
      We say that the cluster percolates if
      it connects at least one face with the opposite one.

      In this way, also the definitions of percolating 
      clusters are different in the two approaches. This 
      certainly influences the results on finite lattices, but
      has no effects on the infinite-volume properties we are interested in.
      In fact, we have seen in Section \ref{patty} that one can have 
      at most a unique spanning cluster above the critical 
      density $p_c$ (in our case below the critical temperature $T_c$).
      Exactly at $p_c$ ($T_c$) there is a finite
      probability to have more than a spanning cluster. 
      So, the two different
      definitions of percolating cluster we have
      adopted can lead to differences between the 
      infinite-volume values {\it only}
      at the critical point $p_c$ ($T_c$). But the critical exponents are, of course,
      not influenced by that, as they are determined by the behaviour
      of the percolation variables {\it near}
      the critical point, not exactly at $p_c$ ($T_c$).

      The second approach has the advantage that it does not
      require a procedure to reduce the configuration 
      of the system to a set of clusters; on the other hand,
      since it gets the information out of a single
      cluster, it requires a higher number of samples
      in order to measure the percolation
      variables with the same accuracy of the first method.
      Nevertheless, the iterations are faster due to the simpler
      measurement of observables, and are less correlated than in the
      first approach, since only a (random) limited region of the lattice  
      is considered in each sample.
      We find that both methods are efficient, and that it is important 
      to be able to compare results obtained in two such different ways.
   
We collected up to $150000$ measurements for 
temperatures close to the critical point. We measured our quantities
every $N$ updates, with $N$ ranging from 
$50$ for the smaller lattice sizes to $100$ for the greater ones: that 
eliminates the correlation of the percolation data.

\begin{figure}[h]
\begin{center}
\epsfig{file=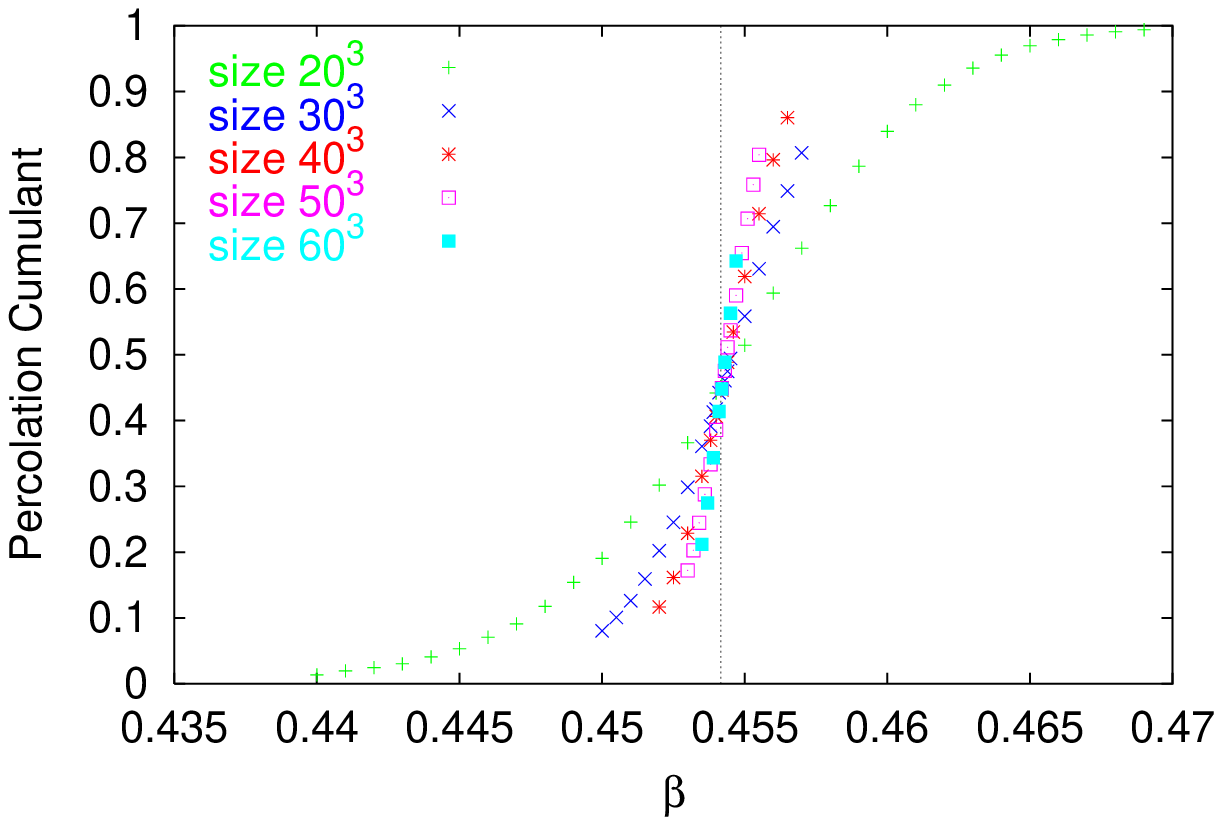,  width=14cm}
\caption[Percolation cumulant as function of $\beta$ for
Wolff clusters in $O(2)$]
{Percolation cumulant as function of $\beta$ for
          $O(2)$ and five lattice sizes. The dashed line indicates the position
          of the thermal threshold \cite{O2}.}
\label{pro2}
\vskip0.4cm
\epsfig{file=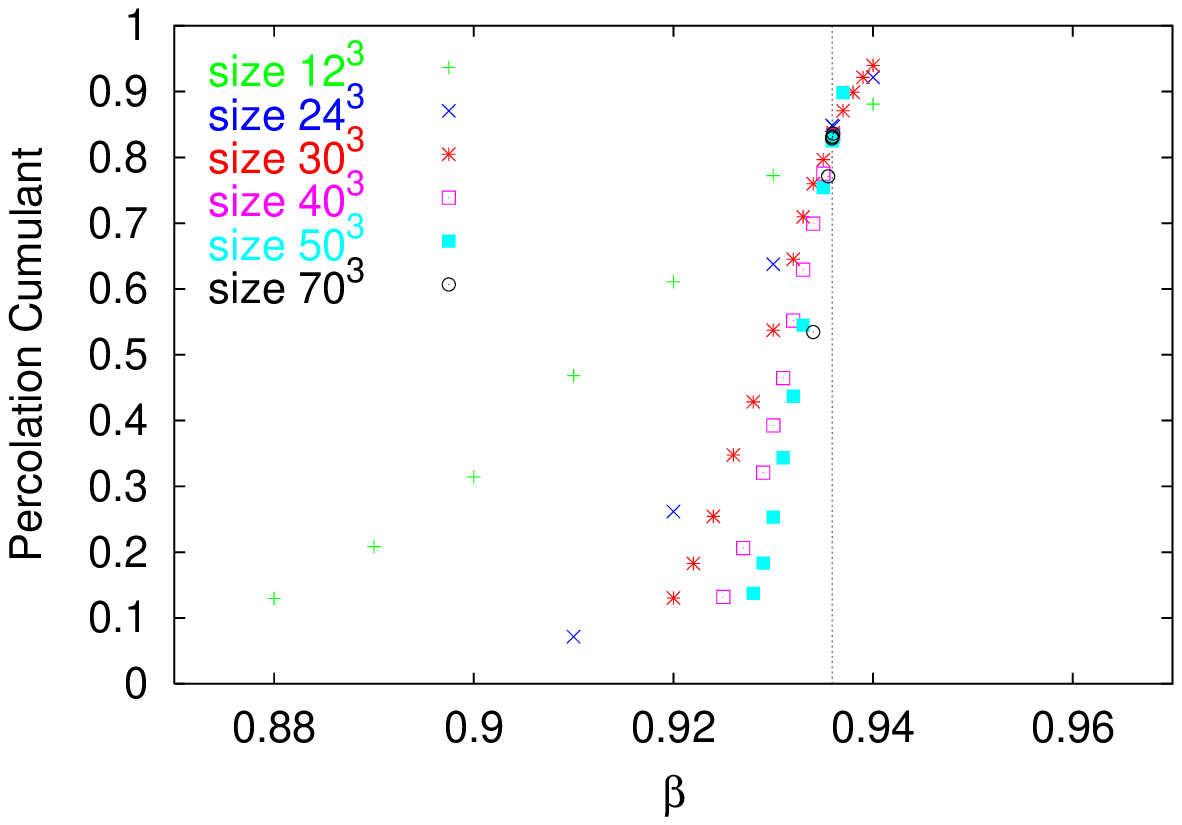,  width=14cm}
\caption[Percolation cumulant as function of $\beta$ for
Wolff clusters in $O(4)$]
{Percolation cumulant as function of $\beta$ for
          $O(4)$ and six lattice sizes. The dashed line indicates the position
          of the thermal threshold \cite{Manf}.}
\label{pro4}
\end{center}
\end{figure}

\begin{figure}[h]
\begin{center}
\epsfig{file=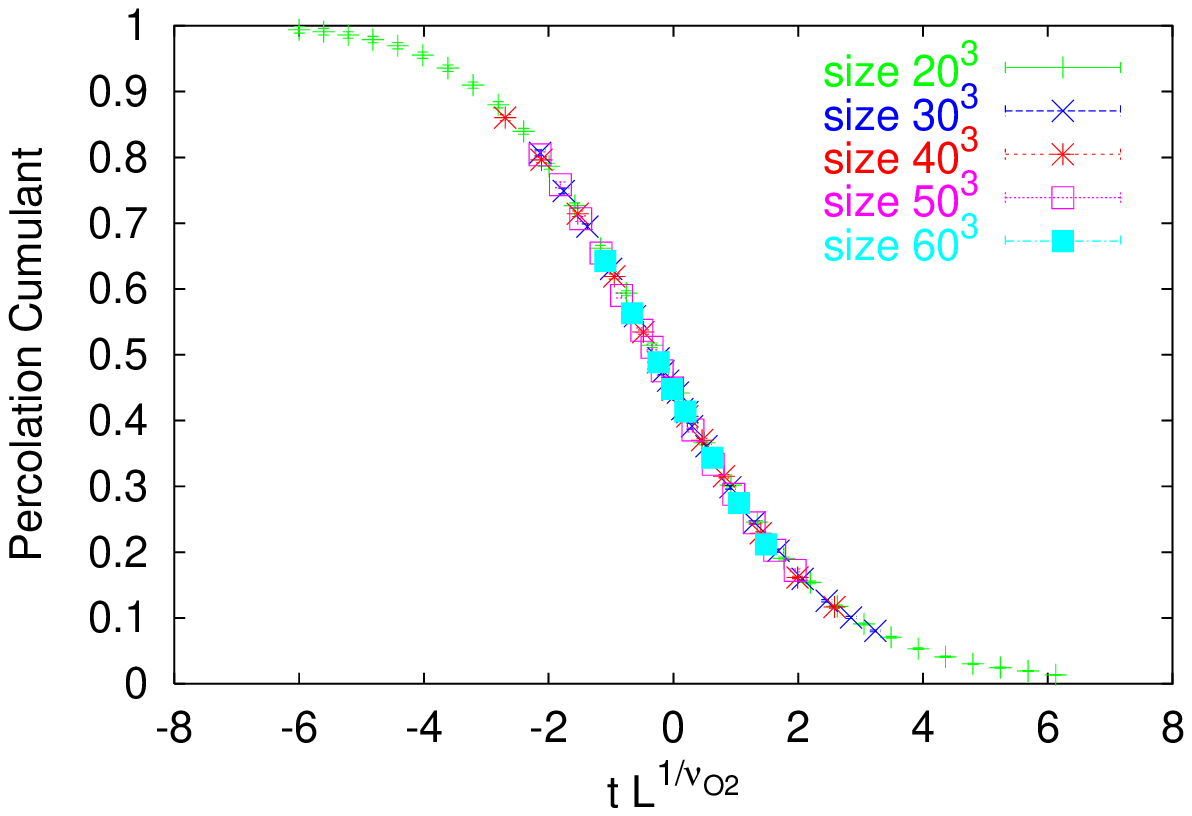,  width=14cm}
\caption[Rescaling of the percolation cumulant with the thermal exponent
for $O(2)$] 
{Rescaled percolation cumulant for $O(2)$
using $\beta_{c}=0.45416$ and the
$O(2)$ exponent $\nu_{O2}=0.6723$.
The values of the thermal critical indices 
are taken from \cite{O2}. The errors on the data points are smaller than the size 
of the symbols in the plot.} 
\label{Scalono2}
\vskip0.4cm
\epsfig{file=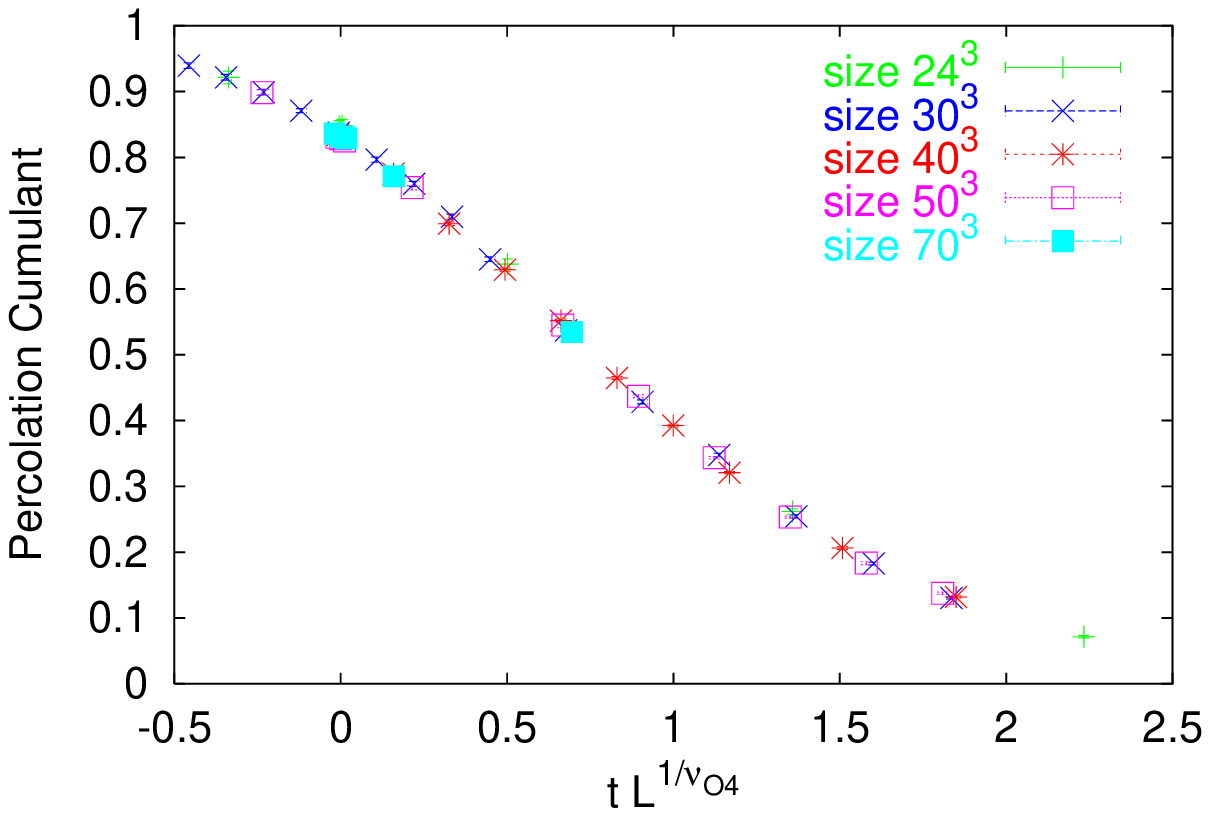,  width=14cm}
\caption[Rescaling of the percolation cumulant with the thermal exponent
for $O(4)$]
{Rescaled percolation cumulant for $O(4)$
using $\beta_{c}=0.9359$ and the
$O(4)$ exponent $\nu_{O4}=0.7479$. The values of the thermal critical indices 
are taken from \cite{Manf} (the threshold) and \cite{O4} (the exponent $\nu_{O4}$).
The errors on the data points are smaller than the size 
of the symbols in the plot.}
\label{Scalono4}
\end{center}
\end{figure}

Figs. \ref{pro2} and \ref{pro4} show percolation cumulant curves for 
$O(2)$ and $O(4)$, respectively. The agreement with
the physical thresholds (dashed lines) is clear.
Successively, we perform the usual scaling tests
to check whether the exponents $\nu_{perc}$ of the
geometrical transitions coincide with 
the ones of the model, $\nu_{O2}=0.6723$
and $\nu_{O4}=0.7479$ respectively, or rather with
the 3-dimensional random percolation exponent 
$\nu_{RP}=0.8765$.
Figs. \ref{Scalono2} and \ref{Scalono4} show that,
by taking the thermal exponents, 
the curves fall on top of each other, confirming
that $\nu_{perc}=\nu_{therm}$ in both cases.

\clearpage

To complete 
our investigation, we 
have determined the critical exponents' ratios 
making use, as usually, of standard finite size scaling techniques.
We list all the critical indices relative 
to the percolation transition for $O(2)$ and $O(4)$
in Tables \ref{o2ta} and \ref{o4ta}, respectively.
In the tables we have reported, for comparison,
the values of the thermal critical indices.
The agreement with the physical values 
in \cite{O2,Manf,O4} is good.
\vskip0.2cm
\begin{table}[h]
\begin{center}
\begin{tabular}{|c||c|c|c|c|}
\hline
& $\beta_{c}$ & ${\beta}/{\nu}$ & ${\gamma}/{\nu}$&$\nu$ \\
\hline\hline
$\vphantom{\displaystyle\frac{1}{1}}$ 
Thermal results \cite{O2}& 0.454165(4)& 0.5189(3)
& 1.9619(5)&0.6723(3)\\
\hline$\vphantom{\displaystyle\frac{1}{1}}$
Percolation results & 0.45418(2)& 0.516(5)
& 1.971(15)& 0.670(4)\\
\hline
\end{tabular}
\caption[Comparison of the thermal and percolation thresholds and
  exponents for $O(2)$]
{\label{o2ta} Comparison of the thermal and percolation thresholds and
  exponents for $O(2)$.}
\end{center}
\end{table}
\vskip0.2cm
\begin{table}[h]
\begin{center}
\begin{tabular}{|c||c|c|c|c|}
\hline
& $\beta_{c}$ & ${\beta}/{\nu}$ & ${\gamma}/{\nu}$&$\nu$ \\
\hline\hline
$\vphantom{\displaystyle\frac{1}{1}}$ 
Thermal results & 0.93590(5)\cite{Manf}& 0.5129(11)\cite{O4}
&1.9746(38)\cite{O4} &0.7479(90)\cite{O4}\\
\hline$\vphantom{\displaystyle\frac{1}{1}}$
Percolation results & 0.93595(3)&  0.515(5)
&1.961(15) & 0.751(6)\\
\hline
\end{tabular}
\caption[Comparison of the thermal and percolation thresholds and
  exponents for $O(4)$]
{\label{o4ta} Comparison of the thermal and percolation thresholds and
  exponents for $O(4)$.}
\end{center}
\end{table}
So far we have presented the results obtained
using the first approach. The results derived using 
the second approach are essentially the same; besides,
we observe an improved quality of the scaling, mainly because 
of the use of periodic boundary conditions, which
reduce considerably the finite size effects.

\begin{figure}[h]
\begin{center}
\epsfig{file=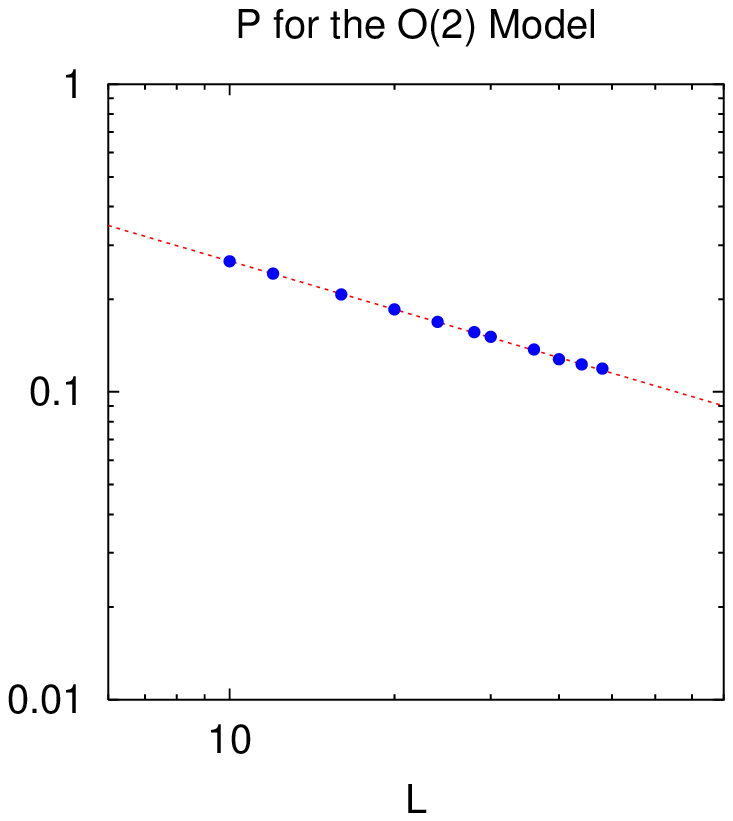,width=7.5cm}
\epsfig{file=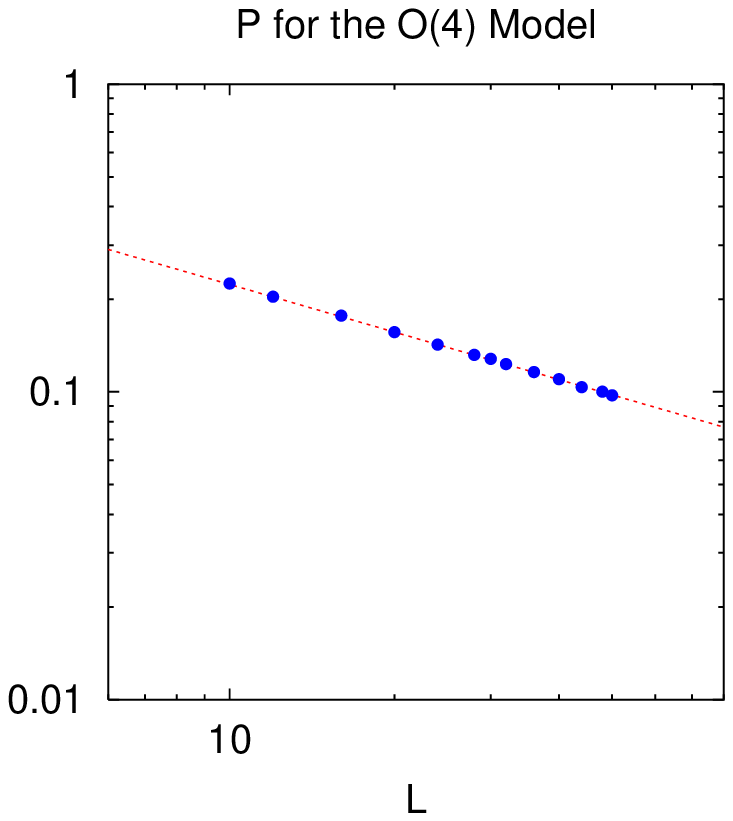,width=7.5cm}
\caption[Finite size scaling plot at $T_c$ of the 
percolation strength $P$ for $O(2)$ and $O(4)$]
{
Finite size scaling plot at $T_c$ for the percolation observable $P$ as
a function of the lattice size $L$. The slopes in the plots correspond
to $\beta/\nu = 0.521(3), 0.513(6)$ respectively for O(2) and O(4).
The errors on the data points are smaller than the size of the points in the plot.}
\label{Popl}
\end{center}
\end{figure}

\begin{figure}[h]
\begin{center}
\epsfig{file=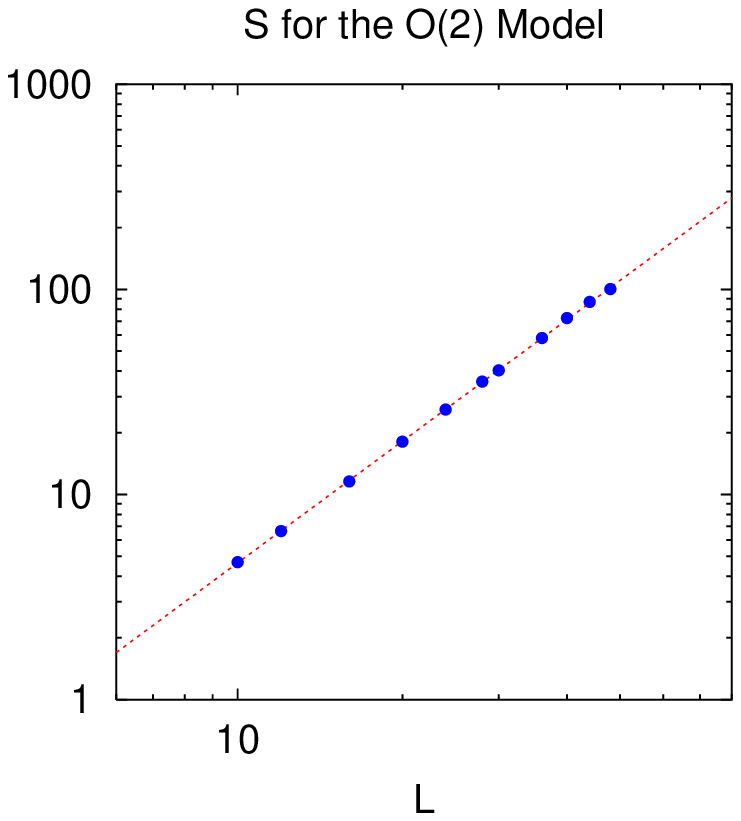,width=7.5cm}
\epsfig{file=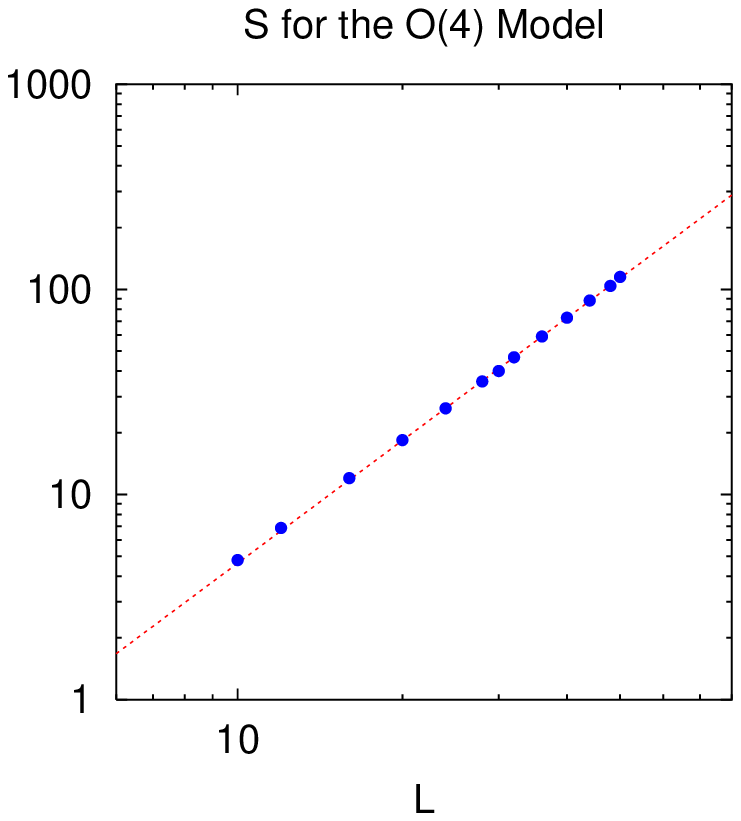,width=7.5cm}
\caption[Finite size scaling plot at $T_c$ of the 
average cluster size $S$ for $O(2)$ and $O(4)$]{
Finite size scaling plot at $T_c$ for the percolation observable $S$ as
a function of the lattice size $L$. The slopes in the plots correspond
to $\gamma/\nu = 1.97(1), 1.99(1)$ respectively for O(2) and O(4).
The two curves look surprisingly similar to each other.
The errors on the data points are smaller than the size of the points in the plot.}
\label{Sopl}
\end{center}
\end{figure}

In particular we show
in Figs. \ref{Popl}, \ref{Sopl} 
the scaling of $P$ and $S$ 
at the thermal thresholds reported in \cite{O2,Manf}.
We observe very small finite size effects (lattices of $L\ge 20$ are
used in the fits), especially for the $O(2)$ case, 
which is in contrast to what 
is observed for thermal observables \cite{engels}.
The slopes 
of the straight lines are in agreement with the 
values of the thermal exponents' ratios ${\beta}/{\nu}$,
${\gamma}/{\nu}$.
\vskip1.3cm 
In conclusion, we have shown that the spontaneous breaking of 
the continuous rotational symmetry for the 3-dimensional $O(2)$ 
and $O(4)$ spin models can be described as percolation of 
Wolff clusters.
In both cases, the number $n$ of components of the spin vectors 
{\bf s} does not seem to play a role; the result is thus likely to be valid
for any $O(n)$ model. 

\chapter{Polyakov Loop Percolation in SU(2) Gauge Theory}\label{Polperc}

\setcounter{footnote}{1}
\thispagestyle{empty}


\def\J{$J/\psi$}
\def\j{J/\psi}
\def\X{$\chi$}
\def\x{\chi}
\def\P{$\psi'$}
\def\p{\psi'}
\def\U{$\Upsilon$}
\def\u{\Upsilon}
\def\C{c{\bar c}}
\def\cg{c{\bar c}\!-\!g}
\def\bg{b{\bar b}\!-\!g}
\def\b{b{\bar b}}
\def\q{q{\bar q}}
\def\Q{Q{\bar Q}}
\def\L{\Lambda_{\rm QCD}}
\def\A{$A_{\rm cl}$}
\def\a{A_{\rm cl}}
\def\N{$n_{\rm cl}$}
\def\n{n_{\rm cl}}
\def\S{S_{\rm cl}}
\def\s{s_{\rm cl}}
\def\bb{\bar \beta}

\def\be{\begin{equation}}
\def\ee{\end{equation}}

\def\lsim{\raise0.3ex\hbox{$<$\kern-0.75em\raise-1.1ex\hbox{$\sim$}}}
\def\gsim{\raise0.3ex\hbox{$>$\kern-0.75em\raise-1.1ex\hbox{$\sim$}}}


\def\CMP{{ Comm.\ Math.\ Phys.\ }}
\def\NP{{ Nucl.\ Phys.\ }}
\def\PL{{ Phys.\ Lett.\ }}
\def\PR{{ Phys.\ Rev.\ }}
\def\PRep{{ Phys.\ Rep.\ }}
\def\PRL{{ Phys.\ Rev.\ Lett.\ }}
\def\RMP{{ Rev.\ Mod.\ Phys.\ }}
\def\ZP{{ Z.\ Phys.\ }}

\section{Finite Temperature SU(N) on the lattice}\label{giorgio}

Finite temperature Quantum
Chromodynamics ($QCD$) has been extensively simulated on the lattice 
over the last two decades, in order to test the hypothesis that, 
at high temperatures and (or) high densities, quark matter should pass
from the (confined) hadronic phase to the (deconfined) phase represented by a 
plasma of quarks and gluons.

$QCD$, like all theories describing
fundamental forces (except gravitation),
is a quantum field theory with local gauge invariance.
The gauge group which rules this invariance
is $SU(3)$: the quarks form a triplet in the fundamental
representation of $SU(3)$, and the gluons, which are
the carriers of the interaction, form an octet in the 
adjoint representation.
    
The non-abelian character of  
the $SU(3)$ group leads to an important 
feature that distinguishes $QCD$ from 
Quantum Electrodynamics ($QED$), which
is ruled by the abelian group $U(1)$: the gluons 
carry a charge and 
can interact among each others, in contrast to
the photons. Therefore, it makes sense
to study systems constitued only by gluons, and to
check whether the interaction gives rise to 
a confinement-deconfinement transition from
a phase in which the gluons are bound in 
glueballs to a phase of free gluons.
This simpler situation is described by the so-called    
{\it pure gauge} $SU(3)$.  

Since any $SU(N)$ group is non abelian, 
the study of the relative pure gauge theories
may be of interest also for $N{\neq}3$.

The Lagrangian density of the $SU(N)$ pure gauge theories is 
\begin{equation}
\label{yamil}
{\cal L}\,=\,-\frac{1}{4}F^{a}_{\mu\nu}(x)\,F^{a\mu\nu}(x), 
\end{equation}
where 
\begin{equation}
\label{field}
F^{a}_{\mu\nu}(x)\,{\equiv}\,\partial_{\mu}A^{a}_{\nu}(x)-\partial_{\nu}A^{a}_{\mu}(x)+
g\,f^{abc}\,A^{b}_{\mu}(x)A^{c}_{\nu}(x).
\end{equation}
Here $A^{a}_{\mu}$ are the gauge fields
($a=1,2,..., N^2-1$), $g$ is the gauge coupling constant
and $f_{abc}$ the structure constants of the $SU(N)$ group. The rightmost
term of (\ref{field}), present because 
for the $SU(N)$ group $f_{abc}\,{\neq}\,0$,
is responsible of the gluon-gluon interaction.

The renormalizability of $SU(N)$ gauge theories \cite{hooft}
assures
the convergence of  
perturbative series expansions: 
$QCD$ was established as 
theory of the strong interaction after 
a great deal of perturbative 
results were confirmed by experiments.
However, the important phenomenon of
confinement lies
well beyond the realm of perturbation theory.

The need of getting predictions 
from the theory in the non-perturbative 
domain led to an alternative calculation pattern,
the {\it lattice regularization}, characterized by a discretization
of space-time which gets rid automatically
of troublesome divergences \cite{wil}.  

With the Lagrangian density (\ref{yamil}) provided,
the formulation of statistical $SU(N)$ theories is,
at least in principle, a well-defined problem. We have to calculate
the partition function
\begin{equation}
\label{part}
{\cal Z}(\beta,V)\,=\,Tr\{e^{-{\beta}H}\}.
\end{equation}
In the trace we have to sum over all physical
states accessible to the system in a spatial volume $V$;
$\beta=1/T$, where $T$ is the physical temperature and
$H$ is the Hamiltonian of the system, which can be expressed
by means of ${\cal L}$.
Once we have ${\cal Z}(\beta,V)$, we can proceed to 
derive all thermodynamic observables. Thus,
\begin{equation}
\label{enden}
\epsilon\,=\,-\frac{1}{V}\Big(\frac{{\partial}\,ln{\cal Z}}{\partial\beta}\Big)_{V}
\end{equation}
is the energy density, and
\begin{equation}
\label{pressure}
P\,=\,\frac{1}{\beta}\Big(\frac{{\partial}\,ln{\cal Z}}{\partial{V}}\Big)_{\beta}
\end{equation}
gives us the pressure.

The lattice formulation of statistical $SU(N)$
is obtained in three steps. First we replace 
the Hamiltonian form (Eq. (\ref{part})) of the partition
function by the corresponding Euclidean functional integral
\begin{equation}
\label{euclide}
{\cal Z}_{E}(\beta,V)\,=\,\int\limits_{A_{\mu}(\beta,\vec{x})=
A_{\mu}(0,\vec{x})}\,{(dA)}\,\exp\Big[-\int_{V}{d^3}x\int^{\beta}_{0}d\tau\,{\cal L}(A)\Big].
\end{equation}
This form involves directly the Lagrangian
density and the sum over states in (\ref{part}) is replaced by the 
integration over the field configurations ${A}$.
The periodicity condition $A_{\mu}(\beta,\vec{x})=
A_{\mu}(0,\vec{x})$ is a consequence of the trace form of 
Eq. (\ref{part}). The spatial integration of (\ref{euclide})
is performed over the whole volume of the system, while in the
imaginary time $\tau\,{\equiv}\,ix_0$, the integration 
runs over a finite slice determined by the temperature. 
The finite temperature behaviour of the partition function thus becomes a finite
size effect in the integration over $\tau$. 

Next, the Euclidean $\vec{x}-\tau$ manifold is replaced
by a discrete lattice, with $N_{\sigma}$ points 
in each space direction and $N_{\tau}$ points 
for the $\tau$ axis. The lattice spacing is $a$.
The overall space volume becomes $V\,=\,(N_{\sigma}a)^3$, the inverse temperature
$\beta^{-1}\,=\,N_{\tau}a$. To ensure the gauge invariance
of the formulation, the gauge fields $A$ must be defined
on the links connecting each pair of adjacent sites.

In the final step, the integration over
the gluon fields is replaced by one over the corresponding gauge group
variables,
or {\it link variables},
\begin{equation}
\label{link}
U_{ij}\,=\,\exp\Big[-ig(x_i-x_j)^{\mu}A_{\mu}\Big(\frac{x_i+x_j}{2}\Big)\Big],
\end{equation}
with $x_i$ and $x_j$ denoting two adjacent lattice sites, so that
$U_{ij}$ is an $SU(N)$ matrix associated to the links between these two sites.

The partition function of finite temperature $SU(N)$ pure gauge theories
takes then the form
\begin{equation}
\label{lapart}
{\cal Z}(N_{\sigma},N_{\tau};g^2)\,=\,\int\prod\limits_{links}\,dU_{ij}\,\exp[-S(U)],
\end{equation}
where $S(U)$ is the {\it Wilson action}
\begin{equation}
\label{plaq}
S(U)\,=\,\frac{2N}{g^2}\sum\limits_{plaq}\Big(1-\frac{1}{N}Re\,Tr\,UUUU\Big).
\end{equation}
The sum is over all the smallest closed paths of the lattice ({\it plaquettes}),
which are formed by four links; $UUUU$ is 
the product of the link variables corresponding to each side
of a plaquette.

By letting the lattice spacing $a$ go to zero, one recovers
the continuum limit (\ref{euclide}). This assures
that, for $a$ small enough, the lattice regularization does not
influence the physical observables and that we can
rely on the results derived by this approach.

\section{Z(N) Symmetry and Deconfinement}\label{gilberto}
    
Pure $SU(N)$ gauge theories have a global symmetry,
resulting from the periodicity of the gauge
fields in the temperature direction, that rules the 
behaviour of the system at finite temperatures.    
Gauge transformations which are compatible with 
the periodicity condition need only 
be periodic up to an element $z$ of the center 
$Z(N)$ of the gauge group $SU(N)$. Thus a gauge transformation
must obey:
\begin{center}
\begin{equation}
\label{zn}
A(\vec{x},0)\,=\,zA(\vec{x},\beta),\,\,\,\,\,\,for\,\,all\,\,\vec{x},
\end{equation}
\end{center}
where $A(\vec{x},\tau)$ is the 
$SU(N)$ matrix associated  
to the gauge field at the point $(\vec{x},\tau)$ and
$z\,\in\,Z(N)$. The $n$-th element of $Z(N)$ is given as
$\exp(2{\pi}in/N)\,\,(n=0,..., N-1)$. 
It is easy to see that, under (\ref{zn}),
the action (\ref{plaq}) remains unchanged. In contrast, 
the {\it Polyakov loop},
\begin{equation}
\label{pollo}
L_{\vec{x}}\,=\,\frac{1}{N}\,Tr\,\prod\limits_{t=1}^{N_{\tau}}\,U_{\vec{x};t,t+1},
\end{equation}
consisting of the product of all
the $U$'s in the temperature direction
taken at a given spatial site $\vec{x}$,
transforms non-trivially under this transformation,
\begin{equation}
\label{potra}
L_{\vec{x}}\,\rightarrow\,z\,L_{\vec{x}}.
\end{equation}
The same relation is valid if we take the average $L={\langle}L_{\vec{x}}\rangle$
over the lattice and over configurations. $L$
is an indicator of the state in which the system finds itself.
It is clear that, if $L\,\neq\,0$,
the transformation (\ref{zn}) will {\it not} leave invariant
the value of $L$, as it would
happen if $L\,=\,0$. 
That means that the state of the system may spontaneously break
the global $Z(N)$ symmetry, just as the ordered phase of the Ising model
breaks the global $Z(2)$ symmetry of its Hamiltonian. 

The quantity $L$ is then the {\it order parameter}
of the phase transition associated to the spontaneous 
breaking of the $Z(N)$ symmetry. Is this transition 
somehow related to deconfinement?

The state of a gluons system can be probed qualitativey by
a heavy test quark. The free energy $F$ of this test quark 
should be infinite in the confinement phase, but finite in the
deconfinement phase. It turns out that
such free energy is related to the lattice average of the
Polyakov loop by the following expression
\begin{equation}
\label{freen}
L\,\propto\,e^{-\beta\,F}.
\end{equation}
In the confinement regime, $F\,=\,\infty$ and Eq. (\ref{freen}) 
implies $L\,=\,0$. If the gluons
are free, $F$ is finite and, consequently, $L\,\neq\,0$.
The (eventual) transition
from the confined to the deconfined state of the 
$SU(N)$ gauge system is thus characterized by
the spontaneous breaking of the global center $Z(N)$ symmetry.

Lattice studies have shown that this phase transition
indeed takes place. 
The first computer 
simulations of finite temperature lattice gauge theories 
were performed in the early $80$'s and
concerned the $SU(2)$ theory \cite{creu},
basically because it is the simplest one and 
the relative simulations are not so lengthy as 
the $SU(3)$ simulations.

\begin{figure}[h]
\begin{center}
\epsfig{file=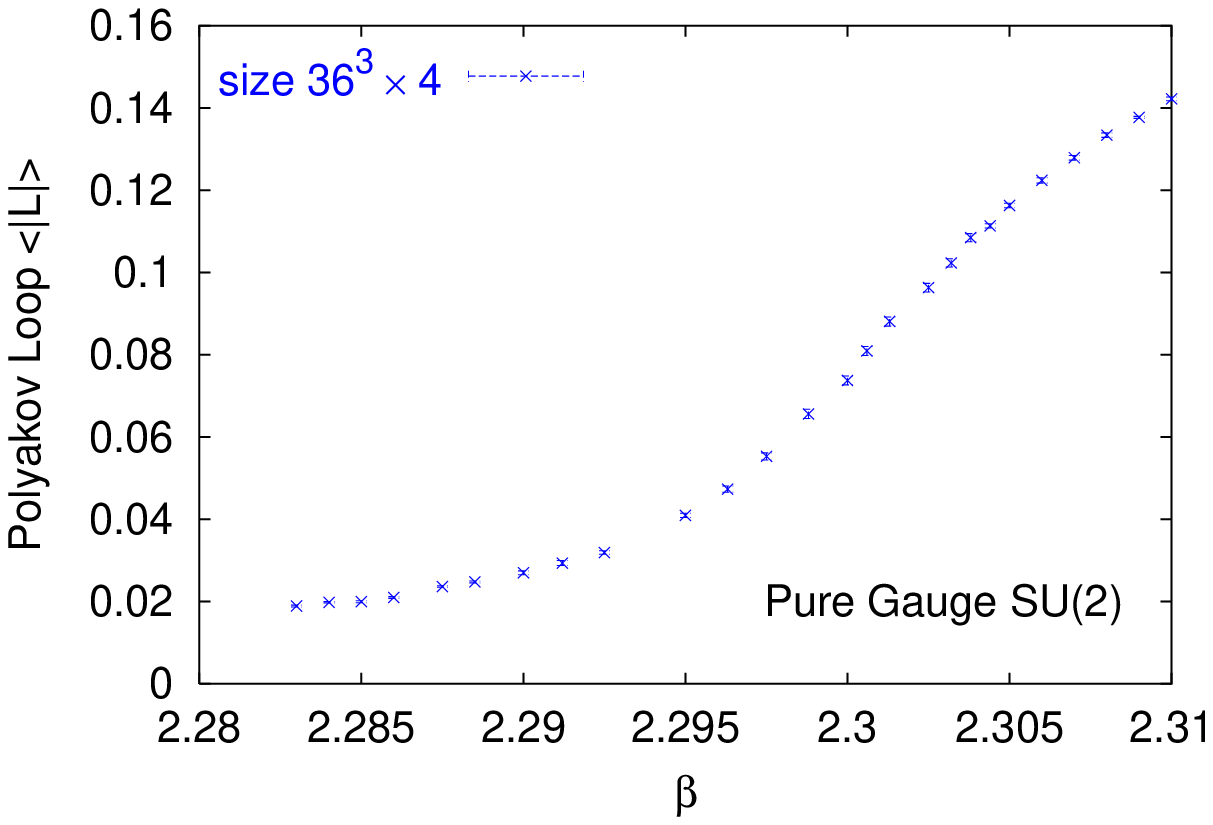,  width=14cm}
\caption[Polyakov loop
as a function of $\beta=4/{g^2}$
for pure gauge $SU(2)$ on a $36^3{\times}4$ lattice]
{Lattice average of the Polyakov loop
as a function of the coupling $\beta=4/{g^2}$
for pure gauge $SU(2)$ on a $N_{\sigma}^3{\times}N_{\tau}$
lattice with $N_{\sigma}=36$ and $N_{\tau}=4$.}
\label{su2data}
\vskip1cm
\epsfig{file=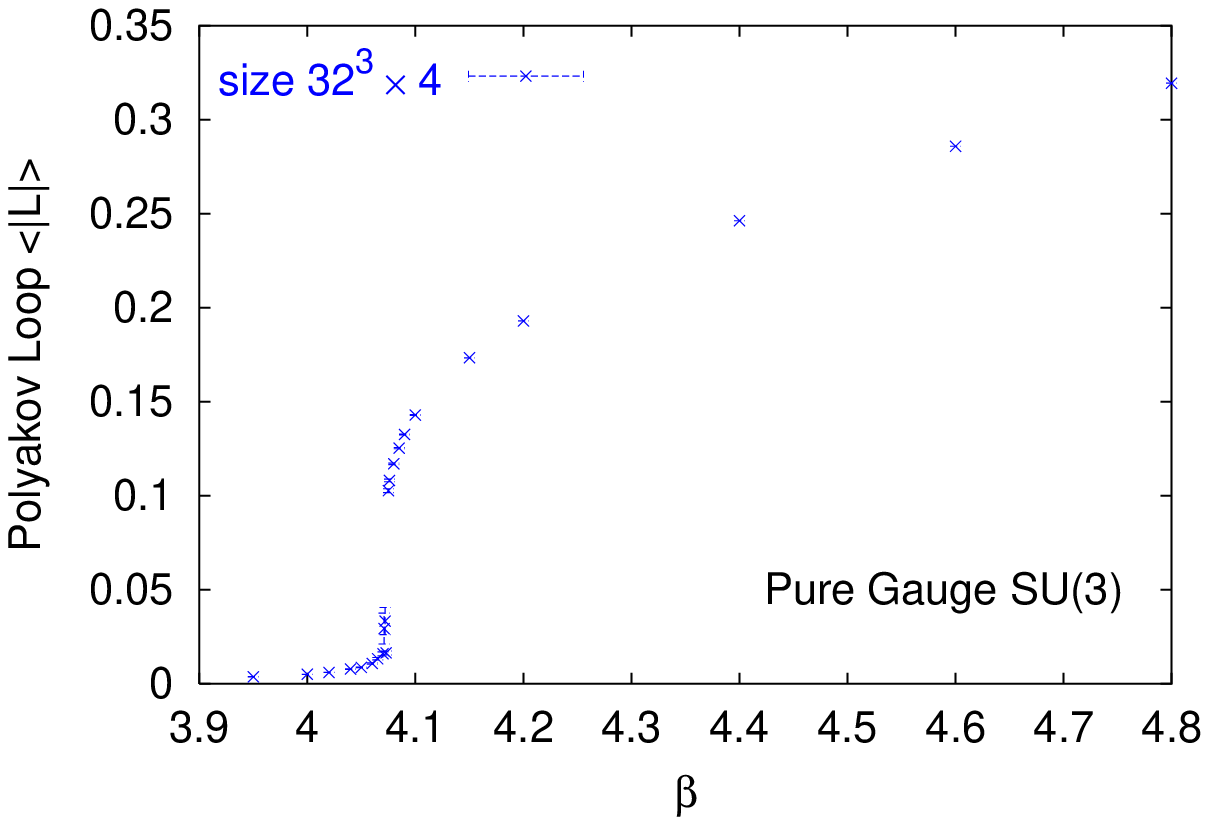,  width=14cm}
\caption[Polyakov loop 
as a function of $\beta=6/{g^2}$
for pure gauge $SU(3)$ on a $32^3{\times}4$
lattice]
{Polyakov loop 
as a function of the coupling $\beta=6/{g^2}$
for pure gauge $SU(3)$ on a $32^3{\times}4$
lattice; 
the data points are taken from \cite{okacz}.}
\label{su3data}
\end{center}
\end{figure}

Fig. \ref{su2data} 
shows $SU(2)$ data relative  
to the Polyakov loop, from which the 
typical behaviour of a second order phase transition is
clearly visible. 

In spite of their higher complexity, $SU(3)$
simulations could be performed shortly after the $SU(2)$ ones.
However, it took a while before
one could be sure to understand what was happening there.
Now it is well established that $SU(3)$ gauge theory undergoes
a (weak) first order confinement-deconfinement phase transition
(Fig. \ref{su3data}).

\clearpage

\section{SU(N) Gauge Theories vs Z(N) Spin Models}\label{buda}

In the previous section we have stressed the 
key role played by
the $Z(N)$ symmetry 
in the 
confinement-deconfinement mechanism of $SU(N)$
gauge theories. 
It is easy to check that 
any $Z(N)$ rotation leaves invariant
the product of link variables 
around each closed path which includes 
spatial links ({\it Wilson loops}). 
However, we have seen that the topologically nontrivial
Polyakov loop $L_{\vec{x}}$ changes after such rotations (see Eq. (\ref{potra})).
If we consider that, from Eq. (\ref{pollo}), $L_{\vec{x}}$ is 
determined {\it only} by the links in the imaginary time/temperature 
direction, we realize that the $Z(N)$ symmetry 
introduces a distinction between 
spatial links and time links: the latter ones alone seem 
to have the control of the phase transition.

It is then natural to try to get rid of the
`irrelevant' degrees of freedom represented
by the spatial links and to express the 
$SU(N)$ action in terms only of the order parameter
field $L_{\vec{x}}$. 
This can be achieved by integrating out the spatial gauge fields;
the resulting effective theory is simpler
than the original model and could facilitate the investigation
of the confinement-deconfinement phase transition. 

Svetitsky and Yaffe \cite{svetitsky} presented a number
of arguments that lead to some 
interesting conclusions about the properties of such
$SU(N)$ effective theories.
Their arguments are essentially based on 
dynamical considerations and simple renormalization group ideas and
allow to deduce, among other things, the 
order of the phase transition and, in case of a continuous 
transition, the exponents which characterize the
critical fluctuations.

The main point of \cite{svetitsky} is the 
hypothesis that, integrating out all degrees of freedom
except the Polyakov loops, one yields an effective theory with short range 
interactions, which is invariant under the center $Z(N)$ symmetry.
In case of a continuous transition, one could in principle
locate the renormalization fixed point which governs the relative
critical behaviour.
If it happens that, in the space of $d$-dimensional theories with short range
interactions invariant under the center symmetry, there exists only a single fixed
point, then the critical behaviour of the original
$(d+1)$-dimensional finite-temperature $SU(N)$ gauge theory
{\it will be the same as that of simple $d$-dimensional
spin models invariant under the same global symmetry $Z(N)$}. 

In many cases theories, related by the same symmetry,
are indeed associated to a single fixed point: this implies
that results for the critical behaviour of simple spin models 
may be used to predict the critical behaviour
in finite-temperature gauge theories.

It is, for example, the case of the $Z(2)$ symmetry, for which
only one fixed point is known. Consequently, 
since $SU(2)$ pure gauge theory undergoes 
a continuous transition, its critical exponents should
fall in the universality class of the $Z(2)$ spin model,
which is the Ising model. Lattice studies have provided
strong evidence that the conjecture is indeed true \cite{engles}. 

As a counter-example, in the space of 
three-dimensional $Z(3)$ symmetric theories,
no stable renormalization group fixed point is known.
That led the authors of \cite{svetitsky} to 
conjecture that $(3+1)$-$d$ $SU(3)$ pure gauge theory undergoes
a first order phase transition, as it was successively
confirmed by lattice simulations (see Section \ref{gilberto}).

\section{Polyakov Loop Percolation}\label{aurelio}

$Z(N)$ spin models have been extensively investigated,
both analytically and numerically. It is 
natural to check to which extent the 
analogy between such models and $SU(N)$ gauge theories
is valid. 

In particular, we know that the critical behaviour of the Ising
model can be equivalently described as percolation of 
well defined clusters of like-sign spins (see Chapter 2).
It is thus spontaneous to ask ourselves whether it is possible
to find a description of critical behaviour 
in terms of percolation also
for the deconfinement transition in $SU(2)$ gauge theory.

This could provide a suggestive view 
of the confinement-deconfinement mechanism.
If we take a typical SU(2) configuration at a certain temperature, there
will be areas where 
the Polyakov loop $L$ takes negative values, and areas where $L$ takes 
positive values. Both the positive and the negative 'islands' can be 
seen as local regions of deconfinement. But as long as there are finite 
islands of both signs, deconfinement remains a local phenomenon and the
whole system is in the confined phase. When one of this islands 
percolates, that
is it becomes infinite, then we can talk of deconfinement as a global phase of the 
system.

The rest of this chapter is devoted
to find a solution to this problem.
The main difficulty is the fact that 
the $SU(2)$ Lagrangian is not directly a function of the 
Polyakov loop, due to the presence
of the spatial gauge fields. 
Moreover, if we integrate out the 
spatial link variables, we yield 
an effective theory that contains non trivial 
combination of operators, which cannot
in general be replaced by suitable 
combinations of Polyakov loops. 

Therefore it is not 
clear how one can extract the expression of the
bond probability which is necessary to build the clusters like 
in the Ising model. 

The only way to face the problem is to
try to approximate $SU(2)$ by means of effective theories
which are easy to handle. This
will allow us to exploit the percolation pictures
for general spin models presented in Chapter 3.

We propose two alternative 
procedures to look for a suitable definition of 
cluster building. 

The {\it first approach} adopts a 
Polyakov loop effective theory derived by means of series 
expansions of the $SU(2)$ partition function 
in the strong coupling limit. 

The {\it second approach}
searches for simple Ising-like spin models 
which approximate the Ising-projected 
Polyakov loop configurations and, even if it 
is more involved than the first one, 
it can also be applied in a case which
approaches 
the weak coupling region.

\section{First Approach: Strong Coupling Expansions}\label{monica} 

\subsection{The Green-Karsch Effective Theory}\label{monaco}

One of the most successful
techniques adopted to 
deduce results from the lattice formulation
of gauge theories is the so-called
{\it strong coupling expansion}, which consists in
expanding quantities like the action, the partition function, etc.,
in powers of the inverse coupling $1/g^2$. 
This procedure, analogous as the  
well known high-temperature expansions in 
statistical physics, allows to obtain 
interesting information about the system.
It was by means of analyses of the 
strong coupling limit that Polyakov \cite{polya}
and Susskind \cite{suss} could show 
for the first time that $QCD$ 
may lose its confining property, if the temperature is
sufficiently high.

We present here a strong coupling expansion
of $SU(N)$ gauge theories derived by
Green and Karsch \cite{Ka}.
Their aim was to perform a 
mean field analysis of the $SU(N)$
deconfinement transition in the presence of 
dynamical quarks, but we will limit 
ourselves to introduce the expressions relative to the 
pure gauge sector, in particular
to pure gauge $SU(2)$, which is the one
we are interested in.

We start from the formula (\ref{lapart}) for the lattice action. 
We can write
\begin{eqnarray}
\label{actplaq}
S(U)\,&=&\,\sum\limits_{P}\,S_{P}(U) \nonumber\\
S_{P}(U)\,&=&\,\frac{4}{g^2}\Big(1-\frac{1}{2}Re\,Tr\,UUUU\Big)
\end{eqnarray}
where the action $S(U)$ is divided in 
the contributions $S_{P}(U)$ coming from each plaquette $P$.
$S_{P}(U)$ can be expanded \cite{gre} in terms of 
the characters $\chi_{r}$ of the $SU(2)$ group ($r$ is an integer which indicates the 
representation of the group)
\begin{equation}
\label{charact}
e^{-S_P}\,=\,{\cal Z}_{0}\Big(\frac{1}{g^2}\Big)\Big[1+\sum_{r}d_{r}z_{r}\Big(\frac{1}{g^2}\Big)\chi_{r}(U_P)\Big].
\end{equation}
In (\ref{charact})
$d_r=r+1$, $z_{r}(1/{g^2})=I_{r+1}(4/{g^2})/I_{1}(4/{g^2})$ and 
${\cal Z}_{0}={g^2}{I_1}(4/{g^2})/2$, where the $I_r$ are
the modified Bessel functions.

Next, we remark that we can neglect
all spacelike plaquettes $P_s$ by 
setting $U_{P_s}=1$ without affecting appreciably the 
critical behaviour of the system,
as long as the coupling $\beta=4/{g^2}$ is small. The
validity of this 
approximation, which corresponds to 
dropping the magnetic term in the hamiltonian of the theory, 
relies on the fact that spacelike plaquettes tend to
decrease the string tension (see \cite{gre}). Hence if a phase
transition is found in the strongly coupled theory,
there is almost certainly one in the full theory.

For little values of $\beta$ we can thus write the 
$SU(2)$ partition function 
\begin{equation}
\label{parchara}
{\cal Z}_{eff}\,=\,\int\,[dU]\,\prod\limits_{P_t}
\Big[1+\sum_{r}d_{r}z_{r}\Big(\frac{1}{g^2}\Big)\chi_{r}(U_{P_t})\Big],
\end{equation}
where the product is exclusively over the timelike plaquettes $P_t$.
Integrating over the spacelike links and 
grouping the timelike links associated to the
same spatial site ${\bf x}$, we easily get
\begin{equation}
\label{parpol}
{\cal Z}_{eff}\,=\,\int\,\prod\limits_{\bf x}\,dW_{\bf x}\,\prod\limits_{{\bf x},{\bf e}}
\Big[1+\sum_{r}z^{N_{\tau}}_{r}\Big(\frac{1}{g^2}\Big)\chi_{r}(W_{\bf x})
\chi_{r}(W^{\dagger}_{{\bf x}+{\bf e}})\Big].
\end{equation}
In the expression above $\{{\bf x},{\bf e}\}$ indicates a link,
$N_{\tau}$ is the number of lattice spacings in the
temperature direction and $W_{\bf x}$
the Wilson line variable
\begin{equation}
\label{willine}
W_{\bf x}\,=\,\prod\limits_{t=1}^{N_{\tau}}\,U_{{\bf x};t,t+1}.
\end{equation}
We stress that the the original $(d+1)$-dimensional lattice 
has now become a simple $d$-dimensional lattice; the 
first product in Eq. (\ref{parpol})
runs over its sites, the second one over its links.

If $\beta$ is small enough, we can keep only the fundamental
$r=1$ term of the expansion, and we finally get
\begin{equation}
\label{finpar}
{\cal Z}_{eff}\,\approx\,\int\,\prod\limits_{\bf x}\,dW_{\bf x}\,
\exp\Big[\beta^{\prime}\sum_{ij}\,L_iL_j\Big],
\end{equation}
with $\beta^{\prime}=4z^{N_{\tau}}_1$ and 
$L_i$ the value of the Polyakov loop at the site $i$
(see Eq. (\ref{pollo})); the sum is over nearest neighbours.
For $\beta$ small,
\begin{equation}
\label{effcou}
z_1\Big(\frac{1}{g^2}\Big)\,=\,\frac{I_2\Big(\frac{4}{g^2}\Big)}{I_1\Big(\frac{4}{g^2}\Big)}
\,\approx\,\frac{1}{g^2}\,=\,\frac{\beta}{4}.
\end{equation}
The final expression for the 
coupling $\beta^{\prime}$ of the effective theory is then
\begin{equation}
\label{effcoupl}
\beta^{\prime}\,=\,4{\Big(\frac{\beta}{4}\Big)}^{N_{\tau}}.
\end{equation}
The partition function ${\cal Z}_{eff}$ of 
Eq. (\ref{finpar}) looks very much like
the one of a spin model with simple nearest-neighbour interactions,
with the Polyakov loop playing the role of the spin variable.
There is, however, an essential difference:
the integration variables
in ${\cal Z}_{eff}$ are not the Polyakov loops $L_{\bf x}$, but
the Wilson line operators $W_{\bf x}$, which are $SU(2)$ matrices.
We know that
\begin{equation}
\label{wilpol}
L_{\bf x}\,=\,\frac{1}{2}Tr\,W_{\bf x},
\end{equation}
but it is not clear whether we can
rewrite the sum in Eq. (\ref{finpar}) as a sum
over Polyakov loops {\it only}.

The properties of the $SU(2)$ 
group may help us to solve the 
problem.
If $U$ is an $SU(2)$ matrix, 
we can use a parametrization in terms of
an angle $\phi$ and a 3-dimensional unit vector $\vec n$:
\begin{equation}
\label{su2param}
U\,=\,e^{i\phi{\vec n}\cdot{{\vec\tau}}/2}\,=\,\Big(\cos\,\frac{\phi}{2}\Big){\bf 1}\,+\,
i\,\Big(\sin\,\frac{\phi}{2}\Big)\,{\vec n}\cdot{{\vec\tau}},
\,\,\,\,\,\,\,\,\,\,\,\,\,0\leq\phi<\,2\pi,\,\, |{\vec n}|=1.
\end{equation}
In (\ref{su2param}), $\vec\tau$ are the Pauli
matrices. According to Eq. (\ref{su2param}), one gets
\begin{equation}
\label{trsu2}
Tr\,U\,=\,2\,\cos{\Big(\frac{\phi}{2}\Big)}
\end{equation}
So, the trace of $U$  
depends only on the angle $\phi$.
With this parametrization, 
the integral over $SU(2)$ matrices can be written as
\begin{equation}
\label{intsu2}
dU\,=\,\frac{1}{4\pi^2}\,d\phi\,d\Omega(\vec n)\,\Big(\sin\,\frac{\phi}{2}\Big)^2,
\end{equation}
where $d\Omega(\vec n)$ is the measure relative
to the angles of $\vec n$. Using Eqs. (\ref{wilpol}), (\ref{trsu2})
and (\ref{intsu2}), we can express ${\cal Z}_{eff}$ 
in the following way
\begin{equation}
\label{finpartsu2}
{\cal Z}_{eff}\,\approx\,\int\,\prod\limits_{\bf x}\,
\frac{1}{4\pi^2}\,d\phi_{\bf x}\,d\Omega(\vec n_{\bf x})\,\Big(\sin\,\frac{\phi_{\bf x}}{2}\Big)^2
\exp\Big[\beta^{\prime}\sum_{ij}\,\cos\,\Big(\frac{\phi_{i}}{2}\Big)\,\cos\,\Big(\frac{\phi_{j}}{2}\Big)\Big],
\end{equation}
The exponential of Eq. (\ref{finpartsu2}) is only a function
of the angles $\phi_i$ associated to the 
Wilson line operators.
The angles of $d\Omega(\vec n)$ can thus be  
integrated out; since $L_i\,=\,\cos({\phi_i}/2)$
and $\sin({\phi_{\bf x}}/2)=\sqrt{1-{L_{\bf x}}^2}$, we reach the final expression
\begin{equation}
\label{finpasu2}
{\cal Z}_{eff}\,\approx\,\int\,\prod\limits_{\bf x}\,dL_{\bf x}\,
\sqrt{1-{L^2_{\bf x}}}\,\exp\Big[\beta^{\prime}\sum_{ij}\,L_iL_j\Big],
\end{equation}
in which we have 
neglected the irrelevant constant
factor due
to the integration over $d\Omega(\vec n)_{\bf x}$.

We stress that, to derive Eq. (\ref{finpasu2}),
we made use of two approximations. We have 
neglected the spacelike plaquettes and 
we have truncated the 
expansion of (\ref{parpol}) to the first 
term.
Both approximations rely on the fact that the coupling $\beta_c$, 
at which the transition occurs, is  
small enough. Since $\beta_c$ shifts to higher values the bigger  
the number of lattice spacings in the time direction, the assumptions
are valid only for small values of $N_{\tau}$.
Green and Karsch showed that 
the mean field analysis of the effective theory
of Eq. (\ref{finpar}) gives results which are compatible with
$SU(2)$ lattice simulations for $N_{\tau}=1,2$ \cite{Ka}.
We decided to concentrate ourselves to the
more interesting case, i. e.  $N_{\tau}=2$. 
  
Eq. (\ref{finpasu2}) is exactly the partition function
of one of the
continuous spin Ising models we have studied in Section \ref{manuele}, 
namely the model whose spin amplitudes $\{\sigma\}$
are distributed according to Eq. (\ref{3eq}). 
From Section \ref{manuele} we know that the critical
behaviour of the continuous Ising models has an equivalent percolation picture;
the clusters are formed by binding nearest neighbouring spins
of the same sign with the probability (\ref{genCK}).
We have also seen that the distribution (\ref{3eq})
does not play a role in the cluster definition.

Assuming that, for $N_{\tau}=2$, the Polyakov loop configurations 
of $SU(2)$ are ruled by the partition
function (\ref{finpasu2}), it is natural to 
test the same definition of clusters of the continuous
Ising model. In our case,
the clusters will be 
then formed by like-signed nearest neighbouring Polyakov
loops, bound with the probability 
\begin{equation}
p(i,j)=1-\exp(-2\beta^{\prime}\,L_i\,L_j).
\label{bowe}
\end{equation}
For $N_{\tau}=2$, from Eq. (\ref{effcoupl}) we get $\beta^{\prime}={\beta^2}/4$,
so that 
\begin{equation}
p(i,j)=1-\exp\Big(-\frac{\beta^2}{2}\,L_i\,L_j\Big).
\label{bowefin}
\end{equation}

With Eq. (\ref{bowefin}), the Polyakov loop percolation problem is
fully defined. We point out that the strong coupling
expansion we have shown is independent on the number
of space dimensions of the system, as long as the corresponding
values of the critical coupling $\beta_c$
remain small. Because of that, we 
decided to investigate $SU(2)$ both in 
$(2+1)$ and in $(3+1)$ dimensions,
to test our cluster definition in two
different cases.

\subsection{Numerical Results for (2+1)-d SU(2)}\label{berlino}

Our analysis is based on four sets of data on
$N_{\sigma}^2{\times}2$ lattices, with $N_{\sigma}$=64, 96, 128 and 160. 
The 
Monte Carlo update consists of one
heat bath and two overrelaxation steps. For the $64^2\times 2$ and
$96^2\times 2$ lattices we evaluated configurations every six updates,
for $128^2\times 2$ and $160^2\times 2$ every eight updates, measuring
in each case the percolation
strength $P$ and the average cluster size
$S$. The percolation variables are essentially uncorrelated. 

\begin{figure}[h]
\begin{center}
\epsfig{file=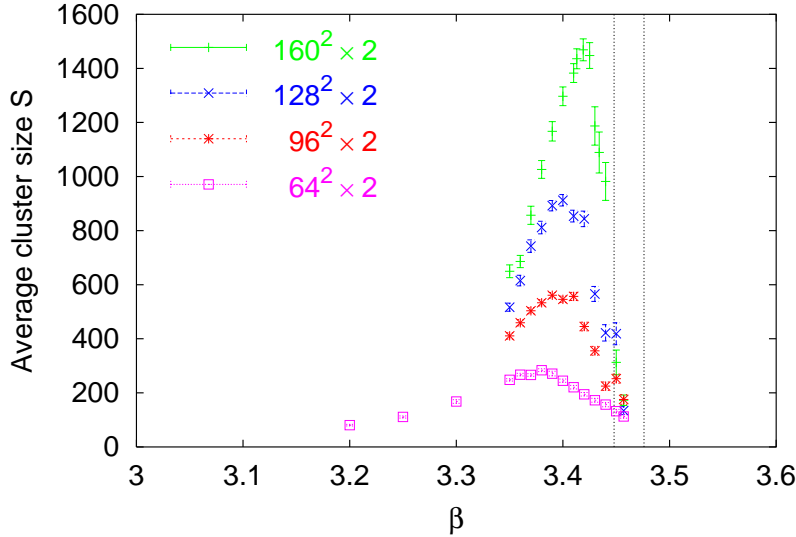,  width=12cm}
\caption[Average cluster size $S$ for 
$(2+1)$-$d$ $SU(2)$, $N_{\tau}=2$: first approach]
{$(2+1)$-$d$ $SU(2)$, $N_{\tau}=2$. 
Average cluster size $S$ as function of 
$\beta$ for four lattice sizes. The curves peak clearly
near the thermal threshold, represented by the dashed
lines (within one standard deviation), 
and tend to approach it the larger the size is.}
\label{SCK2D}
\end{center}
\end{figure}

A first scan for values $3.1 < \beta < 3.5$ leads to the behaviour
of $S$ shown in Fig. \ref{SCK2D}. It is seen that $S$ peaks slightly
below $\beta_c$; with increasing $N_{\sigma}$, the peak moves towards $\beta_c$.
Next, we carried out high-statistics simulations in a
narrower range $3.410 < {\beta} < 3.457$ around the transition. In
general, we performed between 30000 and 55000 measurements per $\beta$
value, with the higher number taken in the region of the interval
closest to the eventual critical point.
The high density of points near the 
threshold allows to determine quite precisely the 
critical indices after the usual finite size scaling analysis
that we have adopted many times in this work.
\begin{figure}[h]
\begin{center}
\epsfig{file=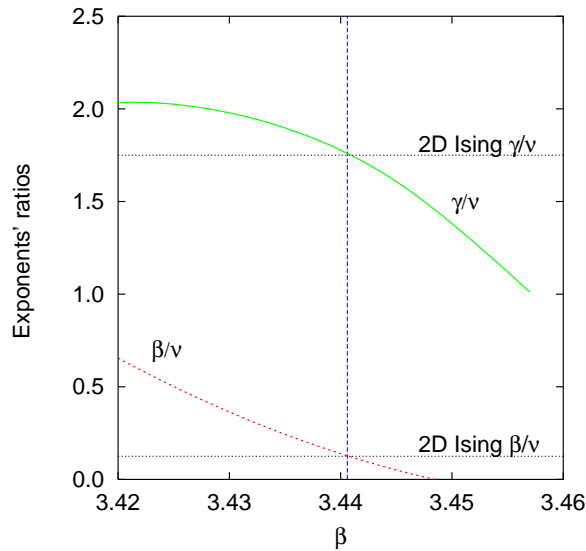,  width=9cm}
\caption[Critical exponents' ratios of finite size scaling fits for $P$
and $S$ in $(2+1)$-$d$ $SU(2)$, $N_{\tau}=2$: first approach]  
{$(2+1)$-$d$ $SU(2)$, $N_{\tau}=2$.
Critical exponents' ratios 
derived by the slope of finite size scaling fits for $P$
(red line) and $S$ (green line), plotted as a function
of the coupling $\beta$. 
The blue line
marks the point at which the $\chi^2$ is minimal.
The corresponding values of the critical exponents' ratios 
${\beta}/{\nu}$ and ${\gamma}/{\nu}$ (vertical axis in
the figure), are in good accord with the 2D Ising values,
represented by the horizontal dashed lines.}
\label{chisu2D}
\end{center}
\end{figure}
\vskip-0.5cm
In Fig. \ref{chisu2D} we present the variation
with $\beta$ of the values of the 
exponents' ratios 
obtained by log-log finite size scaling 
fits for $P$ and $S$. The $\beta$ value
corresponding to the best $\chi^2$ is indicated
by the vertical line in the figure. The values
of ${\beta}/{\nu}$ and ${\gamma}/{\nu}$ around
that point are, within errors,
in the universality class of the 
2D Ising model.  
The final results are reported in Table \ref{resu2D}.
\vskip0.2cm
\begin{table}[h]
  \begin{center}{
      \begin{tabular}{|c||c|c|c|c|}
\hline
  &      Critical point  & $\beta/{\nu}$ & ${\gamma}/{\nu}$ & $\nu$\\ \hline\hline
$\vphantom{\displaystyle\frac{1}{1}}$  Percolation results &
$3.443^{+0.001}_{-0.001}$&
$0.128^{+0.003}_{-0.005}$&$1.752^{+0.006}_{-0.008}$&$0.98^{+0.07}_{-0.04}$\\  
\hline $\vphantom{\displaystyle\frac{1}{1}}$ 
  Thermal results &$3.464^{+0.012}_{-0.016}$ & $1/8\,=\,0.125$&$7/4\,=\,1.75$
&1 \\ \hline
      \end{tabular}
      }
\caption[Thermal and percolation critical indices for 
$(2+1)$-$d$ $SU(2)$, $N_{\tau}=2$: first approach]
{\label{resu2D} 
Thermal and percolation critical indices for 
$(2+1)$-$d$ $SU(2)$, $N_{\tau}=2$. As exponents for the thermal
transition we adopted
the exact 2D Ising exponents, the value of the threshold
is taken from \cite{Teper}.} 
  \end{center}

\end{table}
The critical values of the coupling for the
thermal and the geometric transition are very close,
although they do not overlap within one standard deviation. 
In view of the inevitable approximations
involved by our procedure,
small deviations are not
unexpected. However, the fact that the critical percolation exponents
agree with the Ising values and not
with the 2D random percolation ones shows clearly that our clusters
are, with good approximation, the physical `droplets'
of the system.

\subsection{Numerical Results for (3+1)-d SU(2)}\label{parigi}

As finite temperature $SU(2)$ in $(3+1)$ dimensions 
is more interesting than
in $(2+1)$, because it describes a system
in the `real' $3$-dimensional 
space, we carried on a complete study of the 
model, analysing both 
the thermal and the geometrical transition.

\begin{figure}[h]
\begin{center}
\epsfig{file=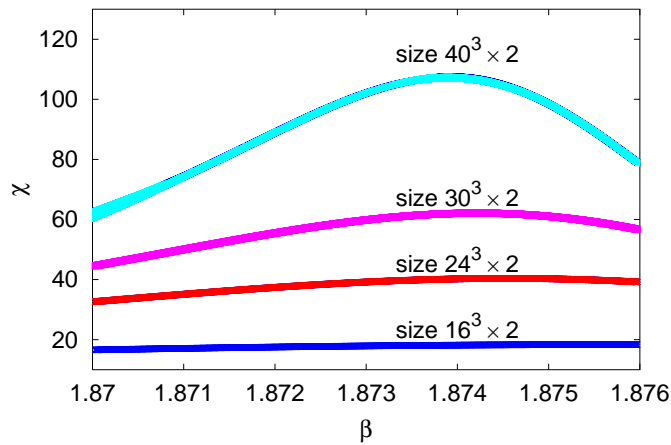,  width=10cm}
\caption[Physical susceptibility $\chi$ as function of $\beta$
for $(3+1)$-$d$ $SU(2)$, $N_{\tau}=2$]
{$(3+1)$-$d$ $SU(2)$, $N_{\tau}=2$.
Physical susceptibility $\chi$ as function of $\beta$ for four 
lattice sizes. For each curve we got 4000 interpolation points.}
\label{chisu23d}
\end{center}
\end{figure}

We
performed four sets of simulations
in correspondence to the following lattice sizes: $16^3\times 2$, $24^3\times 2$,  
$30^3\times 2$, $40^3\times 2$. 
The Monte Carlo update is the same we have used
in the previous case, i. e. it alternates heat bath  
and overrelaxation moves, in the ratio $1:2$. 
We evaluated configurations every ten updates for each lattice size
and value of the coupling $\beta$.
The percolation data are uncorrelated; the thermal 
variables instead show some important
correlation (the autocorrelation time $\tau$ is about $10$
for the magnetization on the $40^3{\times}2$ lattice near criticality).
The number of measurements varies from 10000 to 80000.
We used the density of states method (DSM) \cite{DSM}
to interpolate our data. 
Fig. \ref{chisu23d} shows the results of the 
interpolation for the physical susceptibility
\begin{equation}
\label{chiphysu2}
\chi = V~(\langle{L^2}\rangle-{\langle L \rangle}^2),
\end{equation}

where $L$ is, as usual, the lattice average of the Polyakov loop and V the 
spatial lattice volume.

To find the thermal threshold we used the Binder cumulant\footnote{See footnote
at page 52.}
\begin{equation}
\label{su2gr}
  g_r=\,3-\frac{\langle{L^4}\rangle}{{\langle{L^2}\rangle}^2}~.
\end{equation}
\begin{figure}[h]
\begin{center}
\epsfig{file=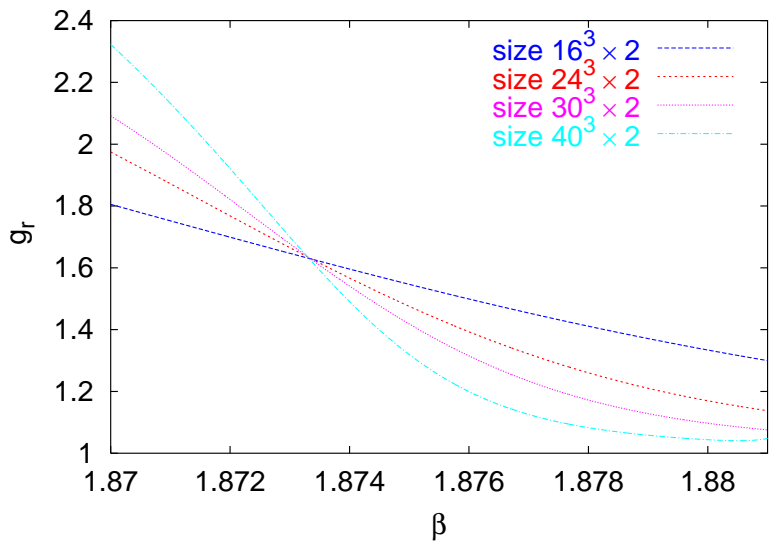,  width=12cm}
\caption[Binder cumulant as function of $\beta$
for $(3+1)$-$d$ $SU(2)$, $N_{\tau}=2$]
{Binder cumulant as function of $\beta$ for $(3+1)$-$d$ $SU(2)$, $N_{\tau}=2$.}
\label{interpgr}
\end{center}
\end{figure}

We put the interpolated curves in Fig. \ref{interpgr}. They clearly cross 
at the same point, around $\beta=1.8734$,  which gives a good idea 
of where the thermal transition takes place. 

To determine precisely this point
we used the $\chi^2$ method \cite{engles}. We applied this method 
to the absolute value of the 
lattice average of the Polyakov loop $|L|$, to the physical susceptibility
$\chi$ and to the derivative of the 
Binder cumulant with respect to $\beta$. In this way 
we could also evaluate the critical exponents 
ratios ${\beta}/{\nu}$, ${\gamma}/{\nu}$ and
$1/{\nu}$. 
Both the threshold and the exponents' ratios are shown in Table \ref{tbsu23d}.

\begin{figure}[h]
\begin{center}
\epsfig{file=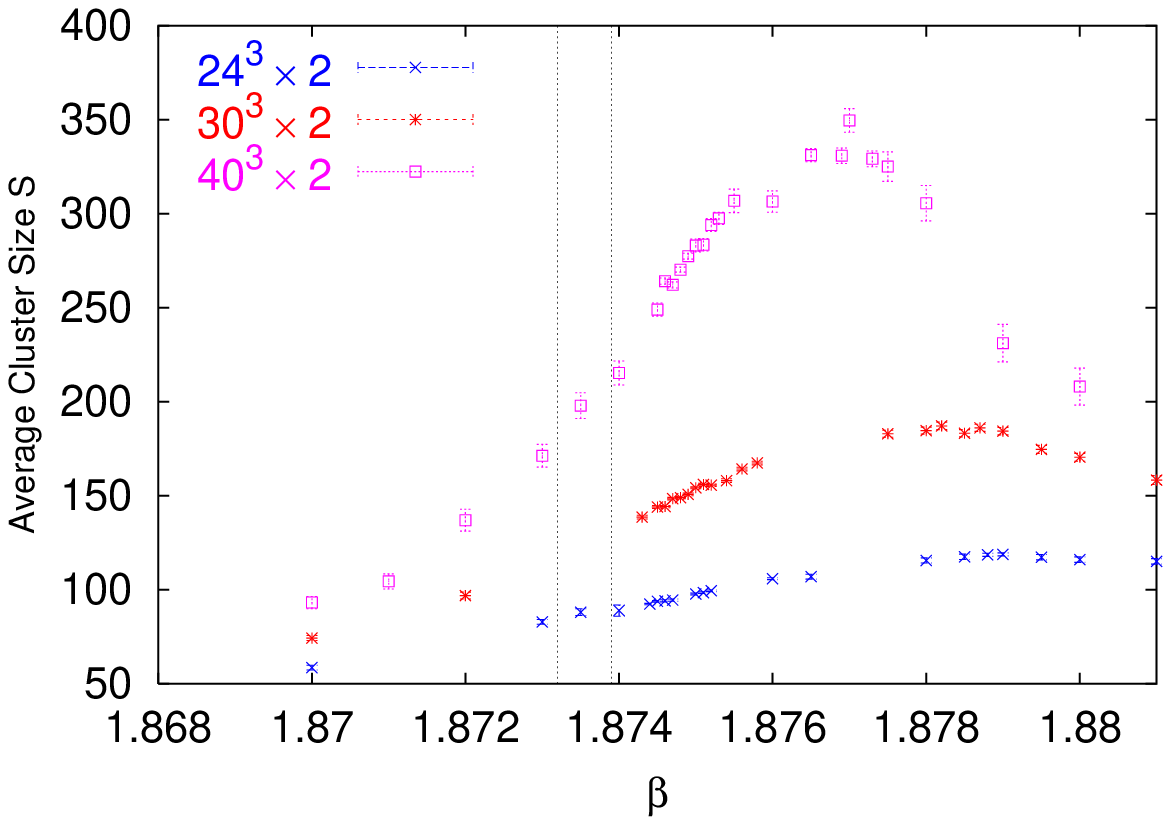,  width=14cm}
\caption[Average cluster size as function of $\beta$
for $(3+1)$-$d$ $SU(2)$, $N_{\tau}=2$: first approach]
{$(3+1)$-$d$ $SU(2)$, $N_{\tau}=2$.
Average cluster size as function of $\beta$ near the thermal threshold
$\beta_{c}$, indicated within one standard deviation by the dashed lines.}
\label{avclusu23d}
\vskip0.7cm
\epsfig{file=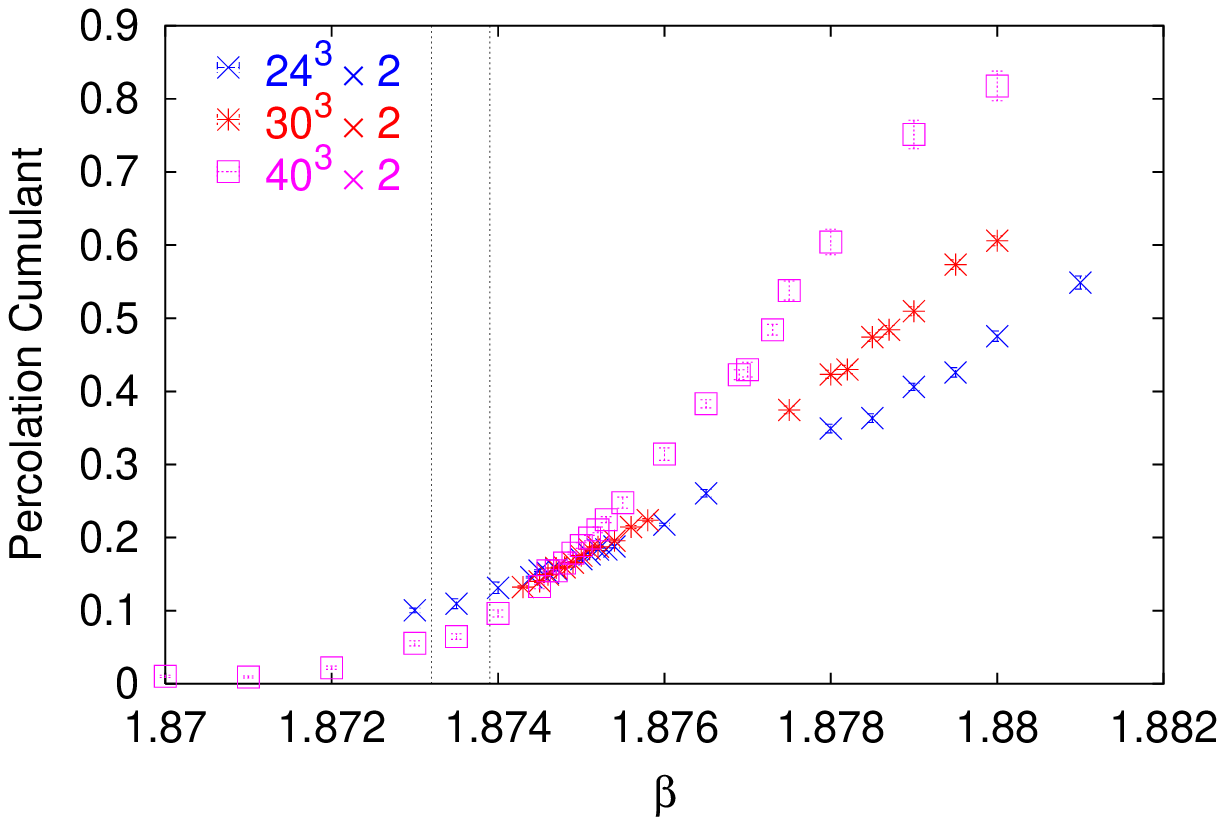,  width=14cm}
\caption[Percolation cumulant as 
a function of $\beta$ for $(3+1)$-$d$ $SU(2)$, $N_{\tau}=2$: first approach]
{$(3+1)$-$d$ $SU(2)$, $N_{\tau}=2$.
Percolation cumulant as function of $\beta$ for three  
lattice sizes. 
The curves cross close to the thermal
threshold (dashed lines).}
\label{alprobsu23D}
\end{center}
\end{figure}

We began our percolation studies performing some test runs 
for different lattice sizes to check the behaviour of our 
percolation variables around criticality.
Fig. \ref{avclusu23d} shows the behaviour of the average cluster size $S$ 
for three lattice sizes, $24^3\times 2$, $30^3\times 2$,
and $40^3\times 2$ respectively. 

To get the critical point of the geometrical transition we made use
of the percolation cumulant. 
In Fig. \ref{alprobsu23D} one can see the percolation cumulant as a function of $\beta$
for $24^3\times 2$,  
$30^3\times 2$ and $40^3\times 2$.
The lines cross at the same point within the errors and 
that restricts further the $\beta$ range  
for the critical threshold.

\clearpage
\begin{figure}[h]
\begin{center}
\epsfig{file=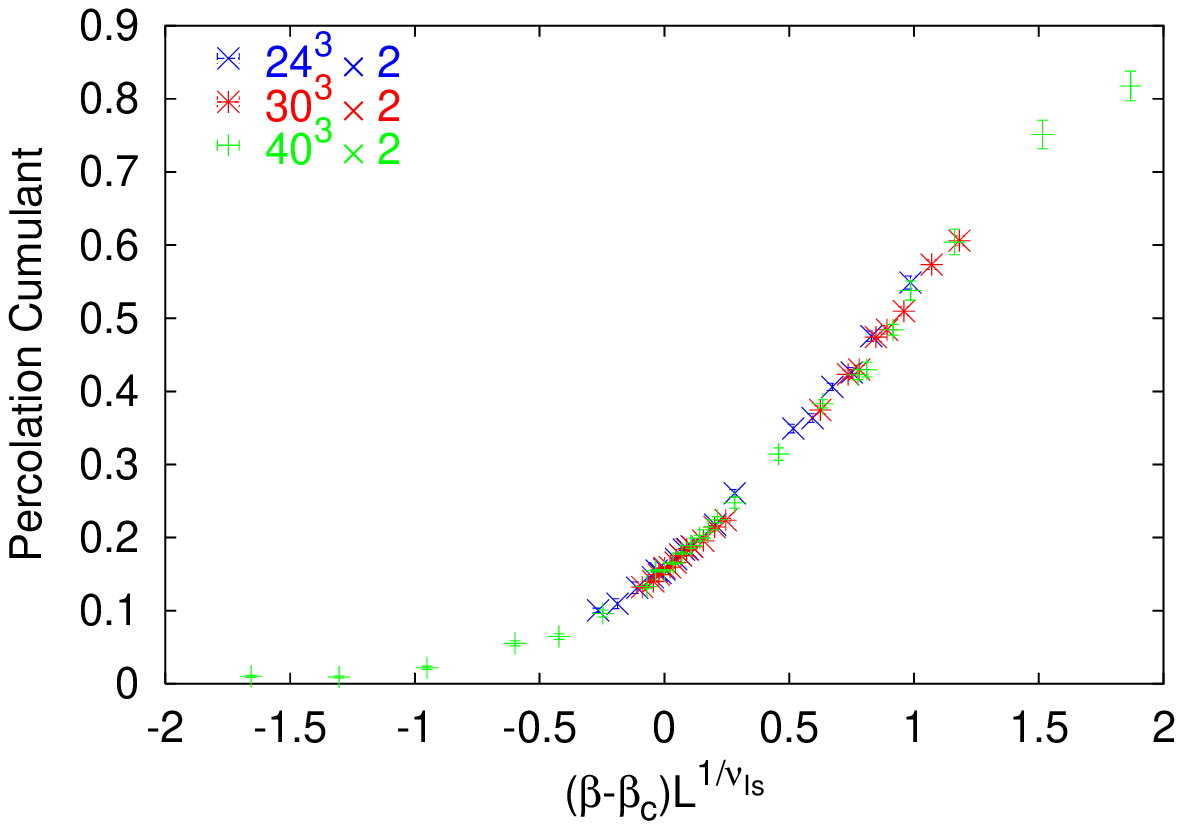,  width=14cm}
\caption[Rescaling of the
percolation cumulant curves of Fig. \ref{alprobsu23D} with the 3D
Ising exponent $\nu_{Is}=0.6294$]
{$(3+1)$-$d$ $SU(2)$, $N_{\tau}=2$.
Rescaling of the percolation cumulant 
curves of Fig. \ref{alprobsu23D} using $\beta_{c}=1.8747$
and the 3-dimensional Ising exponent $\nu_{Is}=0.6294$.}
\label{Scalprobsu23Dis}
\vskip0.7cm
\epsfig{file=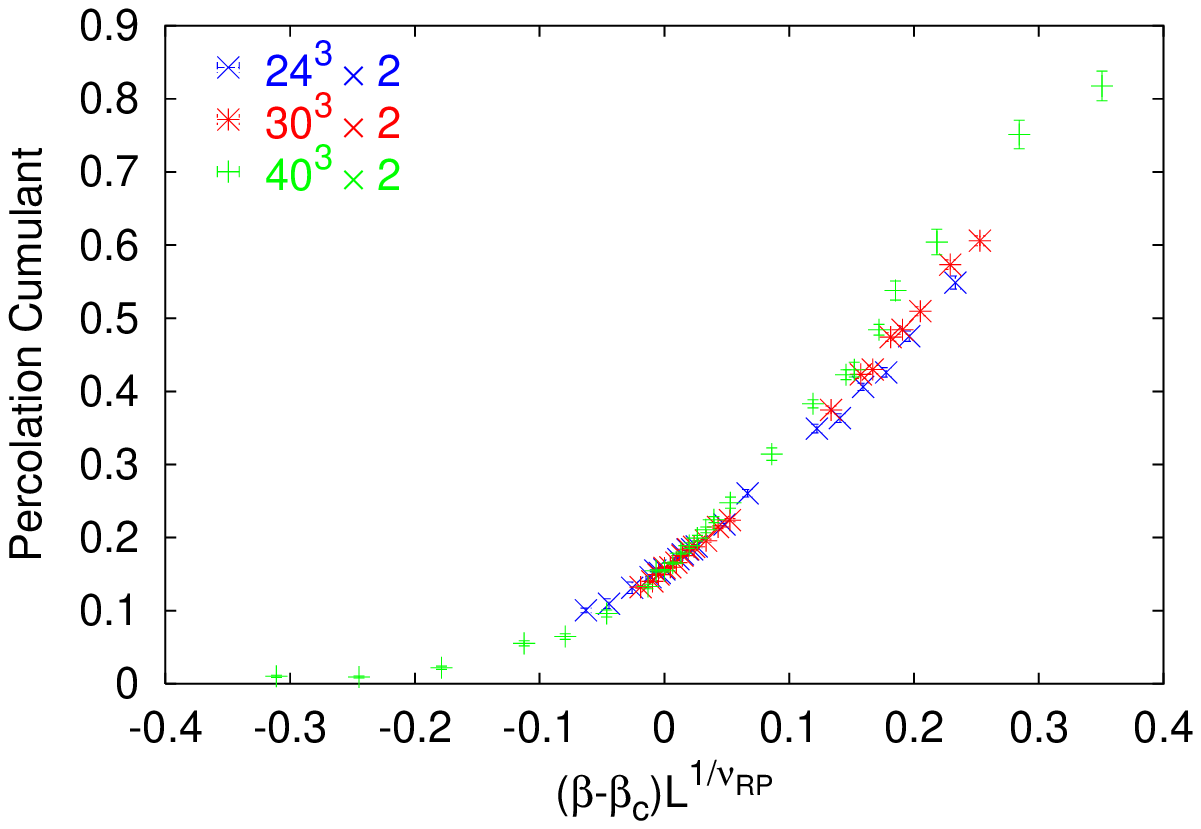,  width=14cm}
\caption[Rescaling of the
percolation cumulant curves of Fig. \ref{alprobsu23D} with the 
3D random percolation exponent $\nu_{RP}=0.8765$] 
{$(3+1)$-$d$ $SU(2)$, $N_{\tau}=2$.
Rescaling 
of the percolation cumulant 
curves of Fig. \ref{alprobsu23D} using $\beta_{c}=1.8747$
and the 3-dimensional random percolation
 exponent $\nu_{RP}=0.8765$.}
\label{Scalprobsu23Drp}
\end{center}
\end{figure}
\clearpage

Figs. \ref{Scalprobsu23Dis} and \ref{Scalprobsu23Drp}
show the rescaled percolation cumulant using $\beta_{c}=1.8747$
and the two major options 
for the exponent $\nu$, respectively the
Ising value and the random percolation one.
From the figures it 
is evident that the 
critical exponent $\nu_{perc}$
of the scaling function is the Ising 
exponent and not 
the random percolation one.

To evaluate the exponents' ratios  $\beta/\nu$ and $\gamma/\nu$ 
we performed high-statistics
simulations in the range where the percolation cumulant curves cross
each other. 
To improve the precision of the scaling fits we considered
several lattice sizes (all even values of the lattice side $L$ 
between $18$ and $40$). 
The number of measurements we took for each 
value of the coupling varies from 50000 to 100000. 
The final results are reported in Table \ref{tbsu23d}.
\vskip0.5cm
\begin{table}[h]
  \begin{center}{
      \begin{tabular}{|c||c|c|c|c|}
\hline
  &      Critical point  & $\beta/{\nu}$ & ${\gamma}/{\nu}$ & $\nu$\\ \hline\hline
$\vphantom{\displaystyle\frac{1}{1}}$  Percolation results &
$1.8747^{+0.0002}_{-0.0002}$&
$0.528^{+0.012}_{-0.015}$&$1.985^{+0.013}_{-0.018}$&$0.632^{+0.010}_{-0.015}$\\  
\hline $\vphantom{\displaystyle\frac{1}{1}}$ 
  Thermal results &$1.8735^{+0.0004}_{-0.0003}$ & $0.523^{+0.010}_{-0.013} $
  &$1.959^{+0.009}_{-0.007} $
&$0.630^{+0.010}_{-0.008}$ \\ \hline$\vphantom{\displaystyle\frac{1}{1}}$
3D Ising Model & &  0.5187(14) 
  & 1.963(7)
&0.6294(10) \\ \hline
      \end{tabular}
      }
\caption[Thermal and percolation critical indices for 
$(3+1)$-$d$ $SU(2)$, $N_{\tau}=2$: first approach]
{\label{tbsu23d} 
Thermal and percolation critical indices for 
$(3+1)$-$d$ $SU(2)$, $N_{\tau}=2$: first approach.} 
  \end{center}
\end{table}

The critical exponents we found, both for the thermal transition
and for the geometrical one, are in good agreement with each other and
with the ones of  
the 3-dimensional Ising model.

Again, the values of the two critical points are very close
but they overlap only within two standard deviations, like in the
$(2+1)$-$d$ case. However, 
our value of the thermal threshold is much more 
precise than the value we adopted for $(2+1)$-$d$ $SU(2)$
(see Table \ref{resu2D}). In this respect, the accord
for $(3+1)$-$d$ is better than the one for $(2+1)$-$d$.
This fact is not a surprise.
In fact, the error induced by the truncation
of the strong coupling expansion to the first term (see Section \ref{monaco})
is smaller the smaller $\beta_c$ is, so that 
the approximation is better 
in the $(3+1)$ dimensional
case ($\beta_c=1.8735$) than in the
$(2+1)$-$d$ case ($\beta_c=3.464$).

\section{Second Approach: Projection on Ising-like Spin Models}\label{dublino}

\subsection{Beyond the Strong Coupling Limit}\label{u2}

On the ground of our previous investigations, we can say that 
the effective theory, derived by means of the
strong coupling expansion of Section \ref{monaco}, 
allows to map the critical behaviour of 
finite temperature $SU(2)$ pure gauge theory
into a geometrical percolation framework.
Our procedure was, necessarily, approximate; nevertheless
the results are encouraging. 

We must point out that a drawback 
of the method is the fact that its
validity is limited to the strong coupling limit
of $SU(2)$ (i.e. for small $N_{\tau}$). If we want to 
address the deconfinement problem more generally,
an investigation of the weak coupling limit
becomes compulsory.

Since, in this case, 
high temperature expansions do not help,
the derivation of an effective theory from
the $SU(2)$ lattice action seems prohibitively
complicated.
We thus decided to try to extract the effective 
theory we need by analysing 
directly the Polyakov loop
configurations. This can be done
by using techniques developed in 
Monte Carlo renormalization group studies \cite{fer, arroyo, deck}.

To simplify the things, we assume 
the $Z(2)$ global symmetry to be
the only relevant feature at the basis
of the critical behaviour of the theory.
This assumption is rather strong but reasonable,
since the $Z(2)$ symmetry seems to be the only unifying
feature of all theories in the Ising universality class.
 
In this way, we can limit ourselves to analyze the
configurations of the {\it signs} of the Polyakov loops,
so that we perform a sort of projection into
Ising-like spin configurations. This approach has been successfully  
applied by 
Okawa to define an effective Hamiltonian for SU(2), in order to
look for the fixed point of the theory by means
of block-spin transformations \cite{okawa}.

The effective Hamiltonian
${\cal H}(s)$ of the signs $\{s_{\bf n}\}$ 
of the Polyakov loop configurations
can be defined through the equation \cite{okawa}
\begin{equation}
\label{effth}
\exp[{\cal H}(s)]\,=\,\int\,[dU]\prod\limits_{\bf n}\,\delta[s_{\bf n}, sgn(L_{\bf n})]
\,\exp(S_{SU2}),
\end{equation}
where $L_{\bf n}$ is, as usual, the 
value of the Polyakov loop at the spatial point {\bf n} and 
$S_{SU2}$ the $SU(2)$ lattice action. 
We stress that we include the factor $-\frac{1}{kT}$ in the definition of the 
Hamiltonian.
Eq. (\ref{effth}) shows that 
all degrees of freedom of the 
original $SU(2)$ field configurations
are integrated out, leaving only the 
distribution of the corresponding
Ising-projected configuration.

The problem is now how to determine 
the expression of ${\cal H}(s)$, starting from the 
original Polyakov loop configurations.

In general, we write
\begin{equation}
\label{effham}
{\cal H}(s)\,=\,\kappa_{\alpha}O^{\alpha},
\end{equation}
in which $\kappa_{\alpha}$ are the couplings,
$O^{\alpha}$ the spin operators and a sum over the 
index $\alpha$ is understood (e.g. in the Ising model
there would be only a single operator $O=\sum_{ij}s_is_j$).
Once we select the 
number and the type of operators, 
to fix the form of ${\cal H}(s)$ we need just 
to calculate the values of the couplings $\kappa_{\alpha}$.

To solve this problem,
Okawa proposed to use Schwinger-Dyson equations,
which are derived by exploiting the $Z(2)$
symmetry of ${\cal H}(s)$ \cite{oka2}.

Suppose to select some point {\bf n} of the 
spatial volume. We can then rewrite
Eq. (\ref{effham}) 
\begin{equation}
\label{effhamsep}
{\cal H}(s)\,=\,\kappa_{\alpha}O^{\alpha}_{\bf n}+\Delta\,H_{\bf n},
\end{equation}
separating the terms
depending on the spin $s_{\bf n}$ at
{\bf n} ($O^{\alpha}_{\bf n}$) from the ones which are independent
of $s_{\bf n}$ ($\Delta\,H_{\bf n}$).
We assume $O^{\alpha}_{\bf n}$ to be linear
in $s_{\bf n}$. 
This assumption is by no means
restrictive, since
all even powers of the spin variables are equal to $1$, and 
consequently any product of spins 
can be reduced to a form where each spin appears at most linearly. 

The thermal average of 
the operator $O^{\gamma}_{\bf n}$ is 
\begin{equation}
\label{thaverg}
\langle O^{\gamma}_{\bf n} \rangle\,=\,\frac{\sum\limits_{\{s\}}\,O^{\gamma}_{\bf n}
\,\exp[{\cal H}(s)]}{\cal Z}
\end{equation}
(${\cal Z}$ is the partition function).
If we perform a
change of variable inside the sum,
`flipping' the spin variable $s_{\bf n}$ to $-s_{\bf n}$,
the operator $O^{\gamma}_{\bf n}$ will change sign and
Eq. (\ref{thaverg}) becomes
\begin{eqnarray}
\langle O^{\gamma}_{\bf n} \rangle\,&=&\,-\frac{\sum\limits_{\{s\}}\,O^{\gamma}_{\bf n}
\,\exp[-\kappa_{\alpha}O^{\alpha}_{\bf n}+\Delta\,H_{\bf n}]}{\cal Z}\nonumber\\
&=&\,-\frac{\sum\limits_{\{s\}}\,O^{\gamma}_{\bf n}
\,\exp(-2\kappa_{\alpha}O^{\alpha}_{\bf n})
\,\exp(\kappa_{\alpha}O^{\alpha}_{\bf n}+\Delta\,H_{\bf n})}{\cal Z}\nonumber\\
&=&\,-\langle O^{\gamma}_{\bf n}\,\exp(-2\kappa_{\alpha}O^{\alpha}_{\bf n}) \rangle.
\label{thavergz2}
\end{eqnarray}
Eq. (\ref{thavergz2}) establishes a relation
between thermal averages of the operators $O^{\gamma}_{\bf n}$ and
the couplings $\kappa_{\gamma}$.
The equations (\ref{thavergz2}) are, however, implicit in the couplings.
They can be solved by means of the
Newton method, which is based on 
successive approximations.
One starts by 
making a guess about the values of the couplings; we indicate by
$\tilde{\kappa_{\gamma}}$
such initial values.
We  
can develop the exponential 
\begin{equation}
\label{expo2000}
\exp(-2\kappa_{\gamma}O^{\gamma}_{\bf n})=\exp\{-2[\tilde{\kappa_{\gamma}}+
(\kappa_{\gamma}-\tilde{\kappa_{\gamma}})]O^{\gamma}_{\bf n}\}\,\approx\,
\exp(-2{\tilde{\kappa_{\gamma}}}O^{\gamma}_{\bf n})[1-2(\kappa_{\delta}-\tilde{\kappa_{\delta}})
O^{\delta}_{\bf n}]
\end{equation}
Combining (\ref{expo2000}) and 
(\ref{thavergz2}), we finally obtain for the first approximation of $\kappa_{\gamma}$
\begin{equation}
\label{expo2001}
\kappa_{\gamma}(1)=\tilde{\kappa_{\gamma}}+\frac{1}{2}\langle{O^{\gamma}_{\bf n}
O^{\alpha}_{\bf n}}\rangle^{-1}[\langle{
O^{\alpha}_{\bf n}}\rangle+\langle{O^{\alpha}_{\bf
  n}}\exp(-2\tilde{\kappa_{\delta}}
O^{\delta}_{\bf n})\rangle].
\end{equation}
From the Polyakov loop configurations we can
calculate the thermal averages of the operator expressions
present in (\ref{expo2001}).
Next, we use the results $\kappa_{\gamma}(1)$
of the first iteration as input values in Eq. (\ref{expo2001})
and we get some values $\kappa_{\gamma}(2)$.
After 
a sufficient number $N$ of iterations,
the series of partial values for the $\kappa_{\gamma}$'s will converge,
i. e. $\kappa_{\gamma}(N+1)\approx\kappa_{\gamma}(N)$ within errors,
$\forall\, \gamma$. The final set of couplings 
is the solution of Eq. (\ref{thavergz2}).

We notice that the general set of equations
(\ref{thavergz2}) refers to a single point 
{\bf n} of the spatial volume. 
The thermal averages are independent of 
the particular point {\bf n}, so it doesn't matter
where we decide to take the averages.
Nevertheless, since we aim
to reduce as much as possible the errors
of the couplings, we chose to 
determine the thermal averages at {\it each point}
of the lattice, and to calculate successively the average value of the 
couplings obtained by solving the equations at any point.
This reduces considerably the effect 
of thermal fluctuations  
and, consequently,
the errors on the final $\kappa_{\gamma}$'s. 
 
So, we have now all necessary tools to derive
an effective theory for $SU(2)$ out of the Polyakov loop configurations.
We still have to specify what kind of spin
operators 
should appear in the expression (\ref{effham}) of the 
hamiltonian ${\cal H}(s)$.
We can in principle choose any 
operator which respects the $Z(2)$ symmetry.
Nevertheless, our choice is bound by 
the condition for the effective spin model to have an 
equivalent percolation formulation.
As far as this is concerned, we know 
that the original Coniglio-Klein picture of the 
Ising model can be extended
to general spin models, as long as the interactions are
spin-spin and ferromagnetic (see Section \ref{maria}).
We proved this result for continuous spin models,
but it remains valid also in the simpler case of Ising spins.

Because of that, we impose that our 
$O^{\alpha}$ are spin-spin operators. 
Our ansatz for the effective Hamiltonian 
${\cal H}(s)$ is thus
\begin{equation}
\label{effans}
{\cal H}(s)\,=\,\kappa_{1}\sum_{NN}s_is_j\,+\,\kappa_{2}\sum_{NTN}s_ks_l\,
+\,\kappa_{3}\sum_{NTNTN}s_ms_n\,+\,etc.\,,
\end{equation}
where the distance between coupled spins increases
progressively starting from the simple nearest-neighbour ($NN$) case
($NTN=$next-to-nearest, $NTNTN=$next-to-next-to-nearest, and so on).

What we have to do is to check whether,
including a sufficient number of operators,
the Hamiltonian (\ref{effans}) can reproduce 
the Ising-projected 
Polyakov loop configurations of 
finite temperature $SU(2)$.
In general, the approximation improves the more operators
we include in (\ref{effans}), because there
will be more parameters. The fact that one must restrict
the choice to some subset of operators involves
an error ({\it truncation error}) in addition to the statistical one. 
The truncation error is, in general, impossible to determine
and can be much bigger than the 
indetermination of the effective theory due to the statistical fluctuations
of the thermal averages.
In this way, the solution 
one finds at the end of the procedure is not necessarily 
a good approximation of the original theory, but only
the closest one belonging to the subspace of theories defined by 
the selected set of operators.
We need thus to establish a criterium to judge how
well the effective theory approximates the original one. 
A good option 
could be to compare average values got from 
the configurations produced by simulating the effective theory with the 
corresponding quantities measured on the
original Ising-projected 
Polyakov loop configurations. We used the lattice average of the
magnetization $m$, 
\begin{equation}
\label{magneff}
m\,=\,\frac{1}{V}\Big|\sum_{i}s_i\Big|\,,
\end{equation}
($V$ is the spatial lattice volume) as test variable for this quality control.

We point out that the approach we have described
is independent of the value of the number
$N_{\tau}$ of 
lattice spacings in the temperature
direction. In this respect, the method is general, although
it is not possible to predict whether it is able to provide 
the required solution in all cases. 

We applied the method to $SU(2)$ in $(3+1)$ dimensions, for two 
different lattice regularizations:
$N_{\tau}=2$ and $N_{\tau}=4$. 
We have already studied the case $N_{\tau}=2$
with the first approach (see Section \ref{parigi}): this gives us 
the possibility to compare the two different procedures.
  
\subsection{Numerical Results for (3+1)-d SU(2), N\boldmath$_{\tau}$\unboldmath= 2}\label{stoccolma}

As we are interested 
in the phase transition of $SU(2)$, 
we focused our attention on
the critical point. The value of the critical coupling
$\beta_c$ was already determined quite precisely 
during the previous investigation; our estimate was
$\beta_c=1.8735^{+0.0004}_{-0.0003}$ (see Table \ref{tbsu23d}).
So, our aim is to check whether,
at $\beta=\beta_c$, we can find a projection
of the theory onto the spin model (\ref{effans}).

We performed a simulation of $SU(2)$ at
$\beta_c$ on a rather large lattice,
$32^3{\times}2$. We chose a large lattice to 
reduce finite size effects. 
The algorithm we used is the same described in
Section \ref{berlino}. We measured 
our quantities every 70 updates, which makes
the analyzed configurations basically uncorrelated; the total number of
measurements is 2000. As usual, the errors were determined 
with the Jackknife method.

We began by making a projection
on a model
with 10 operators. Ten is, 
in fact, the number of spin-spin operators
considered by Okawa in his effective theory of $SU(2)$
\cite{okawa}. However, his
Hamiltonian contains also multispin operators
(products of 4, 6 and 8 spins), which we must exclude.
The average
of the magnetization (\ref{magneff}) of the effective 
theory did not agree with the one of the Polyakov loop
configurations, so that
we progressively
enlarged the set of operators, adding 
further spin-spin interactions, until we
reached a set of 15 couplings.
\begin{table}[h]
  \begin{center}{
      \begin{tabular}{|c|l|c|l|}
\hline
Coupling  & Avg. Value&Coupling  & Avg. Value\\ \hline\hline
$\vphantom{\displaystyle\frac{1}{1}}$  
$\kappa_1$ &\,\,0.1307(1)& $\kappa_9$ &\,\,0.00014(10)
\\  \hline $\vphantom{\displaystyle\frac{1}{1}}$ 
$\kappa_2$ &\,\,0.01905(3)&$\kappa_{10}$ &\,\,0.00058(3)
\\ \hline $\vphantom{\displaystyle\frac{1}{1}}$
$\kappa_3$ &\,\,0.00470(5)&$\kappa_{11}$ &\,\,0.00018(3)
\\ \hline $\vphantom{\displaystyle\frac{1}{1}}$
$\kappa_4$ &\,\,0.0080(1)&$\kappa_{12}$ &\,\,0.00008(1)
\\ \hline $\vphantom{\displaystyle\frac{1}{1}}$
$\kappa_5$ &\,\,0.00192(4)&$\kappa_{13}$ &\,\,0.00001(1)
\\ \hline $\vphantom{\displaystyle\frac{1}{1}}$
$\kappa_6$ &\,\,0.00062(8)&$\kappa_{14}$ &\,\,0.00006(1)
\\ \hline $\vphantom{\displaystyle\frac{1}{1}}$
$\kappa_7$ &\,\,0.00033(2)&$\kappa_{15}$ &-0.00005(3)
\\ \hline $\vphantom{\displaystyle\frac{1}{1}}$
$\kappa_8$ &\,\,0.00007(2)& &
\\ \hline
      \end{tabular}
      }
\caption[Couplings of the effective theory
for the Ising-projected Polyakov loop configurations
of $(3+1)$-$d$ $SU(2)$, $N_{\tau}=2$]
{\label{2couplings} Couplings of the effective theory
for the Polyakov loop configurations
of $(3+1)$-$d$ $SU(2)$ ($N_{\tau}=2$) at the critical coupling $\beta_c=1.8735$.} 
  \end{center}

\end{table}
The relative operators
connect a point ($000$) to ($100$), ($110$), ($111$),
($200$), ($210$), ($211$), ($220$),  ($221$), ($222$),
($300$), ($310$), ($311$), ($320$), ($321$), ($322$).
The final set of couplings is reported in Table \ref{2couplings}.

\begin{figure}[h]
\begin{center}
\epsfig{file=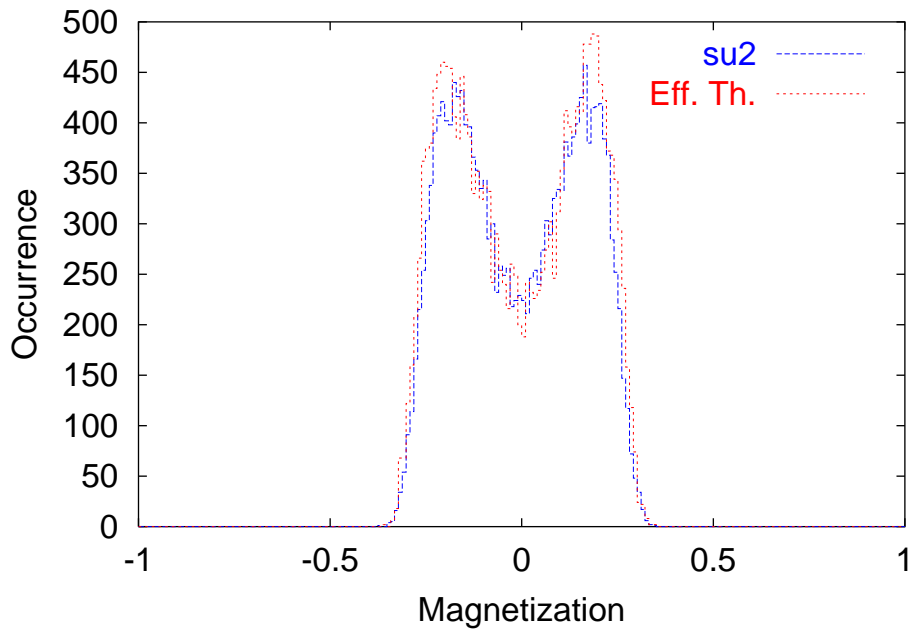,  width=14cm}
\caption[Comparison of the magnetization histograms derived
from the Polyakov loop configurations and the effective
theory: $N_{\tau}=2$]
{$(3+1)$-$d$ $SU(2)$, $N_{\tau}=2$.
Comparison of the magnetization histograms derived
from the Polyakov loop configurations at $\beta_c$ and the effective
theory (\ref{effans}) defined by the couplings of Table \ref{2couplings}.}
\label{histo}
\end{center}
\end{figure}

Fig. \ref{histo} shows a comparison between the 
magnetization distribution of the Polyakov loop configurations
and the one of the effective theory:
the two histograms are very similar. The values  
of the average magnetization $m$ 
are also in agreement: for $SU(2)$, $m=0.091(1)$ and 
for the spin model, $m=0.0923(7)$. 
We notice that all the couplings
in Table \ref{2couplings} are positive, except 
the last one. 
Since the error on $\kappa_{15}$ is of the order of its
average value, we can set $\kappa_{15}=0$ 
without appreciable effects. In this way, we have got
the effective theory we were looking for, with
only ferromagnetic spin-spin interactions.
The values of the couplings can then be used to
determine the bond weights $p^{\alpha}$ of the corresponding percolation model,
according to the usual formula 
\begin{equation}
p^{\alpha}\, =\,1 - \exp(- 2\,\kappa_{\alpha})
\label{ffeqnew}
\end{equation}
($\alpha=1,..., 15$).

The magnetization check indicates that the 
effective theory is a fair approximation of $SU(2)$. 
Anyhow, this does not necessarily imply that 
the two models are very close to each other, so that we 
can conclude that the percolation picture of the effective theory
indeed works for the original Polyakov loop configurations.
The only way to see that is to 
investigate the geometrical transition
of the new clusters in the 
$SU(2)$ configurations. 

Therefore, we performed a percolation analysis 
of $(3+1)$-$d$ $SU(2)$ ($N_{\tau}=2$),
building the clusters according to 
the general definition 
introduced in Section \ref{maria}, with the 
bond probabilities (\ref{ffeqnew}).
We stress that the bond weights are 
temperature-dependent. Our 
effective theory represents a projection
of $SU(2)$ for $\beta=\beta_c$. But,
in order to carry on
our analysis, we need to evaluate the 
percolation variables at different values of $\beta$.
Strictly speaking, for each $\beta_{i}$ at study
we should derive the corresponding 
effective theory, and use the relative
set $\{\kappa\}_{i}$ to 
calculate the bond weights (\ref{ffeqnew}) at $\beta_{i}$. 
But for our analysis the previous consideration
is not important. In fact, we are interested anyhow only
in $\beta$'s which lie near $\beta_c$, so that 
the corresponding couplings of the effective theory will change only
slightly from one to the other extreme of the range. 
In the specific case of our investigations,
it turns out
that the variation is of the order
of the error on the couplings derived by a single projection,
and it is thus irrelevant for our purposes.
Because of that, at each $\beta$,
we shall use the same set of bond probabilities,
namely the set determined by the couplings
of Table \ref{2couplings}.

\begin{figure}[h]
\begin{center}
\epsfig{file=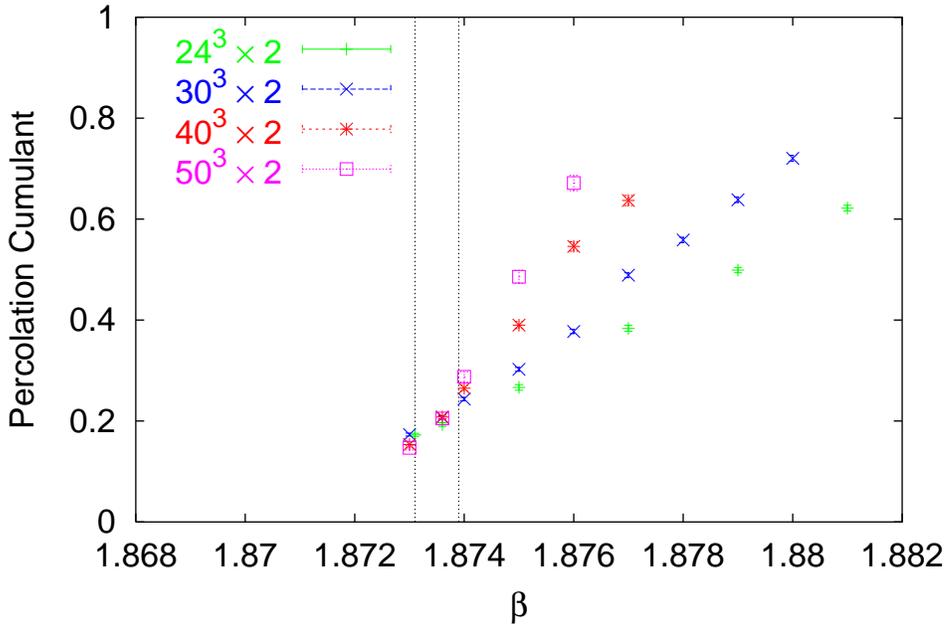,  width=14cm}
\caption[Percolation cumulant as a function of $\beta$ for $(3+1)$-$d$ $SU(2)$, 
$N_{\tau}=2$: second approach]
{$(3+1)$-$d$ $SU(2)$, $N_{\tau}=2$. Percolation cumulant near 
the critical point for four lattice sizes.}
\label{sceffnt2}
\end{center}
\end{figure}

\begin{figure}[h]
\begin{center}
\epsfig{file=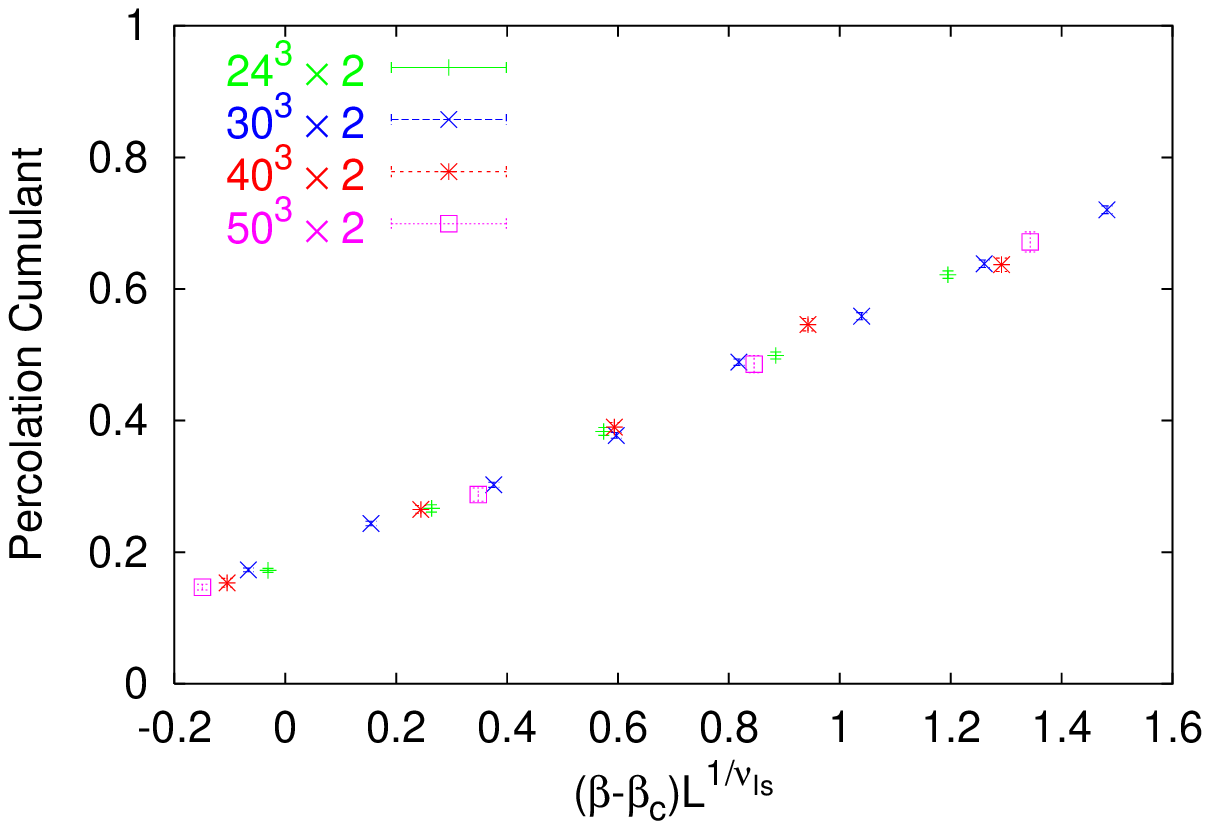,  width=14cm}
\caption[Rescaling of the
percolation cumulant curves of Fig. \ref{sceffnt2} with the 3D
Ising exponent $\nu_{Is}=0.6294$]
{$(3+1)$-$d$ $SU(2)$, $N_{\tau}=2$. 
Rescaling of the percolation cumulant 
curves of Fig. \ref{sceffnt2} using $\beta_{c}=1.8734$
and the 3-dimensional Ising exponent $\nu_{Is}=0.6294$.}
\label{sceffnt2is}
\vskip0.7cm
\epsfig{file=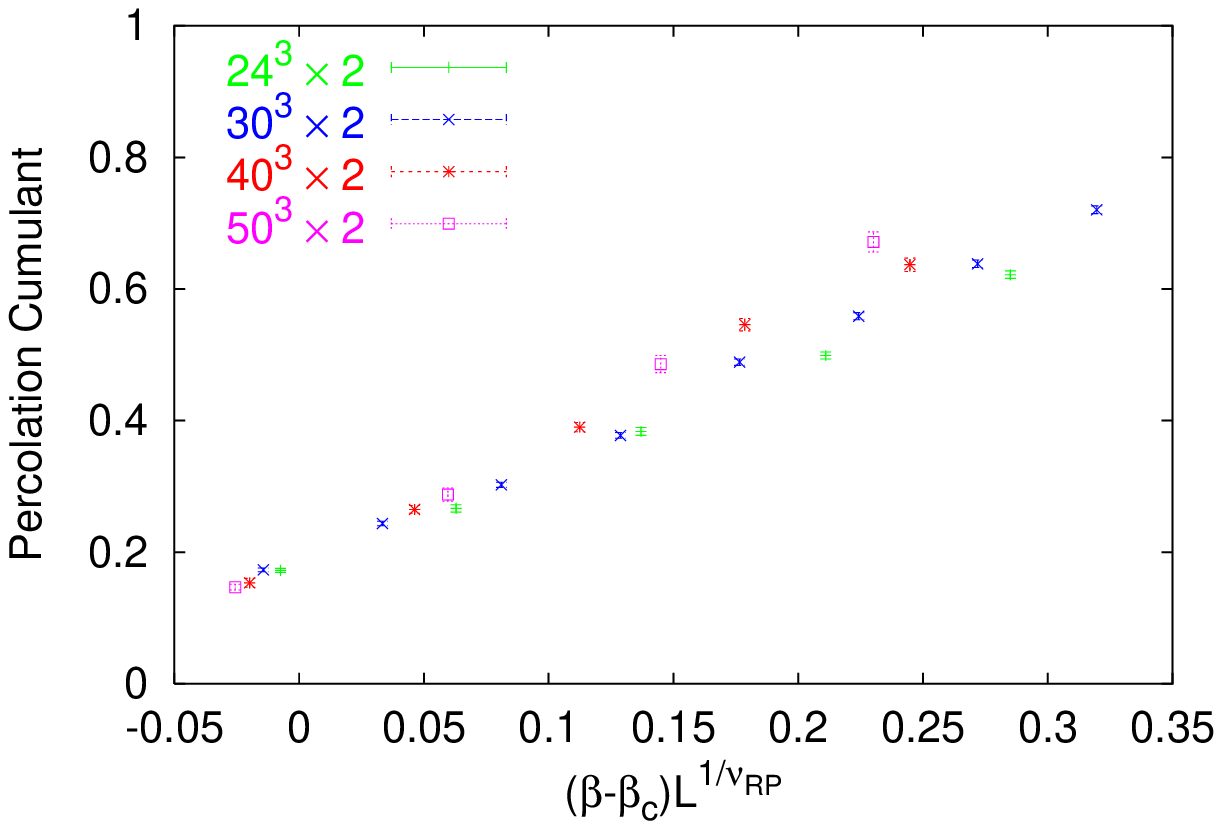,  width=14cm}
\caption[Rescaling of the
percolation cumulant curves of Fig. \ref{sceffnt2} with the 3D
random percolation exponent $\nu_{RP}=0.8765$]
{$(3+1)$-$d$ $SU(2)$, $N_{\tau}=2$.
Rescaling of the percolation cumulant 
curves of Fig. \ref{sceffnt2}
using $\beta_{c}=1.8734$
and the 3-dimensional random percolation
exponent $\nu_{RP}=0.8765$.}
\label{sceffnt2rp}
\end{center}
\end{figure}

We considered four lattice sizes:
$24^3{\times}2$, $30^3{\times}2$, $40^3{\times}2$ and
$50^3{\times}2$. Taking the measurements every $10$ updates,
the percolation data are uncorrelated, even for the 
$50^3{\times}2$ lattice.
Fig. \ref{sceffnt2} shows
the behaviour of the percolation cumulant
as a function of $\beta$. 
\vskip0.5cm
The four curves cross remarkably well at
the same point, within errors,
in excellent agreement with the thermal threshold, indicated within one standard deviation
by the dashed lines.
The rescaling of the percolation cumulant curves indicates
that the percolation exponent $\nu_{perc}=\nu_{Is}$ (Figs. \ref{sceffnt2is} and 
\ref{sceffnt2rp}).

\clearpage
To determine the exponents' ratios $\beta/\nu$ and $\gamma/\nu$,
we used the usual finite size scaling procedure, performing simulations
at criticality of 
many different lattice sizes to improve the quality 
of the scaling fits (we took even values
of the lattice side $L$ between $20$ and $50$). 
Unfortunately,
we could not determine $\beta/\nu$, because of strong 
fluctuations of the percolation strength $P$ around $\beta_c$.
The value of $P$ at criticality is, in general, quite small, and it
suffers more than $S$ the approximations involved by
our procedure.
Consequently, the slopes of the data points
in the log-log scaling fits of $P$ vary wildly, and 
the error of $\beta/\nu$ turns out to be too large.
On the other hand, $\gamma/\nu$
can be evaluated with the usual accuracy (${\approx}\,1\%$)
and its value is in agreement with the one of $SU(2)$ (Table \ref{tabeffnt2}).
\vskip0.5cm
\begin{table}[h]
  \begin{center}{
      \begin{tabular}{|c||l|l|l|}
\hline
  &      Critical point & ${\gamma}/{\nu}$ & $\nu$\\ \hline\hline
$\vphantom{\displaystyle\frac{1}{1}}$  Percolation results &
$1.8734(2)$&$1.977^{+0.011}_{-0.017}$&$0.628^{+0.011}_{-0.009}$\\  
\hline $\vphantom{\displaystyle\frac{1}{1}}$ 
  Thermal results &$1.8735^{+0.0004}_{-0.0003}$
  &$1.959^{+0.009}_{-0.007} $
&$0.630^{+0.010}_{-0.008}$ \\ \hline$\vphantom{\displaystyle\frac{1}{1}}$
3D Ising Model & 
  & 1.963(7)
&0.6294(10) \\ \hline
      \end{tabular}
      }
\caption[Percolation critical indices for 
$(3+1)$-$d$ $SU(2)$, $N_{\tau}=2$: second approach]
{\label{tabeffnt2} 
Percolation critical indices for 
$(3+1)$-$d$ $SU(2)$, $N_{\tau}=2$, with the new cluster definition.
We also put for comparison the thermal results determined in Section \ref{parigi}.
and the 3D Ising values.} 
  \end{center}
\end{table}

In conclusion, in spite of the several approximations we were forced to introduce
to define the percolation picture
with the second approach, for $N_{\tau}=2$ the 
new clusters seem again to follow the behaviour of the thermal 
quantities. Besides, the value of the critical threshold is
better than the one determined by the first approach.

\subsection{Numerical Results for (3+1)-d SU(2), N\boldmath$_{\tau}$\unboldmath= 4}\label{londra}

The case $N_{\tau}=2$, discussed in the previous section, is
important because it shows that the new percolation approach 
can be successfully applied and because 
it confirms the result obtained in Section \ref{parigi},
even if the two types of clusters have apparently nothing 
to do with each other.
However, for $N_{\tau}=4$, the things 
get more interesting, 
since the new method allows us to explore this case,
which is instead unaccessible to the first approach.

As far as the thermal critical behaviour 
is concerned, we adopted as reference values
the results  
of a recent study of Engels et al. \cite{engles}.
In particular, in \cite{engles}
the critical point $\beta_c$
was determined with great accuracy: $\beta_c=2.29895(10)$.
We simulated $(3+1)$-$d$ $SU(2)$ at 
$\beta=2.29895$ and looked for the corresponding 
effective theory. The lattice size was $32^3{\times}4$,
the number of measurements 2000; we evaluated
the configurations every 60 updates to have them
uncorrelated.

We tried first to 
use the same set of $15$ operators 
which worked so well in the $N_{\tau}=2$ case.
Unfortunately, the effective theory we obtained fails
in reproducing the behaviour of the magnetization.
There is, in fact, a clear discrepancy between the 
average values.
This fact is not unexpected: it is known that, by increasing
$N_{\tau}$, longer range interactions  
come into play.
We then enlarged further on the set of 
spin-spin operators. For $19$ operators, we 
got the set of couplings reported in Table \ref{tbeffnt4}.
\vskip0.5cm
\begin{table}[h]
  \begin{center}{
      \begin{tabular}{|c|l|c|l|}
\hline
Coupling  & Avg. Value&Coupling  & Avg. Value\\ \hline\hline
$\vphantom{\displaystyle\frac{1}{1}}$  
$\kappa_1$ &\,\,0.08390(4)& $\kappa_{11}$ &\,\,0.00082(5)
\\  \hline $\vphantom{\displaystyle\frac{1}{1}}$ 
$\kappa_2$ &\,\,0.01839(5)&$\kappa_{12}$ &\,\,0.00055(4)
\\ \hline $\vphantom{\displaystyle\frac{1}{1}}$
$\kappa_3$ &\,\,0.00775(4)&$\kappa_{13}$ &\,\,0.00035(2)
\\ \hline $\vphantom{\displaystyle\frac{1}{1}}$
$\kappa_4$ &\,\,0.00697(1)&$\kappa_{14}$ &\,\,0.00030(4)
\\ \hline $\vphantom{\displaystyle\frac{1}{1}}$
$\kappa_5$ &\,\,0.00343(2)&$\kappa_{15}$ &\,\,0.00013(4)
\\ \hline $\vphantom{\displaystyle\frac{1}{1}}$
$\kappa_6$ &\,\,0.00197(1)&$\kappa_{16}$ &\,\,0.00020(5)
\\ \hline $\vphantom{\displaystyle\frac{1}{1}}$
$\kappa_7$ &\,\,0.00114(1)&$\kappa_{17}$ &\,\,0.00018(3)
\\ \hline $\vphantom{\displaystyle\frac{1}{1}}$
$\kappa_8$ &\,\,0.00083(1)&$\kappa_{18}$ &\,\,0.00017(1)
\\ \hline $\vphantom{\displaystyle\frac{1}{1}}$
$\kappa_9$ &\,\,0.00035(6)&$\kappa_{19}$ &\,\,0.00017(4)
\\ \hline $\vphantom{\displaystyle\frac{1}{1}}$
$\kappa_{10}$ &\,\,0.00105(9)& &
\\ \hline
      \end{tabular}
      }
\caption[Couplings of the effective theory
for the Ising-projected Polyakov loop configurations
of $(3+1)$-$d$ $SU(2)$, $N_{\tau}=4$]
{\label{tbeffnt4} Couplings of the effective theory
for the Polyakov loop configurations
of $(3+1)$-$d$ $SU(2)$ ($N_{\tau}=4$) at the critical coupling $\beta_c=2.29895$.} 
  \end{center}

\end{table}

The new 4 operators connect a point ($000$) to 
($330$) ($\kappa_{16}$), ($331$) ($\kappa_{17}$), ($332$) 
($\kappa_{18}$) and ($333$) ($\kappa_{19}$).
The average value of $m$ from the effective theory
is now $0.121(3)$, in agreement with the 
$SU(2)$ value $0.128(6)$.

We see that all interactions are ferromagnetic.
So, also for $N_{\tau}=4$, there seems to be
a promising effective theory that we can exploit 
to carry on percolation studies.

Next, $SU(2)$ simulations were performed on the following
lattices:  
$24^3{\times}4$, $30^3{\times}4$, $40^3{\times}4$ and
$50^3{\times}4$. To build the clusters we use
the bond weights relative to the set of couplings
of Table \ref{tbeffnt4}, for any value of 
the $SU(2)$ coupling $\beta$ (see Section \ref{stoccolma}).
We took the measurements every $10$ updates for any
coupling and lattice size; in this way the percolation data
are uncorrelated.

\begin{figure}[h]
\begin{center}
\epsfig{file=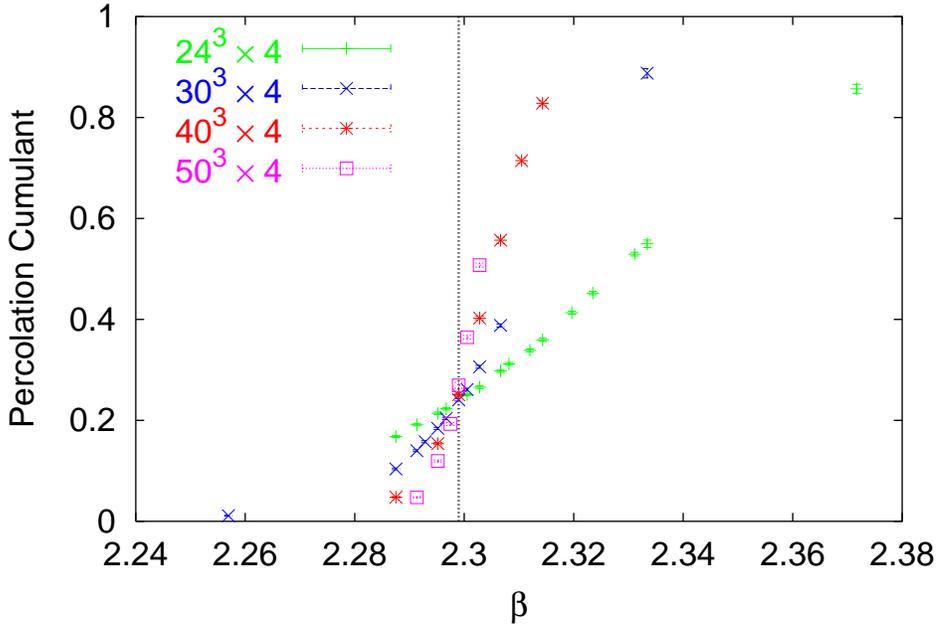,  width=14cm}
\caption[Percolation cumulant as a function of $\beta$ 
for $(3+1)$-$d$ $SU(2)$, $N_{\tau}=4$: second approach]
{$(3+1)$-$d$ $SU(2)$, $N_{\tau}=4$. Percolation cumulant near 
the critical point for different lattice sizes.}
\label{sceffnt4}
\end{center}
\end{figure}
 
\begin{figure}[h]
\begin{center}
\epsfig{file=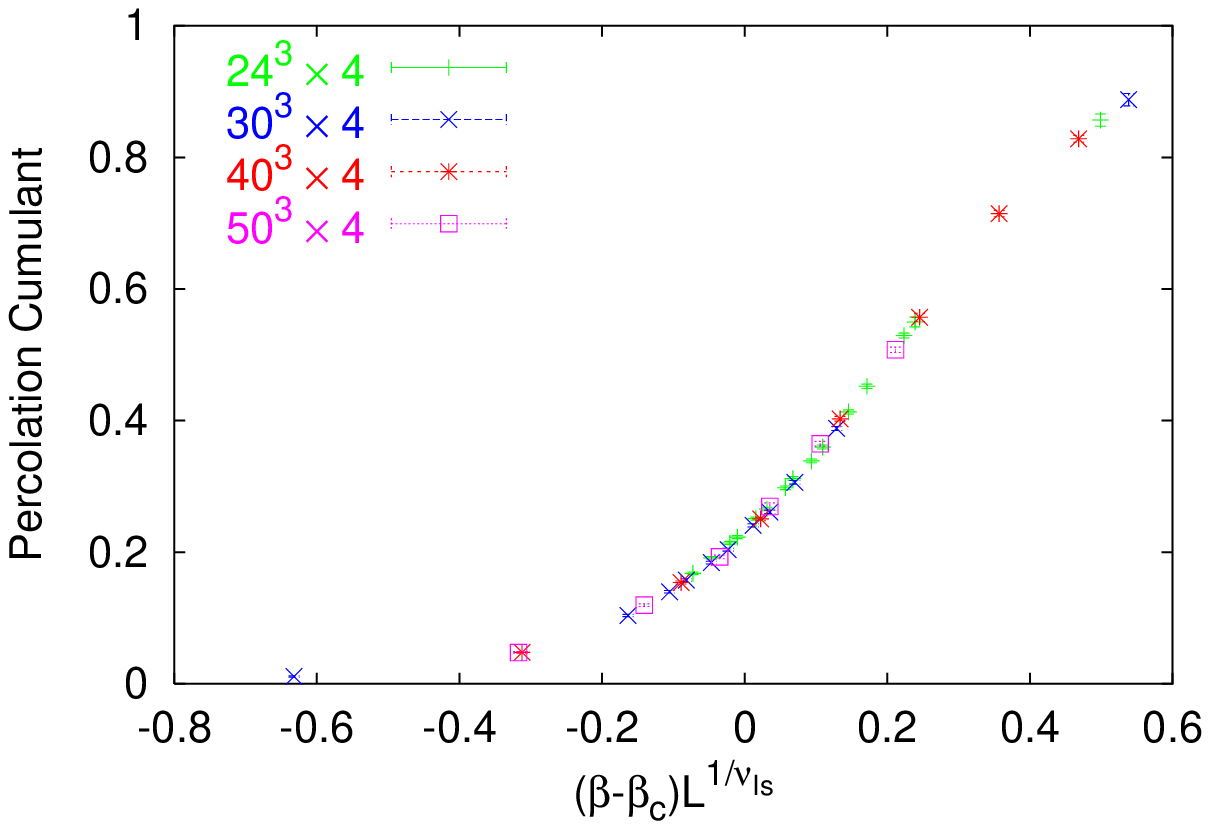,  width=14cm}
\caption[Rescaling of the
percolation cumulant curves of Fig. \ref{sceffnt4} with the 3D
Ising exponent $\nu_{Is}=0.6294$]
{$(3+1)$-$d$ $SU(2)$, $N_{\tau}=4$.
Rescaling of the
percolation cumulant curves of Fig. \ref{sceffnt4}
using $\beta_{c}=2.2991$
and the 3-dimensional Ising exponent $\nu_{Is}=0.6294$.}
\label{sceffnt4is}
\vskip0.7cm
\epsfig{file=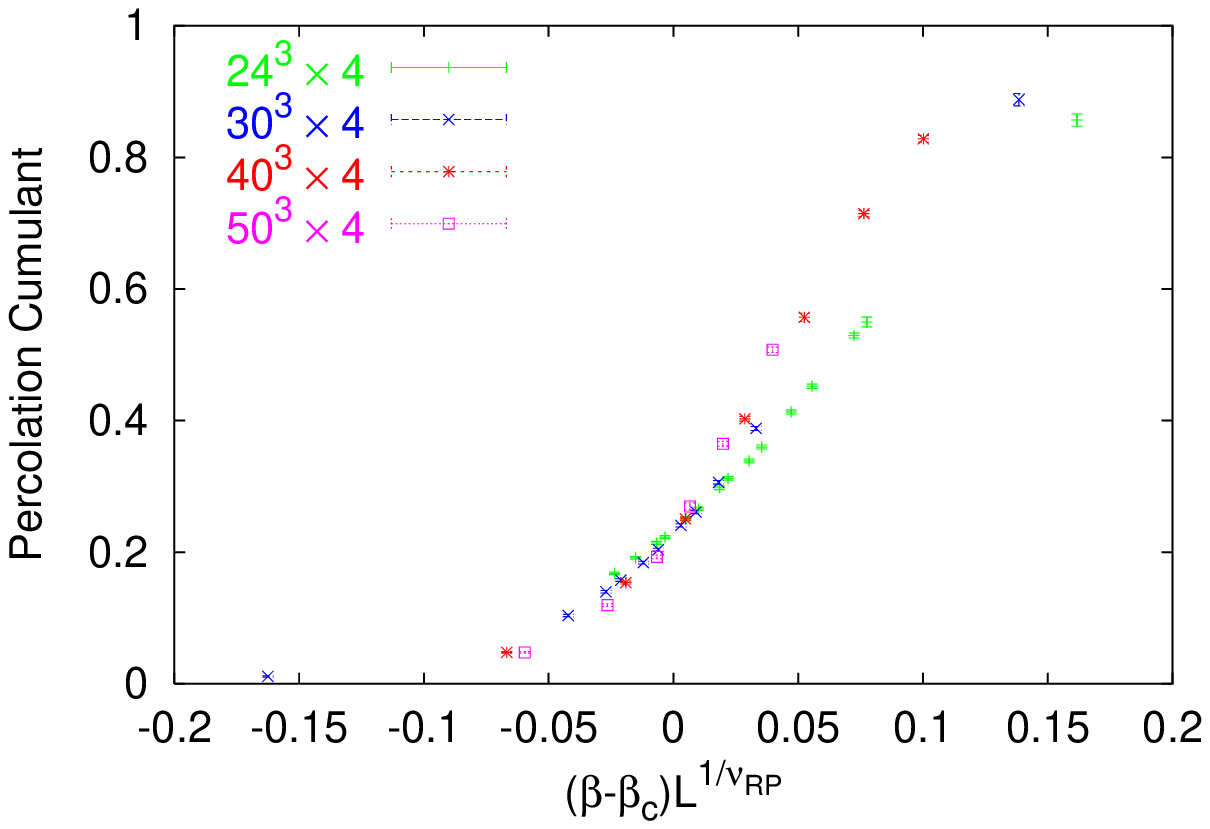,  width=14cm}
\caption[Rescaling of the
percolation cumulant curves of Fig. \ref{sceffnt4} with the 3D
random percolation exponent $\nu_{RP}=0.8765$]
{$(3+1)$-$d$ $SU(2)$, $N_{\tau}=4$. 
Rescaling of the percolation cumulant curves of Fig. \ref{sceffnt4} 
using $\beta_{c}=2.2991$
and the 3-dimensional random percolation
 exponent $\nu_{RP}=0.8765$.}
\label{sceffnt4rp}
\end{center}
\end{figure}

Fig. \ref{sceffnt4} illustrates 
where the geometrical transition takes place:
the crossing point of the percolation cumulant curves
concides with the thermal threshold (dashed line)
within errors.
The scaling analysis of the cumulant curves can be seen
in Figs. \ref{sceffnt4is} and \ref{sceffnt4rp}. Also here
it turns out that $\nu_{perc}=\nu_{Is}$.
The final results of the finite size scaling analysis are 
presented in Table \ref{tabeffnt4}. 
To get better scaling fits we considered again 
several lattice sizes close to the critical point, taking
for the lattice side $L$ all even numbers between $20$ and $50$.
The value
of the exponents' ratio $\beta/\nu$ is missing 
for the same problem stressed in the previous section.
\vskip0.4cm
\begin{table}[h]
  \begin{center}{
      \begin{tabular}{|c||l|l|l|}
\hline
  &      Critical point & ${\gamma}/{\nu}$ & $\nu$\\ \hline\hline
$\vphantom{\displaystyle\frac{1}{1}}$  Percolation results &
$2.2991(2)$&$1.979^{+0.016}_{-0.014}$&$0.629^{+0.007}_{-0.011}$\\  
\hline $\vphantom{\displaystyle\frac{1}{1}}$ 
  Thermal results &$2.29895(10)$
  &$1.944(13)$
&$0.630(11)$ \\ \hline$\vphantom{\displaystyle\frac{1}{1}}$
3D Ising Model & 
  & 1.963(7)
&0.6294(10) \\ \hline
      \end{tabular}
      }
\caption[Percolation critical indices for 
$(3+1)$-$d$ $SU(2)$, $N_{\tau}=4$: second approach]
{\label{tabeffnt4} 
Percolation critical indices for 
$(3+1)$-$d$ $SU(2)$, $N_{\tau}=4$.
They are compared with the thermal results of \cite{engles}
and the 3D Ising values.} 
  \end{center}
\end{table}

We notice that $\gamma/\nu$ is not in accord with the corresponding
$SU(2)$ estimate taken from \cite{engles}.
Nevertheless, it overlaps with the 3D Ising value, although
the agreement is not as good as in the $N_{\tau}=2$ case.
This fact indicates that, for $N_{\tau}=4$,
the effective theory (\ref{effans}) does not approximate 
$SU(2)$ so well as for $N_{\tau}=2$.
The main reason could be the approximation
induced by the condition 
that the theory must contain only spin-spin operators.
As a matter of fact, Okawa showed that, going from
$N_{\tau}=2$ to $N_{\tau}=4$, multispin couplings become important \cite{okawa}.
Besides, for $N_{\tau}>4$, we do not exclude that 
antiferromagnetic couplings may appear, which cannot still be 
handled in a percolation framework.

\clearpage

We stress that our aim 
was to check whether it is possible
to find a percolation picture for $SU(2)$ which works in   
the weak coupling regime as well. 
For $N_{\tau}=4$
the cluster definition
of our approach leads to a percolation
transition which reproduces fairly well the 
thermal counterpart. 
The arguments we have presented above suggest that our method
may fail for $N_{\tau}>4$; this statement should be verified through
numerical simulations.

\chapter*{Summary}
\addcontentsline{toc}{chapter}{Summary}

\thispagestyle{empty}
\markboth{\sc\ \ Summary }{\sc Summary }


\def\J{$J/\psi$}
\def\j{J/\psi}
\def\X{$\chi$}
\def\x{\chi}
\def\P{$\psi'$}
\def\p{\psi'}
\def\U{$\Upsilon$}
\def\u{\Upsilon}
\def\C{c{\bar c}}
\def\cg{c{\bar c}\!-\!g}
\def\bg{b{\bar b}\!-\!g}
\def\b{b{\bar b}}
\def\q{q{\bar q}}
\def\Q{Q{\bar Q}}
\def\L{\Lambda_{\rm QCD}}
\def\A{$A_{\rm cl}$}
\def\a{A_{\rm cl}}
\def\N{$n_{\rm cl}$}
\def\n{n_{\rm cl}}
\def\S{S_{\rm cl}}
\def\s{s_{\rm cl}}
\def\bb{\bar \beta}

\def\be{\begin{equation}}
\def\ee{\end{equation}}

\def\lsim{\raise0.3ex\hbox{$<$\kern-0.75em\raise-1.1ex\hbox{$\sim$}}}
\def\gsim{\raise0.3ex\hbox{$>$\kern-0.75em\raise-1.1ex\hbox{$\sim$}}}


\def\CMP{{ Comm.\ Math.\ Phys.\ }}
\def\NP{{ Nucl.\ Phys.\ }}
\def\PL{{ Phys.\ Lett.\ }}
\def\PR{{ Phys.\ Rev.\ }}
\def\PRep{{ Phys.\ Rep.\ }}
\def\PRL{{ Phys.\ Rev.\ Lett.\ }}
\def\RMP{{ Rev.\ Mod.\ Phys.\ }}
\def\ZP{{ Z.\ Phys.\ }}

We have seen that the Coniglio-Klein percolation picture 
of the paramagnetic-ferromagnetic transition 
of the Ising model can be used, with suitable modifications,
to describe the spontaneous symmetry-breaking of a wide
class of theories.  

In particular, there seems to be a one-to-one correspondence
between spin-spin interactions and 
geometrical bonds connecting the spins involved in the interactions.
The probability for a bond to be active is 
a simple function of the associated coupling strength.
We have found out that the spin variables need not be discrete
like in the Ising model, but they can vary
continuously in a range, and even be $n$-components
vectors, like in $O(n)$ spin models.
The only difference 
in the percolation picture is that the bond probability is 
{\it local}, since it depends on 
the values of the spin variables 
at the sites connected by the bond.

In the more
general case of a theory 
characterized by more interactions, the percolation picture
can be trivially extended if all interactions are  
spin-spin and ferromagnetic. 
For that, one needs just to combine together all
bonds corresponding to each interaction.
It is not clear whether
it is possible to formulate a general percolation 
frame in the presence of 
frustration. Besides, 
if multispin couplings have a geometrical counterpart, 
more complicated 
objects than simple bonds (e.g. plaquettes) may be involved. 
This could lead to a 
highly non-trivial representation, which is still far 
from being established.

The possibility of 
dealing with continuous degrees of freedom 
opens the way, in principle, 
to a possible application of 
percolation models to field theories.
We considered the case of $SU(2)$ pure gauge theory,
because its critical behaviour is identical 
to the one of the Ising model and we hoped that such
connection could simplify our task.

We stressed that 
no rigorous percolation picture a la Coniglio-Klein
is possible as long as we do not
know the exact expressions of the 
interactions between the Polyakov loops. This fact 
forced us to 
approximate $SU(2)$ by means of special effective theories,
for which an equivalent percolation
model exists.
We have followed two different approaches 
to extract a suitable effective theory.

The first approach, based on series expansions
of the $SU(2)$ lattice action,
works rather well but its validity is limited to the
strong coupling limit.

The second approach is more a brute force procedure,
since it aims to find an Ising-like spin model, with 
just spin-spin interactions, which reproduces the 
configurations of the signs of the Polyakov loops.
The new method leads to good results 
both in the strong coupling case we had examined with the 
first approach ($N_{\tau}=2$) and if we move towards the 
weak coupling limit ($N_{\tau}=4$).
However, for $N_{\tau}=4$ the approximation looks worse than
for $N_{\tau}=2$.
More precisely, the value of the 
exponents' ratio $\gamma/\nu$ seems
to shift slightly towards the random percolation
value, even if it is still in agreement 
with the Ising ratio. This can mean that 
the procedure is not reliable for higher values
of $N_{\tau}$. As a matter of fact, 
we have to recognize that our ansatz for
the Hamiltonian of the effective model is 
probably too restrictive, and that multispin couplings 
may become important for big $N_{\tau}$'s.
Moreover, the precision of the method
decreases the more spin-spin operators we introduce. In fact,
if we analyze any time the same number of $SU(2)$ configurations,
the errors on the final couplings of the effective theory
are of the same order, no matter how many
couplings we have. Consequently, the
corresponding uncertainty on the model 
is the greater the more the couplings are.
The Polyakov loop clusters, 
which are built 
by using the bond weights calculated from the couplings of the 
effective theory, become thus less and less defined.
In order not to lose accuracy, one must lower
the error on each single coupling, and that is
possible only if we
increase the number of 
$SU(2)$ configurations to analyze, which can
lead to prohibitively lengthy simulations.

In conclusion, the second approach has certainly
some drawbacks. Nevertheless, it allowed us to
define some Polyakov loop clusters which 
have, with good approximation, 
the properties of the
physical droplets of $SU(2)$ we were looking for, also
in a case which 
approaches the weak coupling limit ($N_{\tau}=4$). 
For this reason, the second approach
is to be preferred to the first one, which strongly 
depends on a special lattice regularization of $SU(2)$.

From our investigations it is 
not possible to argue whether 
the critical behaviour of other 
field theories can be described 
by means of percolation. 
The strict relationship between 
$SU(N)$ gauge theories and $Z(N)$ spin models
can represent a useful tool
to devise suitable percolation pictures
for the gauge theories
starting from results known for the 
simpler spin models. In principle, that is exactly what we have done
in our case, exploiting the analogy between 
$SU(2)$ and the Ising model.
In practice, the task gets more complicated
for $SU(N)$, when $N>2$. For example, 
$SU(3)$ gauge theory is certainly the most 
interesting case of all, because it
involves the "real" gluons. 
In two space dimensions, $SU(3)$ undergoes
a second order phase transition, like
the three states Potts model. Very recently \cite{daniel}
it was shown that the 2-dimensional three states Potts model
admits an equivalent percolation formulation, which could
thus be used for $SU(3)$. However, 
$SU(3)$ in two space dimensions is rather
an academic model.
One is surely more interested
in the realistic 3-dimensional case. The fact that
the $SU(3)$ phase transition in three space 
dimensions is first order 
poses an essential problem concerning the 
relationship between percolation and first order phase transitions.

The situation gets even more involved when one
considers the case of {\it full} $QCD$,
i. e. $SU(3)$ plus dynamical quarks, since 
the transition from confinement to deconfinement
is probably a {\it crossover}, i.e. it  
takes place without any singularity 
in the partition function. We have 
seen in Section 2.6 that 
there are cases 
in which geometrical properties can change abruptly
without a corresponding discontinuity in
the thermal variables. 
This could provide a criterion to {\it define}
different phases and the relative transition
in an extended sense \cite{satz}. Work in this direction is in progress.

We conclude our summary with some general remarks concerning
the method we have chosen to 
study correlated percolation, 
i. e. Monte Carlo simulations.  
There is, in fact, basically no literature about this 
subject, as most of the known results
are based on analytical proofs, and the few 
numerical studies rely on 
series expansions.

We point out the importance of the 
percolation cumulant, from which one can derive 
a precise estimate of the critical point. Besides,
the scaling of the percolation cumulant curves 
allows to get the
value of the critical exponent $\nu$, with 
$4-5\,\%$ accuracy for the lattice sizes we have
considered. 
The accuracy can be increased by analyzing larger lattices.
Anyway, better estimates of $\nu$ can be obtained
by using standard finite size scaling techniques,
like the scaling of the pseudocritical points
(see end of Section \ref{markus}).

We remark that, for equal statistics,
the errors on the percolation variables are much smaller
than the errors on the corresponding thermal variables.
The latter seems to be a general feature 
of site-bond percolation, because the clusters depend
as well on the bonds' distribution. This introduces a further 
random element which contributes to reduce sensibly
the correlation of the 
percolation measurements with respect
to the thermal counterparts, which depend {\it only}
on the spin configurations. 
We found that the data of the percolation strength $P$ are always
more correlated than the corresponding data of the average cluster size $S$.

For a study of the thermal transition variables
like the susceptibility $\chi$ or the Binder cumulant $g_r$
are necessary. Such quantities cannot be determined directly
from measurements on the spin configurations, but
are calculated by means of averages of powers 
of the order parameter. That usually leads to 
big error bars on the final results of $\chi$ and $g_r$.
Instead, the percolation counterparts of $\chi$ and $g_r$,
i.e. the average cluster size and the percolation cumulant,
are calculated directly from the clusters' configurations,
so that their errors are rather small.

Hence, in order to get the same accuracy on the 
average values, the thermal investigation 
of a model would require 
more $CPU$ time than
the relative percolation 
study. Nevertheless we have to point out that 
the errors on the thermal 
variables can be considerably reduced by means
of reweighting techniques like the $DSM$ \cite{DSM}, which
we have often used in our studies, whereas similar
interpolation methods do not exist for correlated percolation 
\footnote{For random percolation a reweighting method was recently
proposed \cite{harris,balle}; the role of the energy
is carried out by the probability of having
a configuration in correspondence of a value $p$ of the density of occupied sites (bonds).}.
In this work we were thus forced to use directly 
the data points in the finite size scaling fits.
We think that the Fortuin-Kasteleyn-Swendsen-Wang model
we have discussed in Section \ref{mancio} could be used to implement
an efficient method for the interpolation of 
Fortuin-Kasteleyn percolation data relative to the $q$-state Potts model.

From the finite size scaling analysis, it turns out
that the scaling behaviour of the percolation variables is rather
pure: that is clearly shown by
the precision of the scaling of the percolation cumulants we have
performed many times in this work. In particular, in all our
analyses, corrections to scaling seem
negligible, and finite size effects disappear
already for relatively small lattice sizes.
This is quite impressive, especially when one makes comparisons
with the thermal variables, which are normally 
strongly affected by such perturbations.
Nevertheless, we have to 
keep in mind that the accuracy 
on our evaluation of the critical exponents
has always been about $1-2\,\%$, which is 
good for our purposes \footnote{We remind that 
for the systems we investigated
we had to check whether the critical exponents
of the percolation transition agree with the thermal
exponents of the system or rather with the ones of random percolation. 
The thermal exponents of all the models we have considered
differ from the random
percolation exponents of about $10-20\,\%$, so that our
accuracy is good enough to distinguish the two cases.}
but not exceptional. 
Moreover, the percolation data of our $SU(2)$ studies are
already affected by the approximations involved
in the determination of the effective theory, which are 
by far more important than eventual corrections to scaling.
On the other hand, if we want to obtain
more accurate estimates of the results
for models which admit an exact
percolation formulation, like the continuous spin models of Chapter 3,
corrections to scaling may become important: in 
high precision numerical studies of random percolation
that seems indeed to be the case \cite{parisi}.
We remind that we have almost always adopted free boundary conditions 
for the cluster identification.
The results on $O(n)$ spin models, however,
suggest that
the situation could be further on improved by
using periodic boundary conditions (see Section 3.3).

\setlength{\parskip}{1mm}
\parskip3.0ex plus1.2ex minus0.7ex

\begin{appendix} 
\chapter{Cluster Labeling}
\thispagestyle{empty}


\def\J{$J/\psi$}
\def\j{J/\psi}
\def\X{$\chi$}
\def\x{\chi}
\def\P{$\psi'$}
\def\p{\psi'}
\def\U{$\Upsilon$}
\def\u{\Upsilon}
\def\C{c{\bar c}}
\def\cg{c{\bar c}\!-\!g}
\def\bg{b{\bar b}\!-\!g}
\def\b{b{\bar b}}
\def\q{q{\bar q}}
\def\Q{Q{\bar Q}}
\def\L{\Lambda_{\rm QCD}}
\def\A{$A_{\rm cl}$}
\def\a{A_{\rm cl}}
\def\N{$n_{\rm cl}$}
\def\n{n_{\rm cl}}
\def\S{S_{\rm cl}}
\def\s{s_{\rm cl}}
\def\bb{\bar \beta}

\def\be{\begin{equation}}
\def\ee{\end{equation}}

\def\lsim{\raise0.3ex\hbox{$<$\kern-0.75em\raise-1.1ex\hbox{$\sim$}}}
\def\gsim{\raise0.3ex\hbox{$>$\kern-0.75em\raise-1.1ex\hbox{$\sim$}}}


\def\CMP{{ Comm.\ Math.\ Phys.\ }}
\def\NP{{ Nucl.\ Phys.\ }}
\def\PL{{ Phys.\ Lett.\ }}
\def\PR{{ Phys.\ Rev.\ }}
\def\PRep{{ Phys.\ Rep.\ }}
\def\PRL{{ Phys.\ Rev.\ Lett.\ }}
\def\RMP{{ Rev.\ Mod.\ Phys.\ }}
\def\ZP{{ Z.\ Phys.\ }}

Suppose we want to perform percolation studies
by means of lattice Monte Carlo simulations. We can 
divide the process in two phases:
\begin{itemize}
\item{a configuration is created specifying, according 
to the percolation model we have chosen,
which sites are occupied and which ones are empty;}
\item{all occupied sites of the configurations are 
set into clusters, following the prescription of the percolation model
(pure site, site-bond, etc.).} 
\end{itemize}
The first phase depends on the type of system we are studying.
In the case of random percolation, for instance, 
one needs just a good {\it random number generator}
to create the required configurations. 
First, one fixes
the value $p$ of the density of occupied sites. 
In general one associates a
random number $r$ between zero and one to
a lattice site and compares it with $p$. If 
$r<p$, the site is occupied, otherwise
it is empty. The procedure is repeated for all
sites of the lattice.
In the case of correlated percolation, the configuration is 
created by means of suitable Monte Carlo algorithms. 
For example, in the Ising model, the spin configurations can be 
produced by standard updates like Metropolis, heat bath, or cluster
algorithms. Anyhow, such procedures will assign 
a value of the spin to each lattice site. Suppose we define 
the sites as occupied if their spins point up, then also
the percolation frame will be 
established.

The delicate point is then represented by the second phase 
of the process that we have mentioned above, 
namely the cluster building.
To identify a cluster configuration one needs essentially
to associate to each site some label $L$, 
which indicates that the site belongs to some 
cluster.
What we would like to have is an algorithm which gives all sites
within the same cluster the same label, and gives
different labels to sites belonging to different clusters.
If this is possible, the search of an
eventual percolation cluster becomes trivial. In fact, it 
suffices to check whether the same label is present in two 
opposite sides/faces of the lattice. Besides, the size $s$
of a cluster is simply determined by counting
how many times a particular label occurs in the lattice.

\begin{figure}[h]
\begin{center}
\epsfig{file=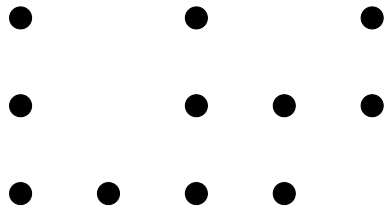,width=5cm}
\vskip0.5cm
\caption[Sample configuration for the cluster labeling]
{Scheme of a $3{\times}5$ lattice with $11$ occupied sites,
which we want to set into clusters according to the pure site percolation
rule.}
\label{hoshe}
\end{center}
\end{figure}

To have an idea of the difficulties to find a 
suitable and efficient algorithm for the cluster labeling,
we take as example the simple configuration we have sketched in
Fig. \ref{hoshe}.
The black points represent the occupied sites of 
the lattice, and we want to 
identify the relative site percolation clusters.

We can ask the computer to start from the top-left site and 
to proceed from left to right within each line, and
then from the top to the bottom. Any time, the 
program will check whether the site ($S$) it analyzes
is occupied or not. Assuming it is,  
one checks its left ($L$) and top ($T$)
neighbours. If none of them is occupied,
$S$ takes a label which has not been
used previously; on the other hand, if at least
one of such neighbours is also occupied, say $L$, 
$S$ will take the label of $L$
(we notice that both $L$ and $T$
have been previously examined and therefore, if occupied, they carry
some label). 

Now we can start our procedure for the 
configuration of Fig. \ref{hoshe}. 
The top-left site is occupied, it therefore receives the label
$1$. The next occupied site is the third from the left:
its left neighbour is empty and above there are no lattice sites, so that
it gets a new label $2$. The same happens for the last site of the
top line, which receives the label $3$.
For the first line we then obtain the following
labeling, where the zeros indicate the empty sites:
\begin{center}
$1\,\,\,\,\,0\,\,\,\,\,2\,\,\,\,\,0\,\,\,\,\,3$
\end{center}

The first site of the second line is occupied, like
its top neighbour, which carries the label $1$. Hence the new site 
takes also the label $1$. For the same reason the 
next occupied site of 
the second line receives the label $2$, like the 
third one. But when we arrive to the last site
of the line, we have a problem: in fact, this site is connected
to both its neighbours, which carry different labels. 
Which label shall we then associate to it? 
The present situation is sketched by the scheme below
\begin{center}
$1\,\,\,\,\,0\,\,\,\,\,2\,\,\,\,\,0\,\,\,\,\,3$\\
$1\,\,\,\,\,0\,\,\,\,\,2\,\,\,\,\,2\,\,\,\,\,?$
\end{center}
The crucial point is that the 
site in question connects two clusters which 
were so far separated. From now on, the labels $2$
and $3$ mark thus one and the same cluster.
We assign to our troublesome site the smaller of the
two labels, but we have to keep in mind that 
all sites marked by $3$ must be finally
switched to $2$. The simplest way of doing that is to ask 
our computer to come back to the top-left lattice site and to
mark with the label $2$ all encountered sites 
which carry the label $3$. But this process would be very time consuming,
since it would force the computer 
to scan the lattice a number of times which 
is of the same order of the 
lattice volume $V$.
The total number of operations
involved by the procedure of cluster labeling would 
thus grow as $V^2$, which makes  
the relative computing time prohibitively large
for big lattice sizes.
We would rather like to have a computing time 
proportional to the lattice volume.

Hoshen and Kopelman \cite{kopelman}
found a smart way to solve the problem.
One needs just to associate to each label $M$ a number,
which we indicate $N(M)$. Such number takes 
into account the relations between 
cluster labels that one finds
while scanning the lattice. So, when we introduce a new label
$M$, we set $N(M)=M$. If, at some stage, 
the cluster $M$ turns out to be
connected to another cluster $P$, with $P<M$,
like in our example ($M=3$, $P=2$), then we reset
$N(M)=P$. 
With this prescription let us proceed with the analysis 
of our configuration.
The label $1$ is obviously fundamental, since it is the first we
have introduced, so that $N(1)=1$. The label $2$ marks
a cluster which has, so far, no connections with $1$, and 
therefore
$N(2)=2$. The same for the third label, before one
examines the crucial case at which 
we interrupted our analysis, so $N(3)=3$.
The site we have indicated through the question mark
obtains now the label $2$. The fact that 
the clusters $2$ and $3$ are the same leads
us to reset $N(3)=2$.
Finally, let us investigate the third and last line of the 
lattice. The first two sites are 
simple to identify: they both receive the label $1$.
The third one establishes a connection between the clusters $1$
and $2$. Because of that, the site takes the label 
$1$ and $N(2)=1$.
We end up with the following situation
\begin{center}
$1\,\,\,\,\,0\,\,\,\,\,2\,\,\,\,\,0\,\,\,\,\,3$\\
$1\,\,\,\,\,0\,\,\,\,\,2\,\,\,\,\,2\,\,\,\,\,2$\\
$1\,\,\,\,\,1\,\,\,\,\,1\,\,\,\,\,1\,\,\,\,\,0$\\
\vskip0.3cm
$N(1)=1,\,\,\,\,\,N(2)=1,\,\,\,\,\,N(3)=2.\,\,\,\,\,$
\end{center}
In this way we need to go through the whole lattice once 
and to store the connections found later in the 
"label of labels" array $N$. To finish
our job, we must assign to each site the right label.
For that we just have to classify 
all labels which have been introduced.
Let us assume that 
we take a label $M$. The first thing to do is to check whether
$N(M)=M$. If it is so, all sites marked 
with $M$ carry the correct label. 
If, otherwise, $N(M)=P<M$, then one has to check
whether $P$ is a fundamental label, i.e. whether $N(P)=P$.
In this case, we reset the label $M$ to
the new value $P$. If $N(P)=L<P$ one repeats the procedure 
until one finds that $N(S)=S$ for some label $S$, which 
becomes the final label of the sites initially marked with $M$.
It is easy to check that 
this iterative procedure 
leads to the correct cluster labeling of the
sample configuration 
of Fig. \ref{hoshe}.

In conclusion, the Hoshen-Kopelman algorithm
is a very efficient
method for the identification of the cluster configurations 
which is necessary to carry on numerical 
percolation studies on a lattice. The algorithm 
requires essentially a single scan of the lattice and the 
label classification we have described above, which can be 
done by means of simple routines in the program.

\end{appendix}

\chapter*{Publications}\label{Literatur}
\addcontentsline{toc}{chapter}{Publications}

\thispagestyle{empty}
\markboth{\sc\ \ Publications}{\sc Publications}

The investigations related to
the original work done in this
thesis are exposed in a number of papers and have been presented 
to the physics community at various conferences.
Below we enclose the complete list of publications.

\begin{itemize}
\item
S. Fortunato, H. Huang, H. Satz,\newline 
{\it Deconfinement and Percolation
in $SU(2)$ Gauge Theory}. \newline
Poster at Quark Matter '99, Torino (Italy).
\item
S. Fortunato, H. Satz, \newline
{\it 
Percolation and Deconfinement in $SU(2)$ Gauge Theory}.\newline
Talk given at the 17th International Symposium on Lattice Field Theory
(LATTICE '99), Pisa (Italy). \newline
Published in Nuclear Physics B Proceedings Supplement {\bf 83}, 452 (2000).\newline
E-Print Archive: hep-lat/9908033.
\item
S. Fortunato, H. Satz, \newline
{\it Polyakov Loop Percolation and Deconfinement
in $SU(2)$ Gauge Theory},\newline
Physics Letters B {\bf 475}, 311 (2000).\newline
E-Print Archive: hep-lat/9911020.
\item
P. Bialas, P. Blanchard, S. Fortunato, D. Gandolfo, H. Satz,\newline
{\it Percolation and Magnetization in the Continuous Spin Ising
  Model}, \newline
Nuclear Physics B {\bf 583}, 368 (2000).\newline
E-Print Archive: hep-lat/0003014.
\item
S. Fortunato, H. Satz, \newline
{\it Percolation and Magnetization for Generalized Continuous Spin
  Models}. \newline
Submitted to Nuclear Physics B.\newline
E-Print Archive: hep-lat/0007005.
\item
S. Fortunato, H. Satz, \newline
{\it Percolation and Deconfinement in $SU(2)$ Gauge Theory}.\newline
Talk given at the $3^{rd}$ Catania Relativistic Ion Studies (CRIS 2000),
Acicastello (Italy).\newline
To be published in Nuclear Physics A Proceedings.\newline
E-Print Archive: hep-lat/0007012.
\item
P. Blanchard, S. Digal, S. Fortunato, D. Gandolfo, T. Mendes, H. Satz,\newline
{\it Cluster Percolation in $O(n)$ Spin Models},\newline
Journal of Physics A: Math. Gen. {\bf 33}, 8603 (2000). \newline
E-Print Archive: hep-lat/0007035.
\item
S. Fortunato, F. Karsch, P. Petreczky, H. Satz,\newline
{\it Percolation and Critical Behaviour in $SU(2)$ Gauge Theory}.\newline
Talk given at the 18th International Symposium on Lattice Field Theory
(LATTICE 2000), Bangalore (India). \newline
To be published in Nuclear Physics B Proceedings.\newline
E-Print Archive: hep-lat/0010026
\item
S. Fortunato, F. Karsch, P. Petreczky, H. Satz,\newline
{\it Effective $Z(2)$ Spin Models of Deconfinement and Percolation in $SU(2)$
  Gauge Theory}.\newline
Submitted to Physics Letters B.\newline
E-Print Archive: hep-lat/0011084.

\end{itemize}

\end{document}